\documentclass[acmtog, authorversion,nonacm]{acmart}
\usepackage[utf8]{inputenc}
\usepackage{natbib}
\usepackage{graphicx}
\usepackage[percent]{overpic}
\usepackage{caption}
\usepackage{subcaption}
\usepackage{paralist}
\usepackage{wrapfig,booktabs}
\usepackage{xcolor,colortbl}
\usepackage{microtype}
\usepackage{enumitem}  
\usepackage{listings}
\usepackage{textcomp}
\usepackage{float}
\usepackage{placeins}
\usepackage{stfloats}
\usepackage[T1]{fontenc}
\usepackage[ruled,vlined]{algorithm2e}
\newcommand{\figref}[1]{Fig.~\ref{#1}}

\newcommand{\smallboldcode}[1]{\begin{small}\textbf{\lstinline{#1}}\end{small}} 
\newcommand{\smallcode}[1]{\begin{small}\lstinline{#1}\end{small}}

\makeatletter
\newcommand{\thesisversion}[1]{%
  \@ifundefined{thesis}{}{#1}%
}
\makeatother

\AtBeginDocument{%
  \providecommand\BibTeX{{%
    \normalfont B\kern-0.5em{\scshape i\kern-0.25em b}\kern-0.8em\TeX}}}

\graphicspath{figures}

\acmJournal{TOG}
\begin{document}

\title{Finding Fast Filters}

\author{Karima Ma}
\affiliation{%
 \institution{Adobe Research}
 \city{New York}
 \state{New York}
 \country{USA}}
\email{kma@adobe.com}
\orcid{1234-5678-9012}

\author{Andrew Adams}
\email{andrew.b.adams@gmail.com}
\affiliation{%
 \institution{Adobe Research}
 \city{San Francisco}
 \state{California}
 \country{USA}
}

\author{Jonathan Ragan-Kelley}
\affiliation{%
 \institution{MIT}
 \city{Cambridge}
 \state{Massachusetts}
 \country{USA}}
\email{jrk@mit.edu}

\begin{abstract}

Processing images, video, and audio often requires running large finite impulse response (FIR) filters with strict performance and latency requirements. 
Prior methods for fast filter approximations are special cases or combinations of a few key techniques: multi-rate and recurrent filtering, and decomposing filters into sums or cascades. We unify these techniques as primitives within a single design language for fast 1D and 2D filters. 
Given a target filter to approximate, we automatically search this program space, fitting continuous parameters with gradient descent, to generate a Pareto frontier of algorithms that trade off performance with quality. Our system produces substantially higher-quality and faster filter approximations than have been previously described for several popular imaging and audio filters. Furthermore we demonstrate how to automatically lower programs in this design space to optimized, vectorized, parallel, C++ code which is fused for data locality.



\end{abstract}


\begin{teaserfigure}
\vspace{-0.3em}
\includegraphics[width=\linewidth]{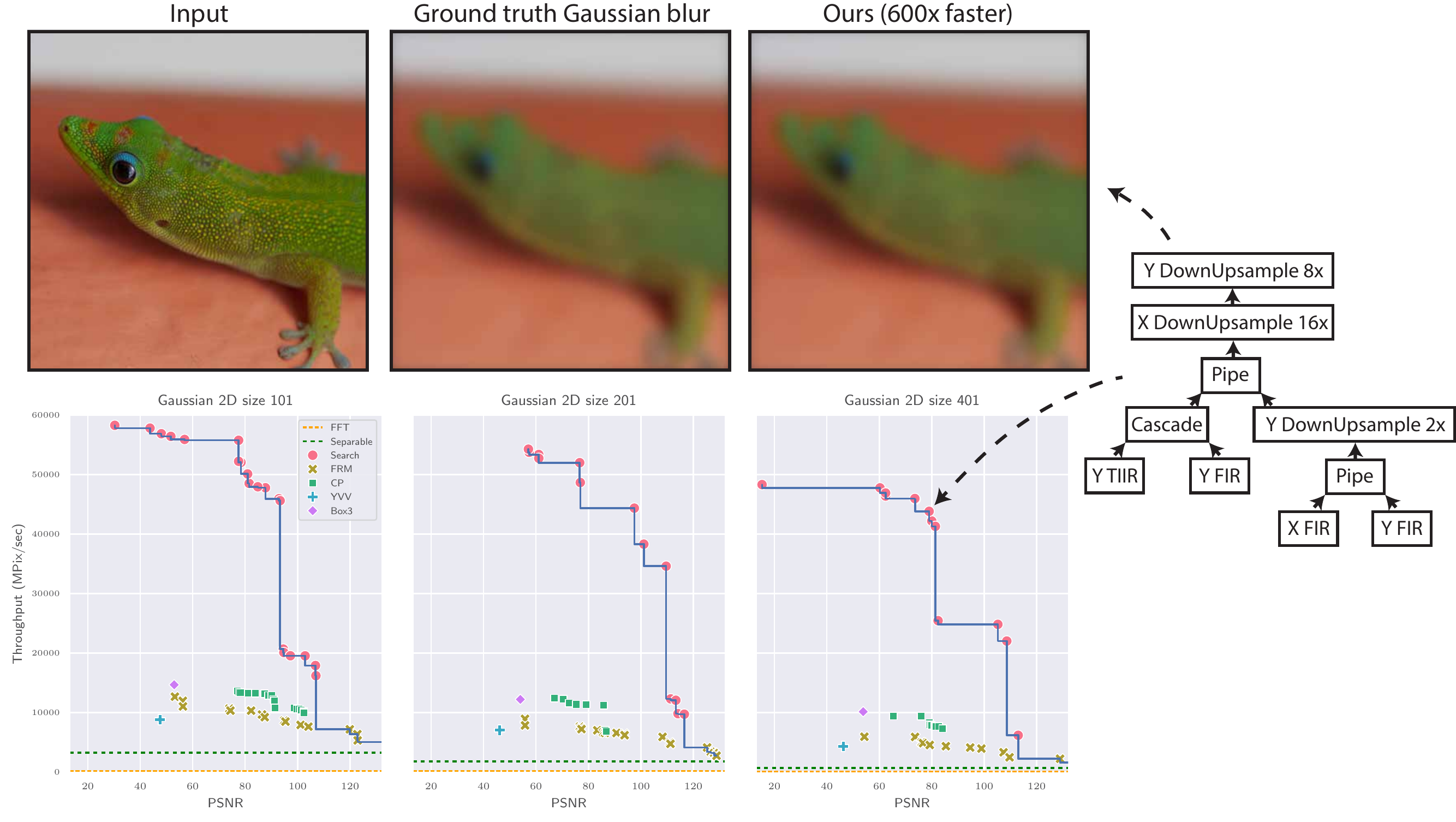}
\vspace{-2.0em}
\caption{We unify fast filter approximation techniques as primitives within a single design language. This allows us to search over programs to find novel acceleration schemes for large 1D and 2D filters automatically. Here we show the output from one our programs for approximating a Gaussian blur with radius 200 ($\sigma = 40$) compared to FFT-based convolution. Our program produces outputs that are indistinguishable from the ground truth while being 600 times faster than the fastest exact filtering method. Our search also found programs that are 3.9 to 4.7x faster than the triple box blur approximation, which is often used in production software, at comparable or higher quality. The bottom row shows the PSNRs and throughputs achieved by designs from our search for Gaussian sizes 101, 201, and 401 compared to existing methods. Our system can represent and generate optimized implementations of many existing acceleration strategies (\cite{FRM2D,convolutional-pyramids,young-gaussian,MCDONNELL1981-triple-box-blur}). However, these designs are significantly Pareto-dominated by the novel designs we find through search. On the right is the program structure that produced our output with PSNR 79dB and throughput 43819 MPix/sec.}
\label{fig:teaser}
\end{teaserfigure}

\keywords{Filter approximation, program synthesis, image processing, signal processing, audio processing, machine learning, high performance computing}

\maketitle
\section{Introduction}
\label{sec:introduction}
Image, video, and audio processing require large-support finite impulse response (FIR) filters for many tasks like denoising, edge detection, binaural sound synthesis, and convolutional reverb. 
These applications often need to provide interactive feedback or meet real-time constraints, sometimes running in highly compute-constrained contexts like mobile phone cameras or wireless headphones.
Direct convolution with large FIRs is prohibitively expensive, because its cost grows linearly or quadratically with the filter size.
Convolution in the Fourier domain is invariant to filter size, but the cost of fast Fourier transforms is often still prohibitive, and the need to process large blocks at a time
introduces potentially unacceptable latency.

In many applications, the best choice is to trade accuracy for performance by \emph{approximating} a filter in ways that reduce the cost of computation. Linear filtering is a well-established field with several cost-saving techniques for approximating expensive FIRs, including recurrent filtering \cite{young-gabor, young-gaussian, TIIR, MCDONNELL1981-triple-box-blur}, multi-rate filtering \cite{multirate-filtering}, and frequency response masking \cite{FRM1D}. 
A more modern approach, Convolution Pyramids \cite{convolutional-pyramids}, extends multi-rate filtering by adding residual FIRs and using gradient descent to fit parameters to the target impulse response.

The claimed performance gains of these methods are based on counting math operations, which does not capture the nuances of efficient parallel implementations on modern hardware, and are often inaccurate. Moreover, they do not provide implementation strategies for their designs on modern CPUs. Proper implementations can be non-trivial and are vital to actually reaping performance benefits. 

More importantly, each strategy alone is not expressive enough to approximate arbitrary targets. Most were created to work for a restricted class of filters, like classical band-pass prototypes, or even narrower classes like Gaussians. These methods often use hand-derived solutions \cite{young-gabor,young-gaussian,TIIR,oppenheim-DTSP}, which provides goodness of fit guarantees, but further prevents them from being applied to any targets. 

Most design methods do not prioritize cost vs. quality exploration. Classical methods for a single FIR or recurrent filter, ask users to specify \emph{both} quality constraints like pass-band ripple (how much the response varies across the desired frequencies) or stop-band attenuation (how much the filter suppresses unwanted frequencies), \emph{and}  computational constraints i.e. the filter order. 
As pointed out by Crochiere et al. \shortcite{design-relationships-between-FIR-parameters}, it is often more useful to control just quality or cost, and let the method determine the other. Some more complex designs like frequency response masking have derived equations for what their cost parameters should be, given band-pass specifications \cite{FRM-analytical-solutions}. Other multi-rate filtering methods require an additional optimization to find the cost parameters \cite{multirate-filtering}. Convolution Pyramids \shortcite{convolutional-pyramids} suggests cost settings for a few target filters, but determining optimal costs in general requires training different structures. In summary, exploring the trade-off space for these designs often requires considerable manual effort. Moreover, because they have fairly restricted variability, i.e. the filter order, stride factor, or number of stages, they have rapidly diminishing returns to quality with increasing cost, as shown in Section~\ref{sec:results}.

To address these issues, we developed a system for automated filter design that unifies traditional approximation methods under a single framework. Through exploring novel structures that utilize all of these cost-saving techniques as composable primitives, we find new approximations for a wide range of FIR targets, that significantly improve upon existing methods in both throughput and quality (\figref{fig:teaser}). Our system generates both the filter designs \emph{and their performance optimized implementations} automatically. 

Design search requires fitting the continuous parameters of each structure in order to assess their approximation quality. We developed differentiable representations for our primitives (Section~\ref{sec:tiir-parameterization}), most notably for recurrent filters, which are known to be difficult to optimize due to non-convexity and exploding gradients \cite{DSP-review}. While we can rely on gradient descent to measure quality, making quality vs. performance trade-offs also requires knowing the throughputs of each program's optimized implementations. We developed a companion lowering system that first algebraically transforms the structure using signal processing identities to expose parallelism and improve locality, and then lowers it to a fast, parallel, vectorized C++ implementation. While it is possible to benchmark these implementations to get ground truth throughputs, we designed a cost model that can accurately predict throughputs instead, allowing us to avoid doing online compilation and benchmarking during the search. After search we use our lowering system to generate fast implementations for the top programs. Exploring discrete structures requires tackling a combinatorial explosion of valid designs. We use knowledge from signal-processing and performance engineering in our enumeration strategies to improve the likelihood of sampling good structures.

Most prior techniques to accelerate filters are special cases or combinations of a few key techniques: multi-rate and recurrent filtering, and decomposing filters into sums or cascades. By using these techniques as primitives in a unified design space, we can express much of the existing literature as individual points in a single differentiable searchable space. We then search this space to find novel acceleration schemes for target filters automatically.

We believe our search system presents a new way to approach filter design for arbitrary frequency response targets. We contextualize well-established cost-saving strategies as primitives within a differentiable domain-specific language for fast image and audio filtering. Our language codifies and extends idiomatic ways of approximating large linear filters into composable abstractions while our compiler systematizes best-practices for fast implementations. Efficient exploration of this design space requires domain knowledge as well. Domain-specific program search, which fuses domain knowledge with techniques from machine learning and program synthesis, can scale traditional signal processing methods both in the degree to which they maximize speed and quality objectives, as well as the breadth of their applicability across design targets. 

\section{Background \& Related Work}
\subsection{Approximate Filtering Techniques}
\subsubsection{Infinite Impulse Response (IIR) Filters}
The most basic cost-saving design is a recurrent infinite impulse response (IIR) filter, which linearly combines input values with previous output values. The recurrence relation produces an infinitely wide support, that either decays, grows, or has constant amplitude over time, while incurring a fixed cost proportional to the number of input and recurrent samples. The infinite response property also means that unlike FIRs, IIRs cannot produce linear phase (symmetric) responses, which is often required for imaging applications. Furthermore, if the target is an arbitrary frequency response, i.e. not a restricted band-pass prototype, classical methods can have issues with accuracy, convergence, stability, and local minima \cite{Lang-WLS, Stoica1981-steiglitz-mcbride, modified-yule-walker, Rabiner-linear-programming}.

Given the limitations of classical IIR design, clever techniques have been developed to approximate specific linear phase FIR filters, such as Gaussians and Gabors \cite{deriche1990fast,young-gaussian,young-gabor,MCDONNELL1981-triple-box-blur}. These designs use cascades of stable (decaying), casual (uses current and past input values) and anti-causal (uses future input values) IIRs to achieve symmetric impulse responses. For an intuition of how this works, note that convolving a negative and positive ramp creates a symmetric tent. The positive ramp is unstable because it grows over time. To avoid numerical issues, it can be implemented as a negative ramp on the \emph{time-reversed} inputs, making it an anti-causal stable IIR. Unfortunately, anti-causal filters require an additional backwards pass over the input, making these designs unsuitable for streaming applications and potentially bandwidth-bound. Furthermore, these methods do not generalize to other target responses, because they are based on mathematical features specific to the target filter. 

\subsubsection{Tail-Canceling Infinite Impulse Response (TIIR) Filters}
Wang et al. \shortcite{TIIR} showed that any IIR can be cascaded with an analytically derived sparse FIR to cancel values after some time step $N$, producing large FIRs with the same computational complexity benefits of IIRs. TIIRs were motivated by the need for fast linear phase narrow-band filters, which require large symmetric support widths. Just like IIRs, a cascade of two TIIRs that are time-reversed versions of each other has linear-phase. TIIRs can use unstable IIRs because their growth is bounded by the truncation. This eliminates the need for a backwards memory pass, but requires twice as much computation and storage as a stable TIIR, to handle numerical precision issues that can prevent unstable tails from getting canceled. 

Both IIRs and TIIRs struggle to efficiently use SIMD vectorization and superscalar hardware because recurrence is inherently sequential. A naive implementation of a third-order IIR reaches only 0.8\% of peak FLOPS on a modern CPU. Transformations like scattered look-ahead \cite{scattered-look-ahead} can expose more parallelism, but increase the arithmetic intensity. In brief, the throughput of recurrent filters cannot be deduced from their operation count. Determining whether a recurrent design is better than other methods with higher arithmetic intensity but more parallelism requires working through implementations.

The method presented by Wang et al. \shortcite{TIIR} only produces linear phase band-pass filters whose source IIRs are derived from classical band-pass methods. However, they note that several popular windowing functions like the box filter are in fact marginally stable TIIRs. The TIIR approach for fast box filtering has been rediscovered 
at least twice. Crow \shortcite{summed-area-tables} is often cited in computer graphics for showing that box filters can be implemented as point-wise differences over integral images. Earlier work in computer vision states that the U.S. Geological Survey had already been implementing box blurs with summed area tables for several years \cite{MCDONNELL1981-triple-box-blur}. 
Box filters are used extensively in graphics applications because they are known to have constant cost regardless of the filter width. But what is less well known is that this is not unique to box filters.  TIIRs contextualize this popular trick as just one member of a large family of computationally efficient filters. We will show that TIIRs can be combined with other cost-saving building blocks to target a broad range of frequency responses. 

\subsubsection{Multi-Rate Filtering}
\label{sec:related-work-multi-rate-filtering}
Narrow-band FIRs can be cheaply approximated with one small FIR sandwiched between a downsampling and upsampling stage \cite{multirate-filtering}. These stages require low-pass and anti-aliasing filtering respectively. Wide-band FIRs with sharp transition widths also require large support. Frequency response masking (FRM) \cite{narrow-band-FRM, FRM1D} can produce cheap approximations of sharp narrow or wide-band filters by combining zero-upsampled FIRs with anti-aliasing interpolators. To see how this works for narrow-band filtering, note that zero-upsampling  an FIR (inserting zeros between samples) by a factor of $S$ produces $S$ shrunken copies of its frequency response, each with $\frac{1}{S}$ the original bandwidth and transition width. The interpolator removes the unwanted aliased copies except one, leaving a sharp narrow-band frequency response. 

Convolution pyramids (CP) \shortcite{convolutional-pyramids} is a more modern approach, and uses multi-rate filtering in a fully differentiable structure. The structural difference between traditional multi-rate filtering and CP is that CP combines the intermediate outputs from each downsampling stage with the corresponding upsampled outputs of equal resolution. Each intermediate downsampling output is convolved with a small FIR, and then summed with the corresponding upsampling output in a skip connection. 

Using multi-rate filtering in an end-to-end differentiable structure allows CP to fit arbitrary frequency responses. Parameters are optimized in the spatial domain by back-propagating the $L2$ loss of the model's output against the target impulse response. One can manually train structures with various computational costs by changing the number of stages and the FIR sizes, but because these designs have limited structural diversity, they are Pareto dominated by designs that use multiple cost-saving techniques (Section~\ref{sec:results}).  

Hierarchical discrete correlation filters \cite{hierarchical-discrete-correlation} were proposed as a way to produce cheap and large Gaussian-like filters. They  have the same design structure as the prefilters and interpolators used by FRM (Section~\ref{sec:primitives}, \ref{sec:differentiable-parameters}), except they use a different heuristic for determining the filter coefficients.

Unlike recurrent filters, multi-rate designs are more amenable to efficient implementation on modern hardware. However, because they rely on striding or downsampling to reduce cost, their throughput scales with the target FIR size.

\subsubsection{Multi-Rank Decompositions for 2D Filters}
\label{sec:multi-rank-methods}
In 2D, the singular vector decomposition decomposes any 2D filter into a weighted sum of linearly separable filters. Certain filters like Gaussians and Gabors are exactly represented by a single term, but in general, 2D filters require multiple or infinite terms. Fortunately, in many cases only a few terms with the largest weights are needed to get a sufficiently accurate approximation. For $k$ terms, this reduces the cost of an $NxN$ filter from $O(N^2)$ to $O(kN)$. For large $N$ however, this approach can still be prohibitively slow. To address this, \cite{Deriche-Abramatic-1982-SVD-IIRs} propose further approximating each 1D term with a single causal higher order recursive filter. As we have discussed above, such designs are limited in their expressive power. Instead, our system considers multi-rank approximations for 2D filters that can use more complex cost-saving structures for its 1D terms and can combine 2D terms with addition as well as convolution. 

\subsubsection{Frequency-domain Convolution}
When the target response cannot be described as a band-pass prototype, and no hand-crafted fast approximations exist, applications use the Fast Fourier Transform (FFT) to apply large convolutions. Unlike the above methods, FFT convolution is mathematically exact, though concrete implementations add a small amount of noise due to floating point precision issues. Tiled FFT convolution has high throughput, but does so by amortizing work across large image tiles or audio windows. This introduces latency proportional to the tile size, which can make it unsuitable for real-time applications. Large tiles can make it challenging to process modest-sized images using multiple cores. FFTs also require complex data movement. Profiling Intel's implementation \cite{IntelIPP} reveals that it spends more time shuffling values in memory than doing multiply-adds. Our system found approximations with outputs visually indistinguishable from the ground truth and significantly higher throughput while meeting real-time latency requirements.

Elboher et al. \shortcite{elboher2011cosine} present a related general-purpose method, which uses cosine-weighted integral images to approximate filters as sums of truncated cosines. It exploits sparsity in the frequency domain, and takes an amount of work proportional to the frequency content of the filter. Notably, as with other integral image methods, this method is capable of handling spatially-varying filters. However, this comes at the cost of substantial work and memory usage ahead of time, and is not suitable for high-performance streaming filters. Interestingly, we note that if spatially-varying behavior is not required, truncated cosine filters can be more efficiently implemented as TIIRs (appendix \ref{appendix:truncated-cosines-are-tiirs}). 

\subsection{Differentiating Recursive Filters}
Design search requires automatically fitting the continuous parameters of each structure to assess their approximation quality. Back-propagating through time (BPTT) to optimize recurrent filters is difficult because outputs can grow exponentially, causing exploding gradients \cite{DSP-review}. To circumvent this, recent music application-focused approaches train neural networks to predict the parameters in the frequency domain, that would produce a target response across a range of sampled frequencies. These parameters are then fed to a differentiable frequency domain representation of the filter to generate an output response on which loss is computed. Simply moving into the frequency domain does not prevent the optimization from struggling with unstable filters. However, parameters in the frequency domain can be constrained to a well-behaved subspace. Several works \cite{Nercessian-neural-eq,synthesize-sound-matching-ddsp,DSP-style-transfer} restrict their IIRs to parametric equalizers so that gradient updates will always produce stable filters with interpretable behavior. Colonel et al. \shortcite{neural-sampling-biquads} notes that this subset may be insufficient for other applications. They show that a neural proxy can predict parameters for a broader range of IIRs given a carefully crafted dataset of target frequency responses. However, these filters are restricted to using complex conjugate poles and prediction performance is sensitive to out of distribution targets. 
 
 Our designs use recurrent filters within larger computational graphs. We cannot apply the frequency sampling approach to our structures because not all of our primitives are linear time invariant (LTI) systems, i.e. there is no closed form equation for the entire structure's frequency response. Using neural proxy methods provides no computational advantage because we would have to train a different network for each design structure. We cannot apply a neural proxy to just the recurrent filters because their desired frequency response is learned jointly during optimization with the behavior of all the primitives in a given structure. We developed a new method for directly optimizing recurrent filter parameters in the time domain, that overcomes the instability of BPTT (Section~\ref{sec:tiir-parameterization}).

\subsection{Fast Filter Implementation Techniques}
Maximizing throughput requires some strategy for taking advantage of multiple CPU or GPU cores. For a given IIR filter, it is possible to parallelize it in small tiles without redundant re-computation at tile boundaries~\cite{Nehab2011,recfilter}. We have not found these techniques necessary for our context, as we focus on CPU implementations, where tiles can be larger because there is less parallelism available. We do go to lengths to make good use of SIMD instructions~(\ref{sec:lowering}) in ways that are similar in spirit to those works. 

The technique we use for vectorizing horizontal IIRs is scattered look-ahead \cite{scattered-look-ahead}, which rewrites an IIR of stride $s$ into an IIR of stride $2s$ and a small FIR of stride $s$. This rewrite can be done repeatedly to increase the stride of an IIR until it is at least the vector size, at which point vectorization is trivial because there are no within-vector data dependencies. We found this superior to other methods for vectorizing IIRs both in performance and in numerical stability.

Fast filter implementations can also be obtained by lowering to a language like Halide~\shortcite{halide}, as was done in Ma et al.~\shortcite{demosaicking}. Halide excels at fusion for locality and parallelization, especially in stencil pipelines. However, it can neither fuse nor vectorize recurrent filters along the recurrent axis, which is critical for our work.

\subsection{Domain-Specific Program Search}
We define domain-specific program search as program search within a domain-specific language (DSL), where expert knowledge is encoded in the language and the search methods. Domain-specific program search has been used across several computational design domains. For example, Zhao et al. \shortcite{robo-grammar} optimize robot designs to traverse different target terrains by searching over compositions of body components defined in their DSL. Zhao et al. \shortcite{carpentry-co-optimization} observe that many carpentry designs and fabrication plans have common substructures. They exploit this with e-graphs to efficiently sample Pareto frontier designs and fabrication plans defined by their carpentry language. These solutions trade-off between competing objectives like material cost, fabrication time, and precision.

It is natural to think of designs as compositions of domain-specific primitives in fabrication contexts. But this idea can be applied more broadly. Programmers in many domains find themselves wrangling large design spaces, often to trade-off between cost and quality. For example, Ma et al. \shortcite{demosaicking} showed that combinations of differentiable image processing building blocks can produce demosaicking and super-resolution programs that dominate existing hand-designed methods in both throughput and quality. We show in this work that combining program synthesis techniques with knowledge from signal processing, performance optimization, and machine learning, produces state of the art fast image and audio filtering programs.

\section{System Overview}
Our search system has four main components: 
\begin{itemize}
    \item a search space defined by composable design primitives
    \item a design sampling strategy
    \item cost and quality evaluation methods
    \item a compiler that generates optimized implementations 
\end{itemize}

\noindent Our goal is to produce a Pareto-frontier of filter approximations. A design is Pareto-dominant if there is no other design with higher throughput at equal or higher quality. The Pareto-frontier is the set of all Pareto-dominant points. Note that we can extend the concept of a Pareto frontier to Pareto tiers, where tier $i$ is the next best Pareto frontier if all points in tier $i-1$ are excluded.

For a given target filter, our system enumerates design structures, fits their parameters with gradient descent to measure quality, and estimates their throughputs with a cost model.
We then extract the first few Pareto tiers and sample permutations on these more promising structures. After this exploitation step we recompute the first few tiers and compile and benchmark them. 
The final Pareto frontier is computed using these profiled throughputs. We will discuss each component of our system in the following subsections.

\subsection{Design Primitives}
\label{sec:primitives}
 Our primitives are unary or binary and 1D or 2D operations. Most have structural parameters that take on discrete values which are sampled during the design search. All nodes have a \smallcode{direction} parameter for the direction(s) it is applied in. 1D nodes have \smallcode{direction = horizontal} or \smallcode{ vertical}. In 1D designs, all nodes are \smallcode{horizontal}. 2D nodes have \smallcode{direction = both} and are created by combining 1D nodes with different directions, two existing 2D nodes,  or a 2D node with a 1D node. A 1D horizontal (vertical) node is applied to a 2D input per row (column) of the input. We use the syntax: \smallcode{primitive_name<parameter_name: parameter_type, ...>} to describe our primitives.

\setlist[itemize]{leftmargin=0.0cm}
\setlist[itemize]{
  leftmargin=0.75em,  
  labelwidth=0.75em,  
  labelsep=-0.2em, 
  align=parleft
}
\begin{itemize}[label=\textbullet]
    \item 1D Leaf operations    
        \begin{itemize}[label=\textendash]
            \item \smallboldcode{Leaf<direction: enum>} (abstract class)
            
            \hspace{0.75em}We use two different leaf operations: dense FIRs and TIIRs.
            \item \smallboldcode{FIR<width: int, direction: enum>}
            
            \hspace{0.75em}A dense FIR of length \smallcode{width} 
            \item \smallboldcode{TIIR<causal: bool, width: int, direction: enum>}
            
            \hspace{0.75em}An order 2 causal or anticausal TIIR with maximum length \smallcode{width}. The TIIR can be fit to use a smaller support (Section~\ref{sec:tiir-parameterization}). 
            We do not use IIRs because back-propagating through time requires evaluating the impulse for a fixed window length anyway. This assumes that the unevaluated values are small enough to ignore which is less accurate than explicitly parameterizing recursive filters as truncated IIRs. 
        \end{itemize}
    \item 1D Binary operations:
        \begin{itemize}[label=\textendash]
            \item \smallboldcode{Cascade<direction: enum>}
             
            \hspace{0.75em}Runs a leaf node on another node's output
        \end{itemize}
    \item 1D Unary operations:
        \begin{itemize}[label=\textendash]        
            \item\smallboldcode{Stride<stride: int, order: int, direction: enum>}
            \hspace{0.75em} Applies its child computation with stride \smallcode{stride}, followed by an interpolator. This primitive is an extension of FRM, where the child computation is not restricted to an \smallcode{FIR}. Interpolators are cascaded small anti-aliasing FIRs, parameterized by a value $\omega$ which indirectly controls which frequencies are completely suppressed \cite{narrow-band-FRM}.  Orders 1-4 of an interpolator with \smallcode{stride=2}, $\omega$\smallcode{=0} are $[1,1]$, $[1,2,1]$, $[1,3,3,1]$, and $[1,4,6,4,1]$. Higher orders have sharper transition widths and less stop-band rippling (there is no pass-band ripping) but higher computational cost. 
        
            \hspace{0.75em}An interpolator must have \smallcode{width} $\ge$ \smallcode{stride} to fill in the sparse FIR. An \smallcode{order=2}, \smallcode{stride=4}, $\omega$\smallcode{=0} interpolator uses two FIRs: $[1,2,1]$ and $[1,0,2,0,1]$ to produce a larger tent: $[1,2,3,4,3,2,1]$. Interpolators use $L=\log_{2}($\smallcode{stride}$)$ FIRs. Each FIR has \smallcode{order+}$1$ non-zero taps strided by $2^l-1$ zeros for levels $l\in[0,L-1]$. We refer to this as a factored tower. A factored tower for \smallcode{order=}$O$, \smallcode{stride=}$S$ has support width $\approx OS$ but only uses $O\log_2(S)$ non-zero taps. If the interpolated FIR has $N$ non-zero taps, the total \smallcode{Stride} support width is $\approx SN+OS$ and uses $N+O\log_2{S}$ non-zero taps, which is $S$ times fewer multiply-adds than a dense FIR.  
        
            \item \smallboldcode{DownUpsample<stride: int, direction: enum>} (abstract class)
            
            \hspace{0.75em}Computation at lower resolutions is performed by subtrees sandwiched between a downsample and upsample. \smallcode{DownUpsample} pre-filters then subsamples the input by \smallcode{stride} before passing it to the child. The child's output is zero-upsampled by \smallcode{stride} and then interpolated. We use two classes that differ in how they prefilter and interpolate.
        
            \item\smallboldcode{DownUpSampleI<stride: int, prefilter_width: int,}\\
            \smallboldcode{interpolator_width: int, direction: enum>}

            \hspace{0.75em}Prefilters and interpolates with FIRs of length \smallcode{prefilter_width} and \smallcode{interpolator_width}. Factored towers only save computation when adjacent output values in higher levels reuse values from lower ones. When downsampling, output values are strided apart and end up sharing no intermediate work. Similarly, when interpolating a zero-upsampled input with an FIR of length $N$, each output value only requires processing $\frac{N}{S}$ non-zero values. When upsampling with a factored tower, only the largest level benefits from the full sparsity as each level creates more non-zero values for the next level to process.
        
            \item\smallboldcode{DownUpSampleII<stride: int, prefilter\_order: int,}\\
            \smallboldcode{interpolator\_order: int, direction: enum>}
                   
            \hspace{0.75em}Pre-filters and interpolates with factored towers, each with their own $\omega$ and \smallcode{order}. Even though they use the same number of operations as a dense FIR, factored towers may still be useful for regularizing the pre-filter and interpolator.
        \end{itemize}
    \item 2D Binary operations:
        \begin{itemize}[label=\textendash]
            \item \smallboldcode{Pipe<direction: enum>}

            \hspace{0.75em}Runs the left child on the output of the right child. Is equivalent to \smallcode{Cascade} when both children are 1D and have the same direction. Children can have different dimensions and directions. It is the only operation that can promote nodes from 1D to 2D.
        \end{itemize}
    \item Point-wise operations:
    \begin{itemize}[label=\textendash]
        \item \smallboldcode{Sum<direction: enum>}:
            
        \hspace{0.75em}Point-wise sums the outputs of two nodes. If they have different widths, the smaller one is centered over the larger output. The search procedure enforces their widths have the same parity. Children must have the same dimensions and directions.
    \end{itemize}
\end{itemize}

\subsection{Design Sampling Strategy}
The main challenge with sampling structures and their parameters is finding \textbf{\emph{valid}} and \textbf{\emph{reasonable}} designs. Valid designs satisfy basic requirements, such as not having any nodes with negative support widths and producing the right output size. Reasonable designs satisfy additional properties that pareto-dominant models should have, based on domain knowledge in high-performance computing and signal processing. For example, any reasonable design should use fewer operations than the target FIR. 

To generate designs we could randomly enumerate grammatically correct structures and their parameters, and then filter for valid and reasonable ones. This approach would be highly inefficient because the raw search space has a combinatorial explosion of options. Instead, our enumeration procedure avoids creating illegal structures by updating constraints through the generation process for a given design and using them to restrict the set of primitives and structural parameters that it samples from. Since the output size is a function of all the nodes in the tree, sampling legal designs also requires performing bounds inference to derive valid support widths for each node. Overly stringent constraints may rule out good designs, but we also want to avoid spending time optimizing non Pareto-dominant models. This type of top-down enumerative search informed by deduction rules to prune the search space is well studied in program synthesis \cite{top-down-synthesis, super-optimizer-progam-search}. We also want to avoid redundantly optimizing programs that are structurally different but semantically equivalent. To do this, we implement normalization rules that transform all semantically equivalent structures into a canonical form. This process is referred to as term rewriting in the programming languages literature \cite{term-rewriting}. 

\subsubsection{Enumerating structures}
\label{sec:enumerating-structures}
Designs in our search space are binary trees. The root node produces the final output. Algorithm \ref{alg:1D-one-res-structure-sampling} samples 1D single-resolution trees using primitives: \smallcode{Stride,Cascade,Sum,} and \smallcode{Leaf}. We first determine the structure, before choosing concrete classes for the \smallcode{Leaf} and \smallcode{DownUpsample} nodes and filling in the discrete structural parameters for all the nodes.  We start generation with a random root primitive. At each recursive step, we select primitives for the current node's children from a set of legal primitives. The set of legal primitives is a function of the enumeration constraints (Section~\ref{sec:search-constraints}) evaluated on the partially generated tree and the location of the node being selected. We then recursively generate the subtrees below the children, stopping at leaf nodes.

To have multi-resolution computations, we use algorithm \ref{alg:1D-one-res-structure-sampling} as a subroutine in algorithm \ref{alg:1D-multi-res-structure-sampling} to fill in lower-resolution subgraphs. Subgraphs can operate at any resolution $r \in R$ where $r=\frac{1}{s}$ and $s$ is an integer from a set of strides $S$.  We pick a top-level resolution $r_1 \le 1$ and start with a \smallcode{DownUpsample} root using stride ${\frac{1}{r_1}}$, unless $r=1.0$, in which case the root is a random primitive excluding \smallcode{DownUpsample} and \smallcode{Pipe}. We then use algorithm \ref{alg:1D-one-res-structure-sampling} to sample the rest of the tree, with one additional legal leaf node type: \smallcode{LowerResolutionTree}, which is a placeholder for an undetermined subtree. This node can be used only once by each call to algorithm \ref{alg:1D-one-res-structure-sampling}, and only when the search deems it possible, $r \ge min(R)$, and useful to do computation at a lower resolution based on a set of conditions on the entire tree (Section~\ref{sec:search-constraints}). If \smallcode{LowerResolutionTree} is used, we recursively perform this procedure again to generate the subtree to replace the \smallcode{LowerResolutionTree}. The recursive call must use resolution $r' < r$.

Algorithm 3 creates 2D multi-resolution computations. It uses algorithm \ref{alg:1D-one-res-structure-sampling}
to generate horizontal and vertical subtrees at different resolutions. A given direction may not be used if the enumeration rules dictate that no more primitives in that direction are allowed, or the procedure randomly chooses to not produce further computation. If neither direction has computation the procedure returns. If both have computation, they are joined by \smallcode{Pipe} to produce the current subtree. If only one does, then it becomes the current subtree. Algorithm \ref{alg:1D-one-res-structure-sampling} does not use \smallcode{LowerResolutionTree} because we want to control lower-resolution computations at the 2D level. We use the same logic as algorithm \ref{alg:1D-multi-res-structure-sampling} to decide if lower-resolutions are allowed. If it is allowed in either direction, we recursively call this procedure to generate a lower-resolution subtree, which we join with the current subtree using either \smallcode{Pipe} or 2D \smallcode{Sum}. When the directionality of the current and lower-resolution subtrees do not match we always use \smallcode{Pipe} because point-wise adding in this case is ill-defined. See the appendix for the algorithms' pseudo-code.

\subsubsection{Sampling structural parameters}
Once the structures have been generated we choose which primitives to use for the \smallcode{Leaf}s and \smallcode{DownUpsample}s and fill in structural parameters. The choices made for the \smallcode{DownUpsample}s and \smallcode{Stride}s inform the valid range of support widths for the leaf nodes below. For example, if the target filter's support width is $N$ and the design is a \smallcode{DownUpsample} by 2x with a \smallcode{FIR} child, then the \smallcode{FIR} must have width $<\frac{N}{2}$, taking into account the widths of the prefilter and interpolator as well. Therefore, to increase the likelihood of sampling legal structures, we make decisions for the \smallcode{DownUpsample} and \smallcode{Stride} nodes first. We enumerate all possible settings for the \smallcode{DownUpsample} types, their prefilter and interpolator widths or orders, and the \smallcode{stride} and \smallcode{interpolator_order} for each \smallcode{Stride}. We randomly select one setting, and do a precursory bounds inference to get upper bounds on the support widths of all the nodes. If any widths are negative, then the current setting is illegal and we keep sampling until a legal setting is found. 

We then sample the leaf node types and their final support widths. Convolutional Pyramids showed that it may be helpful to use small residual FIRs, so we allow designs to have subsets of \smallcode{FIR}s with small support when possible. For example, if a \smallcode{Sum} has a support width of $N$, then only one of its children has to have support $N$, the other can choose any support $w<N$. A \smallcode{Cascade} with support $N$ can have one \smallcode{FIR} child with support $w$ as long as the other child has support $N-w+1$. To determine final support widths, we first enumerate the subsets of \smallcode{FIR}s that can simultaneously have target width-agnostic supports. We then randomly sample one subset and their support widths from a set of discrete options. The support widths of the remaining nodes are inferred top-down from the root, while respecting the already fixed \smallcode{FIR} support widths. The leaf node type choices and the \smallcode{causality} for \smallcode{TIIR}s are constrained by enumeration rules which we discuss below.

\subsubsection{Enumeration Constraints}
\label{sec:search-constraints}
The search procedure uses a list of rules constraining the space of structures that can be enumerated. Each rule affects the search by either limiting the set of primitives that can be used at each step in the recursive tree structure generation or by limiting the range of values for the structural parameters of each node. These rules eliminate: 1) invalid structures, like nodes with negative support width, 2) structures that always have cheaper equivalent alternatives, like a sum of two FIRs, or 3) structures that we deemed unlikely to be competitive based on signal processing intuition. See Appendix \ref{appendix:enumeration-constraints} for the list of constraints.

\subsubsection{Normalizing and accepting sampled structures}
The enumeration process can create trees that are syntactically different but mathematically the same. For example, \smallcode{Sum(A, B)} \smallcode{= Sum(B, A)}. Structures that are mathematically the same will have the same approximation quality (barring optimization issues) so it would be inefficient to optimize both. However, this syntactic difference \emph{does} imply a different order of computation. For example, the left child of a \smallcode{Sum} is evaluated before the right. In some cases, differences in operation order results in implementations with different throughputs. Our search system estimates design throughputs by analyzing their structures with a cost model. (Reasons for using a cost model instead of compiling and benchmarking immediately are discussed in section \ref{sec:cost-model}.) Therefore, it is important that structures are normalized to their lowest cost form. Otherwise, designs can have inaccurately high costs which compromises the accuracy of the Pareto ranking. To address this, we apply normalization passes that transform enumerated structures into their simplest lowest cost form, which we describe below. 

\begin{enumerate}
    \item Flatten adjacent binary nodes into one n-ary operator. For example \smallcode{Cascade(Cascade(A, B), C)} $\xrightarrow{}$ \smallcode{Cascade(A, B, C)}
    \item Sort each subtree's children based on their structures
     \begin{itemize}[label=\textendash]
        \item Flattening adjacent binary operations into one node allows us to easily check if two subtrees with the same n-ary root type are equivalent by sorting and comparing their operands. 
    \end{itemize}
    \item Move the vertical \smallcode{DownUpsample} of adjacent vertical and horizontal \smallcode{DownUpsample}s outermost
    \begin{itemize}[label=\textendash]
        \item applying these in either order are equivalent. However, doing the horizontal \smallcode{DownUpsample} innermost has better throughput because it is more expensive due to vector shuffling. If the vertical \smallcode{DownUpsample} is outermost, the horizontal \smallcode{DownUpsample} is applied to fewer input rows.
    \end{itemize}
    \item Move 1D computation inside orthogonal \smallcode{DownUpsample}s 
    \begin{itemize}[label=\textendash]
       \item In 2D designs, all 1D nodes are joined with a subtree applied in the orthogonal direction, or with a 2D subtree, by a \smallcode{Pipe}. If the root node of one of the subtrees is a \smallcode{DownUpsample} then it would be mathematically equivalent but cheaper to join the orthogonal computation inside the \smallcode{DownUpsample}. If there are multiple \smallcode{DownUpsample}s in a row, the orthogonal computation should be moved as far down as possible. There can be multiple options for which node to move or move into. We use our cost model to pick the one that results in the lowest cost. 
    \end{itemize}
\end{enumerate}

After an enumerated structure is normalized, we apply one additional transformation: if any subtree can be replaced with an \smallcode{FIR} and still cost less than a small fraction of the total cost of the design, we perform the replacement. For a fixed support width, \smallcode{FIR}s are the most expressive primitive in our design space, so it is reasonable to use them wherever it does not significantly contribute to the total cost. The search then sanity checks that none of the subtrees cost more than a \smallcode{FIR}. If it passes the check and has not been sampled already, it is added the the set of structures that will be optimized.

\subsection{Design Cost Model}
\label{sec:cost-model}
Throughputs are obtained by compiling and benchmarking designs with our lowering system. Accurate measurements require benchmarking in an isolated compute environments. This limits our ability to utilize distributed compute systems. However, distributing the parameter optimization is a must because our system samples thousands of designs for a given target. Fortunately, our designs are small and can be optimized on generic modern CPUs. Sharing resources with other processes during optimization is not an issue.
\thesisversion{Benchmarking takes 5-10 seconds per design. Even if each design only required one benchmarking run (ignoring the need to benchmark partial designs), profiling 5k designs would take ~14 hours. Comparatively we had access to 32 distributed nodes each with ~100 CPUs for optimization. Each design used 16 CPUs, so 200 designs could be optimized simultaneously. The median time per job was 1.5 minutes. Fitting 5k designs at that rate takes less than 1 hour.}

Many enumeration constraints and normalization rules rely on assessing the throughputs of multiple subtrees for a given design. If search relied on benchmarking to inform these decisions it would be significantly slower because it has limited access to benchmarking resources. With an accurate cost model, we were able to sample designs without compiling or benchmarking code. The predicted cost is inversely related to throughput and proportional to machine cycles per output sample. Afterwards, a small subset, 2500 out of 150k total designs, that made it to the top 3 Pareto tiers for each target were compiled and benchmarked to produce the final Pareto frontier. Figure \ref{fig:throughputs-vs-cost}
shows the accuracy of our cost model. The correlation coefficients were $0.98$ and $0.97$ for 1D and 2D designs respectively. See the appendix for full cost model details.

\subsection{Differentiable Design Parameters}
\label{sec:differentiable-parameters}
All primitives except \smallcode{Sum}, \smallcode{Cascade}, and \smallcode{Pipe} have continuous parameters that need to be optimized. \smallcode{TIIR} warrants the most exposition due to the challenges with differentiating through recurrent structures. Designs are end-to-end differentiable by construction because we developed differentiable representations for all our primitives. This allows us to optimize their parameters using conventional neural network optimization methods discussed in Section \ref{sec:training}. We first discuss the straight-forward parameterization of all the other primitives and then dive into \smallcode{TIIRs} .

\begin{itemize}[label=\textbullet]
    \item \smallboldcode{FIR<width: int, direction: enum>}\\
    \hspace{0.75em} \smallcode{FIR} is a finite length 1D filter, parameterized by \smallcode{width} weights. We initialize them with uniform random values between -0.1, 0.1. 
    \item \smallboldcode{Stride<stride: int, order: int, direction: enum>}\\
    \hspace{0.75em} \smallcode{Stride} has one learnable parameter, $\omega$, which controls the behavior of its interpolator (Section~\ref{sec:primitives}). \smallcode{stride}  is always a power of 2 and \smallcode{order} $\in [2,3,4]$. If \smallcode{stride=}$2^{L}$, the factored tower interpolator uses $L$ sparse FIRs where the level $l \in [0,L-1]$ FIR has some non-zero taps with $2^l-1$ zeros between them. If \smallcode{order=2}, the non-zero values are $[1,2cos(2^l\omega),1]$.  We call this order 2 because it can be factored into two first order FIRs using Euler's formula:  $[1,2cos(2^l\omega),1] = [e^{(i2^l\omega)}, 1]\ast[e^{(-i2^l\omega)}, 1]$ where $\ast$ is convolution. For all values of $\omega$ except $0$, these first order FIRs are complex valued so we cannot use a first order interpolator. First order interpolators are also low quality. However, when $\omega=0$ the first order FIR is $[1, 1]$, i.e. a box blur, and we can create a third order FIR by convolving the second and first order FIRs: $[1,2,1]\ast[1,1]=[1, 3, 3, 1]$. An \smallcode{order=4} interpolator has non-zero taps $[1,2cos(2^l\omega),1]\ast [1,2cos(2^l\omega),1]$. When $\omega=0$ the interpolator is a low-pass filter. In general, the center of the interpolator's pass-band is at $\omega$ with the exception of some degenerate cases. \thesisversion{DISCUSS THE DEGENERATE CASES} We use $\omega$ to generate the factored tower in the forward pass. If \smallcode{order=3} we fix $\omega=0$. Otherwise, $\omega$ is initialized to a uniform random value in $[0, 0.1]$.
    \item \smallboldcode{DownUpsampleI<stride: int, prefilter_width: int,}\\
    \smallboldcode{interpolator_width: int, direction: enum>}\\
    \hspace{0.75em} \smallcode{DownUpsampleI} uses an FIR parameterized by \smallcode{prefilter_width} weights to prefilter before sub-sampling and another FIR of parameterized by \smallcode{interpolator_width} weights to interpolate after zero-upsampling. We initialize them with uniform random values between -0.1, 0.1. 
    \item \smallboldcode{DownUpsampleII<stride: int, prefilter_order: int,}\\
    \smallboldcode{interpolator_order: int, direction: enum >}\\ \smallcode{DownUpsampleII} uses two factored towers to prefilter and interpolate, each with their own $\omega$.
\end{itemize}

\subsubsection{TIIR background}
There are multiple ways to understand how TIIRs save computation. Readers familiar with signal processing will find the details below. Others may want to skip to Figure \ref{fig:sparsification}, which illustrates the basic idea in the spatial domain with an example.

To discuss our parameterization of TIIRs we first begin with some signal processing background and a summary of the TIIR frequency domain analysis presented in \cite{TIIR}.
An order P linear time invariant (LTI) recursive filter is defined by a difference equation in the spatial domain of the form:
\[y[n] = \sum_{i=0}^{P}b_ix[n-i] + \sum_{j=0}^{P}a_jy[n-j]  \tag{1} \label{eq:basic-difference-equation}\]
where $x[n]$ and $y[n]$ are the input and output signals. This system is equivalent to two systems, where one feeds the other:
\[f[n] = \sum_{i=0}^{P}b_ix[n-i] \quad \text{(convolution with b[i])}\]
\[y[n] = f[n] + \sum_{j=0}^{P}a_jy[n-j] \quad \text{(recurrence relation)}\]
i.e. convolving $x[n]$ with an FIR defined by $b_i$ and a simpler IIR defined by $a_j$. The system response to a given frequency is characterized by its transfer function, which is the ratio of the Z-transforms of $y[n]$ and $x[n]$ and assumes the initial conditions $y[n]$$=$$0$ $\forall n < 0$:
\[H(z)=\frac{Y(z)}{X(z)} = \frac{\sum_{i=0}^{P}b_iz^{-i}}{\sum_{j=0}^{P}a_jz^{-j}} = \frac{B(z)}{A(z)}  \tag{2}\label{eq:iir-transfer-function}\]

\thesisversion{
$z=re^{j\omega}$ is a complex number that represents an exponentially growing or decaying frequency. $X(z)$ and $Y(z)$ show how much of $z$ is in the input and output. The ratio $H(z)$ is a complex or real number that tells us how much this system scales and rotates input frequencies. In the spatial domain, the impulse response $h[n]$ is the analogous characterization. $h[n]$ is the output when $x[n]$ is the Kronecker delta $\delta[n]$, which contains all frequencies.}
\noindent $H(z)$ is also the Z-transform of the system's impulse response $h[n]$:
\[H(z) = Z(h[n]) = h_0 + h_1z^{-1} + h_2z^{-2} + ...
 \tag{3}\label{eq:iir-transfer-function-as-sequence}\]

\noindent LTI filters are called FIRs or IIRs because of their finite or infinite impulse responses. We will use "impulse response" to refer to the output of the filter on $\delta[n]$, and FIR or IIR to refer to the filter itself. Say we want to truncate an infinite impulse response after $N$ samples and use them as the taps for an FIR:
\[y[n] = \sum_{m=0}^{N}h[m]x[n-m]\]
with transfer function:
\[H_{FIR}(z) = h_0 + h_1z^{-1} + ... + h_Nz^{-N}  \tag{4}\]
If $N$ is large, applying this FIR is expensive. Instead, let us define: 
\[z^NH_{FIR} = C(z)=h_0z^N + h_1z^{n-1} + ... + h_N \tag{5}\] 
If we shift the IIR's transfer function  by $z^N$ we get:
\[z^NH(z) = \frac{z^NB(z)}{A(z)} = C(z) + H'(z) \tag{6}\]
where the two equalities follow directly from (\ref{eq:iir-transfer-function}) and (\ref{eq:iir-transfer-function-as-sequence}). The infinite tail we want to remove is:
\[H'(z) = h_{N+1}z^{-1} + h_{N+2}z^{-2} + h_{N+3}z^{-3} + ...  \tag{7}\] 
Note that $z^NB(z)$ is a degree $N+P$ polynomial, $A(z)$ is a degree $P$ polynomial, and $C(z)$ is a degree $N$ polynomial. \\
Therefore, $C(z)$ must be the quotient of $\frac{z^NB(z)}{A(z)}$ and 
\[H'(z)=\frac{B'(z)}{A(z)} \tag{8}\] where $B'(z)$ is the degree $P$ polynomial remainder, i.e. 
\[B'(z) \equiv z^NB(z)\mod A(z) \tag{9} \label{eq:tiir-modulo}\]
It follows from (5) - (8) and simple algebra that:
\[H_{FIR}(z) = \frac{B(z)- z^{-N}B'(z)}{A(z) } =\frac{\sum_{i=0}^{P}b_iz^{-i} -z^{-N}\sum_{j=0}^{P}b'_iz^{-i}}{\sum_{j=0}^{P}a_jz^{-j}} \tag{10}\label{eq:tiir_transfer_function}\]
\begin{figure}
\includegraphics[width=\columnwidth]{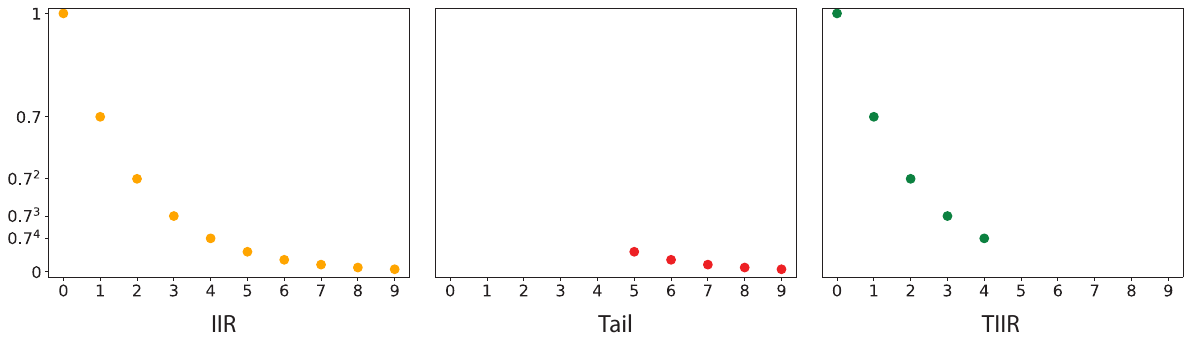}
\captionsetup{skip=2.5pt}
\caption{TIIRs extract FIRs from IIRs. Left: an exponentially decaying first order IIR. Center: the subtracted tail. Right: the remaining finite impulse response. We only plot the first 10 samples of the IIR which is infinite.}
\vspace{-1em}
\label{fig:tiir-explanation}
\end{figure}

\noindent The numerator of (\ref{eq:tiir_transfer_function}) defines a large sparse FIR with $P$ non-zero taps on either end separated by $N-P$ zeros. The denominator defines an IIR with $P$ taps. This representation for $h_{FIR}$ is called a TIIR. The cost of applying the TIIR is $3P$. Typically the progenitor IIR has order $P\ll N$, so this representation requires very few operations.
Intuitively, to remove the tail, we modify the progenitor IIR's non-recursive taps, so that its impulse response starts with the value after the desired region in the original impulse response. The modified IIR is delayed by $N$ samples and subtracted from the progenitor IIR to produce the TIIR. Figure \ref{fig:tiir-explanation} shows a length 5 truncated impulse response extracted from an exponentially decaying, i.e. stable, infinite impulse response produced by the first order IIR: \[y[n] = x[n] + 0.7y[n-1]\] with transfer function \[H(z) = \frac{1}{1 - 0.7z^{-1}}\]
The TIIR difference equation and transfer function are: \[y[n] = x[n] - 0.7^5x[n-5] + 0.7y[n-1]\]
\[H(z) = \frac{1 - 0.7^5z^{-5}}{1-0.7z^{-1}}\]
$-0.7^5x[n-5]$ cancels out $0.7y[n]$ at $n=5$, causing all subsequent outputs of the impulse response to be zero. If the IIR had a recursive tap $>1$  it would be unstable, with an exponentially growing impulse response. A length N box blur as a TIIR is borderline stable: \[y[n] = x[n] - x[n-N] + y[n-1]\]
\[H(z) = \frac{1 - z^{-N}}{1 - z^{-1}}\] 
As written, this is the sliding window box filter. If you decompose it into the recursive term (a sum-scan) followed by the non-recursive terms (a two-tap FIR), this is the summed-area table method. These popular methods are equivalent, and are just a special case of the much more general class of TIIRs.

\subsubsection{Differentiable TIIR parameterization}\label{sec:tiir-parameterization}
TIIRs are parameterized in the spatial domain by the filter taps $b_i$ and $a_j$. Back-propagating through recursive filters suffers from exploding gradients when the IIR is unstable. There is a 1-1 mapping between spatial domain parameters and the frequency domain parameters which are the zeros and poles of $H_{TIIR}$, i.e. the roots of the numerator and denominator polynomials. Unfortunately, we cannot optimize the parameters in the frequency domain for two reasons: 1) it requires differentiating through modulo (see equation (\ref{eq:tiir-modulo})) and 2) we use TIIRs in structures that have no closed form expression for their frequency response  because they use non-LTI primitives: \smallcode{Stride, DownUpsample}.

It can be shown from algebra and Euler's formula on well-known z-transforms that second order IIRs with transfer functions: 
\[\frac{b_0 + b_1z^{-1}}{a_0 + a_1z^{-1} + a_2z^{-2}} \tag{11}\label{eq:our-tiir-transfer-function}\] 
have impulse responses that are weighted sums of two functions defined by the partial fraction decomposition of (\ref{eq:our-tiir-transfer-function}): 
\[\frac{b_0 + b_1z^{-1}}{a_0 + a_1z^{-1} + a_2z^{-2}} = \frac{w_0}{1-p_0z^-1} + \frac{w_1}{1-p_1z^-1} \tag{12}\label{eq:partial-fraction-decomp-of-tiir}\] 
Where $p_0,p_1$ are poles $w_0$ and $w_1$ are real. For real IIRs the  poles are either 1) complex conjugates, 2) distinct reals, or 3) duplicate reals. \\
In case 1, the IIR is a weighted sum of a cosine and sine:
\[h[n] = w_1|p_1|^n\cos(\angle p_1 n) + w_2|p_1|^n\sin(\angle p_1 n)\]
where $|p_1|$=$|p_2|$ and $\mathrm{\angle(p_1)}$= $-\angle(p_2)$. To avoid differentiating through $|\cdot|$ and $\angle$, we observe that any complex number can be written as $re^{i\theta}$=$e^{\ln(r)+i\theta}$, $r \in \mathbb{R}$. Thus $|p_1|=e^{\mathrm{Re}(\ln(p_1))}$ and $\angle p_1 = \mathrm{Im}(\ln(p_1))$.\\
In case 2, the IIR is a weighted sum of real exponentials:
\[h[n] = w_1p_1^n + w_2p_2^n\]
In case 3, the IIR is a weighted sum of a real exponential and the same real exponential times a linear polynomial: 
\[h[n] = w_1p_1^n + w_2(n+1)p_1^n\]
Our differentiable TIIR representation learns its progenitor IIR with the parameters $w_1, w_2, p_1$, and $p_2$. Extending this representation to IIRs with second order numerators i.e. 
\[\frac{b_0 + b_1z^{-1} + b_2z^{-2}}{a_0 + a_1z^{-1} + a_2z^{-2}} = C + \frac{b_0 + b_1z^{-1}}{a_0 + a_1z^{-1} + a_2z^{-2}}, \quad C\in\mathbb{R}\]
is straightforward but we did not find it necessary.

The issue with optimizing TIIRs is that  useful ones typically have poles very close to the unit circle, i.e. they decay slowly. During optimization, poles near the unit circle are likely to stray outside the unit circle, resulting in exploding gradients. In other words, the loss landscape is ill-conditioned; only a small range of pole values result in low error and outside this range the error is catastrophic. \textbf{Our key observation is that we can rescale any range of useful values for $|p_i|$ to take up the entire area of the unit circle}. To do this, we introduce a time step resolution parameter $\gamma< 1$ such that the impulse responses in each case are now:
\begin{enumerate}[label=\arabic*.]
    \item $h[n] = w_1|p_1|^{\gamma n}\cos(\angle p_1 \gamma n) + w_2|p_1|^{\gamma n}\sin(\angle p_1 \gamma n)$
    \item $h[n] = w_1p_1^{\gamma n} + w_2p_2^{\gamma n}$
    \item $h[n] = w_1p_1^{\gamma n} + w_2(\gamma n+1)p_1^{\gamma n}$
\end{enumerate}
\begin{figure}
\includegraphics[width=0.85\columnwidth]{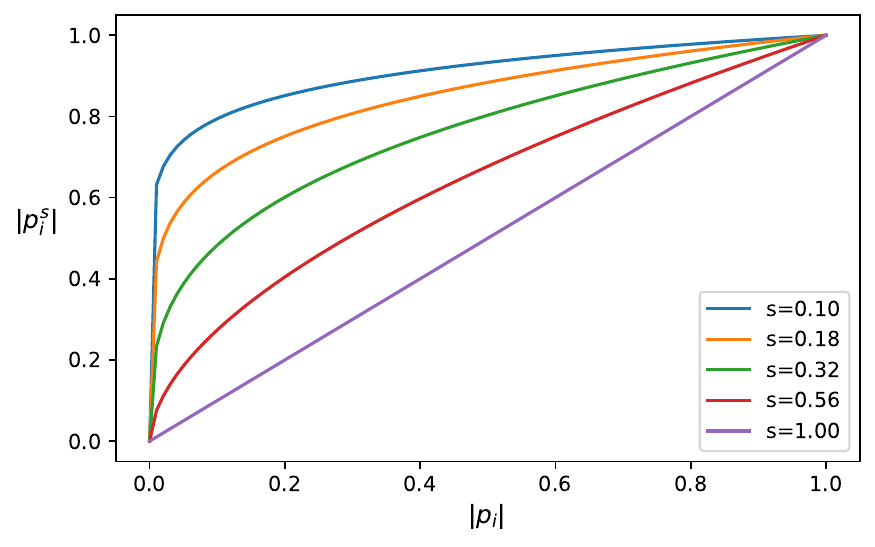}
\captionsetup{skip=1.2pt}
\caption{The time step resolution parameter s rescales any desired range of values for $|p_i|$ to take up the entire range from 0 to 1}
\vspace{-1.25em}
\label{fig:tiir-time-step-resolution}
\end{figure}

\begin{figure}
\includegraphics[width=0.9\columnwidth]{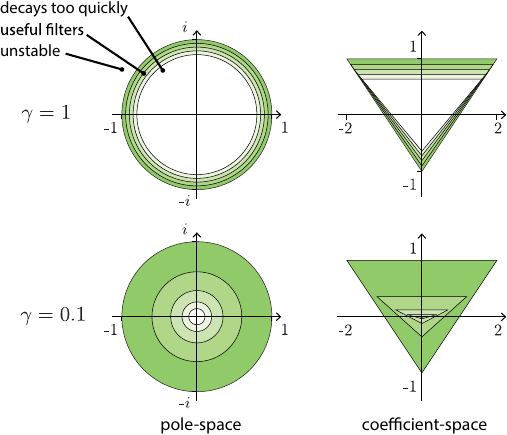}
\captionsetup{skip=4pt}
\caption{The most useful TIIRs use IIRs that decay slowly. In pole-space (top-left), these are in a narrow ring just inside the unit circle (green). Outside this ring, filters are unstable. Inside, filters decay too quickly. Gradient descent in the original pole-space or on the IIR coefficients (top right) is subject to exploding gradients. We ameliorate this by scaling the time axis by $\gamma$$<$$1$. After this reparameterization, the entire unit circle (bottom row) is occupied by IIRs with useful pole magnitudes, and training converges. After training we raise the learned poles to the power $\gamma$ to undo the reparameterization (top left) and then convert from poles to IIR coefficients (top right). }\vspace{-0.2em}
\label{fig:gamma}
\end{figure}
Figures \ref{fig:tiir-time-step-resolution} and \ref{fig:gamma} show the effect of $\gamma$ on the range of useful values for $p_i$. This reparameterization makes it less likely but not impossible for poles to move outside the unit circle, so we project unstable poles back onto the unit circle. After training, to convert back to the original pole space we raise the learned poles to the power $\gamma$.

\begin{figure}
\centering
\vspace{-0.5em}
\includegraphics[width=0.84\columnwidth]{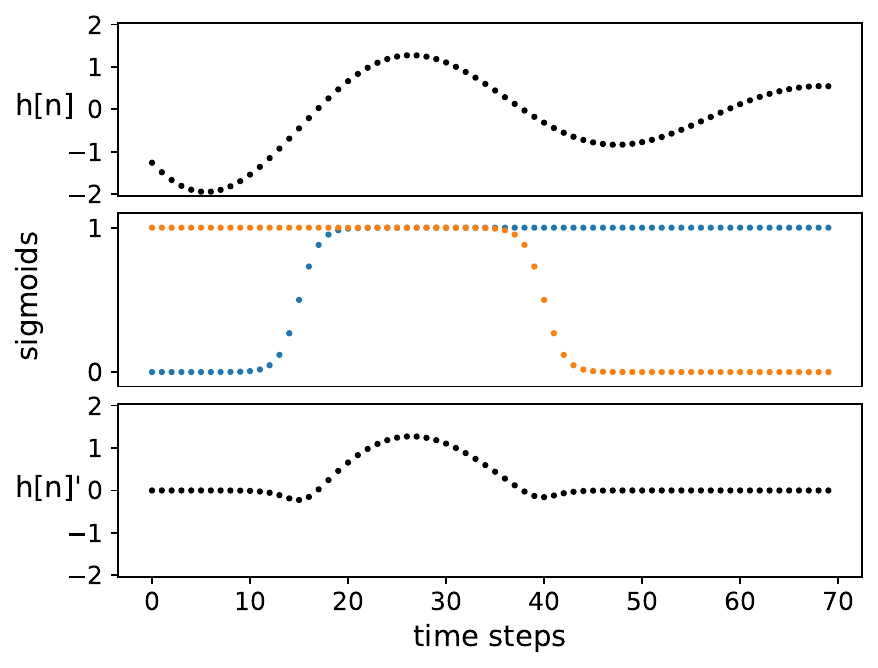}
\captionsetup{skip=1.5pt}
\caption{
Truncation regions for \smallcode{TIIR}s are learned as the inflection locations of increasing and decreasing sigmoids that mask values in the zeros regions. During training we gradually increase the sigmoid temperatures, steepening the transition regions, to encourage integer cutoff locations. From top to bottom: original TIIR $h[n]$, the sigmoids, and the truncated TIIR $h[n]'$.
}
\vspace{-1.0em}
\label{fig:truncation}
\end{figure}

\smallcode{TIIR}s are truncated by their \smallcode{width} parameter since nodes must have a finite width to guarantee the correct output width. This means we sample the progenitor IIR for \smallcode{width} time steps. \smallcode{TIIR}s also learn two truncation parameters $0 \le t_1 < t_2 \le1$, to be able to use a smaller region than \smallcode{width}. $t_1$ and $t_2$ are the inflection points of increasing and decreasing sigmoids with temperature $\tau$:
\[\biggl\{\; \frac{1}{1 + e ^{-\tau(n - t_1)}}\;,\; \frac{1}{1 + e ^{\tau(n-t_2)}} \;\biggr\} \tag{13} \label{eq:sigmoids}\]
The sigmoids are sampled for $n \in [0,$\smallcode{width-1}$]$ and element-wise multiplied with the sampled impulse response (\figref{fig:truncation}). $\tau$ is gradually increased during training to learn discrete cutoffs. In summary, we perform gradient descent on unrolled IIR outputs, but directly relate their frequency domain parameters to their spatial domain impulse responses. This allows us to keep their impulse responses in a well-behaved numerical range during training. We avoid differentiating through modulo to perform truncation by sampling the IIR for a finite number of time steps and using soft windowing. 

\subsubsection{Deriving TIIR coefficients from differentiable parameters}
It remains to explain how we recover the spatial parameters for our TIIRs after training. Given our progenitor IIR (\ref{eq:our-tiir-transfer-function}), the TIIR transfer function (\ref{eq:tiir_transfer_function}) states that our TIIRs have the form:
\[H_{TIIR}(z) = \frac{b_0 + b_1z^{-1} -z^{-(N+1)}(b'_0 + b'_1z^{-1})}{a_0 + a_1z^{-1} + a_2z^{-2}}\]
where $N$ is replaced with $N+1$ because the degree of $B(z)$ is one less than the degree of $A(z)$ in our case. Coefficients $b_i$, $a_j$ belong to the progenitor IIR and are easily derived from $w_0,w_1, p_0,p_1$ by recombining the partial fractions in  (\ref{eq:partial-fraction-decomp-of-tiir}). 

Deriving $b'_0,b'_1$ is more interesting. Recall that we can find them analytically with modulo on polynomials (\ref{eq:tiir-modulo}). However, we can also find them numerically. After training, we know the length $N$ finite impulse response of the TIIR. We know from (\ref{eq:tiir_transfer_function}) that to apply it, we do not naively convolve with the finite impulse response. Instead, we apply a length $N$+$2$ sparse FIR with 2 non-zero taps on each end: 
\[x'[n] = b_0x[n] + b_1x[n-1] - b'_0x[n-N] - b'_1x[n-N-1]\]
and a simple 2-tap IIR:
\[
y[n] = x'[n] + a_1y[n-1] + a_2y[n-2] 
\]
We know $b_i,a_j$, and $N$ but we need to find $b'_i$. We know the finite impulse response is the TIIR output given input $\delta[n]$:
\[x'[n] = b_0\delta[n] + b_1\delta[n-1] - b'_0\delta[n-N] - b'_1\delta[n-N-1]\tag{14}\label{eq:tiir-sparsified}\]
\[h[n] = x'[n] + a_1h[n-1] + a_2h[n-2]\tag{15}\label{eq:tiir-recurrence}\]
We can rearrange (\ref{eq:tiir-recurrence}) to define $x'[n]$ as a function of  $h[n]$ instead:
\[x'[n] = h[n] -a_1h[n-1] - a_2h[n-2] \tag{16}\label{eq:iir-inverter}
\]
\textbf{i.e. we can apply a 3-tap inverter FIR to the TIIR's impulse response to get $x'[n]$.} (\ref{eq:tiir-sparsified}) states that $x'[n]$ is the finite impulse response of the sparse FIR we seek, with non-zero taps $b_i$, $b'_i$.

\begin{figure}
\includegraphics[width=\columnwidth]{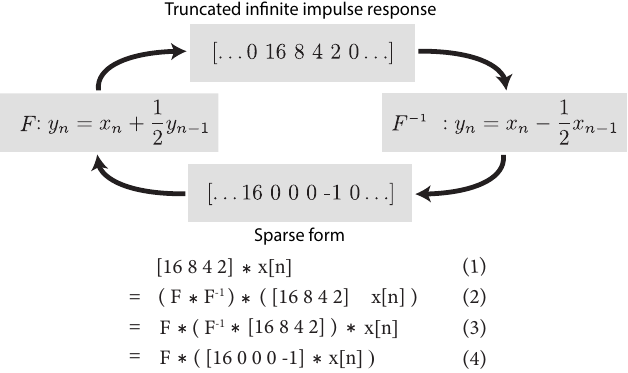}
\captionsetup{skip=2pt}
%
\caption{Convolution by a truncated infinite impulse response of size $n$ can be done with $O(1)$ operations per output. At the top is a truncated impulse response of a first-order IIR: $y[n] = 16x[n] + \frac{1}{2}y[n-1]$. The FIR $F^{-1}$ annihilates this signal except at the boundaries of the truncation region. The truncated dense form is recovered by running the IIR F. Thus, instead of convolving by the truncated impulse response with $O(n)$ operations (1), we can exploit the associative property of convolution (2, 3), and convolve by the sparse form and then run the recurrence $F$ (4), using $O(1)$ operations.}
\vspace{-0.3em}
\label{fig:sparsification}
\end{figure}

Intuitively, when an IIR is run on an impulse, after $P$+$1$ outputs, none of the non-recursive taps $b_i$ overlap the single non-zero input at $n$=0. Thus all subsequent outputs are entirely defined by the recurrence. So if we run the FIR that inverts this relationship on the IIR's impulse response, we will produce zeros, except for the first $P$+$1$ outputs. The relationship also breaks at the end of a truncated infinite impulse response. The inverter will produce non-zero values when only part of its support is in the truncation region. The inverter has support $Q$+$1$, so it will produce $Q$ non-zeros. In our case $P=1, Q=2$ so the sparse output has 2 non-zero values on either end of the truncation region. Obviously to undo the sparsification, we can run the original recurrence (\ref{eq:tiir-recurrence}) on this sparse output. TIIRs encode in a recurrence relation the information needed to reproduce a dense finite impulse response given an input state defined by a sparse FIR. Both have much lower arithmetic intensity compared to the equivalent dense FIR (see figure \ref{fig:sparsification}).

\textbf{Our numerical method has a major advantage over the analytical one: it allows soft truncation windows}. Depending on the truncation locations $t_1,t_2$ and the temperature $\tau$, the truncated impulse response could end up being multiplied by sigmoid values between 0 and 1 at the boundaries. There is no way to account for this with the analytical solution which assumes hard cutoffs. However, our numerical approach generalizes nicely to this setting by producing an extra non-zero value at the boundary for every sampled value of the sigmoid that is not strictly 0 or 1.

\subsection{Optimization}
\subsubsection{Training inputs}
\begin{figure}
\centering
\includegraphics[width=0.88\columnwidth]{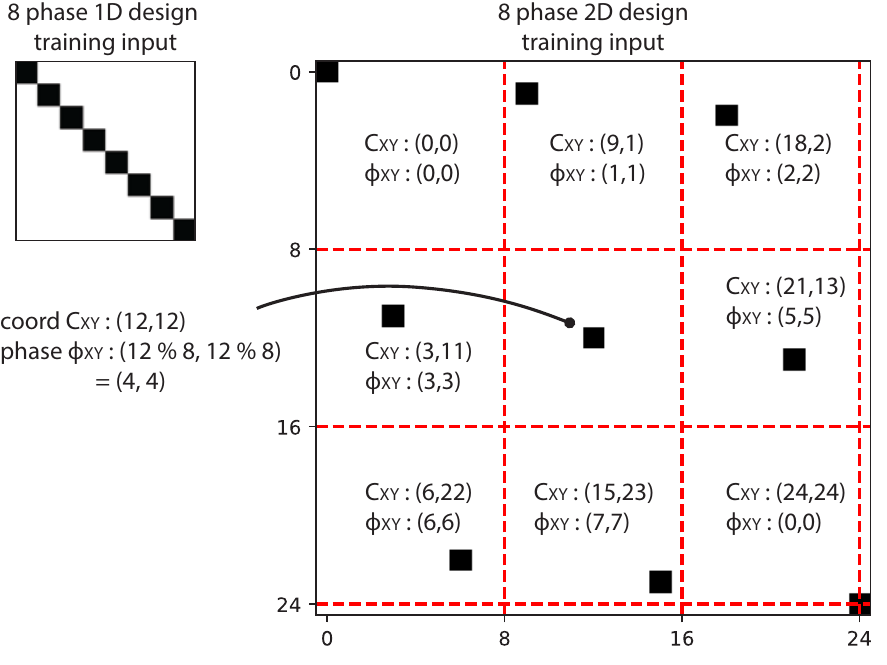}
\captionsetup{skip=1pt}
\caption{Designs that use multi-rate filtering are not shift-invariant. Therefore, training should measure loss on impulse responses at all phases of the design. For 1D designs with $S$ phases, we use an $S$x$S$ training input with impulses along the diagonal. For 2D designs we use an image split into $\lceil \sqrt{S} \rceil$ tiles where each tile has an impulse at a different vertical and horizontal phase.
}
\vspace{-1em}
\label{fig:training-inputs}
\end{figure}
Like CP, we optimize each design with the $L_2$ loss between its response and the target filter's response to some input signal. However, we use different training inputs. CP trains on a ``sum of randomly translated and scaled delta functions" \cite{convolutional-pyramids}. Any filter that upsamples by an integer factor $S$ will produce an output with $S$ phases. Every $S$ output samples will use different subsets of the interpolator's taps after zero-upsampling. In other words, the filter is not shift-invariant. This means that the proper training input for the design should not be randomly placed impulses. Rather, each design's training input should have impulses at each phase of the given design. To compute $S$ we traverse the design's tree and track the maximum total resolution change produced by nested \smallcode{DownUpsample}s. For 2D designs we set $S$ as the maximum resolution change across both directions.  

The training input for 1D designs is an $S$x$S$ image with impulses along the diagonal. During optimization the design is run horizontally across the input so each row is training on a different phase. For 2D designs, we use a square 2D image divided into $S' = \lceil \sqrt{S} \rceil$ tiles of size $D$x$D$. Each tile in row major order has an impulse at vertical and horizontal phase $s_i \in S'^2$. The image is generated by placing an impulse at location $x=t_xD + s_i$, $y=t_yD + s_i$ where $t_x = s_i$ \smallcode{\%} $\sqrt{S'}$ and $t_y = \lfloor \frac{s_i}{\sqrt{S'}} \rfloor$.  $D$ must be a multiple of $S$. The distance between any two impulses is $>D$. For training efficiency we would like $D$ to be as small as possible. $D$ should obviously be less than $W$ rounded up to the nearest multiple of $S$, where $W$ is the support width of the target filter. In practice, we set $D = \frac{W}{2}$ (rounded up to the nearest multiple of $S$). This still produces good training results because the impulse responses for our targets are significantly smaller in the overlapping regions than in the non-overlapping regions. 
See figure \ref{fig:training-inputs} for a visualization of our training inputs. 

\subsubsection{Training Procedure}\label{sec:training}
Unlike deep neural networks, our designs are not over-parameterized and are thus more sensitive to poor initializations \cite{lottery-ticket}. To combat this, we parallelize training many different initializations across the batch dimension. We keep track of the set of parameters with the lowest loss during training and save those as the design parameters. 

To finish optimizing designs within a constrained amount of time, we set maximum training times for each target where larger targets are given more time. We end training early if the design converges or if the impulse response's PSNR against the target exceeds a threshold. We also observed that while bad designs may see steady improvement in PSNR, they never reach a PSNR within the usable range. Empirically, good designs very quickly exceed this threshold during training. We set a minimum PSNR that designs must reach after a set number of iterations to continue training. The training time statistics for our designs (in minutes) are as follows:
\vspace{-0.8em}
\begin{table}[H]
\begin{tabular}{c|c|c}
\hline
 & 1D & 2D \\
\hline
    min & 0.02 & 0.08 \\
    max & 40.32 & 50.5 \\
    median & 1.27 & 2.87 \\
    mean & 2.44 & 7.86\\
\hline
\end{tabular}
\caption{Training time statistics for designs from our search space.}\label{tab:training-time-stats}
\vspace{-2.2em}
\end{table}
\vspace{-0.8em}
\noindent See Appendix \ref{appendix:training_time_statistics} for per-target-filter training time statistics.
\subsubsection{TIIR Hyper-parameters}
The maximum frequency that a stable TIIR can have in its impulse response is bounded by the time step resolution parameter $\gamma$. We set $\gamma \ge$ the maximum frequency in the target impulse response in terms of cycles per sample. As stated in section \ref{sec:tiir-parameterization} TIIRs have poles that fall into one of three cases: 1) complex conjugates, 2) distinct reals, 3) duplicate reals. Using complex parameters for the poles would technically support all three cases, however the chances of the numerical values being exactly real, not to mention exactly duplicate, is essentially zero. Therefore, we sample pole types before training with probabilities $[0.875, 0.1, 0.025]$ respectively. This assumes that truncations of sinusoidal impulse responses will be more useful. The sigmoid temperature $\tau$ starts at 5 and increases at a rate of 1.002 per iteration until it reaches 50. We initialize poles to have random stable magnitudes $0.8<|p|<1$.

\subsection{Lowering}
\label{sec:lowering}

The system uses three intermediate representations (IRs) for filter designs, at progressively lower levels of abstraction. All three IRs represent filter designs hierarchically, as compositions of smaller designs. Search and training is done in the first IR (Section~\ref{sec:primitives}). Optimization passes happen over the second IR (Section~\ref{appendix:optimization_passes}). In this section, we describe the most salient aspects of the final IR.

Our filter designs are ultimately lowered to C++ expression templates, where each primitive is represented by a type, and each type has template parameters that encode the child primitives. Strides and FIR tap locations are also encoded in the type. Knowing strides statically helps generate good SIMD code, and knowing tap locations means that small delay buffers can be promoted out of memory into registers. Any changes to a filter design that change these parameters requires recompilation.

Single-dimensional nodes are compiled to C++ objects that produce and consume SIMD vectors at a fixed rate, maintaining internal state if necessary (e.g. a few previous output vectors, for IIRs). For audio pipelines, this is the only type of node used. In this way, complex single-dimensional pipelines are fused into a single loop, with intermediate values and state held in vector registers instead of going to memory.

For single-dimensional nodes with larger memory requirements (e.g. large strided FIRs), we allocate circular buffers. Circular buffers present a challenge for vectorization, as vector loads might cross the buffer boundary and require loading a wrapped-around vector. We address this by allocating an extra page of virtual memory at the end of the circular buffer that is mapped to the same physical page as the start~\cite{Darkroom}.

Two-dimensional nodes produce and consume entire tiles of image data, which places substantially more strain on the memory hierarchy. A thread-local caching allocator is used to keep scratch memory resident, and any temporarily unused memory (for example the output buffer when computing the first stage of a \emph{Cascade}) is loaned to the allocator to be used as scratch memory when computing child nodes, in a form of scoped "anti-allocation". Two-dimensional nodes support either assigning to the output, or adding to it in-place, which avoids requiring scratch memory for \emph{Sum} nodes.

We compile this C++ with clang 17, targeting the AVX-512 instruction set. Parallelism is handled at the top level, with each thread responsible for computing a single tile of the output.

For our imaging filters, one could also imagine a lowering procedure that generates GPU kernels. Our nodes admit GPU-friendly implementations (e.g. using \cite{recfilter} for IIRs). However, our filters already run at gigapixels per second on CPU, which is competitive with a copy from host memory to a discrete GPU and back, so the image data would have to be already resident on GPU for this to make sense.

\section{Results}
\label{sec:results}
We evaluated our system on 1D and 2D target filters of various sizes. Our 1D filters are Gaussians, Lanczos filters, Head Related Impulse Responses (HRIR), and a telephone effect filter created in Apple Inc's GarageBand \cite{apple_garageband}. Our 2D filters are Gaussians, Gabors, and Lowpass filters. For each target, we cap the number of samples we take and also set a time limit on sampling. The number of samples taken varies given the size and dimensionality of the target. 
Table \ref{tab:samples-per-target} shows the average samples taken per target type.
\begin{table}
    \centering
    \begin{tabular}{c|c}
    \hline
     target class &  avg samples across sizes \\
     \hline
    Gaussian 1D & 4871\\
    Lanczos 1D & 4869\\
    Telephone 1D & 9539\\
    HRIR 1D	& 8147\\
    Gaussian 2D	& 10226\\
    Gabor 2D & 8855\\
    Lowpass 2D & 10740 \\
    \hline
    \end{tabular}
\caption{Design samples taken per target filter}\label{tab:samples-per-target}
\end{table}
\begin{figure}
\centering
\vspace{-1.25em}
\includegraphics[width=\columnwidth]{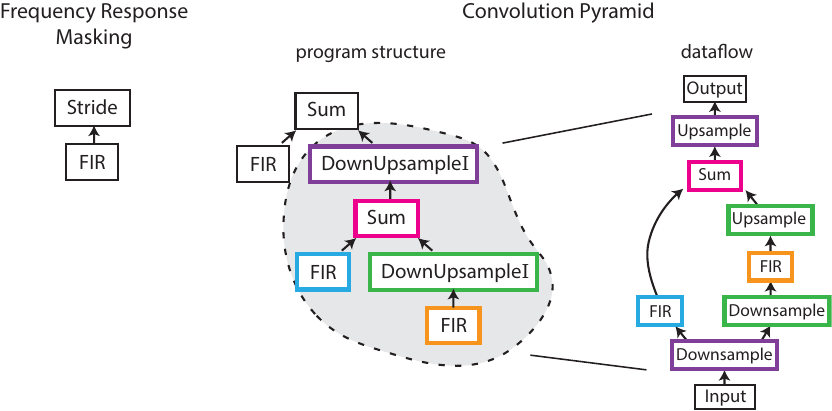}
\captionsetup{skip=1.5pt}
\caption{Our search space subsumes FRM and CP. FRM designs are a single \smallcode{Stride} node with an \smallcode{FIR} child. CP structures are  nested \smallcode{DownUpsampleI}s where each has a lower-resolution subtree that sums the output of an \smallcode{FIR} with the output of another inner lower-resolution subtree. Here we show the binary tree of a CP design with two levels and the dataflow through the highlighted subtree when the design is evaluated.
}
\vspace{-1.25em}
\label{fig:frm-cp-dags}
\end{figure}
For all targets we compare to CP and FRM. We extended these approaches by 1) searching over their design spaces and 2) lowering their designs to fast code with our compiler. Both CP and FRM exist in our search space. FRM designs are a \smallcode{Stride} with an \smallcode{FIR} child. The search space is all legal \smallcode{stride}, \smallcode{prefilter_order}, and \smallcode{interpolator_order} choices. It is small, so we enumerate all options. As discussed in \ref{sec:related-work-multi-rate-filtering}, CP designs are nested \smallcode{DownUpsampleI} nodes that halve the resolution at each level. Each level also convolves the pre-downsampled output with a small \smallcode{FIR}, and adds it to the upsampled output (Figure \ref{fig:frm-cp-dags}). The search space is the cartesian product of the number of levels $L \in [1, \lfloor \log_2(W)\rfloor]$ ($W$ is the width of the target filter), the \smallcode{prefilter_width} and \smallcode{interpolator_width} used in each \smallcode{DownUpsampleI}, and the \smallcode{width} of each \smallcode{FIR}. This space is too large to exhaustively enumerate so we sample the same number of structures as we do for our method. We also perform the exploitation step, with the same number of exploit samples as used for our results. 

Gaussians and Gabors have additional baselines, triple box blur for Gaussians and YVV for both. We do not compare to YVV for 1D Gaussians because anti-causal IIRs cannot be used in streaming scenarios. Our Lowpass filters also have a rank 2 singular vector decomposition (SVD) baseline. To measure quality, we use the PSNR of each design's output on white noise (4 million samples for 1D and 4 megapixels for 2D), versus the target's output. We use white noise because it has equal intensity across all frequencies in expectation.

To measure throughput, we benchmark our fast implementations. We benchmark on an 8-core AMD 9800X3D processor running Ubuntu 24.04. Last-level cache sizes are now large and are growing faster than megapixel counts, so we expect image data to be resident in L3. We therefore size inputs to fit our CPU's cache. For imaging filters we use 4 megapixel single-channel inputs, and for audio filters we use 16k samples. We do not address quantization in this work, so all inputs and outputs are 32-bit floating point. For each target we show the Pareto frontier of designs.

We also show the speed of exact methods: FFT-based convolution, direct convolution in 1D, and linearly separable filtering for 2D Gaussians and Gabors. For 1D FFT convolution, we used a single block (16k samples) which had the fastest throughputs. \textbf{This is an overestimate of realistic throughputs. Practical implementations would use much smaller block sizes since the latency is 2x the block size. Using a single block also does not take into account the cost of merging outputs across FFT blocks}. Naive FFT convolution is not used for real-time applications because it requires a block size $\ge$ 2205 to capture 20Hz, the lowest audible frequency. This corresponds to a latency of 100ms. Real-time applications like live performances require latencies under 5-10ms \cite{latency-for-musicians}, which corresponds to block sizes of 110-220 samples. Our only source of additional latency comes from computing outputs in groups of four SIMD vectors, adding at most 63 samples of latency. For 2D FFT convolution we used the same optimized tile size as for the approximate designs. For all other baselines besides FFT, we use our lowering system to generate fast implementations.

To show qualitative results, for each target we visualize the frequency responses of two Pareto-dominant designs from our method versus baseline designs that are on the frontier within their smaller search spaces. \textbf{All reported speedups are \emph{relative to the fastest exact method}} which is FFT convolution or linearly separable convolution when applicable. 

We will show and discuss some Pareto frontier structures for each target. These descriptions will use the abbreviations: \smallcode{porder}, \smallcode{pwidth}, \smallcode{iorder}, \smallcode{iwidth}, and \smallcode{dir} for prefilter order, prefilter width, interpolator order, interpolator width, and direction respectively.

\subsection{Main Takeaways}\label{sec:main-takeaways}
For 2D filters, our system finds designs that span a wide range of PSNRs with significantly higher throughputs than the baselines. The impulse and frequency response plots we show for Pareto-dominant designs from our search and the baselines illustrate the qualitative difference between designs with different PSNRs. Our 1D designs also significantly Pareto-dominate the baseline methods in 1D. 

HRIR and telephone filters are challenging to approximate because they have wide-band responses. For these targets, FRM and CP struggle to achieve better throughputs than exact filtering at good PSNRs. For these filters, we provide in the supplemental material outputs from our designs and those using CP or FRM at similar (where possible) throughputs on sample audio. For both filters, CP designs cannot produce approximations with competitive throughputs to our search. There are audible artifacts in the approximations produced by FRM and CP for both targets.

For static 8-bit depth images, errors less than 0.5 will not show up in the output image. This corresponds to a PSNR of 54dB. For our 2D targets, CP and FRM have designs with PSNRs that are above or near this threshold. However, they span a narrow and low throughput range. CP designs in 1D also have very low throughputs. FRM designs can achieve higher throughputs but at very low PSNRs. Designs from our search are up to 4.9x faster than the best baseline design with similar PSNR in 1D and up to 8x faster in 2D. 

There are also situations where users may require higher PSNRs. Not all applications use 8 bit images. Raw images from digital cameras typically have 12-14 bit depths. A filter is usually just one stage in a larger processing pipeline. Minimizing error is beneficial because errors compound across stages. Data-analysis may require higher accuracies than what is necessary for aesthetic qualities. For multi-resolution filters, higher PSNRs are correlated with less variance across phases. Finally, artifacts from lower-PSNR approximations can become noticeable when applied to moving objects in videos.

\subsection{Gaussian 1D}
\begin{figure*}
\centering
\includegraphics[width=\textwidth]{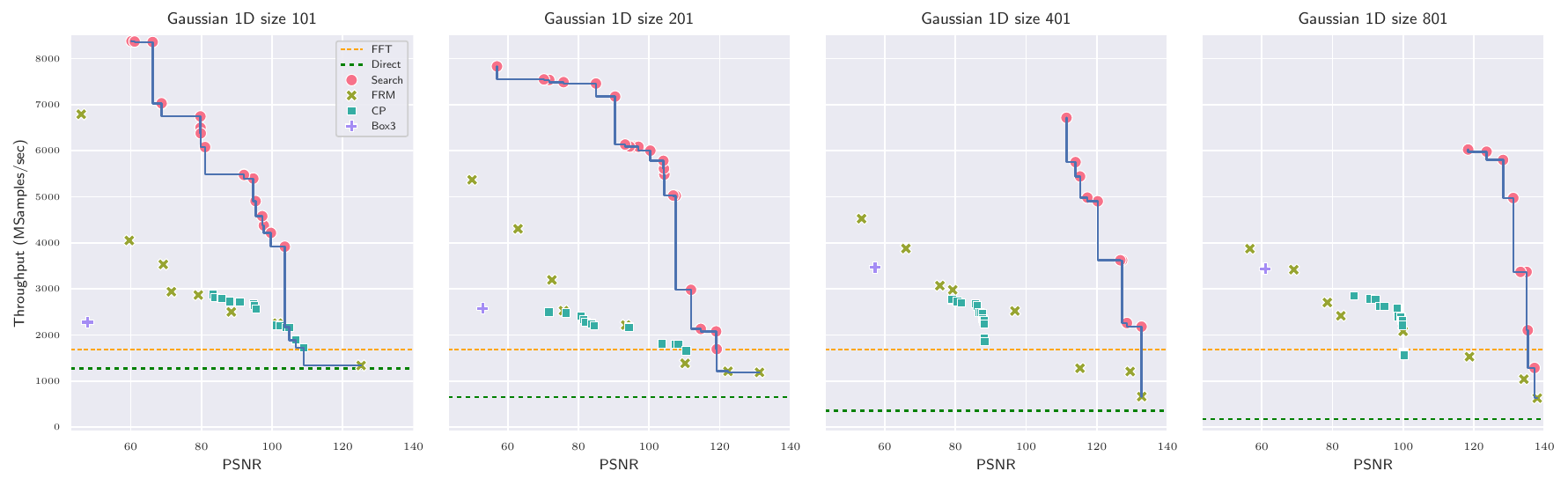}
\captionsetup{skip=1.5pt}
\caption{Throughput vs. quality for approximations of 1D Gaussian sizes 101, 201, 401, and 801. We show the Pareto frontier across all methods (designs on the blue line) as well as the within-class frontiers from the CP and FRM design spaces.
Designs from our search Pareto-dominate the baselines: FRM, CP, and triple box blur, by multiple factors in throughput and quality for all throughputs that are significantly faster than exact filtering via FFT convolution. Our designs are up to 4.25x faster with comparable or higher quality than the closest baseline and up to 74.5dB higher quality at comparable or higher throughputs.
}
\vspace{-0.3em}
\label{fig:gaussian1D-pareto}
\end{figure*}
1D Gaussians are used for tasks like denoising, smoothing and time windowing. Differences of Gaussians serve as linear phase band-pass filters. We targeted sizes 101, 201, 401, and 801 with standard deviation equal to 1/10th of the support width. Figure \ref{fig:gaussian1D-pareto} shows the Pareto plots for all sizes. The exact methods are FFT convolution and direct convolution. We also compare to triple box blur. Figures \ref{fig:gaussian1d-101-freq-plot}-\ref{fig:gaussian1d-801-freq-plot} show for each size, the impulse and frequency responses of two Pareto frontier designs from our system versus the responses of FRM and CP designs and the triple box.

For all sizes, designs from our search significantly Pareto dominate the baselines, up to very high PSNRs $> 100$dB. While they are technically on the Pareto frontier, these baselines have throughputs that are close to or worse than exact methods. Our system avoids sampling low-throughput designs. The Pareto plots show that CP designs span a limited range of low throughputs clustered in the higher PSNR regime with rapidly diminishing returns to lowering the PSNR. FRM designs span a wider range of throughputs but the PSNRs are significantly lower than our designs with comparable or higher throughputs. For size 401 our search found a design 4.25x faster than the closest baseline, both have 115.3dB PSNR. For size 201, our search found a design with 54.4dB higher PSNR (104.3 versus 49.9dB) at comparable throughput (ours is 1.02x faster). Below is a Pareto frontier design for size 801 with PSNR: 123.6dB, throughput 5981 MSamples/sec and 3.54x speed-up over FFT convolution:\\
\noindent\smallcode{DownUpSampleII<stride=8, porder=3, iorder=3, dir=x>}\\
\hspace*{1em}\smallcode{Cascade<dir=x>}\\
\hspace*{2em}\smallcode{DownUpSampleII<stride=2, po=4, io=4, dir=x>}\\
\hspace*{3em}\smallcode{FIR<width=43, dir=x>}\\
\hspace*{2em}\smallcode{FIR<width=5, direction=x>}\\
It applies a cheap FIR at the lowest resolution, upsamples by 2x, improves the upsampling interpolator by cascading with a small FIR, before  upsampling to the full resolution. 
For large Gaussians, Pareto-dominant designs do most of the computation at a low resolution. A Pareto frontier design for size 101 with PSNR 97.2dB, throughput 4580 MSamples/sec, and a 2.7x speedup has the structure: \\
\begin{minipage}{\columnwidth}
\vspace{0.35em}
\noindent\smallcode{DownUpsampleII<stride=2, porder=3, iorder=4, dir=x>}\\
\hspace*{1em}\smallcode{DownUpsampleI<stride=2, pwidth=5, iwidth=5, dir=x>}\\
\hspace*{2em}\smallcode{Sum<dir=x>}\\
\hspace*{3em}\smallcode{FIR<width=7, dir=x>}\\
\hspace*{3em}\smallcode{Stride<iorder=4, stride=2>}\\
\hspace*{4em}\smallcode{FIR<width=9, dir=x>}\\
\vspace{-0.5em}
\end{minipage}
The design downsamples by 4 in two stages, then sums an FIR with an FRM structure before upsampling back to the full resolution. For smaller sizes, high PSNR designs cannot downsample as aggressively, but they can still save computation with \smallcode{Stride} nodes.

\subsection{Lanczos 1D}
\begin{figure*}
\centering
\includegraphics[width=\textwidth]{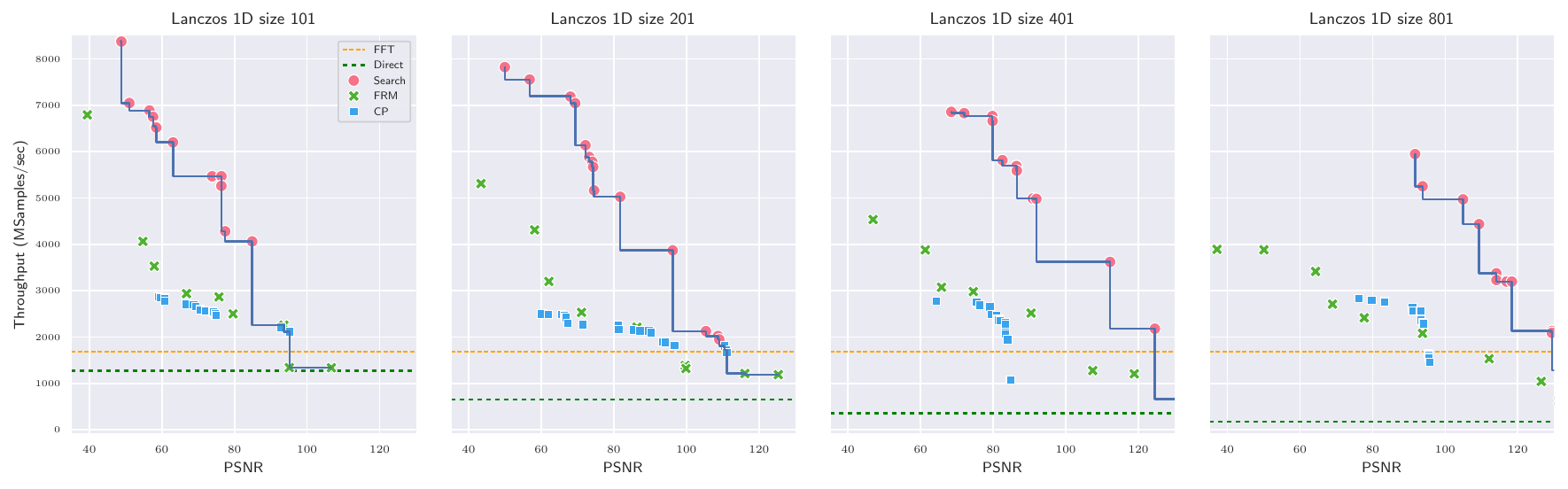}
\captionsetup{skip=1.5pt}
\caption{Throughput vs. quality for approximations of 1D Lanczos filter sizes 101, 201, 401, and 801. We show the Pareto frontier across all methods (designs on the blue line) as well as the within-class frontiers from the CP and FRM design spaces.
Designs from our search Pareto-dominate the baseline FRM and CP designs by multiple factors in throughput and quality across all throughputs that are significantly faster than exact filtering via FFT convolution. Our designs are up to 3.2x faster with comparable or higher quality than the closest baseline and up to 72.2dB higher quality at comparable or higher throughputs.
}
\vspace{-0.5em}
\label{fig:lanczos1d-pareto}
\end{figure*}
Normalized sinc filters, which have infinite support, are the theoretically optimal reconstruction filters because their frequency domain representation is a box filter. In practice resampling methods use finite approximations of it like the Lanczos filter, defined as: 
\[\frac{\sin(x\pi)}{x\pi} * \frac{\sin(ax\pi)}{ax\pi}\quad x\in [-\pi, \pi]\]
We target this filter for $a=3$ at sizes 101, 201, 401, and 801. The exact baselines are FFT and direct convolution. For all sizes, designs from our search significantly Pareto-dominate the baselines across all throughputs that are significantly faster than exact filtering and across a wide range of PSNRs. CP designs span a narrow range of low throughputs. FRM designs reach higher throughputs than CP but at low PSNRs. Figures \ref{fig:lanczos1d-101-freq-plot}-\ref{fig:lanczos1d-801-freq-plot} show the impulse and frequency responses for two Pareto-dominant designs from our search versus designs from the baselines. For size 801, our search found a design 3.2x faster than the closest baseline design with 9.4dB higher PSNR (104.9 versus 95.5dB) and another with 72.2dB higher PSNR (109.3 vs 37.2dB) at 1.14x higher throughput. This Pareto frontier design for size 801 has PSNR 114dB, throughput 3378 MSamples/sec and a 2x speedup over FFT convolution: 
\vspace{0.3em}\\
\begin{minipage}{\columnwidth}
\smallcode{DownUpsampleII<stride=2, porder=3, iorder=3, dir=x>}\\
\hspace*{1em}\smallcode{DownUpsampleII<stride=2, porder=4, iorder=3, dir=x>}\\
\hspace*{2em}\smallcode{Stride<stride=2, iorder=3, dir=x>}\\
\hspace*{3em}\smallcode{FIR<width=97, dir=x>}\vspace{-0.7em}\\
\end{minipage}
Lanczos filters have side lobes, so high PSNR designs downsample less aggressively than for Gaussians, in this case by 4x instead of 16x. Lower PSNR models downsample more aggressively. For example, this Pareto frontier design for size 101 with PSNR 58.5dB, throughput 6521 MSamples/sec, and a 3.86x speedup has the structure:
\vspace{0.3em}\\
\begin{minipage}{\columnwidth}
\smallcode{DownUpsampleI<stride=4, pwidth=7, iwidth=9, dir=x>}\\
\hspace*{1em}\smallcode{Sum<dir=x>}\\
\hspace*{2em}\smallcode{DownUpsampleI<stride=2, pwidth=5, iwidth=5, dir=x>}\\
\hspace*{3em}\smallcode{FIR<width=8, dir=x>}\\
\hspace*{2em}\smallcode{Stride<stride=4, iorder=2, dir=x>}\\
\hspace*{3em}\smallcode{FIR<width=5, dir=x>}\vspace{-0.7em}\\
\end{minipage}
It downsamples by 8x, which is aggressive for the target size, but it compensates for the low resolution computation by adding a slightly higher resolution strided computation. 

\subsection{HRIR 1D}
\begin{figure}
\centering
\includegraphics[width=0.52\columnwidth]{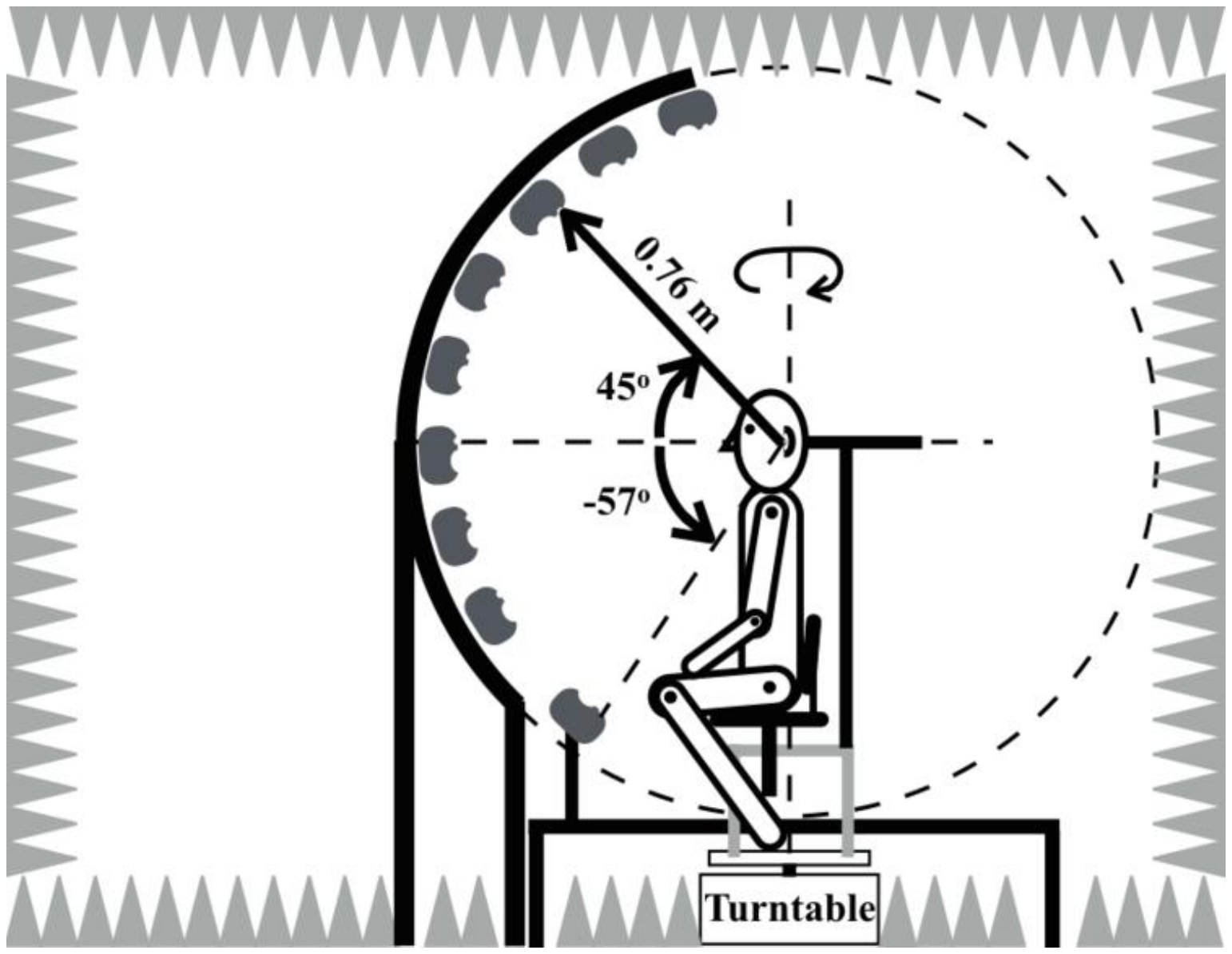}
\captionsetup{skip=1.5pt}
\caption{Measurement setup for the HRIR dataset. The subject is seated in an anechoic chamber in front of a vertical arc of 9 loudspeaker emitters. Binaural microphones in the subject's ears measure HRIRs for each emitter. The subject is rotated in 5 degree increments for a total of 648 measurements.
}
\vspace{-1.5em}
\label{fig:hrir-dataset-measurment}
\end{figure}\label{fig:hrir-measurement}
\begin{figure*}
\centering
\includegraphics[width=0.86\textwidth]{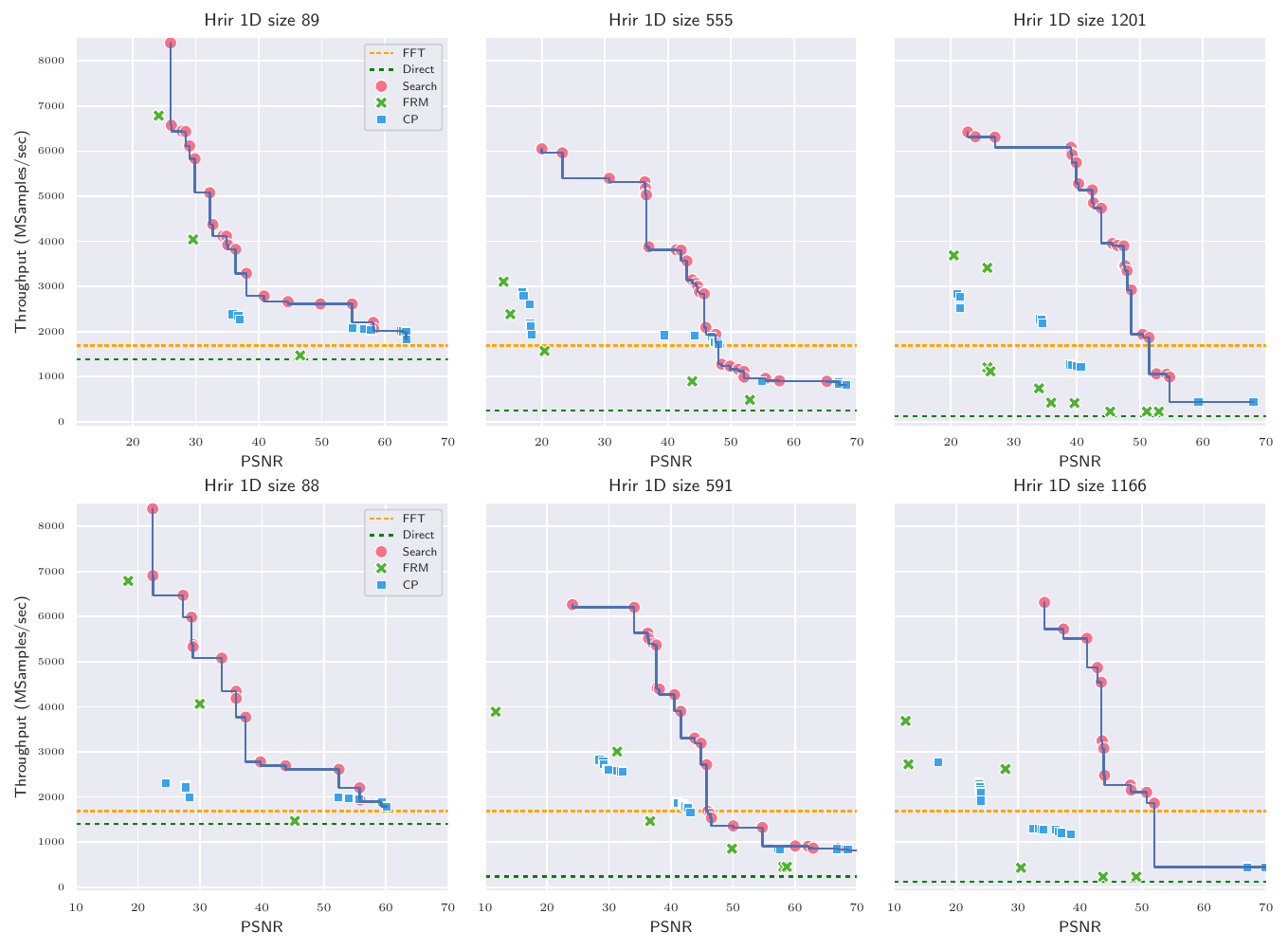}
\captionsetup{skip=1.5pt}
\caption{Throughput vs. quality for approximations of each HRIR section for the right (top row) and left (bottom row) ear. We show the Pareto frontier across all methods (designs on the blue line) as well as the within-class frontiers from the CP and FRM design spaces. Designs from our search Pareto-dominate the baseline FRM and CP designs for all throughputs that are significantly faster than exact filtering via FFT convolution. Our designs are up to 4.9x faster with comparable or higher quality than the closest baseline and up to 31.5dB higher quality at comparable or higher throughputs.
}
\label{fig:hrir1D-pareto}
\end{figure*}
\label{fig:hrir-pareto}
Head related impulse responses (HRIR) capture how an individual's body and ear shape filters directional audio. HRIRs are measured for each ear for different impulse emitters at various points in space around the individual's head. Applying a pair of HRIRs to two channels of non-spatial audio will produce the effect of the sound coming from the relative location of HRIR's emitter. 

We target a pair of HRIRs from the 3D3A Lab Head-Related Transfer Function Database \cite{hrir_dataset}. We use measurement 1 from subject 1 with source location  $5^\circ$ to the right, $57^\circ$ below, and 0.76 meters from the center of the subject's head (Figure \ref{fig:hrir-dataset-measurment}). Most of the energy is in the head of the HRIR, which has a long tail of decaying values. To prevent the optimization from ignoring smaller values, we separate the HRIRs into three sections with values within the same order of magnitude. Figure \ref{fig:hrir1D-pareto} shows the Pareto-plots for each section. The first row is the right ear, the second is the left. Figures \ref{fig:hrir1d-89-freq-plot}-\ref{fig:hrir1d-1166-freq-plot} show the impulse and frequency responses of two of our Pareto frontier designs and CP and FRM designs for each section. Achieving high PSNR and throughput for HRIRs is challenging because they have non-smooth wide-band frequency responses. Hand-designed approximations for convolutional reverb-like filters use direct convolution for the head and coarsely approximate the tail with IIRs. Our search found  nuanced versions of this approach, for example, for the head of receiver 1, this structure:\\
\smallcode{DownUpsampleI<stride=2, pwidth=5, iwidth=5, dir=x>}\\
\hspace*{1em}\smallcode{FIR<width=46, dir=x>}\\
takes advantage of the lack of energy in the upper half of the spectrum (Figure \ref{fig:hrir1d-88-freq-plot}) by downsampling. It has PSNR 52.4dB, throughput 2611 MSamples/sec, and a 1.55x speedup over FFT convolution. 

For the longest tail section of receiver 1, our search found a Pareto frontier design with PSNR 43.4dB, throughput 4542 MSamples/sec, and 2.69x speedup that predominantly depends on TIIRs:\vspace{0.3em}\\
\begin{minipage}{\columnwidth}
\smallcode{DownUpsampleI<stride=2, pwidth=5, iwidth=5 dir=x>}\\
\hspace*{1em}\smallcode{DownUpsampleI<stride=4, pwidth=9, iwidth=9 dir=x>>}\\
\hspace*{2em}\smallcode{Sum<dir=x>}\\
\hspace*{3em}\smallcode{Cascade<dir=x>}\\
\hspace*{4em}\smallcode{FIR<width=21, dir=x>}\\
\hspace*{4em}\smallcode{TIIR<causal=True, width=123, direction=x>}\\
\hspace*{3em}\smallcode{Cascade<dir=x>}\\
\hspace*{4em}\smallcode{FIR<width=5, dir=x>}\\
\hspace*{4em}\smallcode{TIIR<causal=True, width=139, direction=x>}
\end{minipage}\vspace{0.3em}
It uses two wide-support TIIRs, each cascaded with a small FIR and summed together. This is done at 1/8 resolution to save computation, sacrificing accuracy in the higher frequencies that have significantly lower magnitudes (Figure \ref{fig:hrir1d-1166-freq-plot}). Designs from our search Pareto dominate CP and FRM up to the low-throughput, high PSNR regime where speed-ups are not significantly better than FFT convolution. 

We created a full binaural HRIR approximation, with 2.37x total speedup over FFT convolution with a 16k block size, by combining Pareto frontier designs from each section. We also created CP-only and FRM-only approximations using Pareto frontier designs from each baseline. The fastest CP combined design is only 1.55x faster than FFT convolution. For FRM, we created a design with a 2.33x speedup. Note that a 16k block size is abnormal for latency and cache size reasons so we also compare to block size 256. Table \ref{tab:hrir-speedups} summarizes the speedups for each  method over the FFT convolutions and direct convolution. We ran all three designs on a percussion audio clip from Freesound \cite{freesound-drums} and include them in the supplemental material along with the ground truth output. Our design's output is indistinguishable from the ground truth. Both the CP and FRM outputs have noticeably less high frequency content and sound muffled, with the FRM errors being more severe.
\begin{table}[h]
    \centering
    \begin{tabular}{c|c|c|c}
    \hline
    design method & FFT 16k & FFT 256 & direct convolution \\
     \hline
    Ours & 2.37 & 8.66 & 17.68\\
    CP & 1.55 & 5.66 & 11.57\\
    FRM & 2.33 & 8.54 & 17.3\\
    \hline
    \end{tabular}
\caption{Speedups of design methods for HRIR filters over exact filtering}\label{tab:hrir-speedups}
\end{table}

\subsection{Telephone effect filter 1D}
\begin{figure*}
\centering  \includegraphics[width=\textwidth]{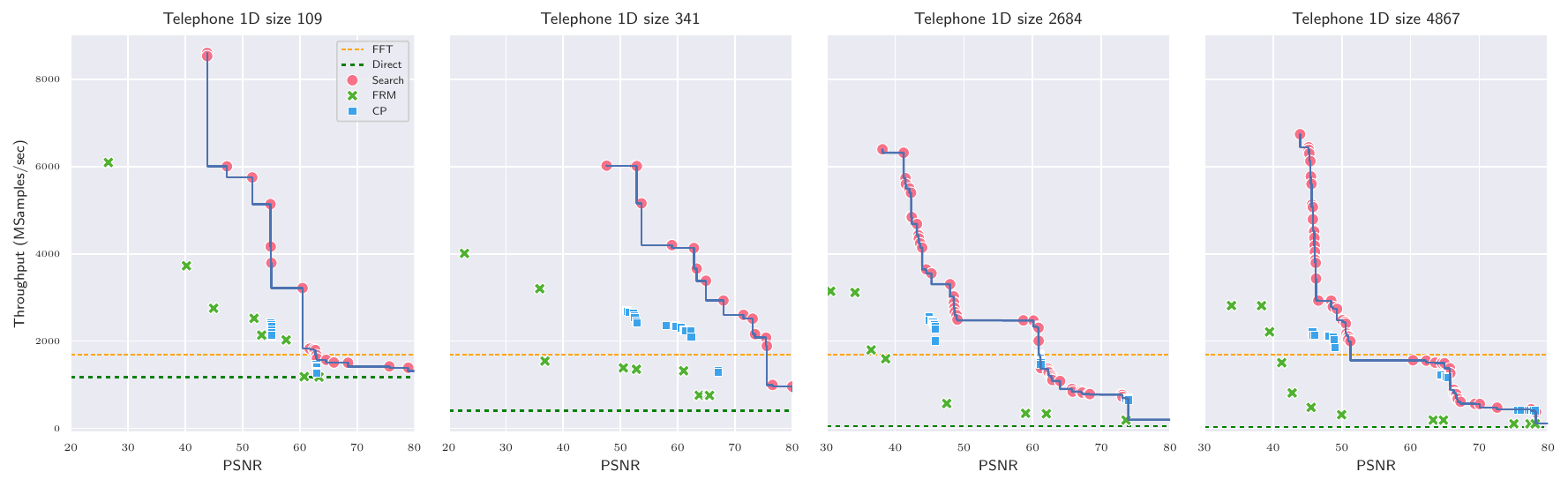}
\captionsetup{skip=1.5pt}
\caption{Throughput vs. quality for approximations of each 1D Telephone filter section. We show the Pareto frontier across all methods (designs on the blue line) as well as the within-class frontiers from the CP and FRM design spaces.
Designs from our search Pareto-dominate designs from the baseline methods: FRM and CP, across all throughputs that are faster than exact filtering via FFT convolution. Our designs are up to 3x faster at comparable or higher quality than the closest baseline design and up to 40.1dB higher quality with comparable or higher throughputs.}
\label{fig:telephone1d-pareto}
\end{figure*}
Telephone filters simulate the audio quality of old speakers and are used in sound mixing for stylistic effect. They suppress low frequencies at around 300 Hz and high frequencies at around 3500 Hz. We created a telephone filter with slight reverb. Like the HRIRs, these are long filters, in our case 8k samples long, with a high energy head and a long decaying tail. We also separated this filter into 4 sections with values roughly within the same order of magnitude. Figure \ref{fig:telephone1d-pareto} shows the Pareto plots for each section. Figures \ref{fig:telephone1d-109-freq-plot}-\ref{fig:telephone1d-4867-freq-plot} show the impulse and frequency responses of two of our Pareto frontier designs versus a CP and FRM design for each section. Designs from our search Pareto-dominate the baselines for all sections.

The Pareto frontier designs for the first section share similar structures to those for Gaussians and Lanczos filters because this section is smooth. TIIRs are useful when their support widths are large enough to overcome the overhead of making their recursive structure vectorizable. Designs for smooth filters with small sizes like 109, can use more aggressive \smallcode{DownUpsample}s. The lower-resolution support widths are small enough to not use TIIRs. The other sections are much larger and make use of TIIRs. Section 2 has a Pareto frontier design with PSNR 62.8dB, throughput 4132 MSamples/sec, and speedup 2.45x over FFT convolution:
\vspace{0.3em}\\
\begin{minipage}{\columnwidth}
\noindent\smallcode{DownUpsampleII<stride=2, porder=3, iorder=3, dir=x>}\\
\hspace*{1em}\smallcode{DownUpsampleII<stride=2, porder=3, iorder=3, dir=x>}\\
\hspace*{2em}\smallcode{Sum<dir=x>}\\
\hspace*{3em}\smallcode{DownUpsampleII<stride=2, porder=4, iorder=3, dir=x>}\\
\hspace*{4em}\smallcode{FIR<width=39, dir=x>}\\
\hspace*{3em}\smallcode{TIIR<causal=True, width=84, dir=x>}\vspace{-0.5em}\\
\end{minipage}
It adds a lower-resolution FIR to a wide-support TIIR. This Pareto frontier structure for the last and longest tail section uses 3 TIIRs:\vspace{0.4em}\\
\begin{minipage}{\columnwidth}
\noindent\smallcode{DownUpsampleI<stride=8, pwidth=15, iwidth=21, dir=x>}\\
\hspace*{1em}\smallcode{Cascade<dir=x>}\\
\hspace*{2em}\smallcode{FIR<width=153, dir=x>}\\
\hspace*{2em}\smallcode{TIIR<causal=True, width=152, dir=x>}\\
\hspace*{2em}\smallcode{TIIR<causal=True, width=152, dir=x>}\\
\hspace*{2em}\smallcode{TIIR<causal=True, width=153, dir=x>}\vspace{-0.5em}\\
\end{minipage}
It has PSNR 48.4dB, throughput 2928 MSamples/sec, and speedup 1.73x. The FIR contributes wide-band frequency content while each TIIR cheaply extends the support width of the filter by adding narrow-band middle to high frequency content. 

We created two full approximations using our Pareto frontier designs, with 1.96x and 2.74x speedup over FFT convolution (block size 16k). The fastest combined approximation we could generate using CP designs has a 1.45x speedup. The fastest FRM design has a 2.05x speedup. We also show speedups over FFT convolution with block size 256 and direct convolution (Table \ref{tab:telephone-speedups}). 
\begin{table}[h]
    \centering
    \begin{tabular}{c|c|c|c}
    \hline
    design method & FFT 16k & FFT 256 & direct convolution \\
     \hline
    Ours 1 & 1.96 & 21.94 & 46.97\\
    Ours 2 & 2.74 & 30.6 & 65.52\\
    CP & 1.45 & 16.24 & 34.77\\
    FRM & 2.05 & 22.97 & 49.18\\
    \hline
    \end{tabular}
\caption{Speedups of designs for the telephone filter over exact filtering}\label{tab:telephone-speedups}
\vspace{-2em}
\end{table}
\noindent We include in the supplemental materials the outputs from both our designs and the CP and FRM designs on recorded human speech and the ground truth. The CP output is audibly missing high frequencies and has a slight ringing in the mid frequencies. The FRM output has a very noticeable undesirable resonance in the mid frequencies and fails to capture the reverb. Our 2.74x speedup design, which is faster than both baselines, does not have artifacts like ringing but it has slightly less high frequencies than the ground truth and does not capture the reverb entirely. Our 1.96x speedup design, which is significantly faster than the CP design and has approximately the same speed as the FRM design, sounds indistinguishable from the ground truth.

\subsection{Gaussian 2D}
\begin{figure*}
\centering
\includegraphics[width=0.86\textwidth]{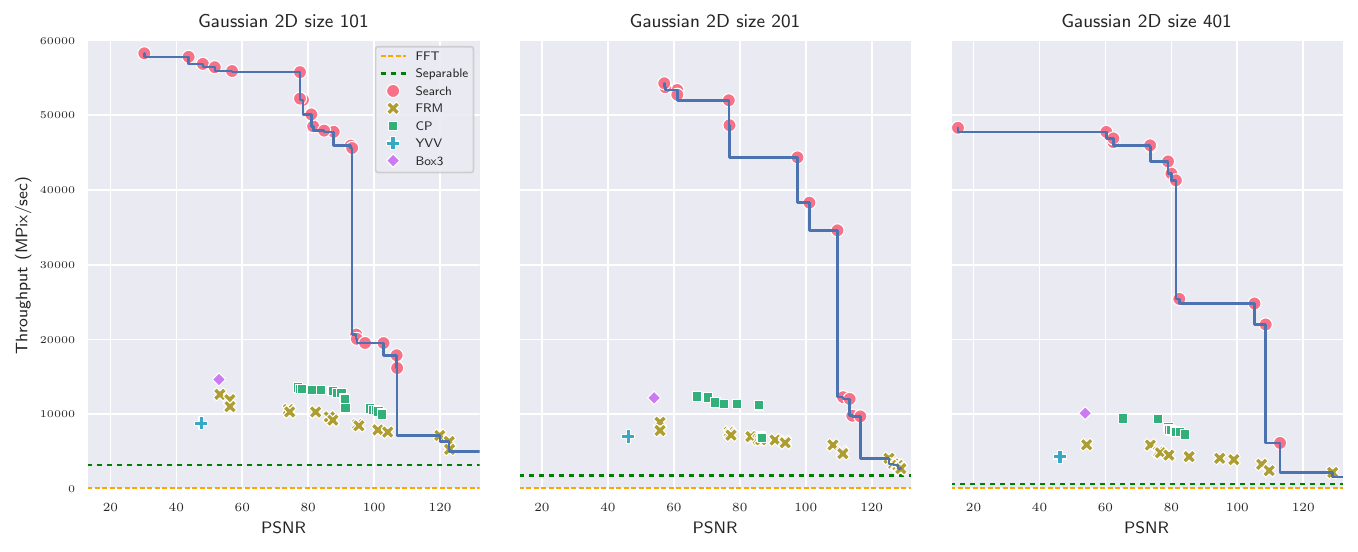}
\captionsetup{skip=1.5pt}
\caption{Throughput vs. quality for 2D Gaussian approximations for sizes 101, 201, and 401. We show the Pareto frontier across all methods (designs on the blue line) as well as the within-class frontiers from the CP and FRM design spaces. Designs from our search Pareto-dominate the baselines: FRM, CP, triple box blur, and YVV by multiple factors in throughput and quality for all throughputs that are significantly faster than exact filtering. Our designs are up to 7.2x faster with comparable or higher quality than the closest baseline and up to 44.3dB higher quality at comparable or higher throughputs. Our search found designs that are 3.9 to 4.7x faster than triple box blur, which is a commonly used Gaussian approximation in production software, with comparable or higher quality.
}
\vspace{-0.6em}
\label{fig:gaussian2D-pareto}
\end{figure*}
2D Gaussians are used for tasks ranging from denoising, general blurring, to depth of field effects. Figure \ref{fig:gaussian2D-pareto} shows the Pareto plots for sizes 101, 201, and 401, with standard deviation equal to 1/10th the support width. We compare to the exact methods FFT and linearly separable convolution. For all sizes, linearly separable convolution is faster. We also compare to FRM, CP, triple box blur, and YVV. Triple box performs better than YVV. We show in figures \ref{fig:gaussian2d-101-freq-plot}-\ref{fig:gaussian2d-401-freq-plot} the impulse and frequency responses of two of our Pareto frontier designs compared to an FRM and CP design and the triple box blur for each size. The baselines are significantly Pareto-dominated by our designs up till the low-throughput regime. 

CP and FRM do not provide much of a trade-off space. For example, at size 101, a CP design with PSNR 80.9dB has throughput 12760 MPix/sec. Moving to a lower quality CP design with PSNR 62.8dB, only improves the throughput by 3\%. In 1D FRM designs reach higher throughputs than CP (at much lower quality), but in 2D FRM is strictly worse than CP because the throughput scales by $\frac{1}{\text{stride}}$ per dimension instead of $\frac{1}{\text{stride}^{2}}$ when using downsampling. For size 201, our search found a design at 97.5dB PSNR that is 7.2x faster than the closest baseline design which has 3.7dB lower quality. Our search also found a design with 44.3dB higher quality than the closest baseline at about the same throughput (12280 vs 12374 MPix/sec). 

In 2D, vertical TIIRs are easily vectorized horizontally (orthogonal to the recurrence), so they can reap efficiency gains at smaller support widths. The fastest size 201 design has the following structure:\vspace{0.3em}\\
\begin{minipage}{\columnwidth}
\noindent\smallcode{DownUpsampleI<stride=8, pwidth=15, iwidth=17, dir=y>}\\
\hspace*{1em}\smallcode{DownUpsampleI<stride=8, pwidth=15, iwidth=17, dir=x>}\\
\hspace*{2em}\smallcode{Pipe<dir=both>}\\
\hspace*{3em}\smallcode{DownUpsampleII<stride=2, porder=3, iorder=3, dir=y>}\\
\hspace*{4em}\smallcode{Pipe<dir=both>}\\
\hspace*{5em}\smallcode{FIR<width=4, dir=y>}\\
\hspace*{5em}\smallcode{FIR<width=23, dir=x>}\\
\hspace*{3em}\smallcode{TIIR<causal=True, width=11, dir=y>}\vspace{-0.5em}\\
\end{minipage}
It has PSNR 57.1dB, throughput 54271 MPix/sec, and a 30.5x speedup over linearly separable convolution. Half of the support width of the vertical gaussian is approximated by a causal TIIR with a hump-like impulse response that skews left. The other half is approximated by a small upsampled FIR that has a hump-like right-skewed impulse response. When convolved together they produce a centered approximation of a 1D Gaussian. Because TIIRs are more expensive horizontally, the entire horizontal Gaussian is approximated by a larger FIR at the lowest resolution.

This Pareto-frontier structure for size 401 also uses TIIRs to approximate the vertical Gaussian:\vspace{0.5em}\\
\begin{minipage}{\columnwidth}
\noindent\smallcode{DownUpsampleI<stride=8, pwidth=15, iwidth=17, dir=y>}\\
\hspace*{1em}\smallcode{DownUpsampleI<stride=8, pwidth=15, iwidth=17, dir=x>}\\
\hspace*{2em}\smallcode{Pipe<dir=both>}\\
\hspace*{3em}\smallcode{DownUpsampleII<stride=2, porder=3, iorder=3, dir=x>}\\
\hspace*{4em}\smallcode{Pipe<dir=both>}\\
\hspace*{5em}\smallcode{DownUpsampleII<stride=2,porder=4,iorder=3, dir=y>}\\
\hspace*{6em}\smallcode{Pipe<dir=both>}\\
\hspace*{7em}\smallcode{FIR<width=10, dir=x>}\\
\hspace*{7em}\smallcode{FIR<width=10, dir=y>}\\
\hspace*{5em}\smallcode{Cascade<dir=y>}\\
\hspace*{6em}\smallcode{TIIR<causal=True, width=12, dir=y>}\\
\hspace*{6em}\smallcode{TIIR<causal=True, width=13, dir=y>}\\
\hspace*{3em}\smallcode{Stride<stride=2, iorder=3, dir=x>}\\
\hspace*{4em}\smallcode{FIR<width=11, dir=x>}\vspace{-0.5em}\\
\end{minipage}
It has PSNR 62.4dB, throughput 46899 MPix/sec, and a 70.2x speedup. Similar to the size 201 design, the two causal TIIRs produce a slightly left-skewed hump, which is adjusted by cascading with the right-skewed hump from the lower-resolution FIR. In this case the horizontal gaussian is composed of a small FIR at 1/16 resolution and another strided FIR at 1/8 resolution.              


It is perhaps surprising that downsampling by 16x for a size 201 Gaussian can achieve good PSNRs. But looking at the Gaussian’s frequency response, we can see that frequencies larger than $\sim$1300 Hz are entirely suppressed. The theoretical band limit of downsampling by 16 is $\sim$2750 Hz. This is wide enough for the design to capture the response of the Gaussian. 
\subsection{Gabor 2D}
\begin{figure*}
\centering
\includegraphics[width=0.86\textwidth]{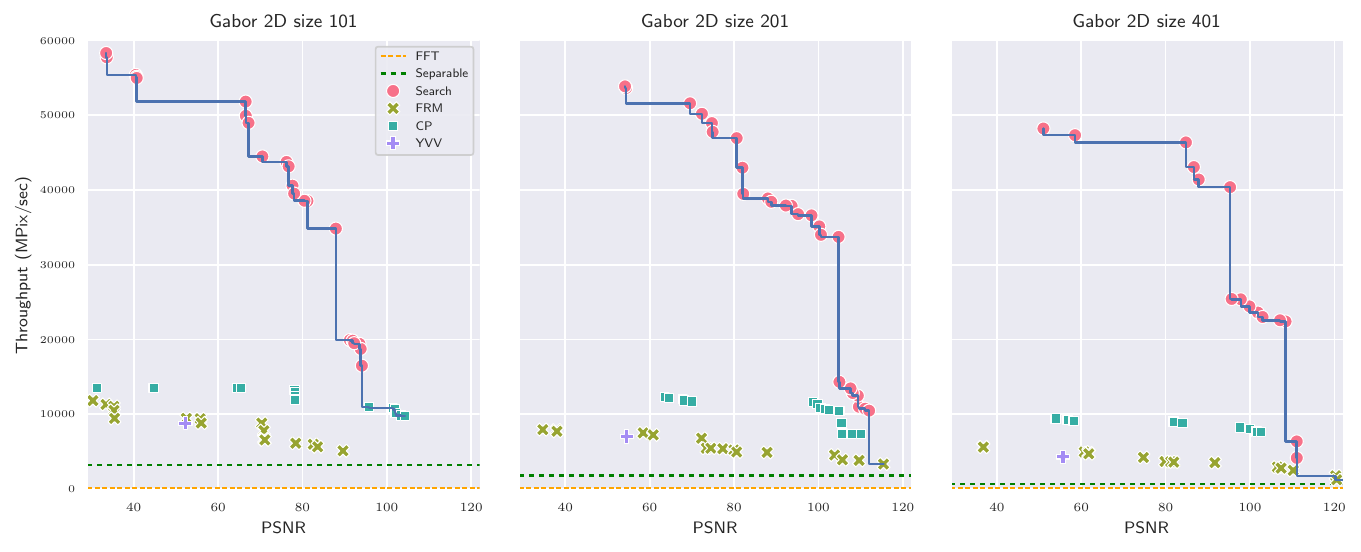}
\captionsetup{skip=1.5pt}
\caption{Throughput vs. quality for 2D Gabor filter approximations for sizes 101, 201, and 401. We show the Pareto frontier across all methods (designs on the blue line) as well as the within-class frontiers from the CP and FRM design spaces.
Designs from our search Pareto-dominate designs from the baselines: FRM, CP, and YVV by multiple factors across all throughputs that are significantly faster than exact filtering, and across a wide range of PSNRs. Our designs are up to 8x faster with comparable or higher quality than the closest baseline design and up to 62.8dB higher quality with comparable or higher throughputs. 
}
\vspace{-0.65em}
\label{fig:gabor2D-pareto}
\end{figure*}
Gabor filters are directionally varying band-pass filters used for tasks like edge detection and feature extraction. We target linearly separable Gabor filters of sizes 101,201, and 401 with harmonic function frequencies $0.075, 0.075/2, 0.075/4$ respectively and standard deviations in x and y equal to 1/10th of the support width. The orientation is 0, i.e. the harmonic is in the x direction. Figure \ref{fig:gabor2D-pareto} shows the Pareto plots for all sizes. We compare to exact methods FFT and linearly separable convolution. YVV for Gabors uses complex-valued IIRs, which use more operations than real IIRs. Since complex IIRs are not in our lowering space, we over-estimate the YVV throughput with the YVV throughput for 2D Gaussians. 

Figures \ref{fig:gabor1d-101-freq-plot}-\ref{fig:gabor1d-401-freq-plot} show the impulse and frequency responses of two of our Pareto frontier designs compared to the CP and FRM designs as well as YVV for all sizes. Our designs significantly Pareto-dominate all the baselines up till the low throughput regime across a wide range of qualities and throughputs. For size 401, our search found a design 8x faster than the closest baseline design, both with PSNR 107dB, and another with 54.5dB higher PSNR and 2.4x faster than the closest baseline design at PSNRs 108.4 vs 53.9dB. 

The following Pareto frontier structure for size 401 has PSNR 84.8dB, throughput 46363 MPix/sec, and a 69.4x speedup over linearly separable convolution:\vspace{0.25em}\\
\begin{minipage}{\columnwidth}
\noindent\smallcode{DownUpsampleI<stride=8, pwidth=15, iwidth=17, dir=y>}\\
\hspace*{1em}\smallcode{DownUpsampleII<stride=8, porder=3, iorder=4, dir=x>}\\
\hspace*{2em}\smallcode{Pipe<dir=both>}\\
\hspace*{3em}\smallcode{DownUpsampleII<stride=2, pwidth=4, iwidth=3, dir=y>}\\
\hspace*{4em}\smallcode{Pipe<dir=both>}\\
\hspace*{5em}\smallcode{FIR<width=10, dir=y>}\\
\hspace*{5em}\smallcode{FIR<width=45, dir=x>}\\
\hspace*{3em}\smallcode{Cascade<dir=y>}\\
\hspace*{4em}\smallcode{FIR<width=7, dir=y>}\\
\hspace*{4em}\smallcode{TIIR<causal=True, width=18, dir=y>}\vspace{-0.6em}\\
\end{minipage}

\noindent It downsamples less horizontally than vertically, because the high frequency oscillation is horizontal. Vertically, the target is a smooth Gaussian, and the design saves cost by using a larger \smallcode{TIIR} with two smaller \smallcode{FIR}s. This strategy of downsampling less horizontally and using more computation horizontally is consistent across Gabor filter sizes. For example, this size 101 Pareto frontier approximation:\vspace{0.4em}\\
\begin{minipage}{\columnwidth}
\noindent\smallcode{DownUpsampleII<stride=4, porder=3, iorder=3, dir=y>}\\
\hspace*{1em}\smallcode{DownUpsampleII<stride=4, porder=4, iorder=4, dir=x>}\\
\hspace*{2em}\smallcode{Pipe<dir=both>}\\
\hspace*{3em}\smallcode{DownUpsampleI<stride=2, pwidth=5, iwidth=5, dir=y>}\\
\hspace*{4em}\smallcode{Pipe<dir=both>}\\
\hspace*{5em}\smallcode{FIR<width=3, dir=y>}\\
\hspace*{5em}\smallcode{FIR<width=20, dir=x>}\\
\hspace*{3em}\smallcode{Stride<stride=2, iorder=2, dir=y>}\\
\hspace*{4em}\smallcode{FIR<width=5, dir=y>}\vspace{-0.5em}\\
\end{minipage}
\noindent downsamples by 8x vertically and 4x horizontally. It uses a relatively large FIR horizontally, but vertically, it only uses a small 3-tap FIR at the lowest resolution and another 5-tap strided FIR at 1/4 resolution. This design has PSNR 67.2dB, throughput 48983, and a 15.2x speedup over linearly separable convolution. 
                                
\subsection{Lowpass 2D}
\begin{figure*}
\centering
\includegraphics[width=0.86\textwidth]{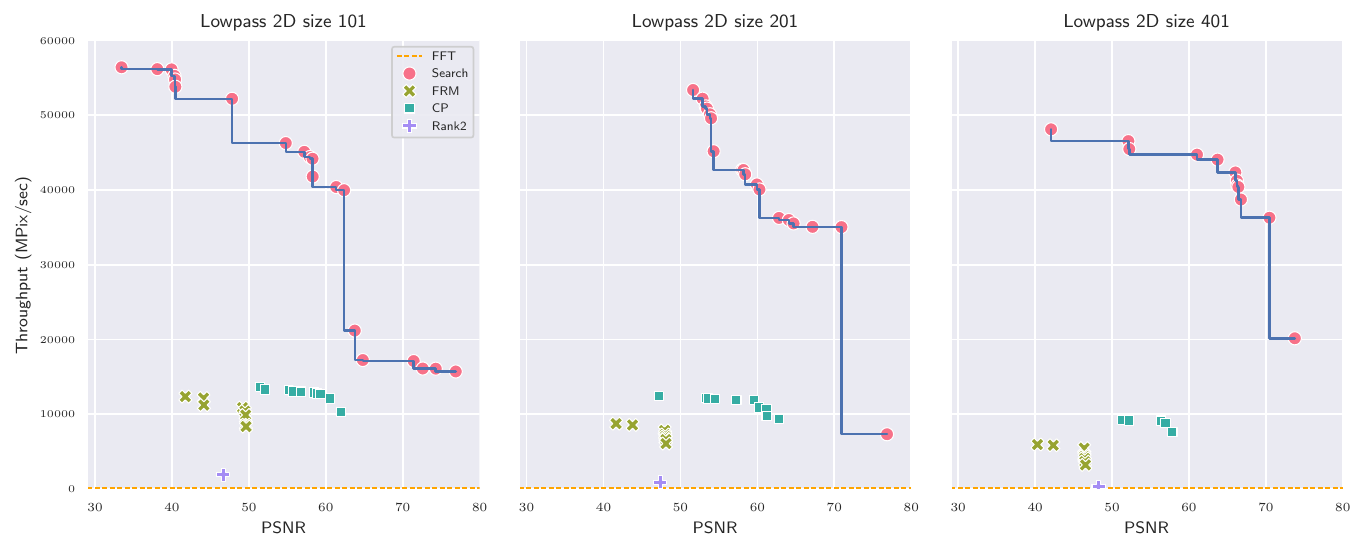}
\captionsetup{skip=1.5pt}
\caption{Throughput vs. quality for 2D Lowpass approximations for sizes 101, 201, and 401. We show the Pareto frontier across all methods (designs on the blue line) as well as the within-class frontiers from the CP and FRM design spaces. Designs from our search Pareto-dominate the baselines: CP, FRM, and a linearly separable rank 2 SVD approximation, across a wide range of quality and throughputs. Our designs are up to 5.9x faster with comparable or higher quality and up to 25.5dB higher quality at comparable or higher throughputs.
}
\vspace{-0.6em}
\label{fig:lowpass2D-pareto}
\end{figure*}
Our final 2D targets are low-pass filters of sizes 101, 201, and 401. They are the inverse FFT of a radius 4 disk in the frequency domain multiplied by a  Blackman window. The pass-band gets narrower as the filter size increases. Unlike Gaussians, these low-pass filters have a flatter pass-band and a steeper transition band. These filters are also not linearly separable so we only compare to FFT convolution as an exact method and add a rank 2 SVD approximation baseline.  

Figure \ref{fig:lowpass2D-pareto} shows the Pareto plots for all sizes. Figures \ref{fig:lowpass2d-101-freq-plot}
-\ref{fig:lowpass2d-401-freq-plot} show the impulse and frequency responses of two Pareto frontier designs from our search and a CP and FRM design. Our designs Pareto-dominate all the baselines across all sizes. For size 101, our search found a design 3.9x faster than the closest baseline design with comparable PSNRs (61.4 vs 61.9dB), and another with 25.5dB higher PSNR (76.9 versus 51.4dB) and 1.15x higher throughput. 

These targets have rank $>2$, so we searched over designs that summed two 2D structures. This Pareto frontier structure for size 101 has PSNR 54.8dB, throughput 46275 MPix/sec, and a 317x speedup over FFT convolution:\vspace{0.5em}\\
\begin{minipage}{\columnwidth}
\noindent\smallcode{DownUpsampleI<stride=4, pwidth=7, iwidth=9, dir=y>}\\
\hspace*{0.7em}\smallcode{DownUpsampleII<stride=4, porder=3, iorder=3, dir=x>}\\
\hspace*{1.4em}\smallcode{Sum<dir=both>}\\
\hspace*{2.1em}\smallcode{DownUpsampleII<stride=2, porder=3, iorder=3, dir=y>}\\
\hspace*{2.8em}\smallcode{Pipe<dir=both>}\\
\hspace*{3.5em}\smallcode{FIR<width=9, dir=y>}\\
\hspace*{3.5em}\smallcode{Stride<stride=2, iorder=3, dir=x>}\\
\hspace*{4.2em}\smallcode{FIR<width=10, dir=x>}\\
\hspace*{2.1em}\smallcode{DownUpsampleII<stride=2, porder=3, iorder=3, dir=x>}\\
\hspace*{2.8em}\smallcode{Pipe<dir=both>}\\
\hspace*{3.5em}\smallcode{DownUpsampleII<stride=2, porder=3, iorder=4, dir=y>}\\
\hspace*{4.2em}\smallcode{Pipe<dir=both>}\\
\hspace*{4.9em}\smallcode{FIR<width=7, dir=y>}\\
\hspace*{4.9em}\smallcode{FIR<width=9, dir=x>}\\
\hspace*{3.5em}\smallcode{FIR<width=5, dir=y>}
\end{minipage}
Inspecting the frequency responses of these subtrees shows that for a given direction, one of the terms fits the higher frequencies and captures the low-pass filter's side lobes. The other term fits the lower frequencies and its impulse response looks more like a smooth hump. Essentially this design learned that a good approximation separates the higher frequency oscillations in each direction across the two terms. 

This Pareto frontier design for size 401 learned a similar decomposition with a different structure:\vspace{0.3em}\\
\begin{minipage}{\columnwidth}
\noindent\smallcode{DownUpsampleI<stride=8, pwidth=15, iwidth=17, dir=y>}\\
\hspace*{0.7em}\smallcode{DownUpsampleII<stride=8, porder=3, iorder=3, dir=x>}\\
\hspace*{1.4em}\smallcode{Sum<dir=both>}\\
\hspace*{2.1em}\smallcode{DownUpsampleII<stride=2, porder=3, iorder=4, dir=x>}\\
\hspace*{2.8em}\smallcode{Pipe<dir=both>}\\
\hspace*{3.5em}\smallcode{DownUpsampleI<stride=2, pwidth=5, iwidth=5, dir=y>}\\
\hspace*{4.2em}\smallcode{Sum<dir=both>}\\
\hspace*{4.9em}\smallcode{Pipe<dir=both>}\\
\hspace*{5.6em}\smallcode{FIR<width=7, dir=y>}\\
\hspace*{5.6em}\smallcode{FIR<width=6, dir=x>}\\
\hspace*{4.9em}\smallcode{Pipe<dir=both>}\\
\hspace*{5.6em}\smallcode{FIR<width=9, dir=y>}\\
\hspace*{5.6em}\smallcode{FIR<width=20, dir=x>}\\
\hspace*{3.5em}\smallcode{TIIR<causal=True, width=24, dir=y>}\\
\hspace*{2.1em}\smallcode{Pipe<dir=both>}\\
\hspace*{2.8em}\smallcode{DownUpsampleII<stride=2, porder=3, iorder=4, dir=x>}\\
\hspace*{3.5em}\smallcode{DownUpsampleI<stride=2, pwidth=5, iwidth=5, dir=y>}\\
\hspace*{4.2em}\smallcode{Pipe<dir=both>}\\
\hspace*{4.9em}\smallcode{FIR<width=21, dir=y>}\\
\hspace*{4.9em}\smallcode{FIR<width=18, dir=x>}\\
\hspace*{2.8em}\smallcode{FIR<width=5, dir=x>}\vspace{0.5em}
\end{minipage}
It has PSNR 61.dB, throughput 44729 MPix/sec, and a 613.4x speedup. The first term in the \smallcode{Sum} is itself a rank 2 structure cascaded with a vertical \smallcode{TIIR}. The second term is a rank 1 structure cascaded with a horizontal \smallcode{FIR}. 

\section{Limitations and Future Work}

Our design space works with primitives that have continuous parameters or parameters that can be made continuous (i.e. the truncation locations of TIIRs). Some primitives like comb filters, have discrete parameters like the delay parameter, that are non-differentiable. Figuring out how to include these primitives in our design space would further improve its expressivity. Audio filters with reverb effects like the Telephone filter may be more efficiently approximated with some combination of our existing primitives and comb filters. 

Analyzing the frequency response of the target filter to inform the search constraints can improve sample efficiency. For example, the pass-band of the target can inform which downsampling factors to allow in the search. Pareto frontier structures are sometimes shared across different sizes of a given target filter. Search efficiency can be further improved by using more sophisticated search strategies like genetic search, and seeding the search with Pareto frontier structures from smaller sizes which are cheaper to optimize. 

Our element-wise sum operators always center their inputs before summing. Deeper inspection of the frequency responses learned by subtrees inside designs reveal that allowing variable offsets instead of always centering could further improve the Pareto frontier.

Another limitation is the cost of training an approximation per filter. For a specific filter, this is a one-time expense, but it is common to use parameterized families of filters, e.g. Gaussians of many different sizes. It is possible to dynamically resize certain primitives of a given design to achieve smaller or larger target sizes. For example, the interpolators, prefilters, and scaling factors of \smallcode{DownUpsampleII} and \smallcode{Stride} nodes can be adjusted. \smallcode{TIIRs} can also be stretched or contracted using the same theory we applied for our time step resolution parameter (Section \ref{sec:tiir-parameterization}), and by scaling the truncation region locations. Fine-tuning these parameters may be necessary to achieve the best possible quality after dynamic resizing.

Our figures in Section \ref{sec:frequency-response-gallery} show that PSNR is a good measure for design quality. However, certain use-cases may prefer designs with lower PSNR (i.e. more absolute frequency response error) than other designs, in exchange for other properties like smoother error behavior or narrower error band-widths. Our system design is orthogonal to the choice of loss function and can be used with any user-preferred error metric.  

Finally, it may be desirable to ensure that the frequency response behavior smoothly interpolates across designs for a given filter family. For example, interactive audio applications like Digital Audio Workstations allow users to dynamically adjust filter parameters like EQ settings. If each setting corresponds to a different approximation design, audible artifacts from discontinuities in behavior across these designs should be avoided. 


\section{Conclusion}

In this paper we have generalized and unified techniques from the literature into a design space of fast filter approximations that make the most of modern CPUs. We have shown how to search this space automatically, and how to use gradient descent on the points within it to tune the continuous parameters of fully differentiable structures that utilize diverse fast filter approximation strategies. We have shown how to lower these representations to fast, fused, vectorized, and parallelized implementations. Finally, we have used our system to automatically generate novel state-of-the-art approximations for a variety of common image and audio filters. The Pareto-dominant designs from our system cover a wide range of quality versus throughput trade-offs, to satisfy a broad range of user preferences. 


\clearpage
\section{Appendix}
\SetKwComment{Comment}{/* }{*/} 
\SetKwComment{ShortComment}{// }{} 
\SetCommentSty{small}

\subsection{Structure sampling algorithms}
\begin{algorithm}
\caption{Sampling 1D trees using a single resolution}\label{alg:1D-one-res-structure-sampling}
\DontPrintSemicolon
\SetKwFunction{Genstruct}{gen1D}
\SetKwProg{Fn}{Function}{:}{}
\Fn{\Genstruct{\smallcode{tree, parent, P, dir, left=True, right=True}}}{
    \Comment{\smallcode{tree}: root of the entire tree}
    \Comment{\smallcode{parent}: parent of the current subtree}
    \Comment{\smallcode{P}: design primitives}
    \Comment{\smallcode{dir}: direction of computation}
    \Comment{\smallcode{left}: if true fill in left child of binary nodes}
    \Comment{\smallcode{right}: if true fill in right child of binary nodes}

    \lIf{\smallcode{parent} is leaf} {
    \KwRet
    }
    $P' \gets$ \smallcode{filterPrimitives(tree, parent, P)}\\
    \uIf{\smallcode{parent} is unary} {
        \smallcode{parent.child} = $\gets$ random pick from $P'$\\
        \smallcode{gen1D(tree, parent.child, P)}\\
    } 
    \uElseIf{\smallcode{parent is binary and left}}{
          \smallcode{parent.leftChild} $\gets$ random pick from $P'$\\
          \smallcode{gen1D(tree, parent.leftChild, P, dir)}\\
    }
    \uElseIf{\smallcode{parent is binary and right}}{
        $P'' \gets$ \smallcode{filterPrimitives(tree, parent, P)}\\
        \smallcode{parent.rightChild} $\gets$ random pick from $P''$\\
        \smallcode{gen1D(tree, parent.rightChild, P, dir)}\\
    }
    \KwRet\\
}
\smallcode{tree} $\gets$ random pick from \smallcode{P}\\
\smallcode{gen1D(tree, tree, P, horizontal)}\\
\end{algorithm}

\begin{algorithm}
\caption{sampling 1D trees using multiple resolutions}\label{alg:1D-multi-res-structure-sampling}
\DontPrintSemicolon
\SetKwFunction{GenstructII}{multiRes1D}
\SetKwProg{Fn}{Function}{:}{}
\Fn{\GenstructII{\smallcode{R, tree, parent, P, dir}}}{
    \Comment{\smallcode{R}: set of allowed resolutions}
    \Comment{\smallcode{tree, parent, P, dir}: same definition as algorithm 1}
    
    \uIf{\smallcode{lowerResOK(R, tree, dir)}}{
    \smallcode{P'} $\gets$ \smallcode{P}$+\{$\smallcode{LowerResolutionTree}$\}$\\   
    }
   \smallcode{gen1D(tree, DUNode, P', dir)}\\   
   \smallcode{dummy = findLowerResPlaceholder(tree)}\\
    \If{\smallcode{dummy != null}}{
        \Comment{gen1D created a tree with a lower-res placeholder. Fill it.}
        \smallcode{r} $\gets$ random pick from \smallcode{R}\\
        \smallcode{lowerDU} $\gets$ \smallcode{DownUpample<1/r, dir>}\\ 
        \smallcode{R'}$\gets \{$\smallcode{r'|r'} $\in$ \smallcode{R} $\land$ \smallcode{r'< r}$\}$ \tcp{new legal resolutions}
        \smallcode{dummy.replaceWith(lowerDU)} \tcp{update the tree}
     \smallcode{multiRes1D(R', tree, lowerDU, P, dir)}\\
    }    
    \KwRet\\
}
\smallcode{r} $\gets$ pick from \smallcode{R}\\
\eIf{\smallcode{r == 1}}{
    \smallcode{tree} $\gets$ random pick from \smallcode{P} 
}{
\smallcode{tree} $\gets$ \smallcode{DownUpsample<1/r, dir>}
}
\smallcode{multiRes1D(R, tree, tree, P, horizontal)}
\end{algorithm}

\begin{algorithm}
\caption{sampling 2D trees using multiple resolutions}\label{alg:2D-multi-res-structure-sampling}
\SetKwFunction{GenstructIII}{multiRes2D}
\SetKwProg{Fn}{Func}{:}{}
\DontPrintSemicolon
\Fn{\GenstructIII{\smallcode{R,tree,dX,dY,parent,grandParent,P}}}{
    \ShortComment{\smallcode{R}: allowed horizontal, vertical resolutions. \smallcode{Tree,P}: see alg. 1}
    \ShortComment{\smallcode{dX,dY}: Make lower res. horiz. or vert. computation when True}
    \ShortComment{\smallcode{parent}: parent of computation at this resolution}
    \ShortComment{\smallcode{grandParent}: joins resolutions, null if no lower resolution}
    \vspace{0.1em}
    \uIf{\smallcode{parent is Pipe}}{ 
        \smallcode{gen1D(tree, parent, P, horizontal, left=True)}\\ 
        \smallcode{gen1D(tree, parent, P, vertical, right=True)}\\  
    }
    \lElse {
        \smallcode{gen1D(tree, parent, P, parent.direction)}
    }
    \vspace{0.15em}
    \smallcode{rx,ry} $\gets$ random pick from \smallcode{R}\\
    \smallcode{R'}$\gets\{$\smallcode{rx',ry'} $\in$ \smallcode{R} $\land$ \smallcode{(rx'< rx)} $\land$ \smallcode{(ry'< ry)}$\}$\\    
    \smallcode{dX' = lowerResOK(R'.x,tree,dir)} \smallcode{and !randNoOp()}\\
    \smallcode{dY' = lowerResOK(R'.y,tree,dir)} \smallcode{and !randNoOp()}\\
    \smallcode{grandParent'= pickBinary2D(dX, dY, dX', dY')}\\
    \vspace{0.05em}
    \ShortComment{join current resolution with lower resolution}
    \uIf{\smallcode{dX and dY}}{
        \smallcode{DU = createNestedDownUpsamples(rx, ry)}\\
        \eIf{\smallcode{grandParent' != null}}{
            \smallcode{DU.child = grandParent'}\\
            \smallcode{grandParent'.leftChild = Pipe}\\
        }{
            \smallcode{DU.child = Pipe}
        }
    }
    \uElseIf{\smallcode{dX or dY}}{
        \lIf{\smallcode{dX}}{\smallcode{DU = DownUpsample<1/rx, horizontal>}}
        \lElse{\smallcode{DU = DownUpsample<1/ry, vertical>}}
        \eIf{\smallcode{grandParent' != null}}{
            \smallcode{DU.child = grandParent'}\\
            \smallcode{grandParent'.leftChild} $\gets$ random pick from \smallcode{P}
        }{
            \smallcode{DU.child} $\gets$ random pick from \smallcode{P}
        }
    }
    \lElse{\KwRet}
    \smallcode{grandParent.rightChild = DU}

    \uIf{\smallcode{grandParent' == null}}{
        \smallcode{parent' = DU.child}
    }
    \Else{
        \smallcode{parent' = grandParent'.leftChild}
    }    \smallcode{multiRes2D(R',tree,dX',dY',parent',grandParent',P)}\\
    \KwRet
}
\vspace{0.1em}
\smallcode{rx,ry} $\gets$ pick from \smallcode{R}\\
\smallcode{R'}$\gets\{$\smallcode{rx',ry'} $\in$ \smallcode{R} $\land$ \smallcode{(rx'< rx)} $\land$ \smallcode{(ry'< ry)}$\}$\\    
\smallcode{dX = lowerResOK(R'.x,tree,dir)} \smallcode{and !randNoOp()}\\
\smallcode{dY = lowerResOK(R'.y,tree,dir)} \smallcode{and !randNoOp()}\\
\smallcode{grandParent} $\gets$ \smallcode{pickBinary2D(True, True, dX', dY')}\\
\eIf{\smallcode{rx == 1 and ry == 1}}{\smallcode{root = grandParent}}{
    \uIf{\smallcode{rx < 1 and ry < 1}}{
        \smallcode{root = createNestedDownUpsamples(rx, ry)}\\
    } 
    \uElseIf{\smallcode{rx < 1}} {
        \smallcode{root} $\gets$ \smallcode{DownUpample<1/rx, horizontal>}
    }
    \lElse{
        \smallcode{root} $\gets$ \smallcode{DownUpample<1/ry, vertical>}
    }
    \smallcode{root.child = grandParent}\\
}

\smallcode{grandParent.leftChild = Pipe}\\
\smallcode{multiRes2D(R, root, dX, dY, grandParent.leftChild,}\smallcode{grandParent, P)}
\end{algorithm}

\begin{algorithm}
\caption{Pick Binary2D node to join multi-resolutions}\label{alg:2D-binary-node-selection}
\SetKwFunction{pickBinaryNode}{pickBinary2DNode}
\SetKwProg{Fn}{Function}{:}{}
\DontPrintSemicolon
\Fn{\pickBinaryNode{\smallcode{dX, dY, dX', dY'}}}{
    \ShortComment{\smallcode{dX,dY}: True if current res. has horiz., vert. computation}
    \ShortComment{\smallcode{dX',dY'}: True if lower res. has horiz., vert. computation}
    \vspace{0.05em}
    \eIf{\smallcode{dX and dY}}{
       \smallcode{currentResDim = 2}
    }{
        \smallcode{currentResDim = 1}
    }
    \uIf{\smallcode{dX' and dY'}}{
       \smallcode{lowerResDim = 2}
    }
    \uElseIf{\smallcode{dX' or dY}'}{
        \smallcode{lowerResDim = 1}
    } 
    \Else{
        \smallcode{lowerResDim = 0}
    }
    \uIf{\smallcode{lowerResDim == 0}}{
        \smallcode{node = null}
    }
    \uElseIf{\smallcode{currentResDim == lowerResDim}}{
        \smallcode{node =} random pick from $\{$\smallcode{Pipe, Sum}$\}$
    }
    \Else{
        \smallcode{node = Pipe}
    }
    \KwRet node
}
\end{algorithm}
\FloatBarrier

\subsection{Design Enumeration Constraints}
\label{appendix:enumeration-constraints}

\setlist[enumerate]{
  leftmargin=1.5em,  
  labelwidth=1.3em,  
  labelsep=0.5em, 
  align=parleft
}

\begin{enumerate}
    \item the support width for all nodes must be $>0$
    \begin{itemize}[label=\textendash]
        \item This is the definition of a legal structure
    \end{itemize}
    \item the support width for non \smallcode{Leaf} nodes must be $>$ \smallcode{N}
    \begin{itemize}[label=\textendash]
        \item \smallcode{n} must be $>0$ for constraint 1 to hold. Furthermore, for small enough support widths, an FIR is cheap enough that more complex structures will not provide better throughputs due to implementation overhead. We do bounds inference as we generate the tree to determine when a \smallcode{Leaf} must be inserted.
    \end{itemize}
    \item a design can use at most \smallcode{L} \smallcode{Leaf} nodes
    \begin{itemize}[label=\textendash]
        \item The tree size must be limited to ensure that sampling terminates. This assumes there are good structures with $<$ \smallcode{L} leaves.  
    \end{itemize}
    \item a design cannot use a resolution lower than \smallcode{R}
    \begin{itemize}[label=\textendash]
        \item No prefilter is perfect, so there will be some aliasing of frequencies above the Nyquist limit when downsampling. The amount of aliasing in the pass-band of the target filter increases with the decimation factor, so there is a limit to how much a design can decimate and still produce good approximations.
    \end{itemize}
    \item all subtrees under \smallcode{DownUpsample} should have support width $\ge$ \smallcode{D}
    \begin{itemize}[label=\textendash]
        \item The support width has to be $\ge0$ for the design to be legal, and we can use a larger value to ensure that all \smallcode{DownUpsample}s do some work besides decimation and interpolation. For narrow-band filters, it is cheaper to do the frequency domain shaping inside the lower-resolution subtree.
    \end{itemize}
    \item no \smallcode{Cascade} has an \smallcode{Stride} child
    \begin{itemize}[label=\textendash]
        \item Cascading a \smallcode{Leaf} with a \smallcode{Stride} is equivalent to using a more complex interpolator than the factored tower. This could be useful, but we assume that the interpolator prescribed by \cite{narrow-band-FRM} is sufficient.
    \end{itemize}
    \item no designs have nested \smallcode{Stride}s 
    \begin{itemize}[label=\textendash]
        \item The pass-band width of two nested \smallcode{Stride}s with strides $a$ and $b$ is equivalent to that of a single \smallcode{Stride} with stride $a+b$. It is reasonable to assume that the center frequencies of the nested \smallcode{Stride} pass-bands would align, in which case the single \smallcode{Stride} with the same center frequency has the same pass-band. The computation used to do frequency response shaping in the pass-band across two different stride levels can instead be used at a single stride with lower cost.
    \end{itemize}
    \item All \smallcode{Stride}s cost $<x\%$ the cost of a \smallcode{FIR} with equal support width
    \begin{itemize}[label=\textendash]
        \item Designs should only use \smallcode{Stride} if it costs less than an \smallcode{FIR} since it is strictly less expressive. There is also overhead to performing the more complicated computation of \smallcode{Stride} compared to an \smallcode{FIR}. This threshold accounts for the overhead. 
    \end{itemize}
    \item Trees can have at most \smallcode{S} \smallcode{Sum}s and \smallcode{C} \smallcode{Cascade}s 
    \begin{itemize}[label=\textendash]
        \item The limit \smallcode{L} on \smallcode{Leaf}s means that trees have $<$ \smallcode{L-1} \smallcode{Sum}s or \smallcode{Cascade}s. \smallcode{S} $<$ \smallcode{L-1} or \smallcode{C} $<$ \smallcode{L-1} biases how much designs rely on adding versus multiplying intermediate frequency responses. 
    \end{itemize}
    \item \smallcode{Sum} cannot have two \smallcode{FIR} children
    \begin{itemize}[label=\textendash]
        \item A sum of two \smallcode{FIR}s with maximum support $n$ is equivalent to one length $n$ \smallcode{FIR} and is more expensive
    \end{itemize}
    \item \smallcode{Cascade} cannot have two \smallcode{FIR} children
        \begin{itemize}[label=\textendash]
        \item A cascade of two \smallcode{FIR}s with support widths $n,m$ is equivalent to one length $n+m-1$ \smallcode{FIR} and is more expensive
    \end{itemize}
    \item A \smallcode{TIIR} must cost less than an \smallcode{FIR} with equal support width
    \begin{itemize}[label=\textendash]
        \item A \smallcode{TIIR} is less expressive than an \smallcode{FIR} with the same support width, so it can only be advantageous when it costs less.
    \end{itemize}
    \item Within a resolution \smallcode{Sum}s are adjacent and above \smallcode{Cascade}s
    \begin{itemize}[label=\textendash]
        \item We assume \smallcode{Sum}s should be used to combine multi-band decompositions of a target response.
    \end{itemize}
\end{enumerate}

\subsection{Cost model}
\begin{figure}[H]
\centering
\includegraphics[width=0.9\columnwidth]{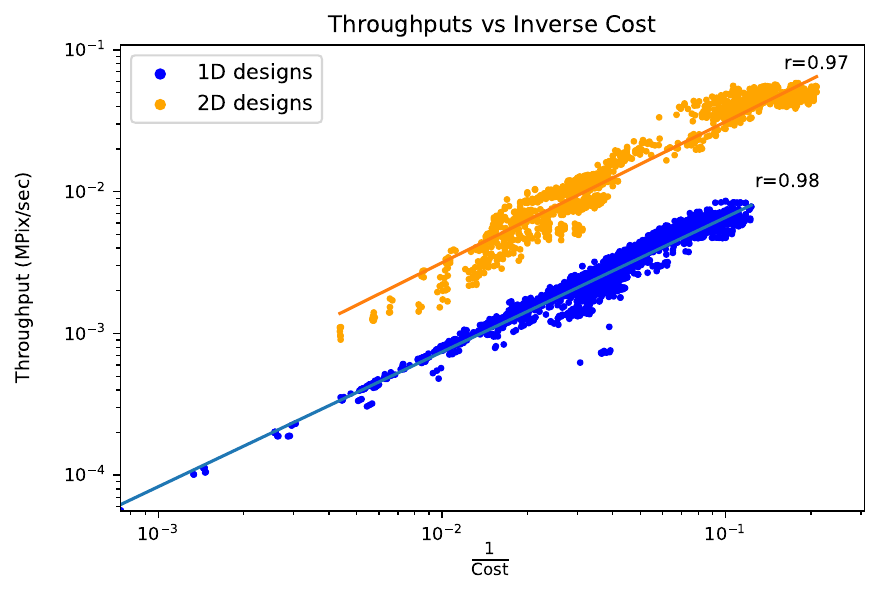}
\captionsetup{skip=0.5pt}
\caption{Given a design, our cost model predicts machine cycles spent per output sample, so cost is inversely related to throughput. Here we plot benchmarked throughputs in megapixels per second versus 1/cost for the top 3 Pareto tiers from our search and all baseline designs on a log-log scale.  
}
\label{fig:throughputs-vs-cost}
\end{figure}
The cost of a design is inversely related to throughput and is proportional to machine cycles per output sample. Our model evaluates cost recursively. The cost of a node includes all operations below. Here are the cost formulas for each primitive:
\setlength{\abovedisplayskip}{0pt}
\setlength{\belowdisplayskip}{2pt}
\begin{enumerate}
    \item \smallcode{cost(FIR)=} 
        \[\begin{cases}
          \text{\smallcode{width}}, & \text{\smallcode{width}} < 32\\
          \text{\smallcode{width}} *\frac{0.5}{0.8}, & 32\le \text{\smallcode{width}} < 64\\
          \text{\smallcode{width}} *\frac{0.5}{0.9}, & \text{otherwise}
        \end{cases}\]
     \begin{itemize}[label=\textendash]
        \item  Larger FIRs reach a higher percentage of peak FLOPs, because they have better reuse of loaded values. This has an effect on throughput once \smallcode{width} $\ge$ two SIMD vectors. The constants were determined from benchmarking \smallcode{FIR}s. Note that the footprint of a \smallcode{Leaf} with a \smallcode{Stride} parent is \smallcode{stride}$*$\smallcode{(width-1)}+1.
    \end{itemize}
    \item \smallcode{cost(TIIR)=lookup_table[pole_type, stride]}
    \begin{itemize}[label=\textendash]
        \item All nodes under \smallcode{Stride} are evaluated with the appropriate stride. \smallcode{TIIR}s with larger strides are cheaper because they are more amenable to vectorization (Section~\ref{sec:lowering}). Vertical TIIRs are even cheaper because they are vectorized horizontally and their cost does not depend on stride. Their cost is set to the lowest stride 16 cost. We derived the lookup tables for each case by benchmarking second-order TIIRs for each stride. They are assigned the cost of the FIR size with equivalent throughput. The TIIR costs per stride are:

    \begin{tabular}{c|c}
    stride & cost \\ 
    \hline
    1 & 24 \\ 
    2 & 19 \\ 
    4 & 16 \\ 
    8 & 13 \\ 
    16 & 9 \\ 
    \end{tabular}\label{appendix:tiir-costs}\\
    \end{itemize}
    \item \smallcode{cost(Cascade)=} sum of children's costs
    \item \smallcode{cost(Stride)=} cost of child + $\log_2($\smallcode{stride}$)*$\smallcode{order}
     \begin{itemize}[label=\textendash]
        \item The second term is the interpolator cost (Section~\ref{sec:primitives}) 
    \end{itemize}
    \item \smallcode{cost(DownUpsampleI)=} \\
    (child cost + \smallcode{prefilter_width + interpolator_width} ) / \smallcode{stride} \\
    \smallcode{+ stride/2}   if \smallcode{direction == horizontal}
    \begin{itemize}[label=\textendash]
        \item We measure cost in terms of the work per output sample at the full resolution. The child, prefilter, and interpolator costs are divided by \smallcode{stride} because we do not compute prefiltered values for discarded samples, every \smallcode{stride-1} taps of the interpolator is multiplied by a zero, and the child computation is done at 1/\smallcode{stride} resolution. There last term accounts for the vector shuffle operations required when downsampling and upsampling along the vectorized dimension.  
    \end{itemize}
    \item \smallcode{cost(DownUpsampleII)=}\\
    \smallcode{child cost/stride + prefilter_order + interpolator_order}\\
    \smallcode{+ stride/2} if \smallcode{direction == horizontal}
    \begin{itemize}[label=\textendash]
        \item As discussed in section \ref{sec:primitives} only the highest level of a factored tower can be evaluated at just the retained values during downsampling. During upsampling, each level benefits from less sparsity. It is straightforward to show that the total number of operations required in to prefilter and interpolater for each output sample is equal to the \smallcode{order} of the factored tower.  
    \end{itemize}
    \item \smallcode{cost(Pipe)=} left child cost + right child cost
    \item \smallcode{cost(Sum)=} sum of children's costs + the arity - 1 
\end{enumerate}

\subsection{Optimization passes}
\label{appendix:optimization_passes}

A number of optimization passes that exploit signal processing identities are run during lowering, in order to make faster implementations possible. This is done in a optimization IR, designed for this purpose. To map from the representation used for search into this IR, we do the following:

\begin{itemize}
\item \smallboldcode{Stride} nodes are removed, and instead, recursively, all the FIRs and IIRs below them are given a stride parameter. Rather than being fed a strided subsample of the input, they are now fed the entire input, and are expected to operate in a strided manner. This supports the vectorization that will occur.
\item \smallboldcode{TIIR} nodes are decomposed into their sparse FIR and IIR. In addition to being strided, FIRs may now be sparse. This decomposition happens both to simplify the underlying implementation, but also so that FIRs may be fused and reordered as we see fit.
\item \smallboldcode{Pad} nodes are introduced, which pad the output of their child with zeros on either side in some direction. These are used on the children of Sum nodes to make their shapes match, and on the output of \emph{DownUpSample} nodes to account for quantization effects on size. The inverse - the \smallboldcode{Trim} node - also exists in this IR, but is not yet used.
\item \smallboldcode{ForEachRowOrColumn} nodes are introduced wherever a 1D node is combined with a node that operates along another direction. These nodes lift an algorithm from 1D to 2D, by wrapping a for loop over rows or columns around it. After these nodes are introduced, the distinction between \smallboldcode{Pipe} and \smallboldcode{Cascade} can be dropped, leaving only 1D and 2D \smallboldcode{Cascade} nodes. 1D nodes must be either causal or anticausal, and can only be summed or cascaded with nodes of the same causality, so \smallboldcode{ForEachRowOrColumn} nodes are also used to make causal/anticausal pairs. They have a single causal 1D child, and a single anticausal 1D child, one of which may be a \smallboldcode{NoOp} node, which just returns its input. A vertical \smallboldcode{ForEachRowOrColumn} node has an optional stride parameter that interleaves some number of columns of the input before feeding them to the children, in order to support vectorization of vertical passes.
\end{itemize}

Once in this IR, we run the following optimization passes in the order listed. Optimizations that are critical for performance are marked with a $\dagger$. 

\begin{itemize}

\item{\textbf{Factor third-order IIRs to support baselines.} Third-order IIRs are factored into a second-order and a first-order term by finding a real root with Newton's method. Our search does not generate third-order IIRs, so this pass exists to express baselines that use third-order IIRs in our system (~\cite{young-gaussian}).}

\item{$\dagger$ \textbf{Fuse row-wise or column-wise operations for locality.} All of our operations are linear and therefore associative. This means that sub-trees of \smallboldcode{Cascade} nodes can be re-nested arbitrarily. Also, all operations that do not use \smallboldcode{DownUpSample} commute. We now identify all sub-trees of \smallboldcode{Cascade} nodes, recursively extract the non-\smallboldcode{Cascade} leaves, and sort each commutative subsequence so that all children that operate purely horizontally come before all children that operate purely vertically. This places \smallboldcode{ForEachRowOrColumn} nodes that operate in the same direction next to each other. Recall that \smallboldcode{ForEachRowOrColumn} nodes have a causal child followed by an anticausal child. For the cases where the anticausal pass in the first commutes with the causal pass in the second, we can fuse these nodes into a single \smallboldcode{ForEachRowOrColumn} node, where the causal child is a 1D \smallboldcode{Cascade} of the causal children of the originals, and likewise for the anti-causal child. 1D \smallboldcode{Cascade}s are run entirely fused in the manner of synchronous data-flow, so by turning a 2D \smallboldcode{Cascade} into a 1D \smallboldcode{Cascade} we pass intermediate values in registers instead of spilling to some level of the cache hierarchy (typically L2). The sorted and fused leaves are then reassembled back into a tree of \smallboldcode{Cascade} nodes}

\item{$\dagger$ \textbf{Vectorize horizontal IIRs using scattered look-ahead}. All horizontal \smallboldcode{IIR}s with stride $s$, where $s$ is less than the SIMD width (16 in our benchmarks), are rewritten into a \smallboldcode{Cascade} of an \smallboldcode{IIR} with stride $2s$, and a small FIR of stride $s$. Given a second-order \smallboldcode{IIR} with recursive $\alpha$, $\beta$, the new \smallboldcode{IIR} has coefficients $\alpha^2 - 2\beta$, $-\beta^2$, and the FIR has coefficients $1, \alpha, -\beta$ (\cite{scattered-look-ahead}). The first-order equivalent can be derived by setting $\beta = 0$. This is done repeatedly until the stride is at least the SIMD width, at which point it is possible to run the IIR using vector instructions, because there are no within-vector dependencies. We found that this also makes low-pass IIR filters more numerically stable.}

\item{\textbf{Optimize padding}. 1D sub-trees are run as synchronous data-flow. Each node is a stateful unit that produces and consumes data at a fixed rate. In this context, right-padding a causal pass is almost free - it just requires ensuring the tail of the output buffer contains zeros. Left-padding, however, is equivalent to a delay line, and requires buffering data. \smallboldcode{FIR} nodes also require buffering previous inputs, so we fuse \smallboldcode{Pad} nodes with each other, and also with \smallboldcode{FIR} nodes, to cut down on the number of buffers required. In a causal/anti-causal pair, we convert any left-padding on the causal pass into right-padding in the anticausal pass, and vice-versa.}

\item{\textbf{Fuse FIRs with matching stride}. Similarly to the procedure described above for fusing \smallboldcode{ForEachRowOrColumn} nodes, we sort all commutative subsequences of 1D \smallboldcode{Cascade}s so first we have the sparse FIRs corresponding to the numerators of the TIIRs, then we have the IIRs, and finally we have all strided FIRs in decreasing order of stride. Any back-to-back FIRs with matching stride are then fused by convolving their coefficients. Each such fusion saves a single multiply-add and one buffer of intermediate results.}

\item{$\dagger$ \textbf{Vectorize vertical passes.} By default, a \smallboldcode{ForEachRowOrColumn} operating vertically passes a single column of data into a 1D causal and anticausal pipeline. Lowered naively, this would require vector gather and scatter instructions to pass data to and from the child nodes. A cleaner use of SIMD vectors is to process $n$ columns in parallel. We therefore assign these nodes stride $n$, and feed their child nodes $n$ columns interleaved, which can be assembled using aligned vector load instructions. The children have their strides recursively multiplied by $n$ in order to operate correctly on the interleaved stream.}

\item{\textbf{Do low-resolution operations early.} In a \smallboldcode{Sum} node, the input is accessed by both children. If one stage uses less intermediate memory than the other, it is better to do it first, so that the input has a chance of still being in cache when the second child accesses it. Low-resolution sub-trees (i.e. those within a \smallboldcode{DownUpSample} node) naturally use less intermediate memory, so we reorder the children of Sum nodes to do these first. This was primarily done to optimize the skip connection in Convolution-Pyramid-like structures. Note that it is not possible to similarly reorder \smallboldcode{DownUpSample} nodes in \smallboldcode{Cascade}s, because they are not commutative.}

\item{\textbf{Fuse nested DownUpSamples}. If one \smallboldcode{DownUpSample} is nested immediately around another, we fuse them into a single larger \smallboldcode{DownUpSample} node, combining their prefilters and interpolators by striding the outer filter by the downsampling factor of the outer node, and then convolving the two. This saves a small amount of math and intermediate data.}

\end{itemize}

We parallelize in tiles, and produce all non-zero outputs, so corner and edge tiles are responsible for producing more values than interior tiles. This affects how they should be optimized. Edge tiles on the left are also distinct from edge tiles on the right, because a filter may contain, e.g. only causal TIIRs, so we do not have left/right or top/bottom symmetry in the algorithm, even if the target filter is symmetric. So after the passes above, the filter is split into nine copies for the nine types of tile, with each copy wrapped in the appropriate \smallboldcode{Trim} node to slice out just the portion of the output needed for that tile. Naively this implies a lot of values are computed and then discarded. We use two mechanisms to ameliorate this. First, we run more optimization passes on these nine copies that aim to reduce the amount of wasted work. Second, in the C++ implementation, pipeline stages can be asked to only produce a region-of-interest of the output, and are expected to use this to skip unnecessary work. The second set of optimizations are:

\begin{itemize}

\item{$\dagger$ \textbf{Trim at low-resolution}. Trimming the output of a \smallboldcode{DownUpSample} can be partially converted to trimming the output of its child before upsampling, taking into account the size of the interpolator and the upsampling factor. This optimization is critical for Convolution-Pyramid-like structures.}

\item{\textbf{Cancel trimming against padding.} Padding followed by trimming can be replaced with either a smaller pad or a smaller trim.}

\item{\textbf{Move \smallboldcode{Trim} inside \smallboldcode{Sum}}. Instead of trimming the output of a summation, we can trim the output of each child. This is done mostly to push \smallboldcode{Trim} nodes further in to find new optimization opportunities.}

\item{\textbf{Move \smallboldcode{Trim} nodes early as possible}. In a 2D \smallboldcode{Cascade}, trimming in one direction commutes with any operations in the other direction, and so the trimming should be done first, to reduce the amount of data fed to the other operation.}

\end{itemize}

\subsection{Truncated cosines as TIIRs}\label{appendix:truncated-cosines-are-tiirs}
Below is the proof that the truncated cosines used by \cite{elboher2011cosine} and implemented with integral images can instead be implemented as truncated IIRs:\\
A real valued cosine with frequency $\omega$ is the sum of two complex sinusoids (by Euler's formula):
\[\cos(\omega t) = \frac{1}{2}(e^{j\omega t} + e^{-j\omega t})\]
each term can be represented as an IIR with the transfer function:
\[2H(\cos(\omega t)) =\frac{1}{1-e^{j\omega} z^{-1}} + \frac{1}{1-e^{-j\omega} z^{-1}}\]
Because each term is an exponential function, we know that the value at time $T$ is just the scalar value $e^{j\omega(T)}$ times the value at time 0 for the first term, and $e^{-j\omega(T)}$ for the second term. So to truncate the cosine after $T$ time steps, we just need to subtract the scaled up value from $T$ time steps ago from the current value: 

{\small
\[2H(\cos_{T}(\omega t)) = \frac{e^{j\omega} - e^{j\omega T}z^{-T}}{1-e^{j\omega}z^{-1}} + \frac{e^{-j\omega} - e^{-j\omega T}z^{-T}}{1-e^{-j\omega} z^{-1}}
\]}
Combining terms:
{\footnotesize
\[= \frac{(e^{j\omega} + e^{-j\omega}) - 2z^{-1} -(e^{j\omega T}+e^{-j\omega T})z^{-T} + (e^{j\omega(T-1)}+e^{-j\omega(T-1)})z^{-(T+1)}}{1 - (e^{j\omega} + e^{-j\omega})z^{-1} + z^{-2}}\]
}
{\small
\[= \frac{2\cos(\omega) - 2z^{-1} - 2\cos(\omega T)z^{-T} + 2\cos(\omega(T-1))z^{-(T+1)}}{1 - 2\cos(\omega)z^{-1} + z^{-2}}\]
}
This means convolving with truncated cosine can be implemented as the TIIR with the system equation:
{\small
\[y[n] = \cos(\omega)y[n-1] -\frac{1}{2}y[n-2] + \]}{\small\[\cos(\omega)x[n] - x[n-1] -\cos(\omega T)x[n-T] + \cos(\omega(T-1))x[n-(T+1)]\]}
\subsection{Training Time Statistics}
\label{appendix:training_time_statistics}
Training time statistics in CPU minutes per target filter:
\begin{table}[H]
\begin{tabular}{|c|c|c|c|c|c|}
\hline
Target Filter & min & max & mean & median & samples \\
\hline
Gaussian 1D 101 & 0.17 & 1.43 & 0.48 & 0.43 & 4,228\\
Gaussian 1D 201 & 0.10 & 3.47 & 0.86 & 0.77 & 4,574\\
Gaussian 1D 401 & 0.10 & 7.25 & 1.65 & 1.47 & 4,816\\
Gaussian 1D 801 & 0.02 & 15.53 & 3.00 & 2.60 & 5,864\\
\hline
Lanczos 1D 101 & 0.03 & 1.00 & 0.33 & 0.32 & 4,171\\
Lanczos 1D 201 & 0.05 & 3.48 & 0.54 & 0.48 & 4,299\\
Lanczos 1D 401 & 0.03 & 5.10 & 0.73 & 0.57 & 4,744\\
Lanczos 1D 801 & 0.02 & 15.02 & 2.68 & 2.27 & 6,260\\
\hline
Telephone 1D 109 & 0.12 & 2.30 & 0.51 & 0.47 & 5,886\\
Telephone 1D 341 & 0.13 & 5.98 & 1.08 & 0.92 & 7,488\\
Telephone 1D 2684 & 0.18 & 40.32 & 5.75 & 4.62 & 12,429\\
Telephone 1D 4867 & 0.92 & 38.67 & 8.20 & 7.47 & 12,352\\
\hline
Hrir 1D 88 & 0.12 & 1.70 & 0.75 & 0.77 & 5,272\\
Hrir 1D 89 & 0.10 & 2.37 & 0.83 & 0.77 & 5,305\\
Hrir 1D 555 & 0.23 & 2.67 & 1.15 & 1.13 & 9,854\\
Hrir 1D 591 & 0.15 & 3.82 & 1.13 & 1.10 & 8,657\\
Hrir 1D 1166 & 0.38 & 9.87 & 2.11 & 1.98 & 9,741\\
Hrir 1D 1201 & 0.28 & 14.35 & 2.29 & 2.10 & 10,051\\
\hline
Gaussian 2D 101 & 0.08 & 10.82 & 1.09 & 0.78 & 10,801\\
Gaussian 2D 201 & 0.10 & 20.42 & 5.55 & 3.62 & 10,598\\
Gaussian 2D 401 & 0.83 & 40.55 & 18.26 & 16.12 & 9,279\\
\hline
Gabor 2D 101 & 0.10 & 8.50 & 0.68 & 0.52 & 7,980\\
Gabor 2D 201 & 0.32 & 20.43 & 3.54 & 2.18 & 9,985\\
Gabor 2D 401 & 1.43 & 35.32 & 17.04 & 13.63 & 8,600\\
\hline
Lowpass 2D 101 & 0.13 & 13.38 & 1.20 & 0.80 & 10,439\\
Lowpass 2D 201 & 0.38 & 25.58 & 5.28 & 2.60 & 9,935\\
Lowpass 2D 401 & 1.20 & 50.77 & 16.29 & 8.63 & 11,846\\
\hline
\end{tabular}
\label{tab:detailed-training-time-stats}
\vspace{-0.5em}
\end{table}

\subsection{Gallery of Impulse and Frequency Responses}\label{sec:frequency-response-gallery}
In figures 18 - 44, for each target we show the impulse responses of two Pareto frontier models from our search as well as a CP and FRM design. The CP and FRM designs are chosen from the Pareto frontier of only CP or FRM models respectively. 
Models are sorted by decreasing throughput. We report the PSNRs and throughputs of each and their speedup over the fastest exact filtering method, which is always FFT convolution except for 2D Gaussians which can use linearly separable filtering. For Gaussians we also plot the triple box responses (which Pareto dominate YVV) and for Gabors we plot the YVV responses. 

The first column of each plot shows the impulse responses compared to ground truth. In 2D we show contour plots of the impulse response with a full scale view on the left and a zoomed-in view on the right of the center values which have higher magnitude. The second column compares the magnitude of the frequency responses of the model to the ground truth. For all of our linear phase targets, the third column compares the phase responses of the target to the approximation. We plot the wrapped phase because 1) numerical errors in phase unwrapping distort the actual error 2) deviations from the ground truth are not visible when phases are unwrapped over a wide range of frequencies. For the audio filters, HRIR and Telephone, we plot the group delay instead, which is more informative of filter behavior. For 2D filters we plot the frequency response of the horizontal cross sections of the target and approximation.

For a point of reference, we draw a solid red line on each plot at the frequency beyond which both the ground truth filter and the approximation have more than -50dB attenuation from the peak magnitude response. Generally speaking, the magnitude response for frequencies beyond this point are low enough that errors in the magnitude or phase response can be ignored.

\begin{figure*}[htbp]
\centering
\includegraphics[width=0.9\textwidth]{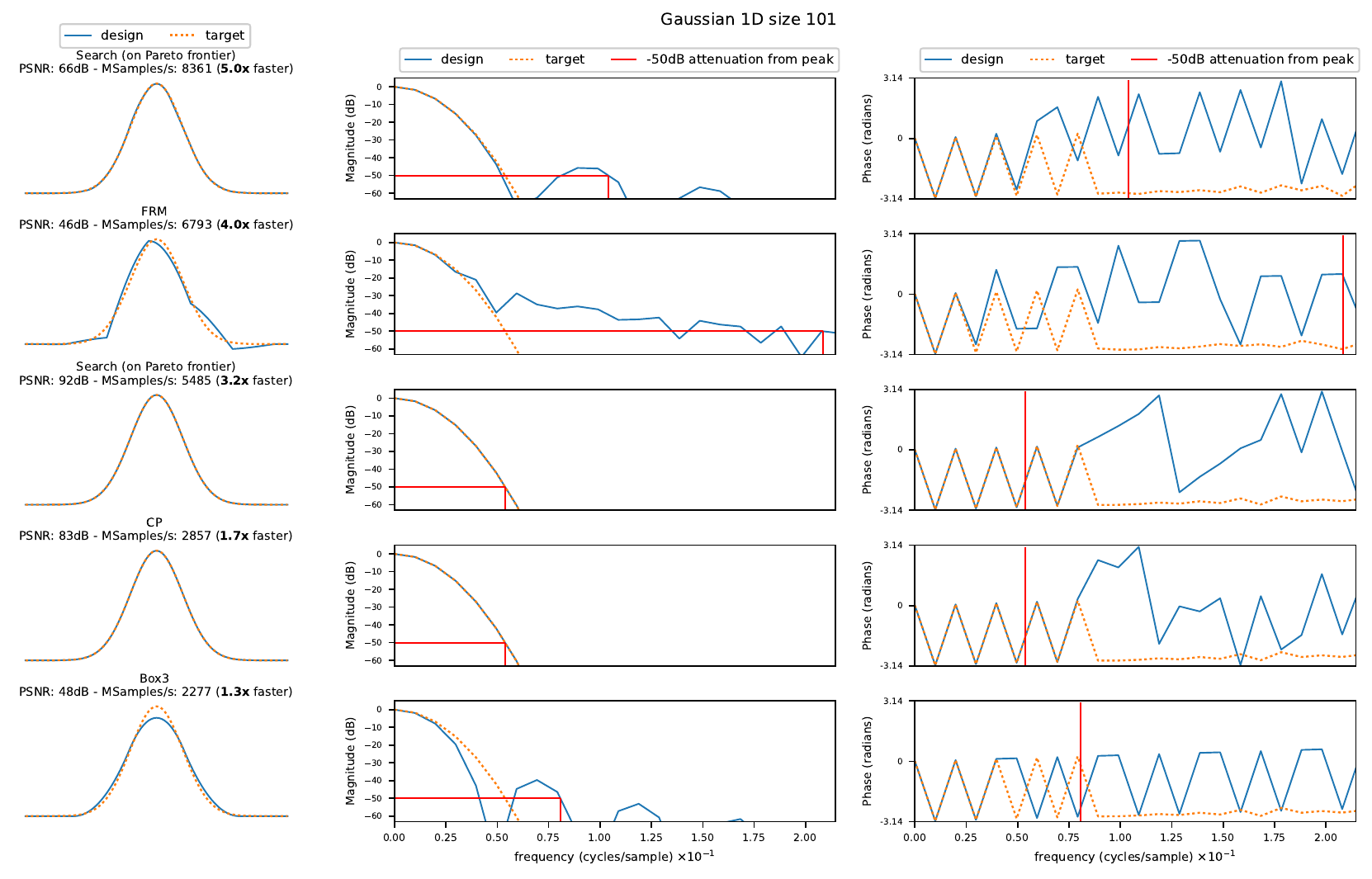}
\captionsetup{skip=1.5pt}
\caption{Figures 18-21 show the responses for selected models across all 1D Gaussian sizes. For each size we show a lower and higher quality Pareto frontier model from our search compared to a CP and FRM design and the triple box blur (Box3). For all sizes, the triple box has ringing. Its kurtosis causes its magnitude response to dip below the target at $\sim$0.025 cycles/sample. However, a benefit of the triple box blur is that it has linear phase (its phase plot does not exactly match the target due to numerical precision issues). The FRM designs have large errors and are not linear phase. For sizes 401 and 801, both of our designs and the CP designs are essentially indiscernible from ground truth but ours are multiple times faster. For sizes 101 and 201, our faster designs are exact up to -40dB and the higher quality ones are exact beyond -50dB. }
\label{fig:gaussian1d-101-freq-plot}
\end{figure*}
\begin{figure*}[htbp]
\centering
\includegraphics[width=0.89\textwidth]{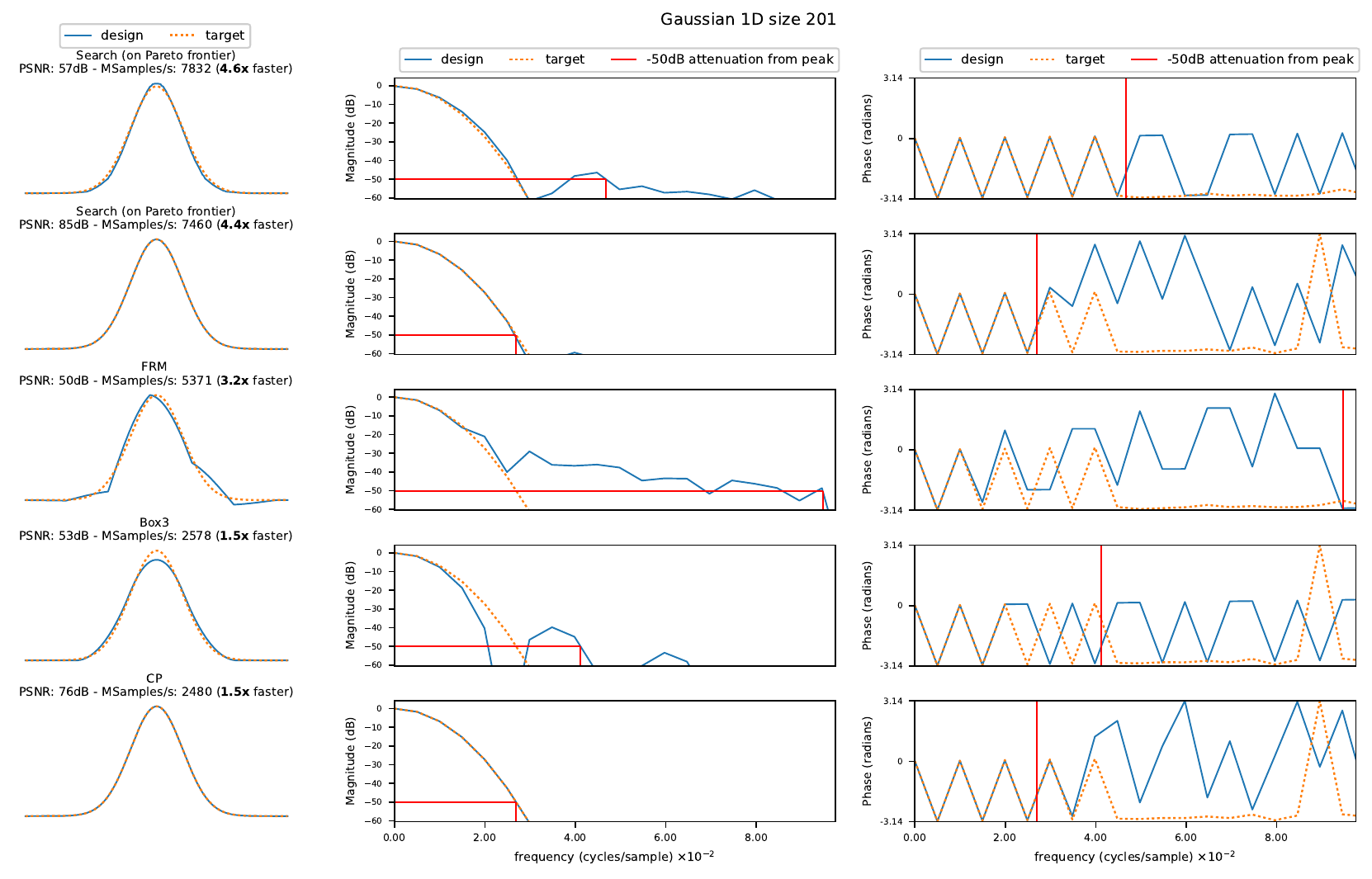}
\captionsetup{skip=1.5pt}
\caption{
}
\label{fig:gaussian1d-201-freq-plot}
\end{figure*}
\begin{figure*}[htbp]
\centering
\includegraphics[width=0.89\textwidth]{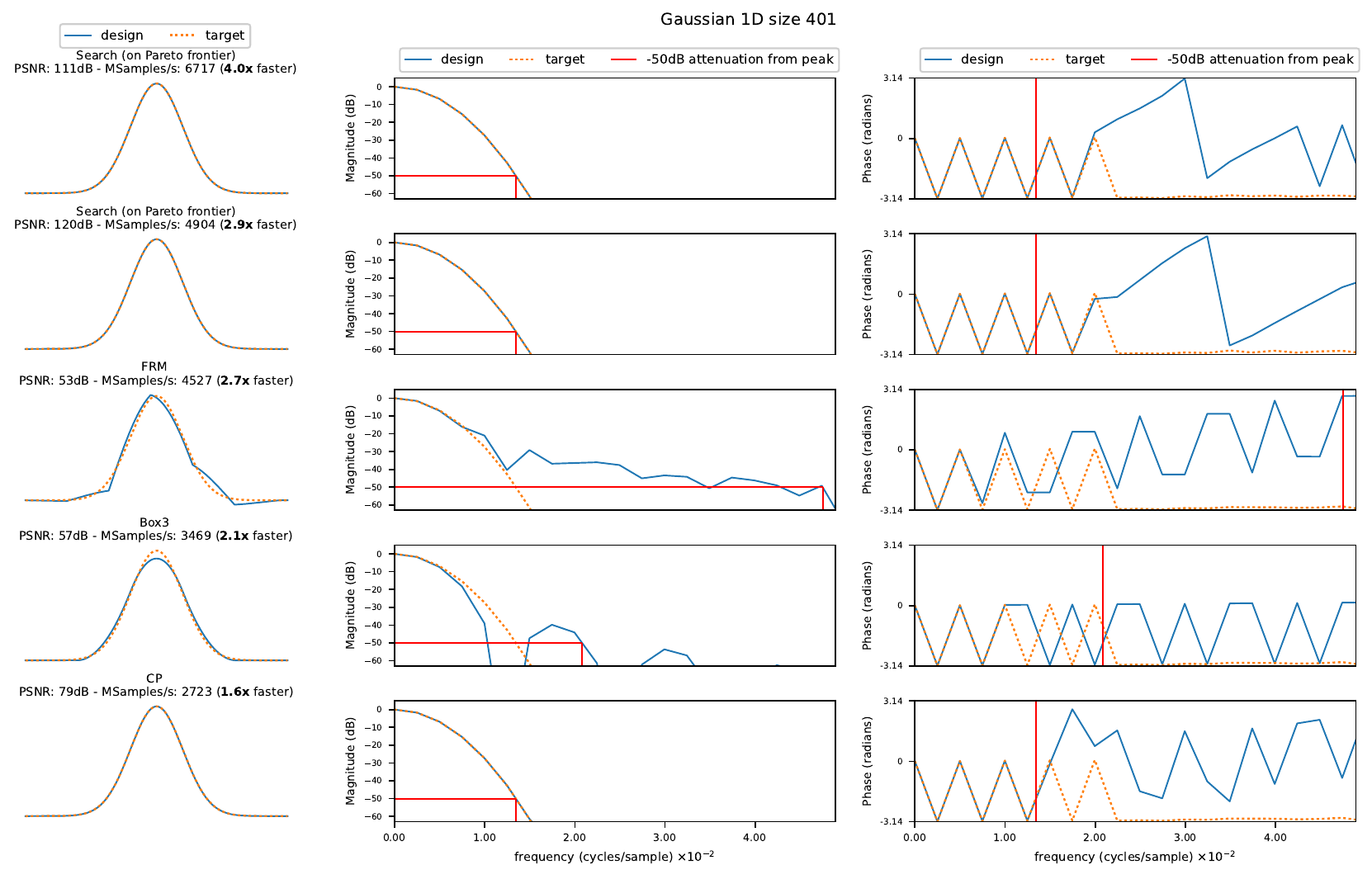}
\captionsetup{skip=1.5pt}
\caption{
}
\label{fig:gaussian1d-401-freq-plot}
\end{figure*}
\begin{figure*}[htbp]
\centering
\includegraphics[width=0.89\textwidth]{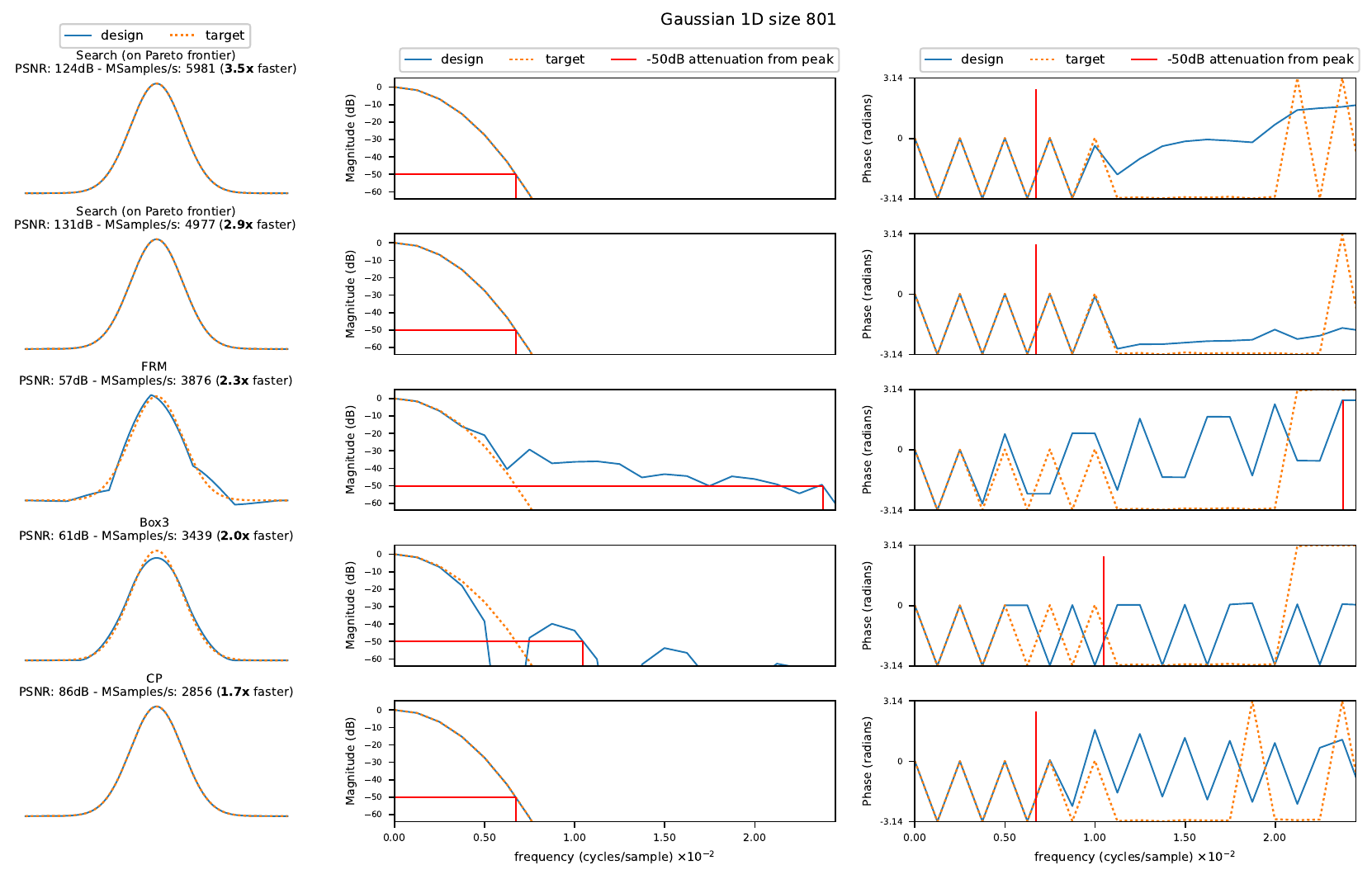}
\captionsetup{skip=1.5pt}
\caption{
}
\label{fig:gaussian1d-801-freq-plot}
\end{figure*}

\begin{figure*}[htbp]
\centering
\includegraphics[width=0.95\textwidth]{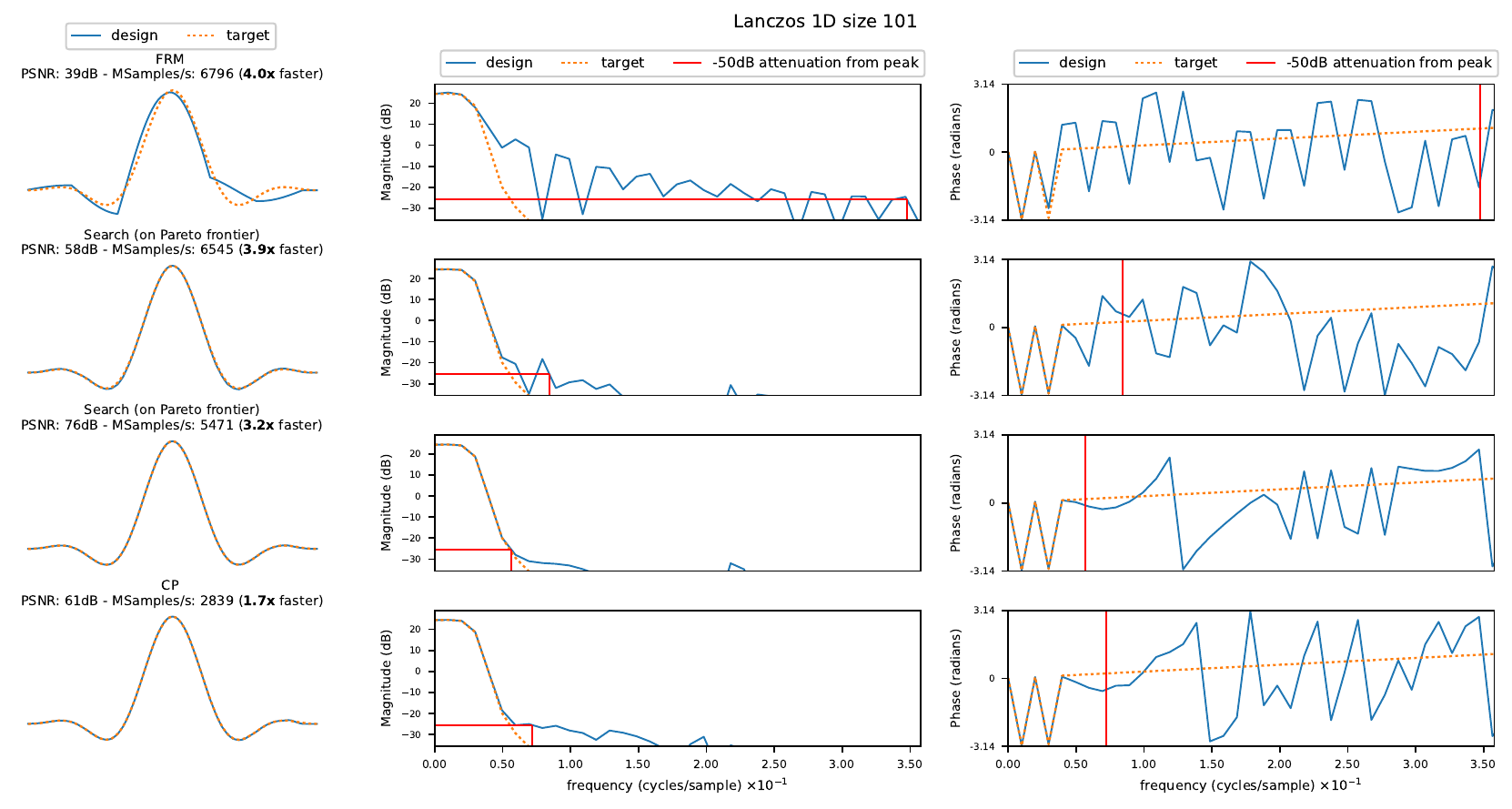}
\captionsetup{skip=1.5pt}
\caption{Figures 21-24 are the response plots for 1D Lanczos filters sizes 101, 201, 401, and 801. The takeaways are similar to those for 1D Gaussians. FRM designs have high error across all sizes and are not linear phase. For all sizes, both of our designs are significantly faster than the CP designs and strictly Pareto dominate FRM or have significantly higher quality. Our designs are indiscernible from ground truth except for sizes 101 and 201 where the faster designs begin to show some errors for frequencies with magnitude responses below -40dB.
}
\label{fig:lanczos1d-101-freq-plot}
\end{figure*}
\begin{figure*}[htbp]
\centering
\includegraphics[width=0.95\textwidth]{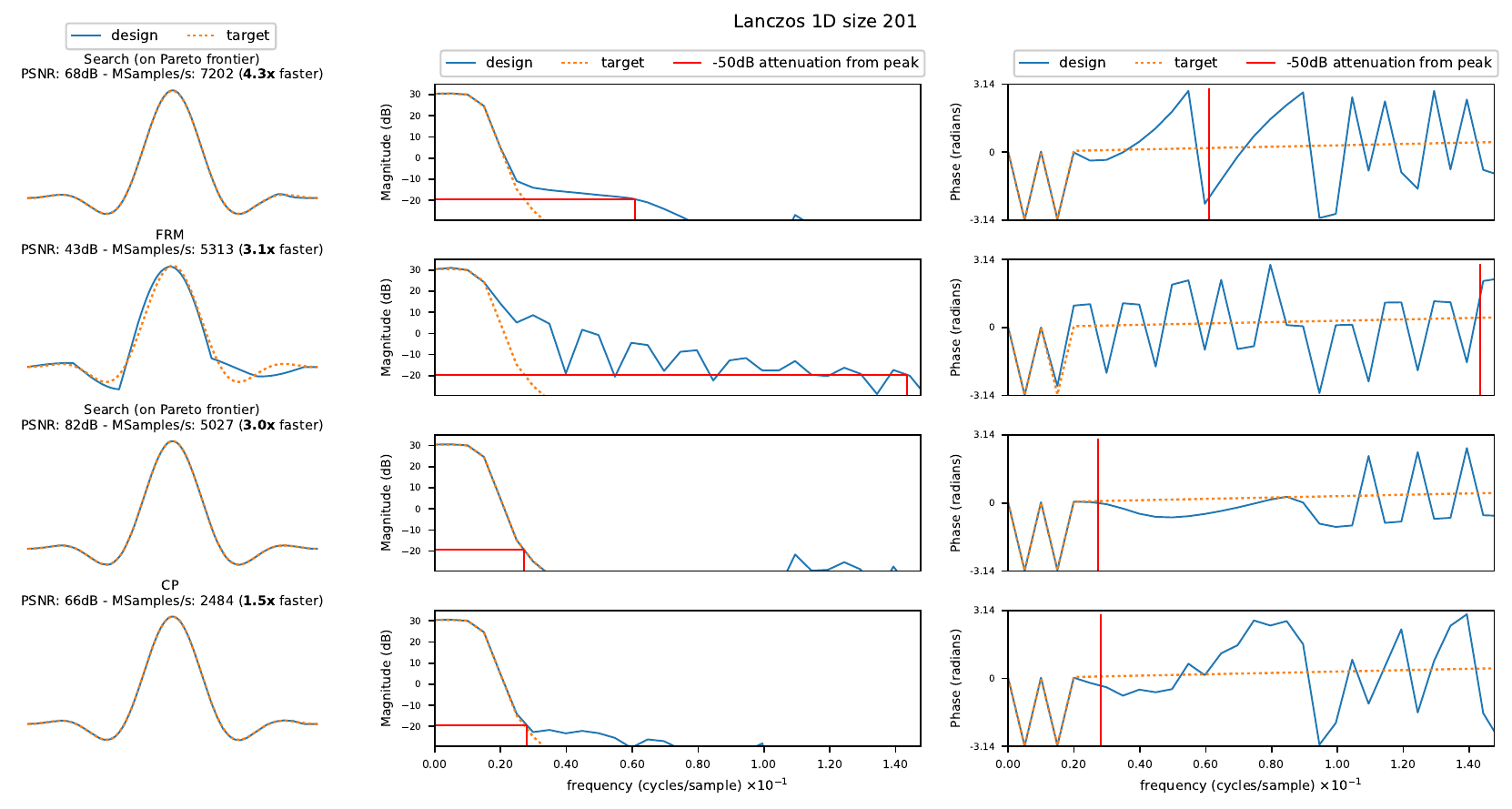}
\captionsetup{skip=1.5pt}
\caption{
}
\label{fig:lanczos1d-201-freq-plot}
\end{figure*}
\begin{figure*}[htbp]
\centering
\includegraphics[width=0.95\textwidth]{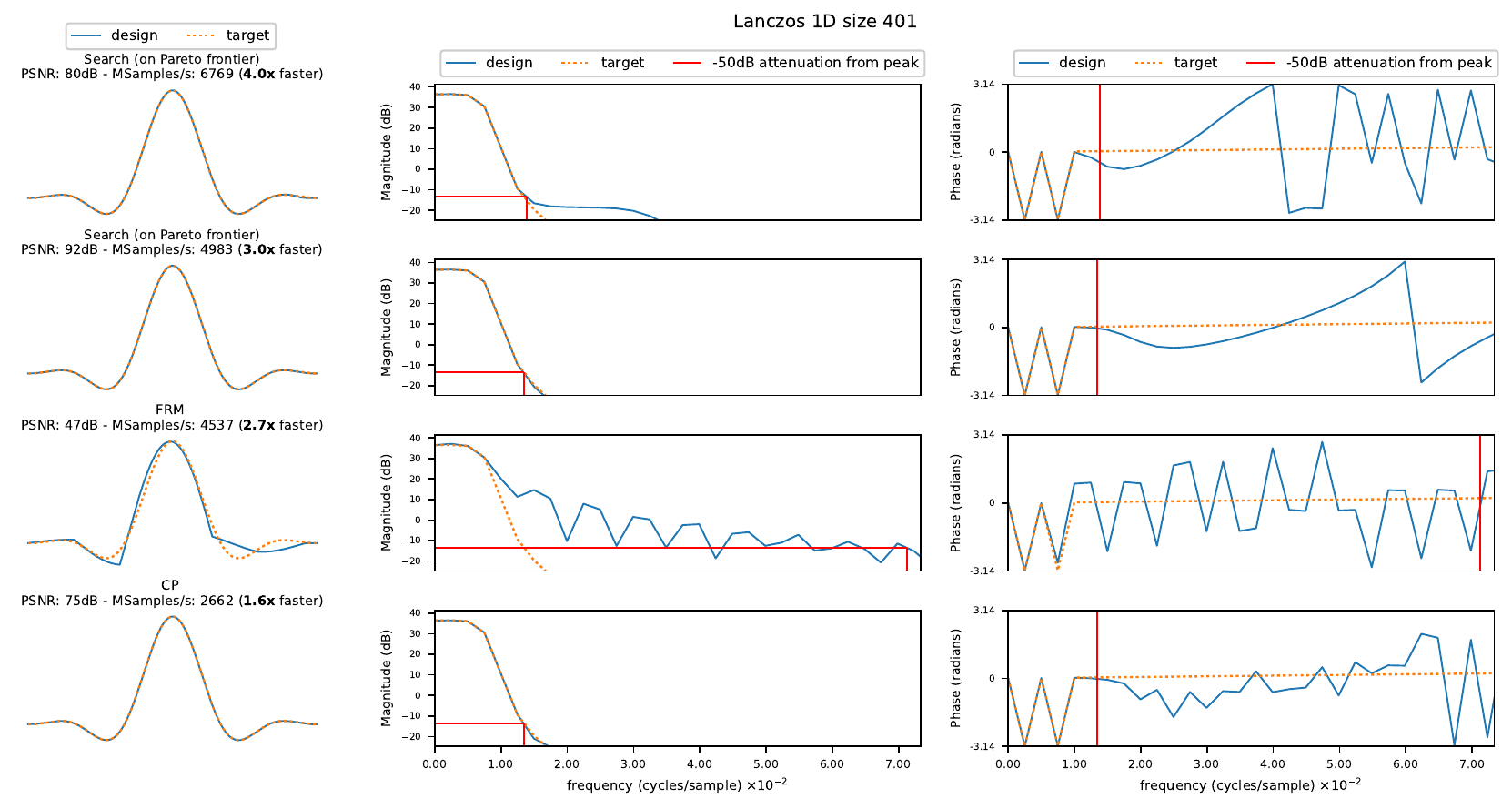}
\captionsetup{skip=1.5pt}
\caption{
}
\label{fig:lanczos1d-401-freq-plot}
\end{figure*}
\begin{figure*}[htbp]
\centering
\includegraphics[width=0.95\textwidth]{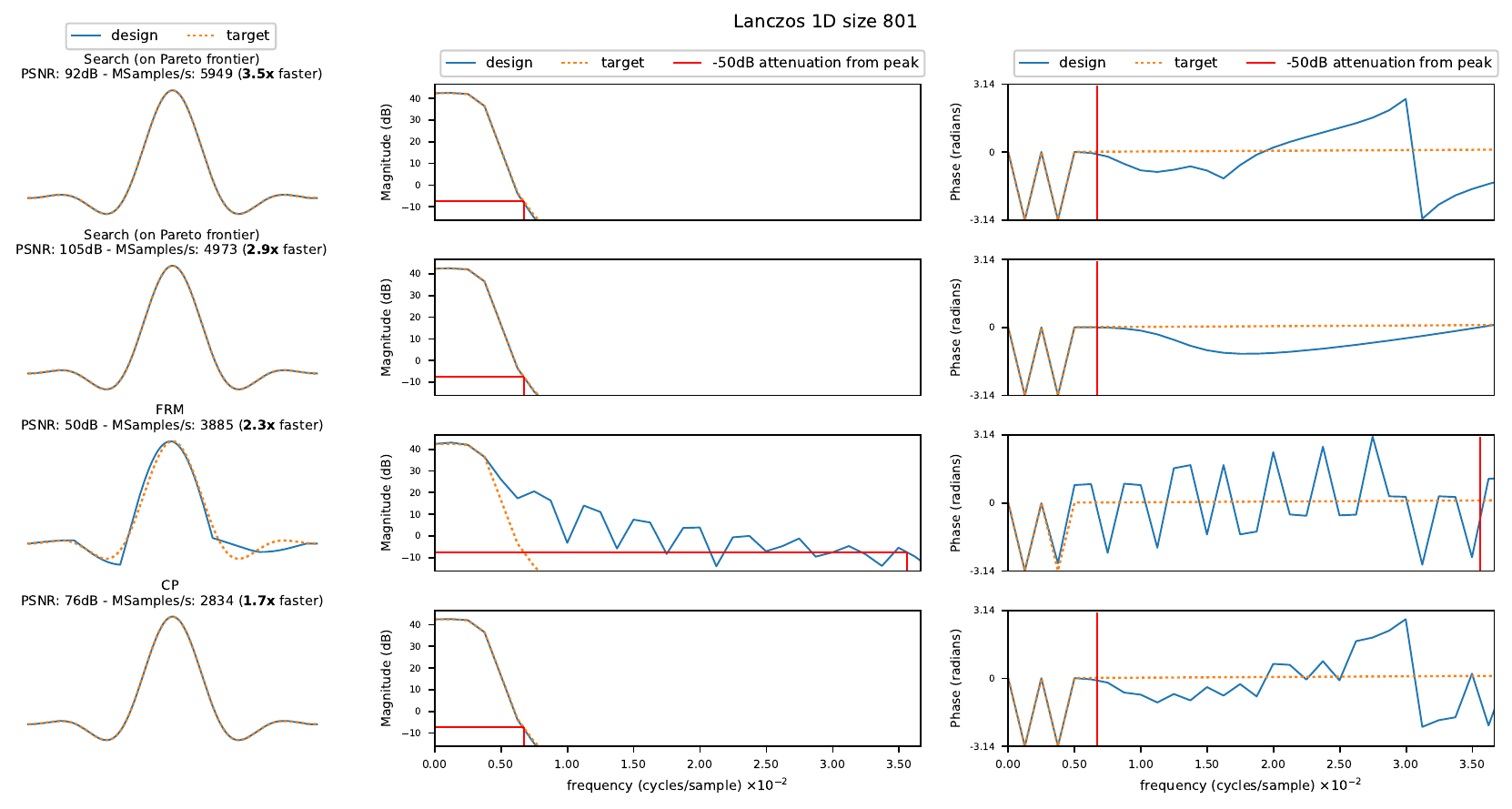}
\captionsetup{skip=1.5pt}
\caption{
}
\label{fig:lanczos1d-801-freq-plot}
\end{figure*}

\begin{figure*}[htbp]
\centering
\includegraphics[width=0.99\textwidth]{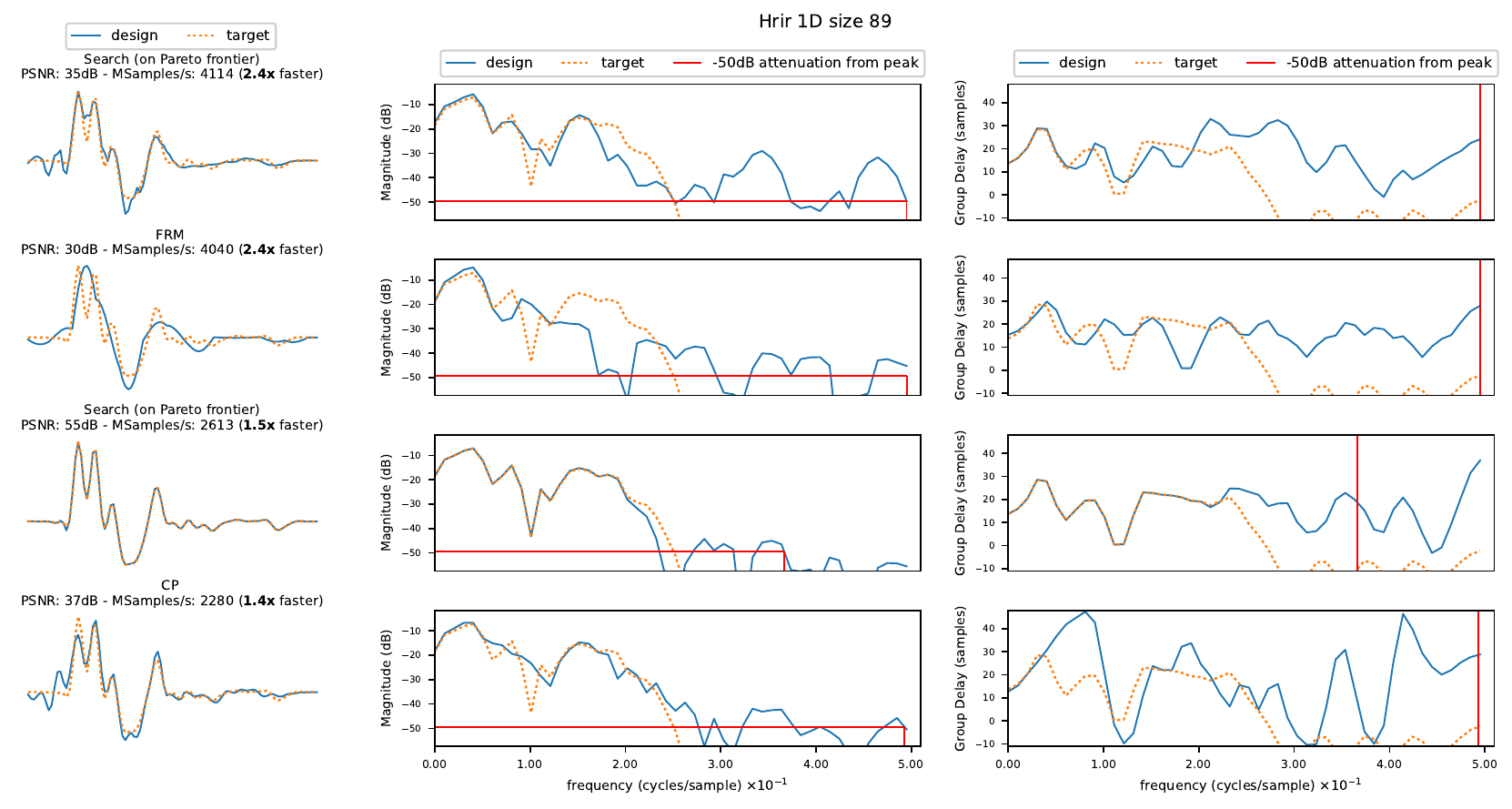}
\captionsetup{skip=1.5pt}
\caption{Figures 25-27 show the responses for each section of the right ear's HRIR. Figures 28-30 show the responses for the left ear's sections. The -50dB line is drawn relative to the peak magnitude response of the combined sections for the given ear. Because HRIRs are not linear phase filters, we show group delay in the third column, which is more meaningful for analyzing audio filters. HRIR filters are more challenging to fit with high accuracy because they have much wider pass-bands than Lanczos or Gaussian filters with non-uniform responses across the pass-band. For all sections our slower designs achieve significantly higher quality than FRM and CP at comparable speeds to the CP designs. Our faster designs have  throughputs that are competitive with or faster than FRM with significantly higher quality. While the FRM designs are faster than the CP ones, they often deviate drastically from the target response.}
\label{fig:hrir1d-89-freq-plot}
\end{figure*}
\begin{figure*}[htbp]
\centering
\includegraphics[width=0.99\textwidth]{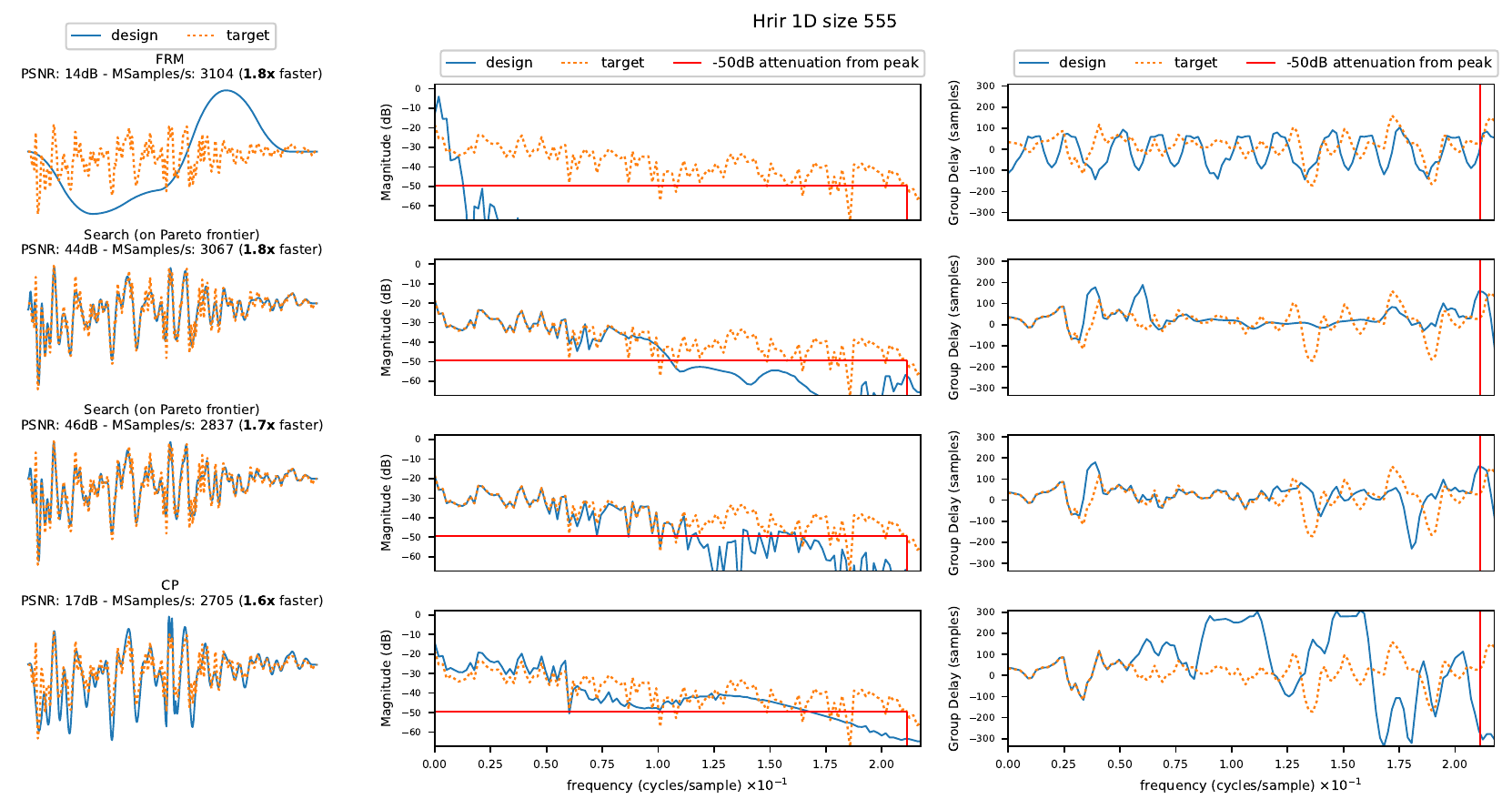}
\captionsetup{skip=1.5pt}
\caption{
}
\label{fig:hrir1d-555-freq-plot}
\end{figure*}
\begin{figure*}[htbp]
\centering
\includegraphics[width=0.99\textwidth]{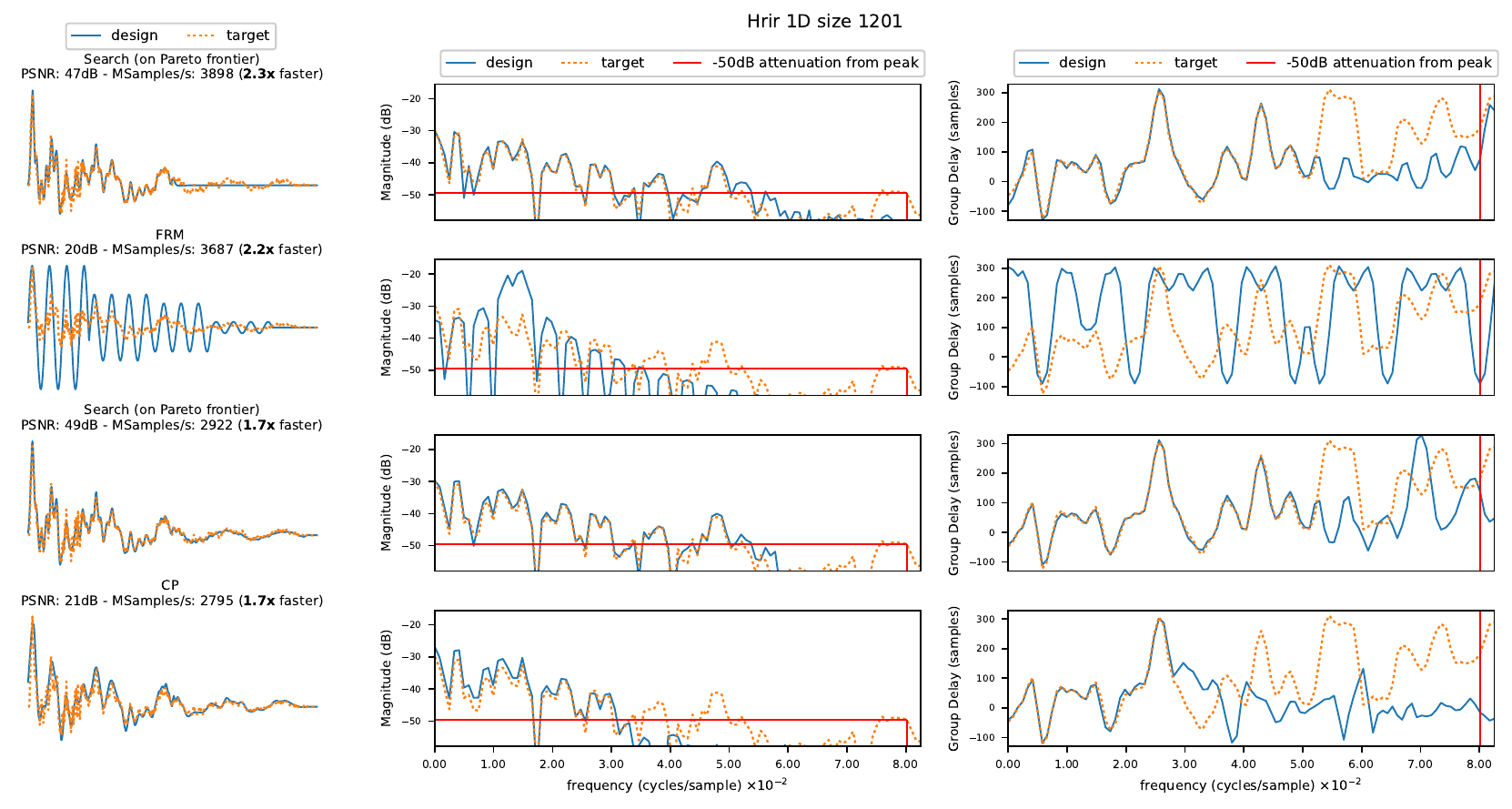}
\captionsetup{skip=1.5pt}
\caption{
}
\label{fig:hrir1d-1201-freq-plot}
\end{figure*}
\begin{figure*}[htbp]
\centering
\includegraphics[width=0.99\textwidth]{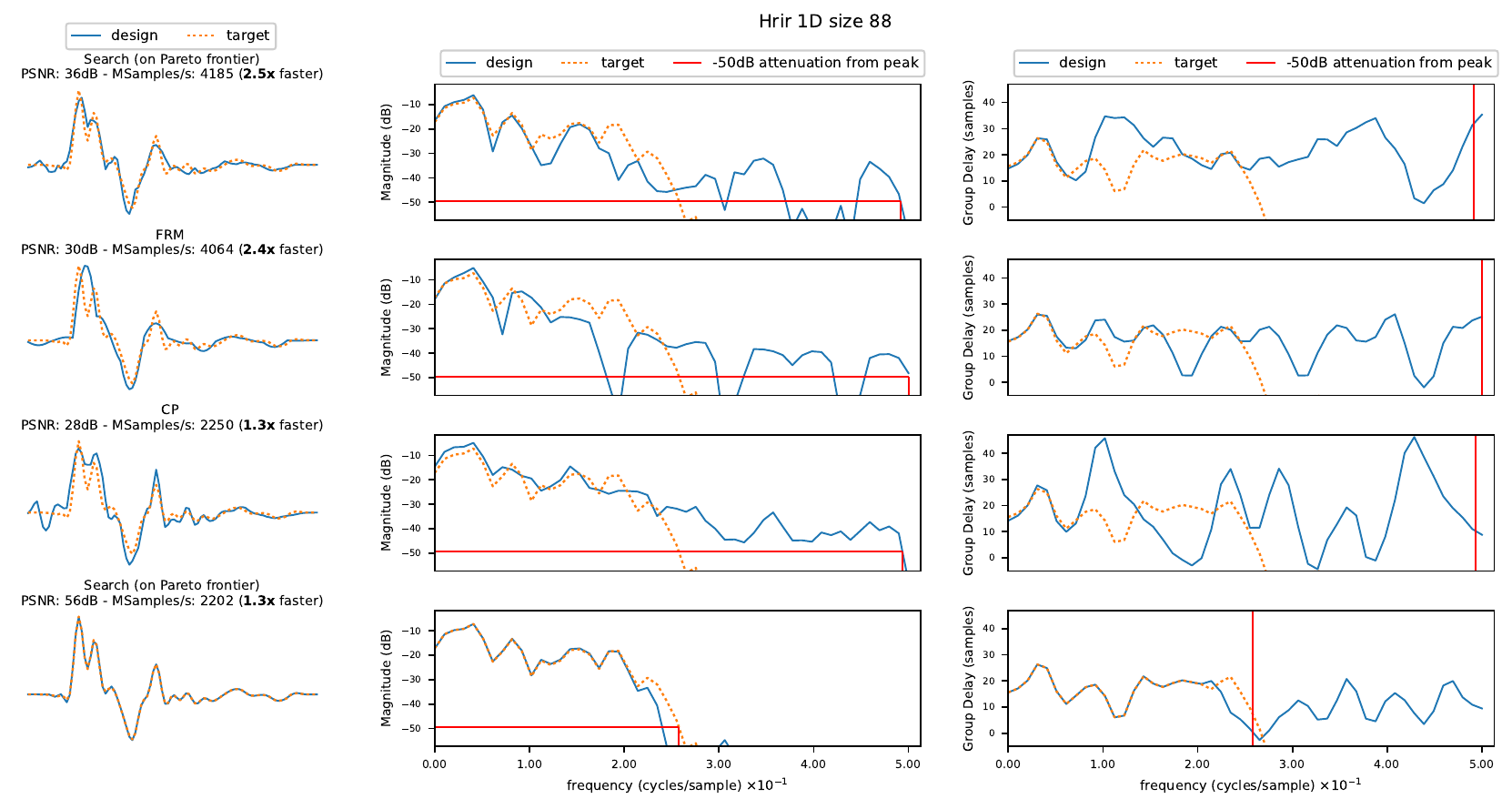}
\captionsetup{skip=1.5pt}
\caption{
}
\label{fig:hrir1d-88-freq-plot}
\end{figure*}
\begin{figure*}[htbp]
\centering
\includegraphics[width=0.99\textwidth]{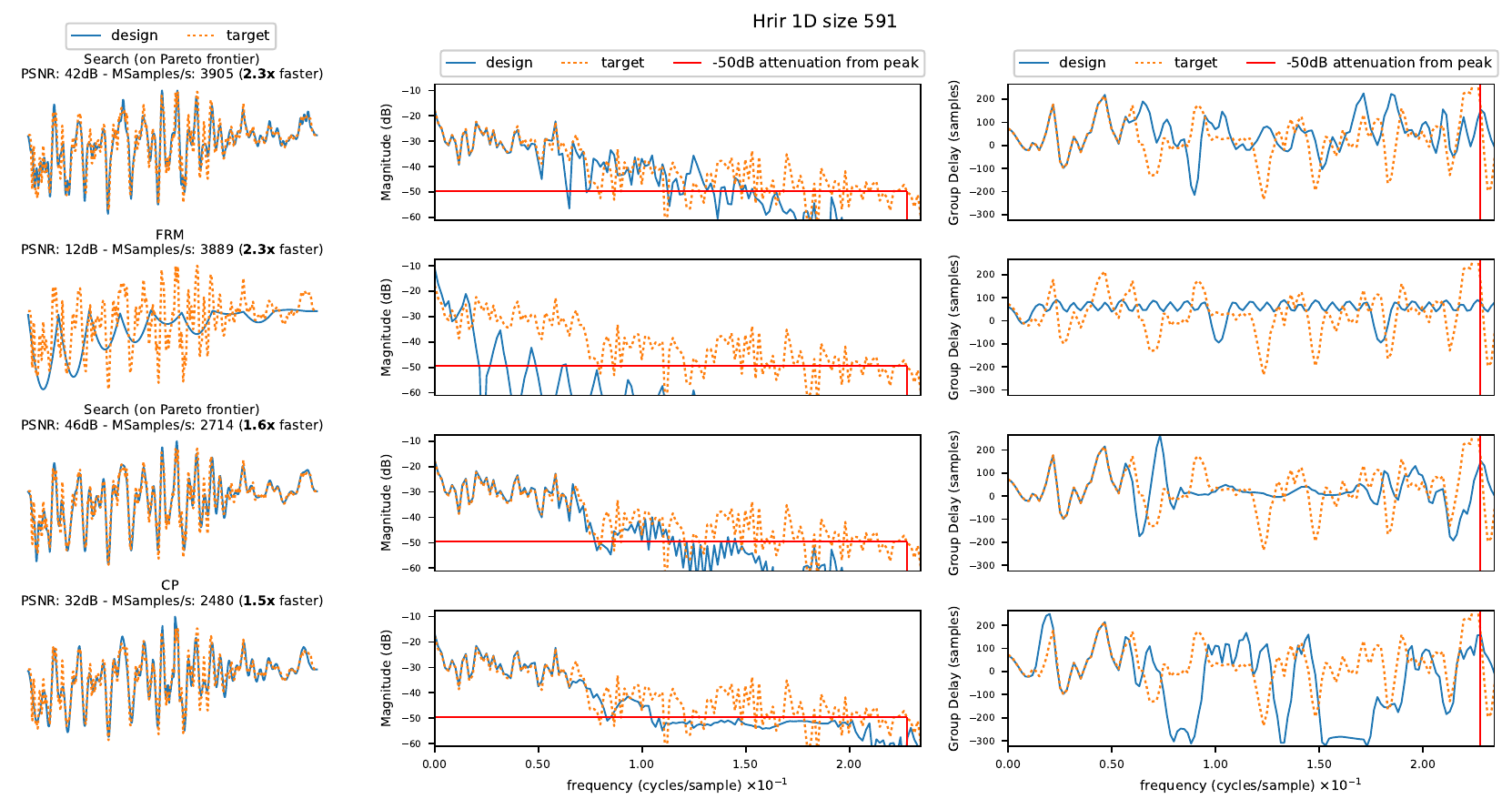}
\captionsetup{skip=1.5pt}
\caption{
}
\label{fig:hrir1d-591-freq-plot}
\end{figure*}
\begin{figure*}[htbp]
\centering
\includegraphics[width=0.99\textwidth]{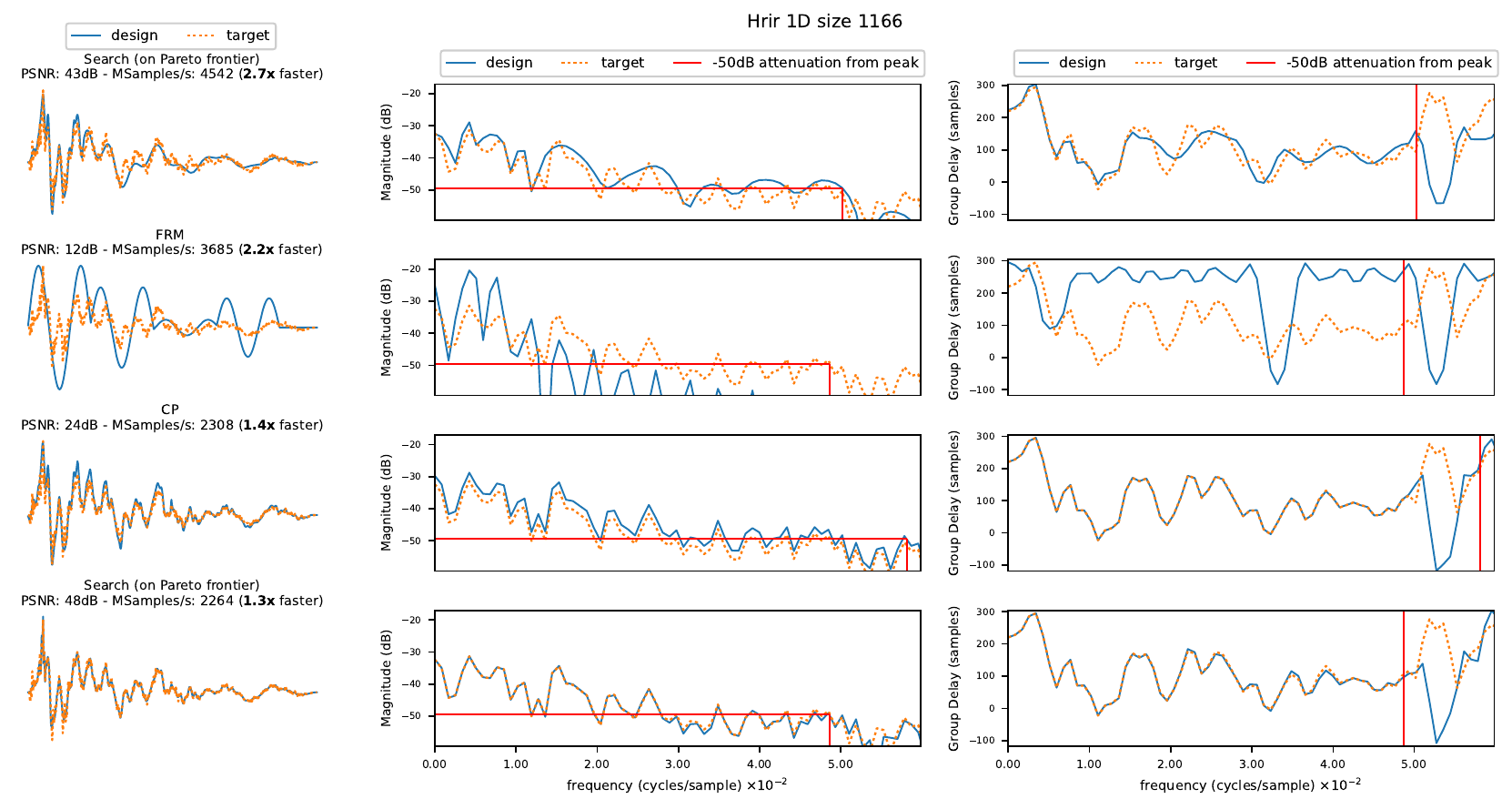}
\captionsetup{skip=1.5pt}
\caption{
}
\label{fig:hrir1d-1166-freq-plot}
\end{figure*}

\begin{figure*}[htbp]
\centering
\includegraphics[width=0.99\textwidth]{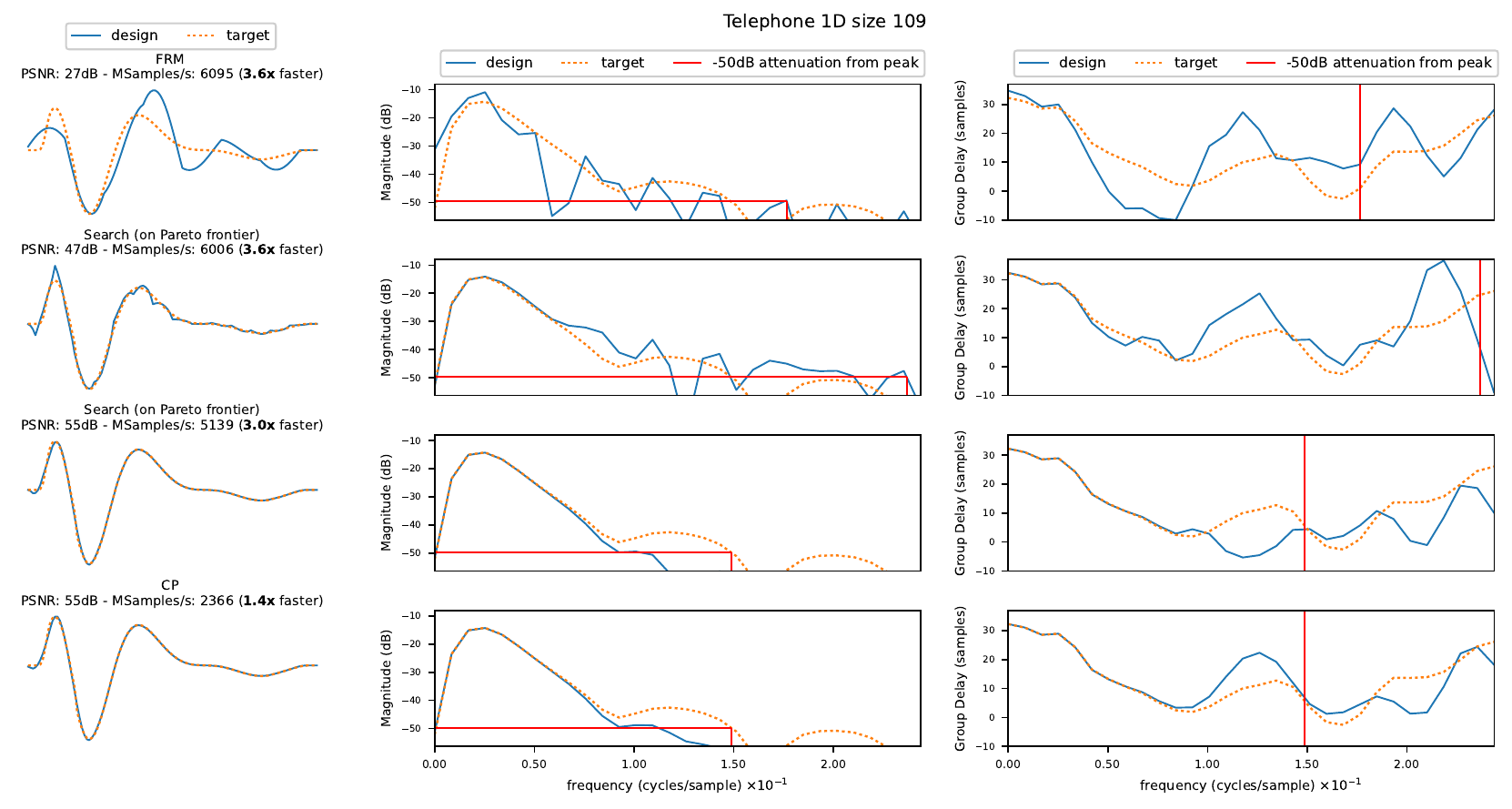}
\captionsetup{skip=1.5pt}
\caption{Figures 31-34 show the responses for each section of the Telephone filter. We show the group delay instead of the phase response because this is an audio filter and not a linear phase filter. The takeaways are similar to those for HRIR. Our faster designs have higher quality than the CP designs are comparable or higher throughputs, except for the shortest section where our higher quality design is >2x faster than CP at a similar quality. Our faster designs have throughputs that are competitive with or higher than FRM with significantly higher quality. In many cases, the FRM designs deviate drastically from the target response because high-throughput designs (ones that have large strides) result in narrow pass-bands.}
\label{fig:telephone1d-109-freq-plot}
\end{figure*}
\begin{figure*}[htbp]
\centering
\includegraphics[width=0.99\textwidth]{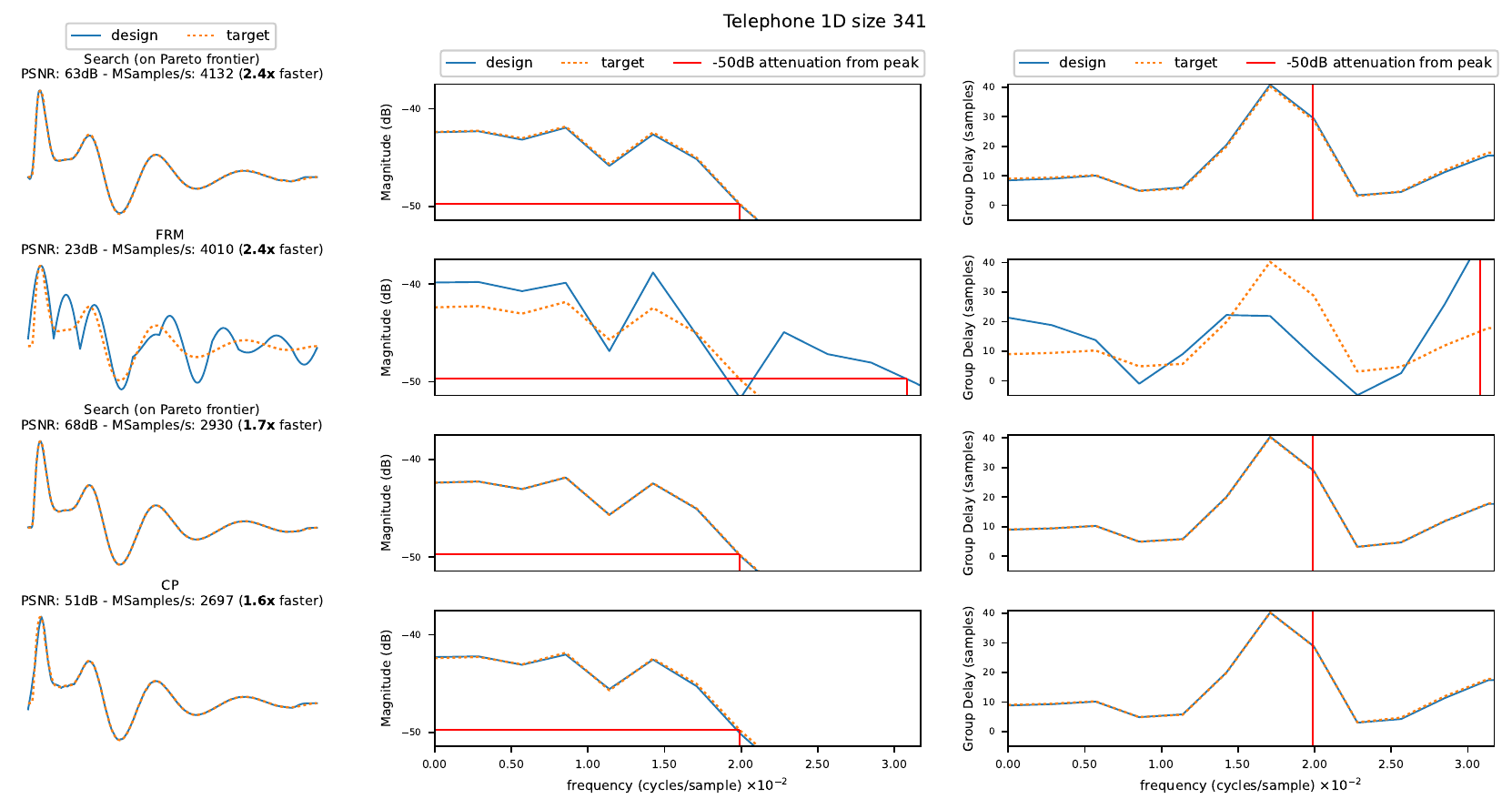}
\captionsetup{skip=1.5pt}
\caption{
}
\label{fig:telephone1d-341-freq-plot}
\end{figure*}
\begin{figure*}[htbp]
\centering
\includegraphics[width=0.99\textwidth]{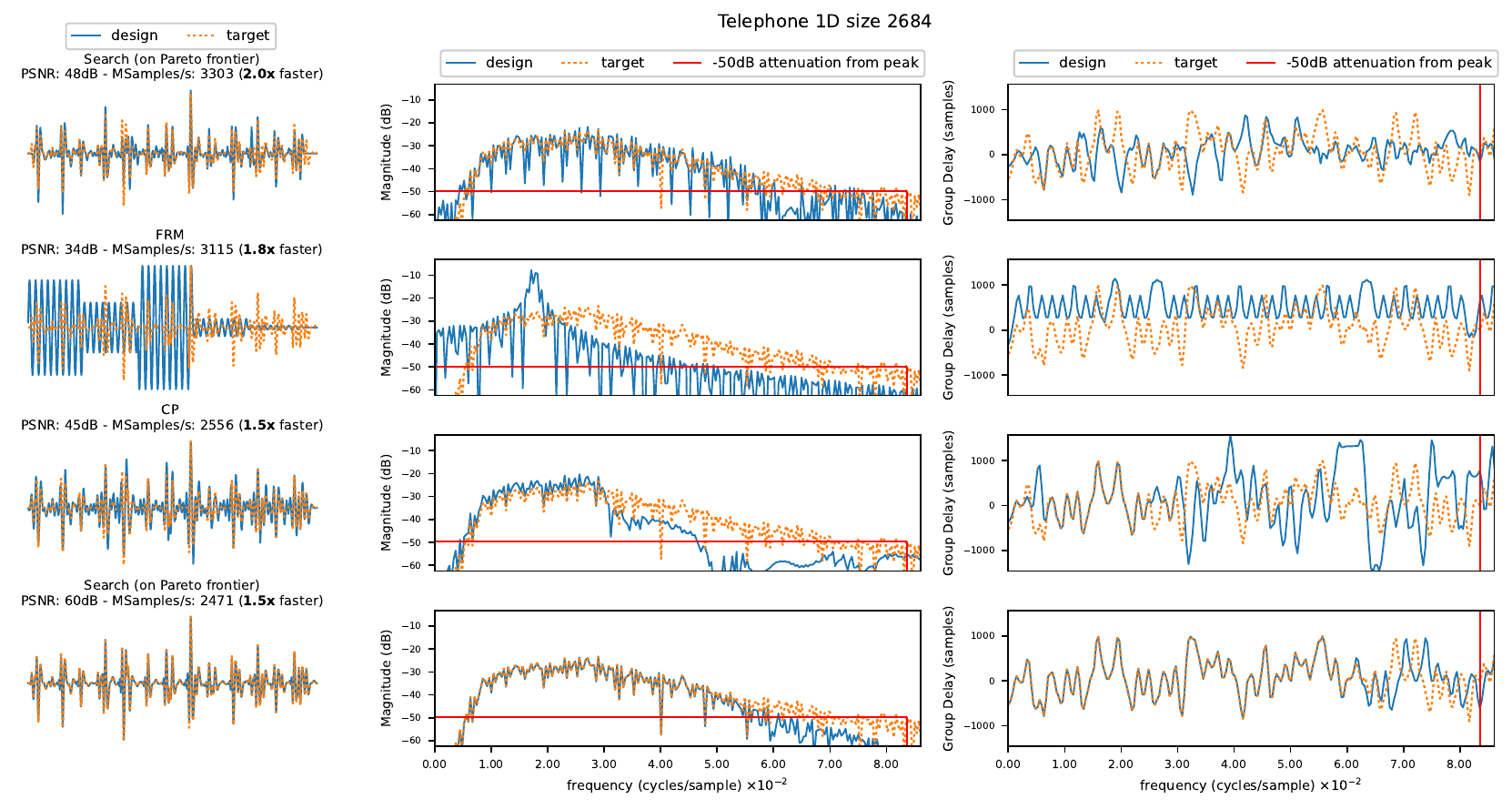}
\captionsetup{skip=1.5pt}
\caption{
}
\label{fig:telephone1d-2684-freq-plot}
\end{figure*}
\begin{figure*}[htbp]
\centering
\includegraphics[width=0.99\textwidth]{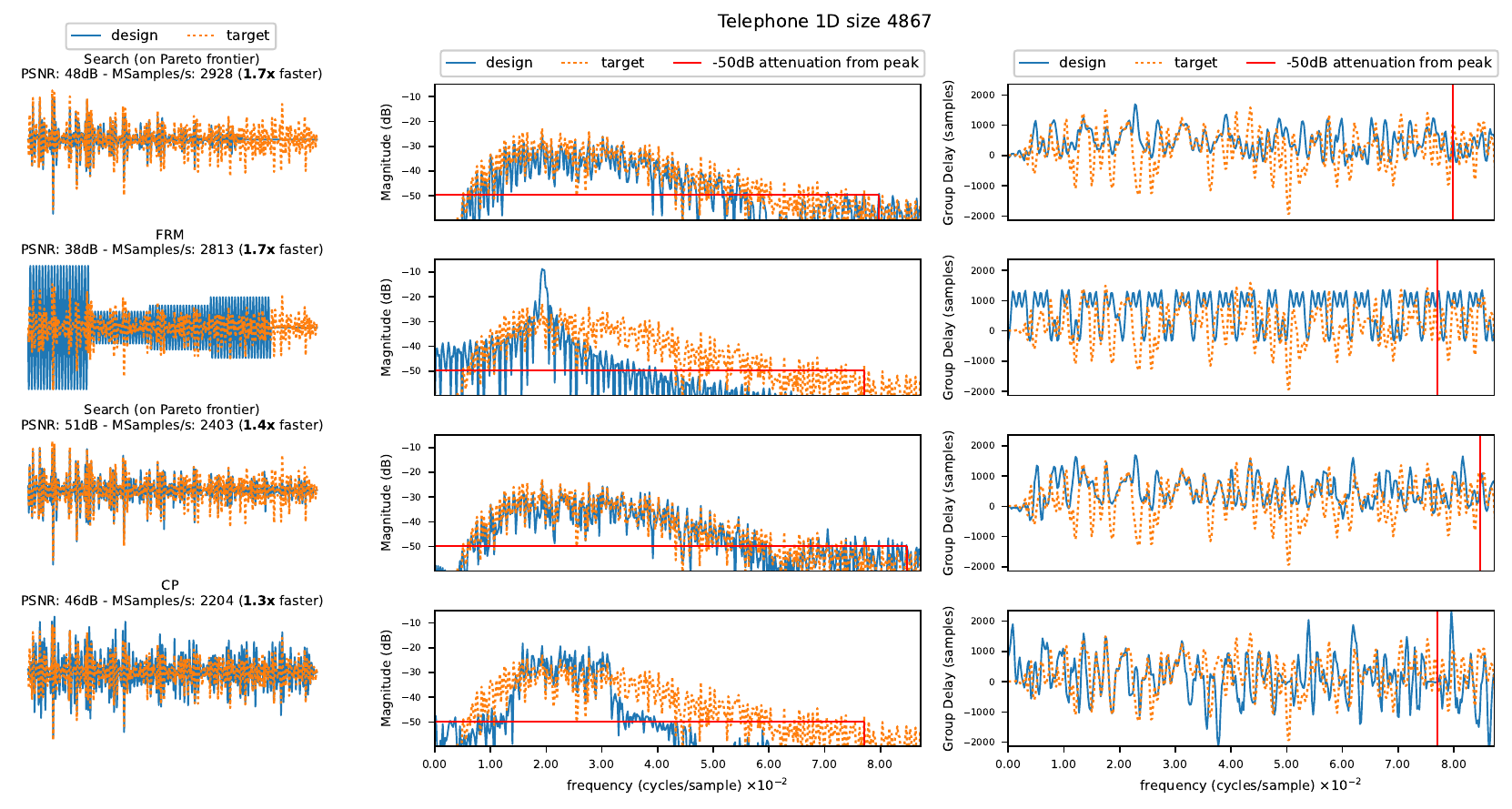}
\captionsetup{skip=1.5pt}
\caption{
}
\label{fig:telephone1d-4867-freq-plot}
\end{figure*}

\begin{figure*}[htbp]
\centering
\includegraphics[width=1.0\textwidth]{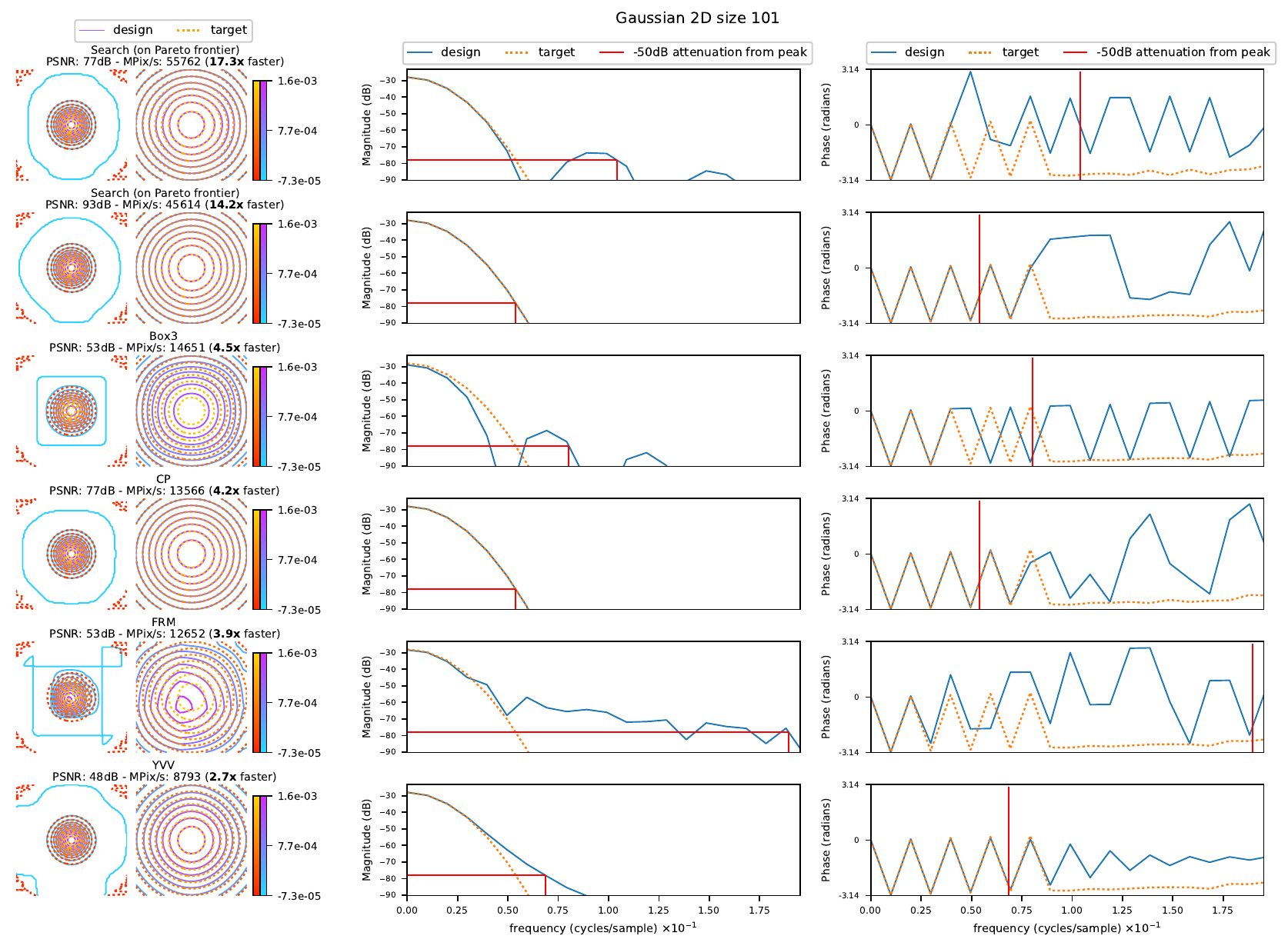}
\captionsetup{skip=1.5pt}
\caption{Figures 35-37 show the responses for 2D Gaussian sizes 101, 201, and 401. For all sizes, triple box blur has ringing and non-circular impulse response. It also has significantly less frequency content at the higher frequency region of the gaussian pass-band. For all sizes, FRM has large errors and is not linear phase. FRM in 2D is also no longer even competitive in terms of throughput because striding only divides the cost by the stride factor in each dimension. For all sizes, the CP designs are indistinguishable from the ground truth but they are up to 5x slower than our designs with higher quality than CP.}
\label{fig:gaussian2d-101-freq-plot}
\end{figure*}
\begin{figure*}[htbp]
\centering
\includegraphics[width=1.0\textwidth]{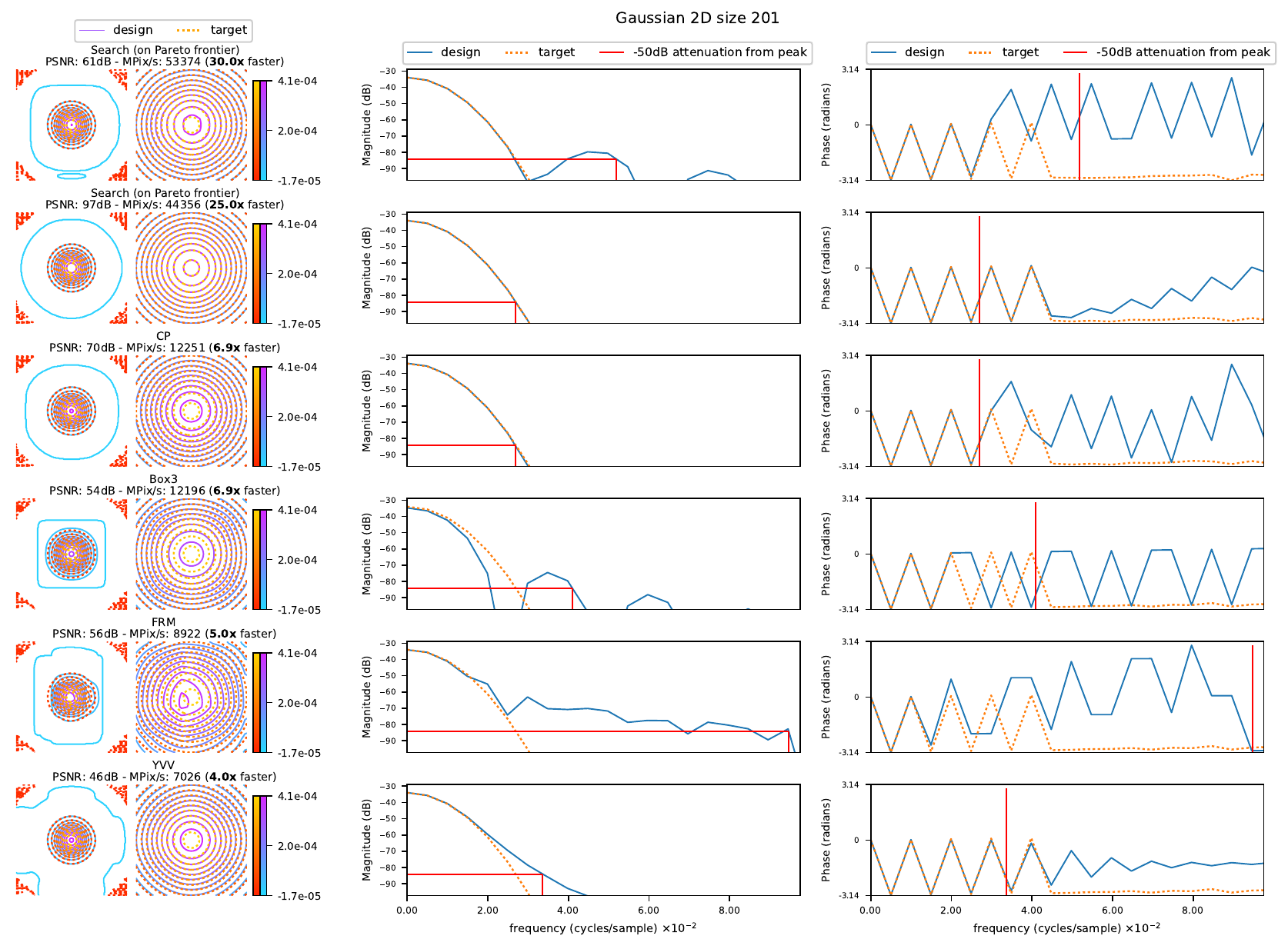}
\captionsetup{skip=1.5pt}
\caption{
}
\label{fig:gaussian2d-201-freq-plot}
\end{figure*}
\begin{figure*}[htbp]
\centering
\includegraphics[width=1.0\textwidth]{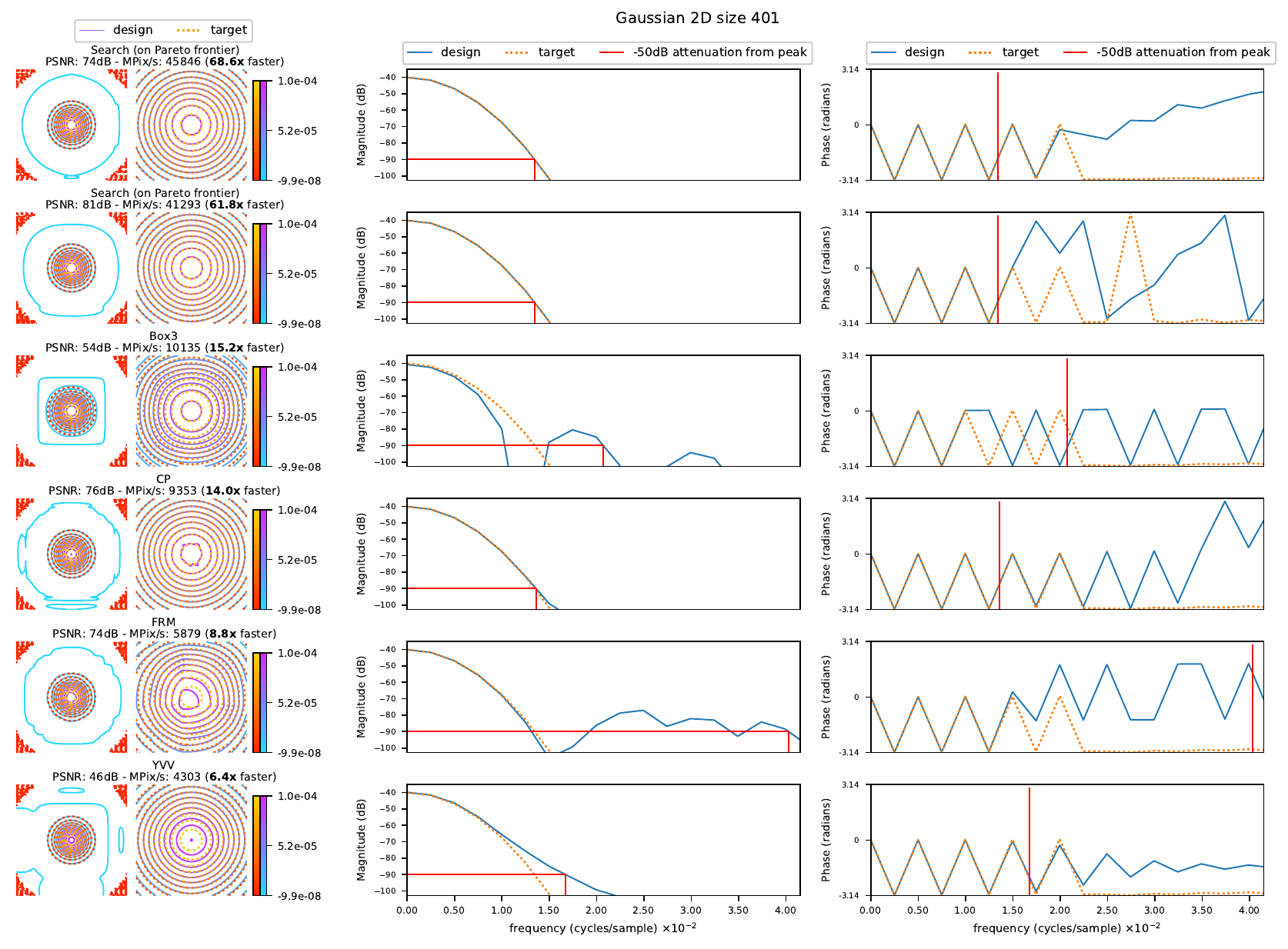}
\captionsetup{skip=1.5pt}
\caption{
}
\label{fig:gaussian2d-401-freq-plot}
\end{figure*}

\begin{figure*}[htbp]
\centering
\includegraphics[width=0.89\textwidth]{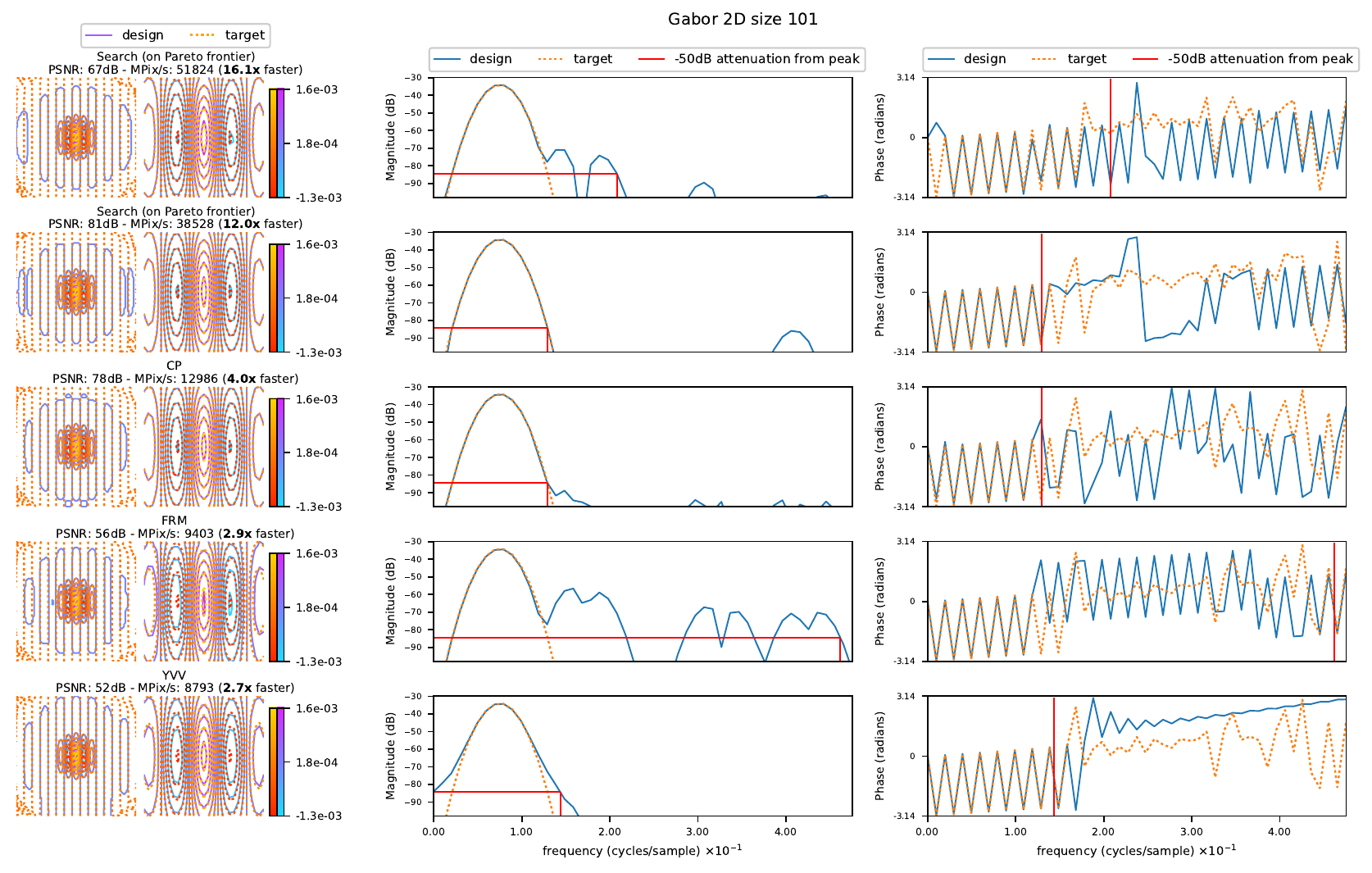}
\captionsetup{skip=1.5pt}
\caption{Figures 38-40 are the response plots for 2D Gabor sizes 101, 201, and 401. CP struggles to get perfect approximations up to -50dB because Gabor filters have oscillations that cannot be aggressively downsampled. For all sizes at least one of our designs is indistinguishable from ground truth beyond -50dB while being up to $\sim$ 5x faster than CP. All the FRM designs have noticeably non-linear phase and ringing. They have worse throughput than CP because FRM throughputs only scale down by 1/stride per dimension. YVV has the lowest throughput because it does not use multi-resolution filtering and requires anti-causal IIRs. Note that these throughputs are over-estimates because Gabor filters require complex-valued IIRs, but we used the timings for real-valued IIRs since complex IIRs are not currently supported by our lowering system. YVV has lower PSNRs than FRM and CP but it may have preferable error behavior compared to these other baselines since it has linear phase, smooth errors, and the error does not extend significantly beyond the desired pass-band.}
\label{fig:gabor1d-101-freq-plot}
\end{figure*}
\begin{figure*}[htbp]
\centering
\includegraphics[width=0.89\textwidth]{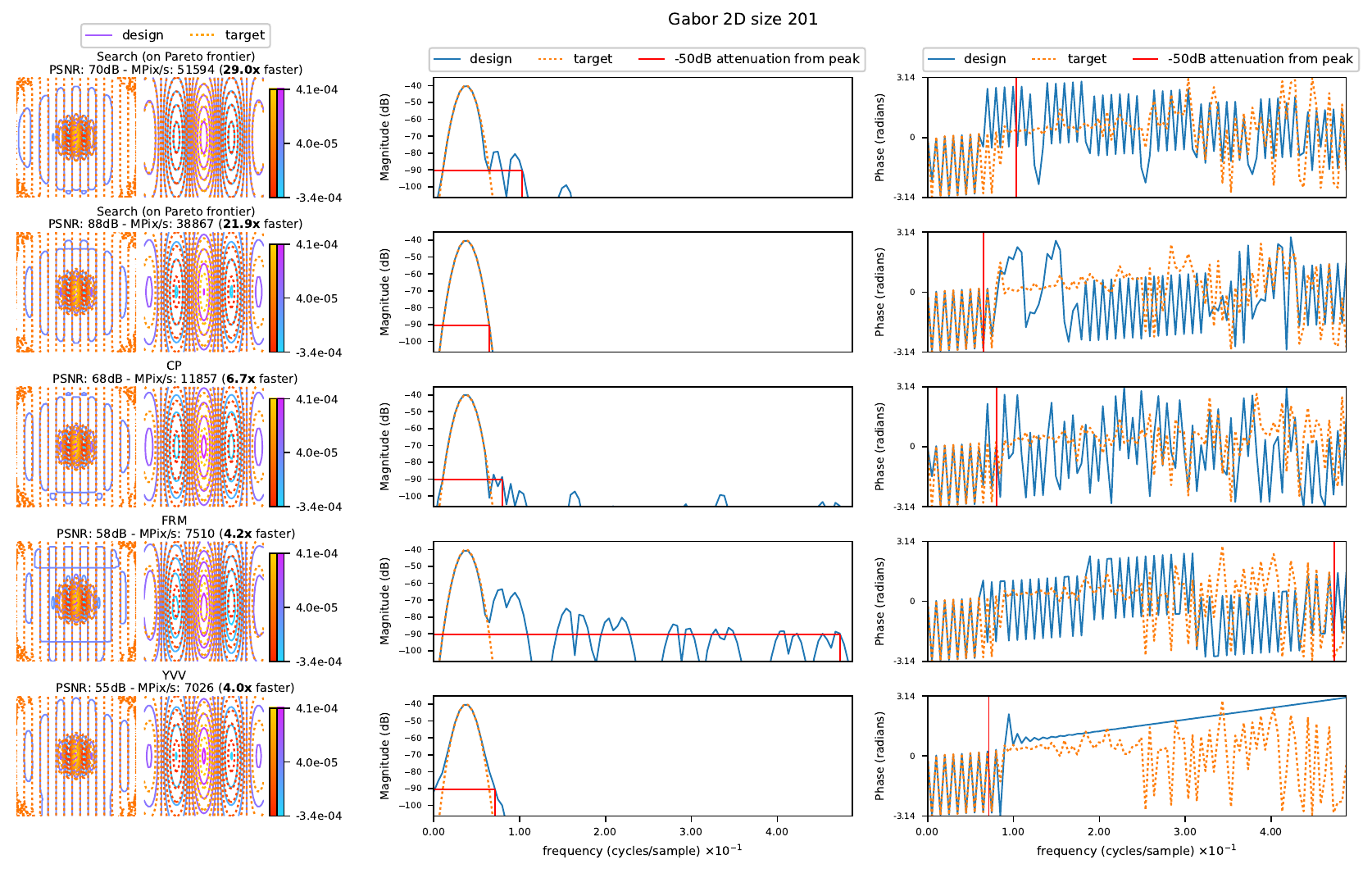}
\captionsetup{skip=1.5pt}
\caption{
}
\label{fig:gabor1d-201-freq-plot}
\end{figure*}
\begin{figure*}[htbp]
\centering
\includegraphics[width=0.89\textwidth]{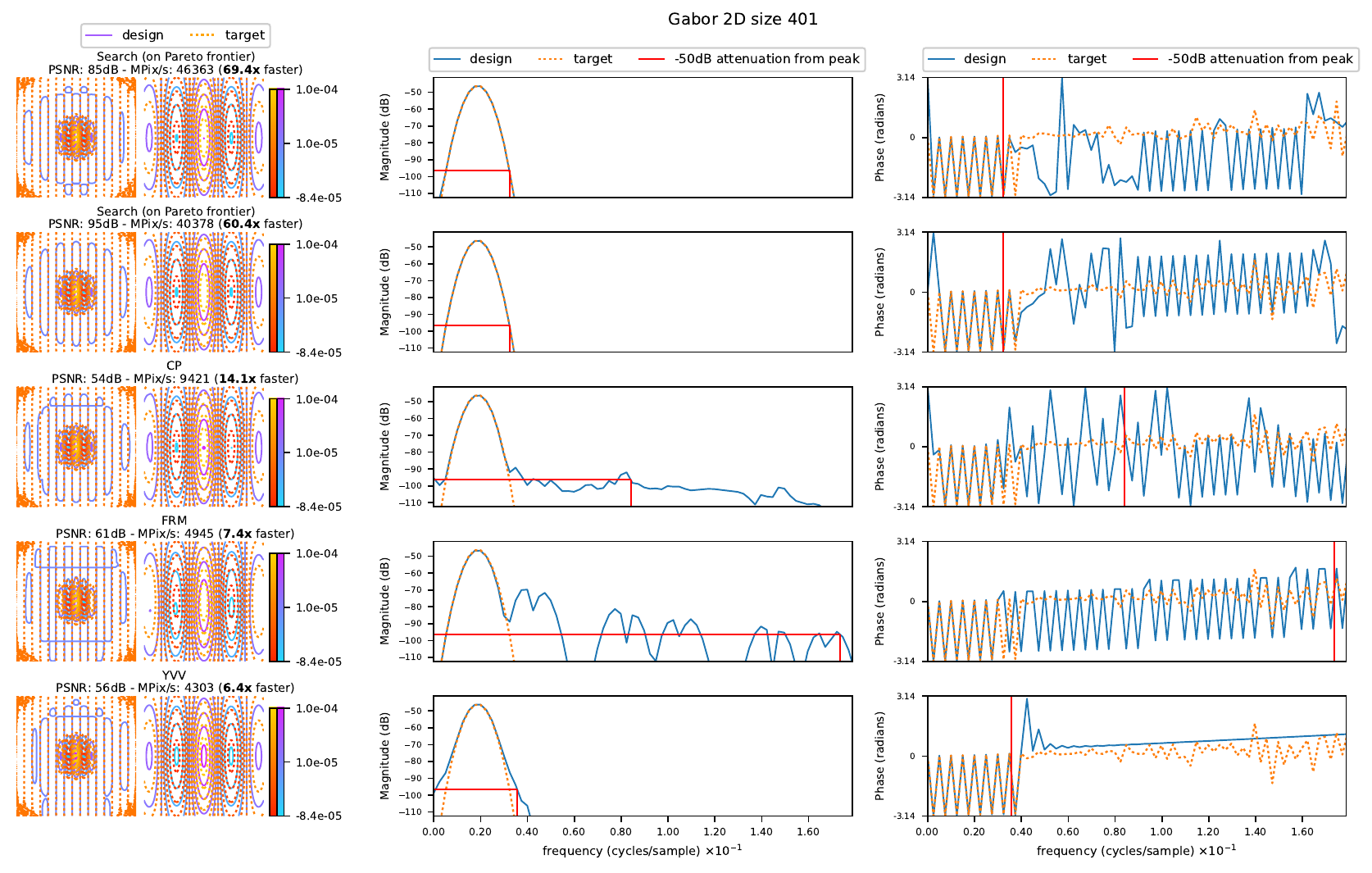}
\captionsetup{skip=1.5pt}
\caption{
}
\label{fig:gabor1d-401-freq-plot}
\end{figure*}

\begin{figure*}[htbp]
\centering
\includegraphics[width=0.98\textwidth]{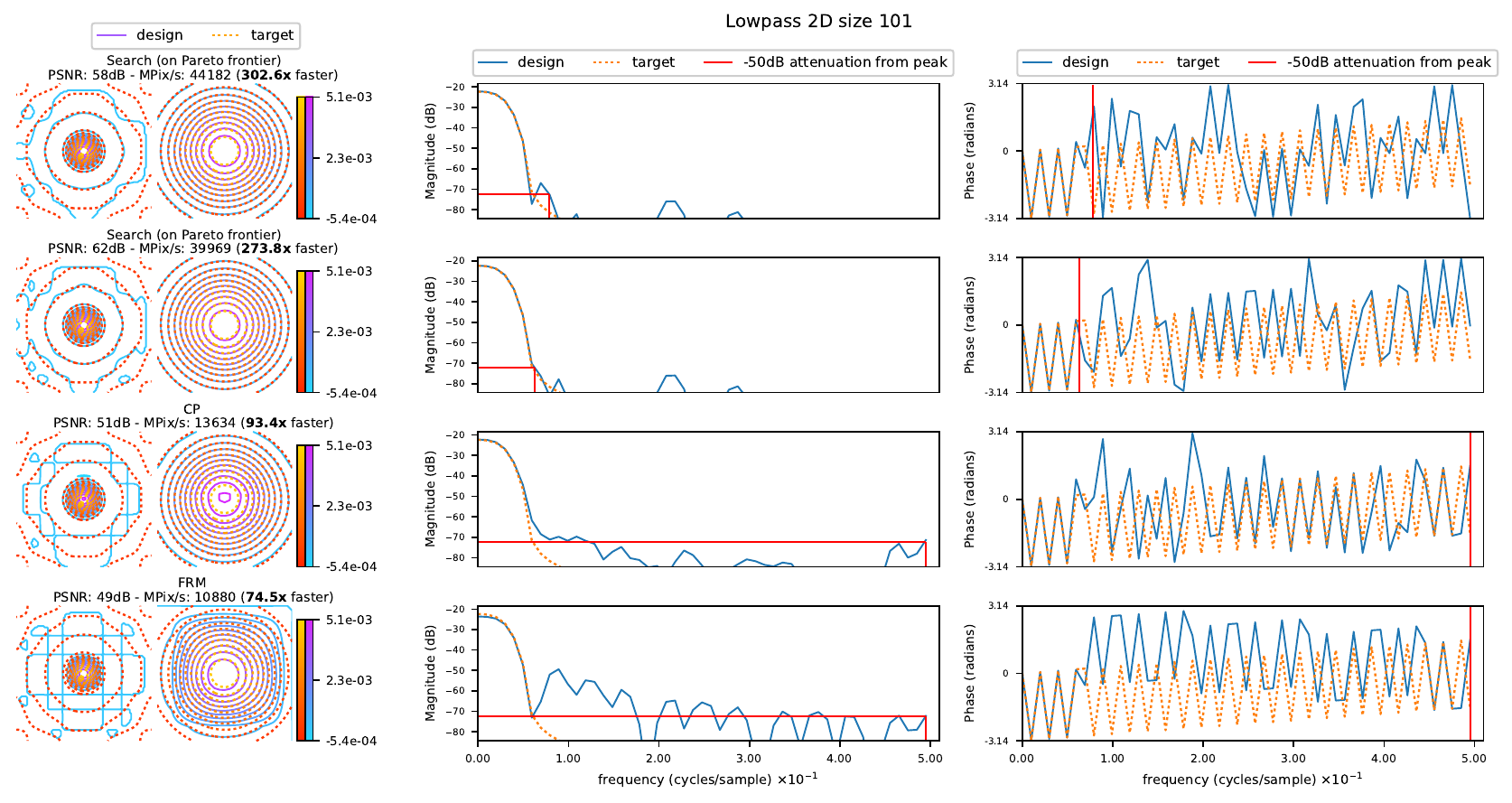}
\captionsetup{skip=1.5pt}
\caption{Figures 41-43 show the response plots for 2D Low-pass filters sizes 101, 201, and 401. For all sizes both of our designs have comparble or higher quality than CP while being up to 4.8x faster. Our higher quality designs are indistiguishable from the ground truth with linear phase up to or beyond -50dB suppression. The CP designs have long tails of unwanted high frequency content, because our Low-pass filters have lobes, which are hard to perfectly capture with aggressive downsampling. The FRM designs again have the lowest quality and throughputs. They are not linear phase and have high magnitude ringing. 
}
\label{fig:lowpass2d-101-freq-plot}
\end{figure*}
\begin{figure*}[htbp]
\centering
\includegraphics[width=0.98\textwidth]{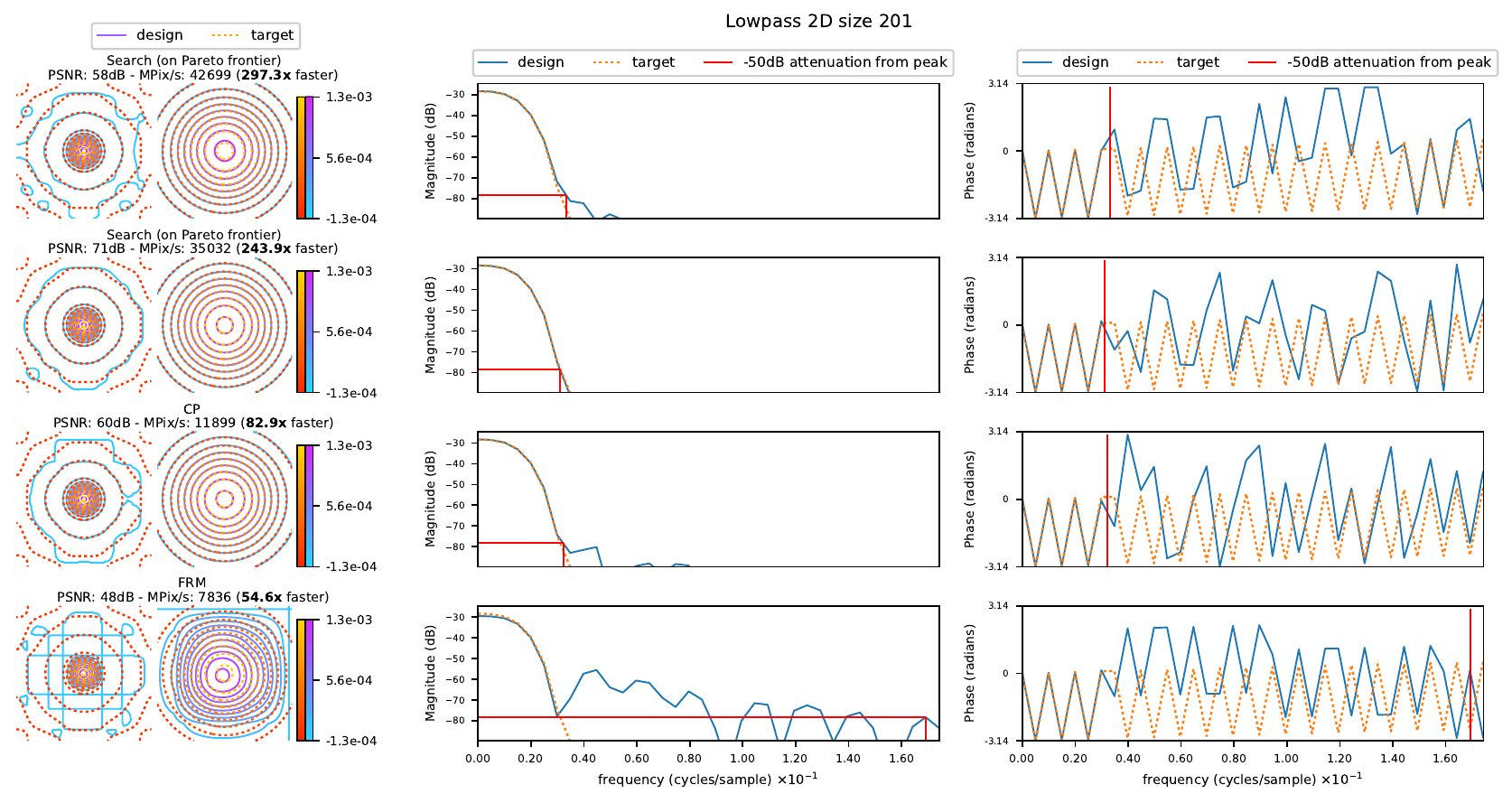}
\captionsetup{skip=1.5pt}
\caption{
}
\label{fig:lowpass2d-201-freq-plot}
\end{figure*}
\begin{figure*}[htbp]
\centering
\includegraphics[width=0.98\textwidth]{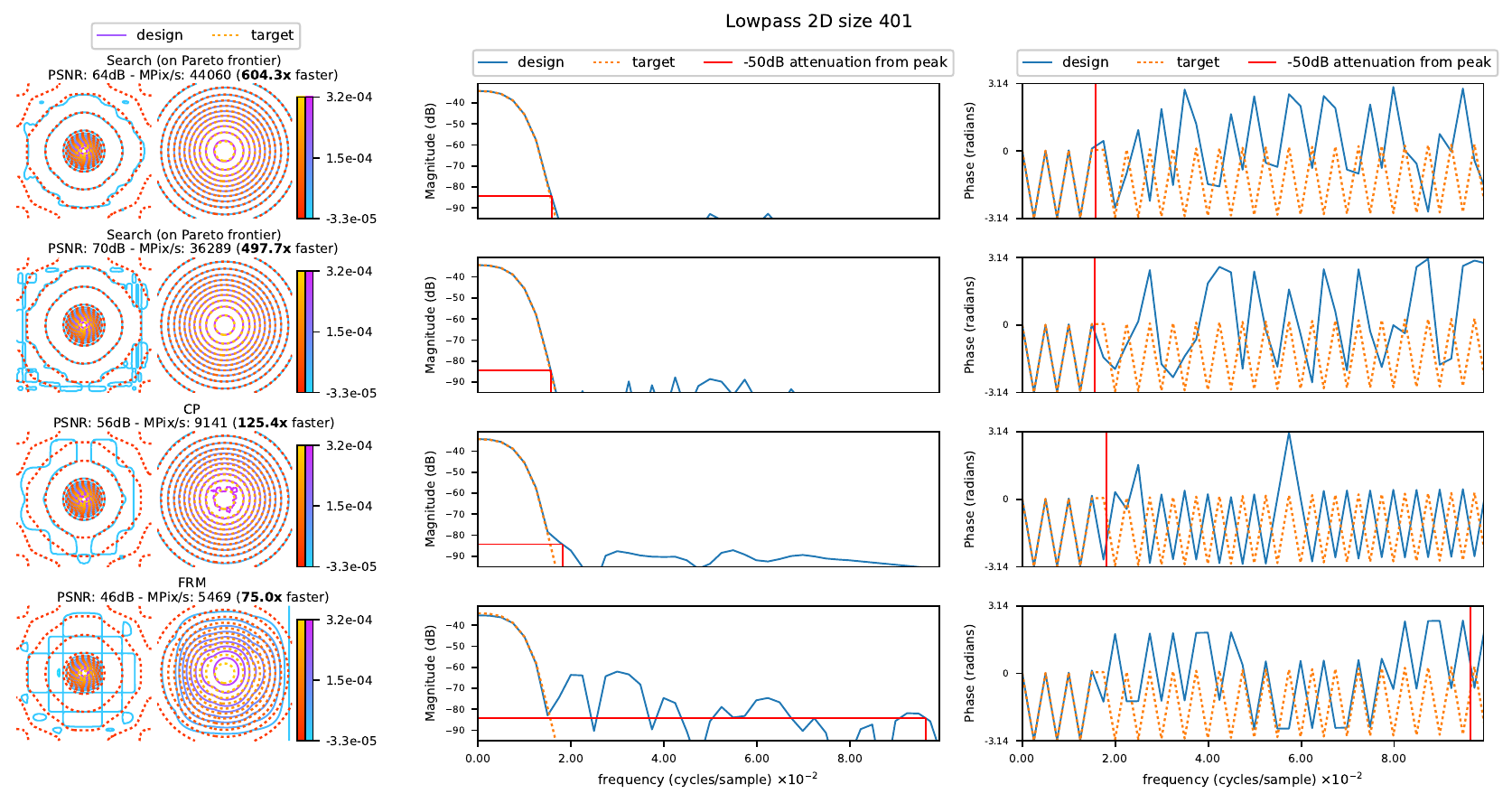}
\captionsetup{skip=1.5pt}
\caption{
}
\label{fig:lowpass2d-401-freq-plot}
\end{figure*}

\subsection{Qualitative outputs on image patches}
Figures 45 - 61 show the outputs of our Pareto-dominant models compared to the baselines for each 2D target filter and size on different image patches taken from the authors' iPhone 13 Pro camera library or raw photographs from the PIXLS.US raw database \cite{pixlsraw} demosaicked with the AMaZE algorithm from RawTherapee \cite{rawtherapee}. 

Across all targets, FRM produces noticeable ringing artifacts. For 2D Gaussians, the triple-box blur over-blurs and YVV under-blurs the input. CP's design structure is most-suited for approximating low-pass filters with gradually decaying responses, so for 2D Gaussians, both CP's outputs and our approximations produce outputs that are visually indistinguishable from ground truth. However, our approximations are several factors faster than all the baselines, ranging from 4.1x faster to 4.9x faster than CP and 3.8x to 4.5x faster than triple-box blur depending on the filter size.

Unlike 2D Gaussians, the 2D Lowpass targets have a more shelf-like frequency response. CP has a harder time approximating filters with steep response drop-offs and produces subtle thin banding artifacts. Our approximations' outputs are visually indistinguishable from the ground truth and are 2.9x up to 4x faster than CP and 4.1x to 6.6x faster than FRM.  

For the Gabor targets, YVV under-blurs but the effect is not visibly noticeable. FRM's ringing artifacts appear as diagonal or vertical bands. The Gabor frequency response is a translated Gaussian, with its peak frequency located halfway between 0 and the band-limit, instead of at 0. This means that the frequency response must drop off much more steeply (like with the lowpass targets), which is harder to do with CP's small prefilters and FIRs. Larger target sizes require more pyramid levels and the error from suboptimal prefiltering compounds with  each level. The CP ringing artifacts are most noticeable in the size 401 outputs as vertical bands. 

\begin{figure*}[p]\centering
\begin{tabular}{cc}
\begin{minipage}[t]{2.500in}\centering \includegraphics[interpolate=false,height=2.500in]{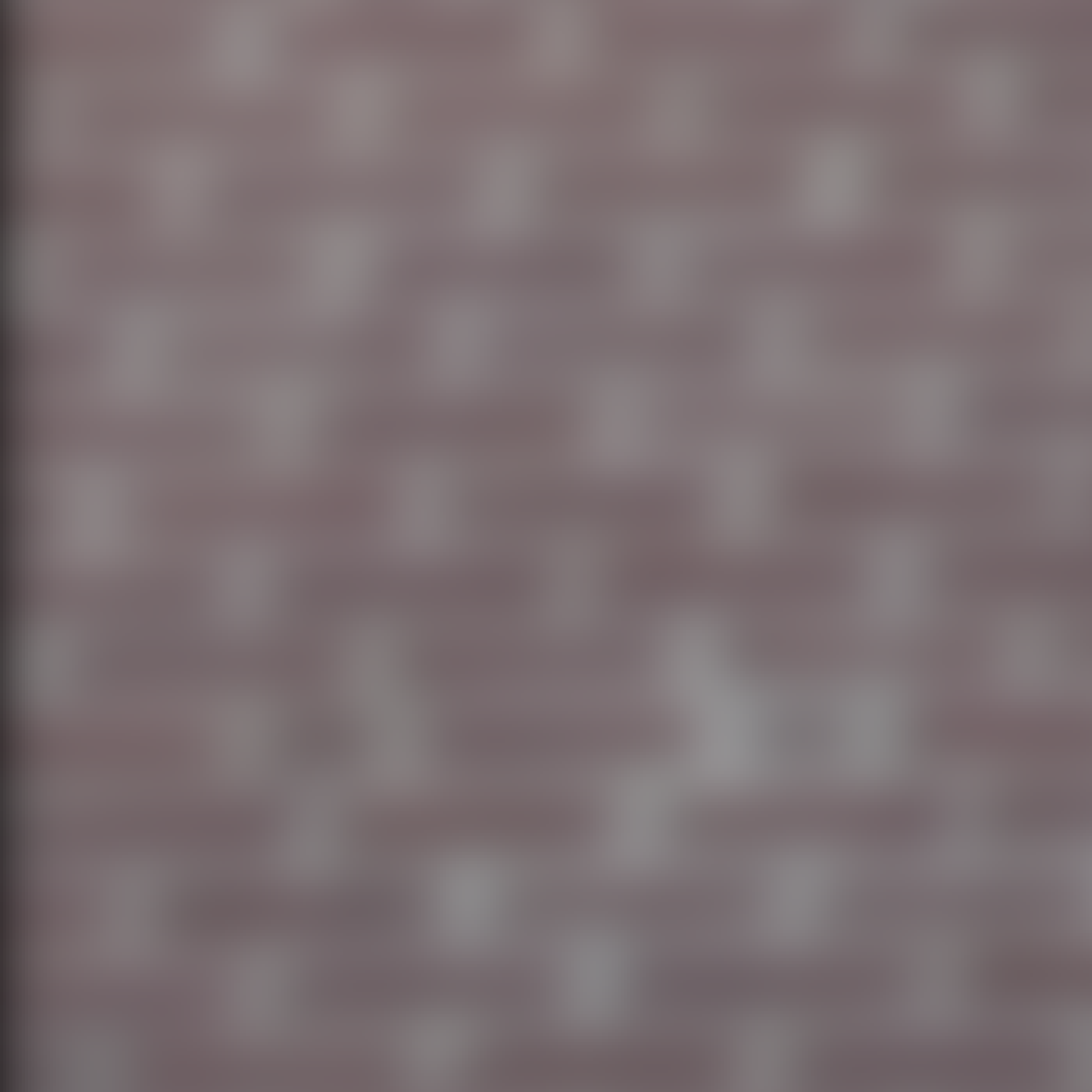}\\[1pt]
\mbox{\fontsize{9pt}{11pt}\selectfont Ground truth}\end{minipage} & \begin{minipage}[t]{2.500in}\centering \includegraphics[interpolate=false,height=2.500in]{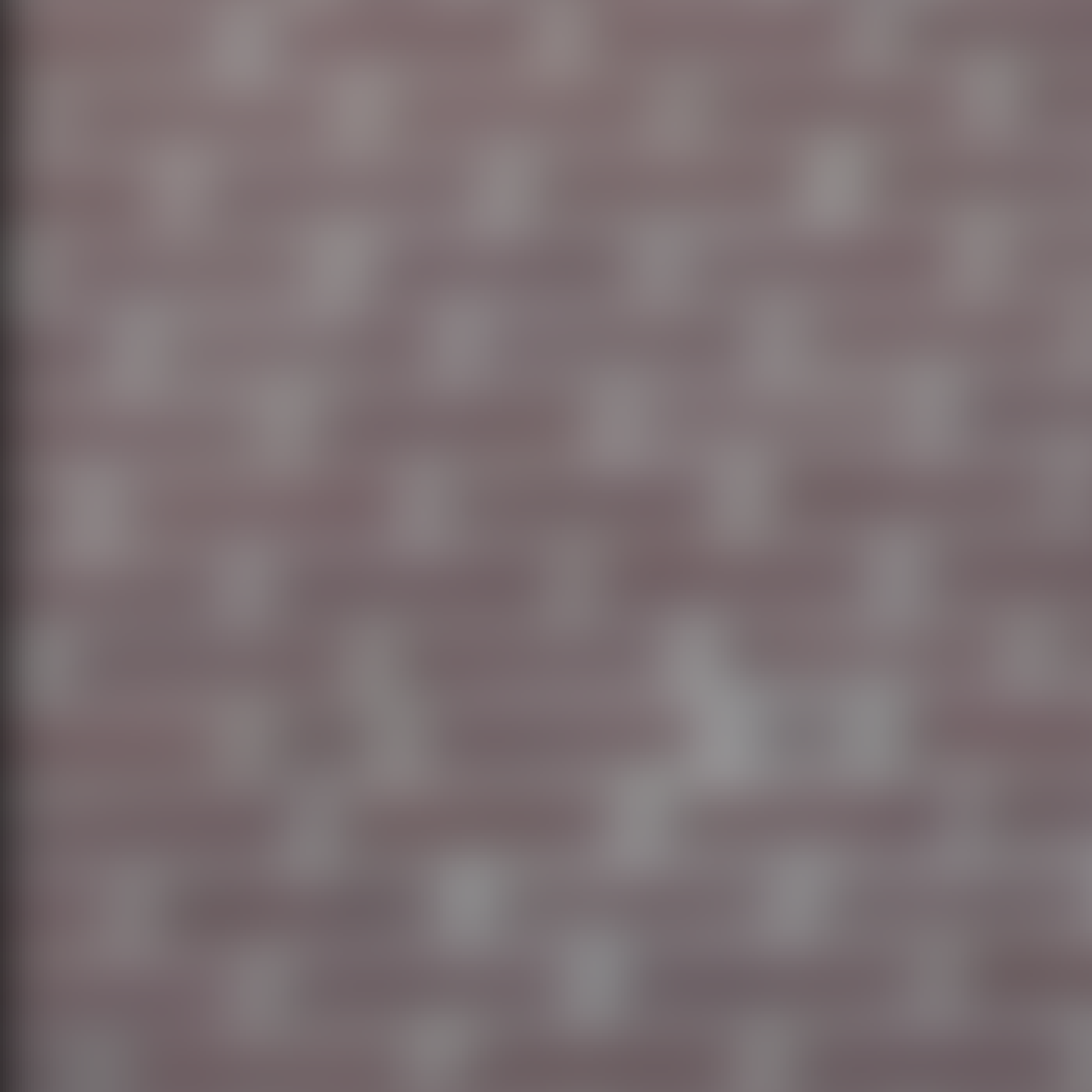}\\[1pt]
\mbox{\fontsize{9pt}{11pt}\selectfont Ours\,/\,71.9\,dB $\cdot$ 55,762 Mpix/s $\cdot$ fastest}\end{minipage} \\[12pt]
\begin{minipage}[t]{2.500in}\centering \includegraphics[interpolate=false,height=2.500in]{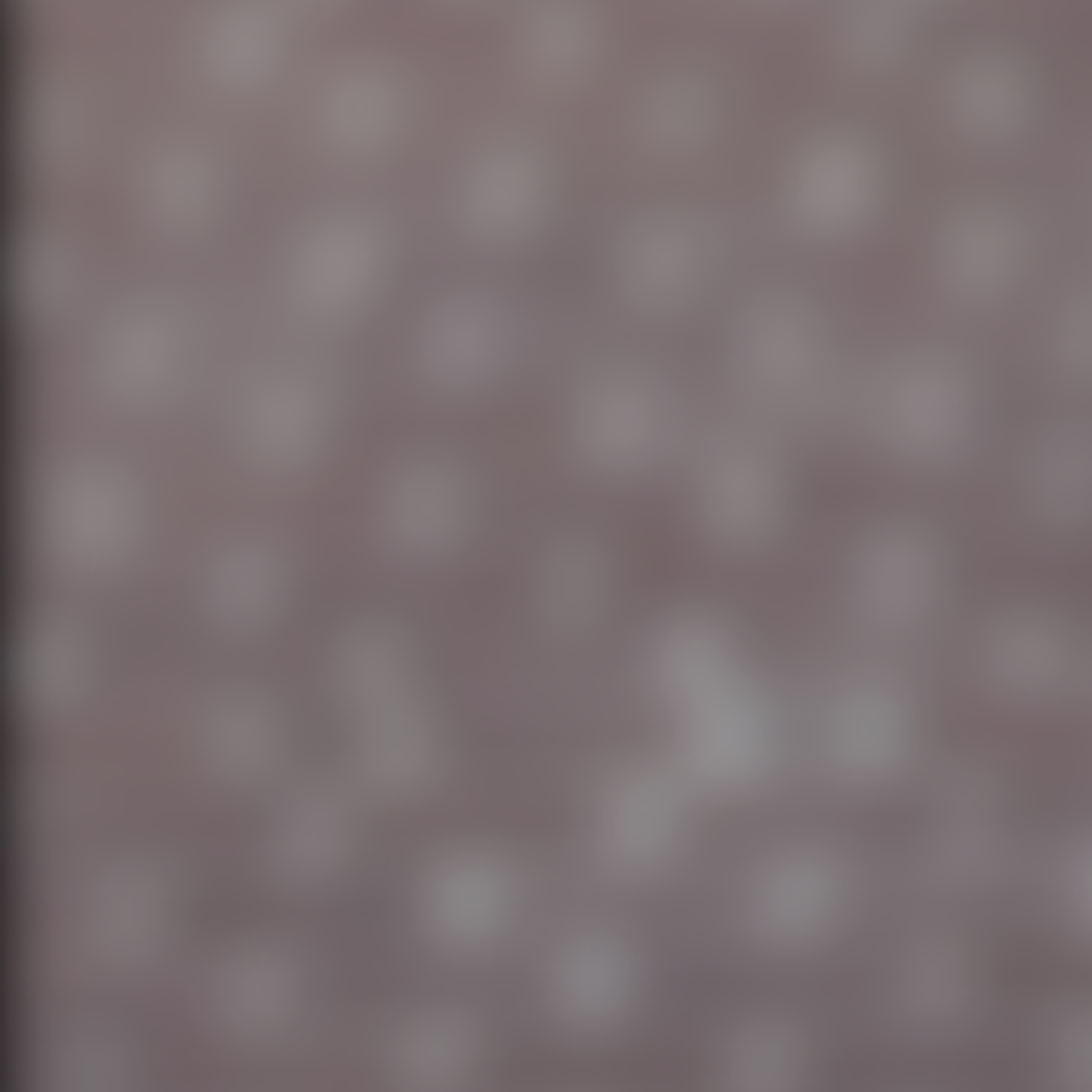}\\[1pt]
\mbox{\fontsize{9pt}{11pt}\selectfont Box3\,/\,44.6\,dB $\cdot$ 14,651 Mpix/s $\cdot$ 3.8$\times$ slower}\end{minipage} & \begin{minipage}[t]{2.500in}\centering \includegraphics[interpolate=false,height=2.500in]{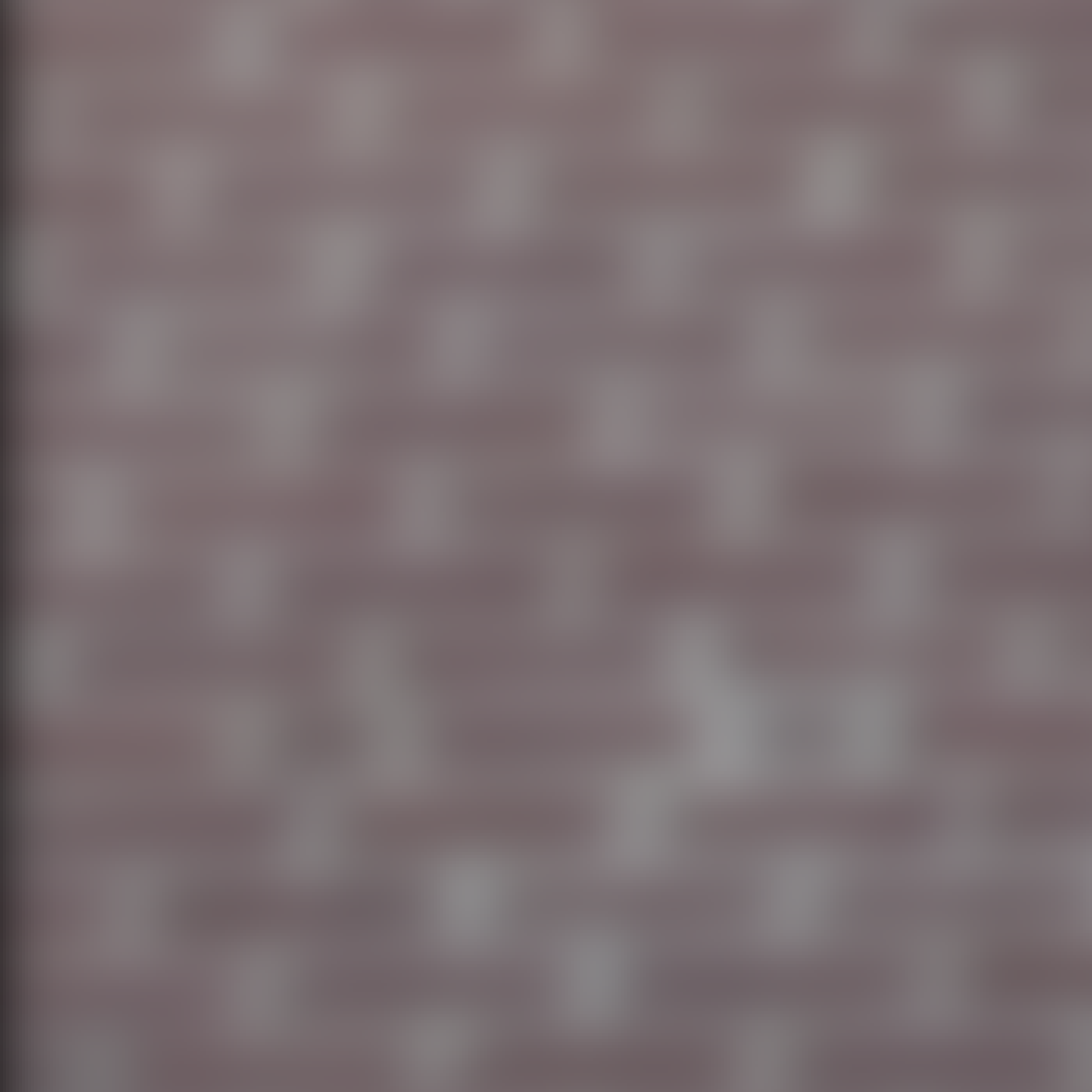}\\[1pt]
\mbox{\fontsize{9pt}{11pt}\selectfont CP\,/\,77.5\,dB $\cdot$ 13,566 Mpix/s $\cdot$ 4.1$\times$ slower}\end{minipage} \\[12pt]
\begin{minipage}[t]{2.500in}\centering \includegraphics[interpolate=false,height=2.500in]{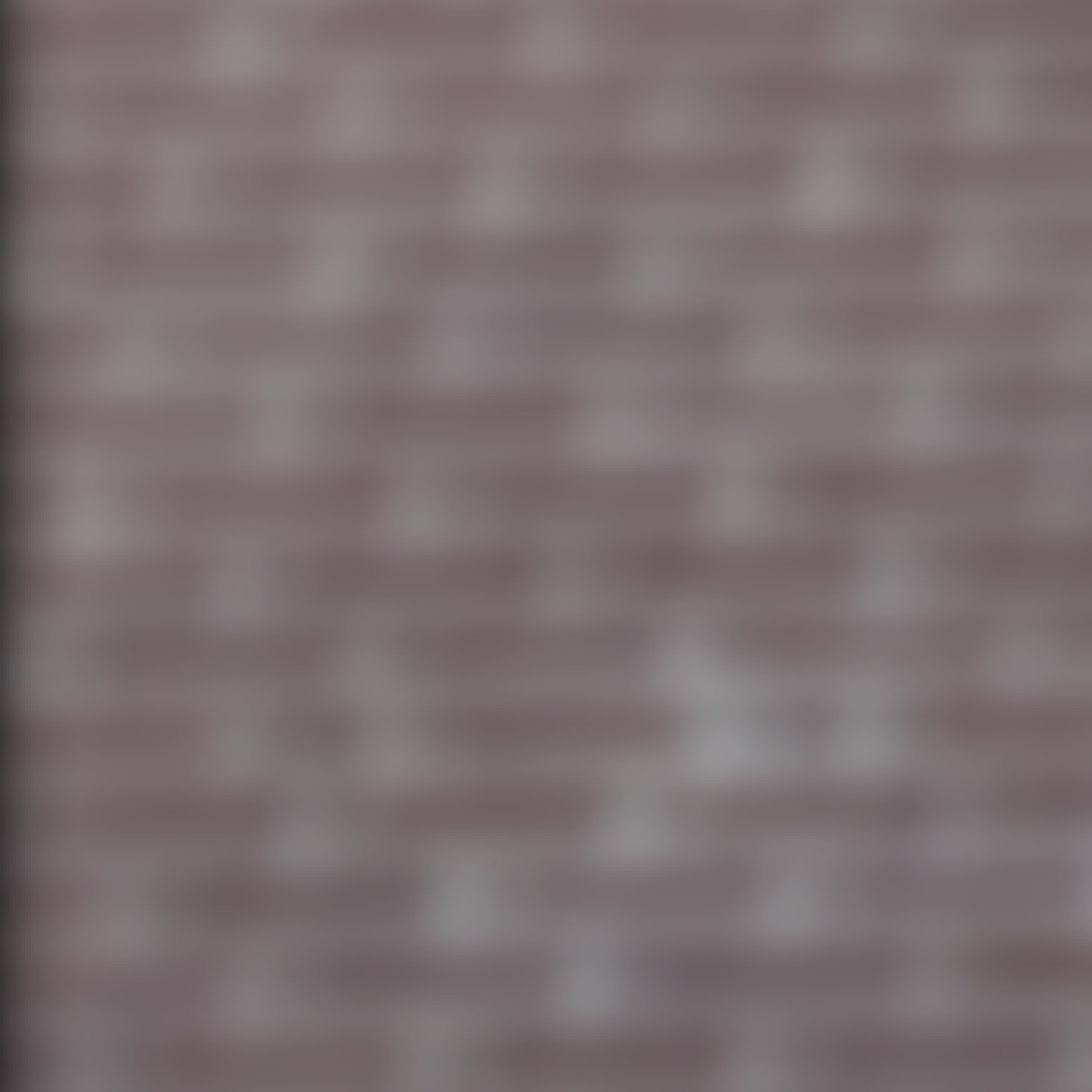}\\[1pt]
\mbox{\fontsize{9pt}{11pt}\selectfont FRM\,/\,39.9\,dB $\cdot$ 12,652 Mpix/s $\cdot$ 4.4$\times$ slower}\end{minipage} & \begin{minipage}[t]{2.500in}\centering \includegraphics[interpolate=false,height=2.500in]{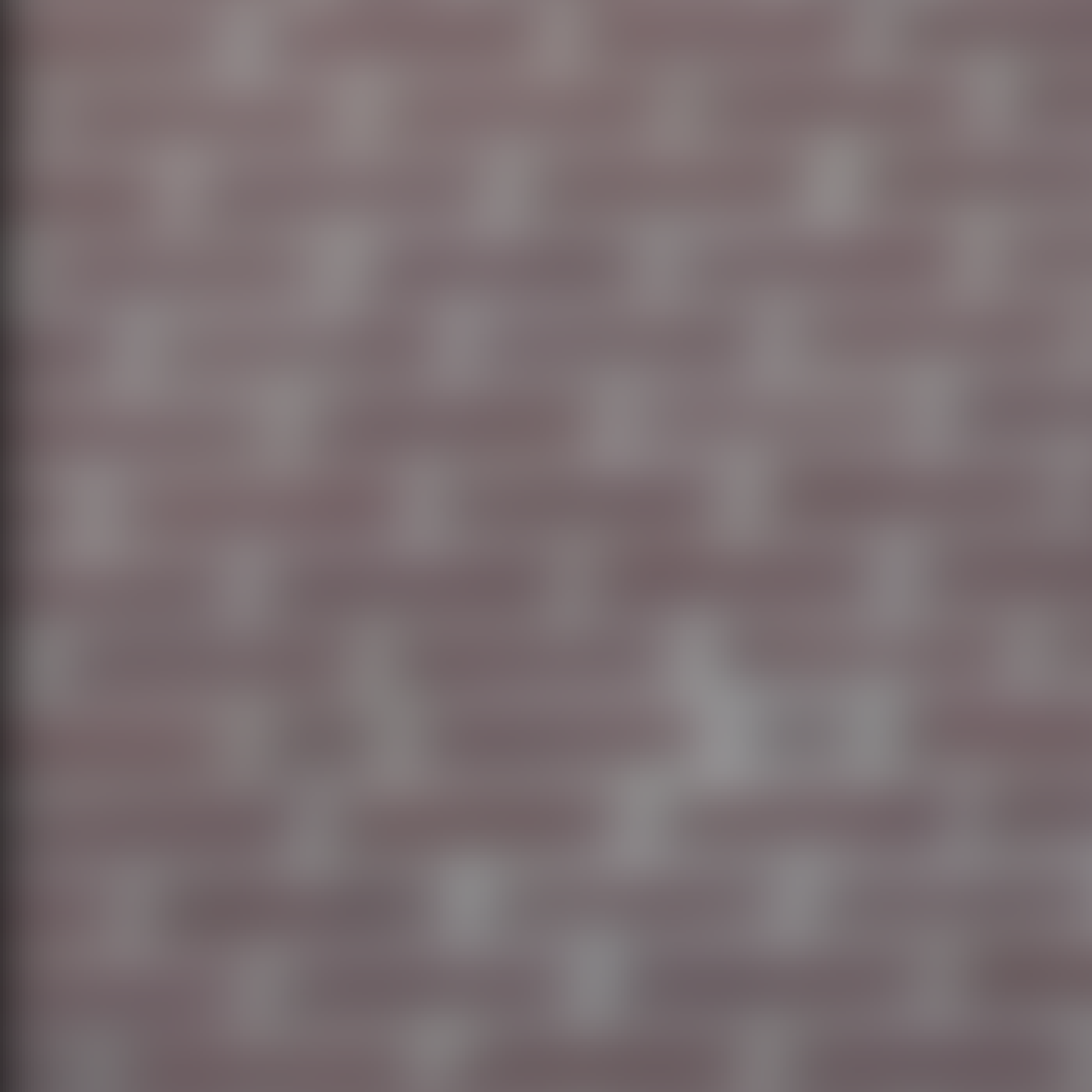}\\[1pt]
\mbox{\fontsize{9pt}{11pt}\selectfont YVV\,/\,47.4\,dB $\cdot$ 8,793 Mpix/s $\cdot$ 6.3$\times$ slower}\end{minipage} \\[12pt]
\end{tabular}
\vspace{-0.9em}
\caption{Figures 45 - 47 show outputs for 2D size 101 Gaussian filter approximations, on patches from different images taken with the authors' iPhone 13 Pro. Our model and CP produce visually indistinguishable outputs from ground truth, but our model is the fastest approximation and is 4.1x faster than CP. Triple box blur over-blurs the image and FRM produces ringing artifacts. YVV under-blurs. These qualitative observations are confirmed by the frequency response plots of these approximations shown in figures \ref{fig:gaussian2d-101-freq-plot},\ref{fig:gaussian2d-201-freq-plot}, and \ref{fig:gaussian2d-401-freq-plot}}
\label{fig:patch_gaussian2D_101_IMG_5700_y2850_x50}
\end{figure*}

\begin{figure*}[p]\centering
\begin{tabular}{cc}
\begin{minipage}[t]{2.500in}\centering \includegraphics[interpolate=false,height=2.500in]{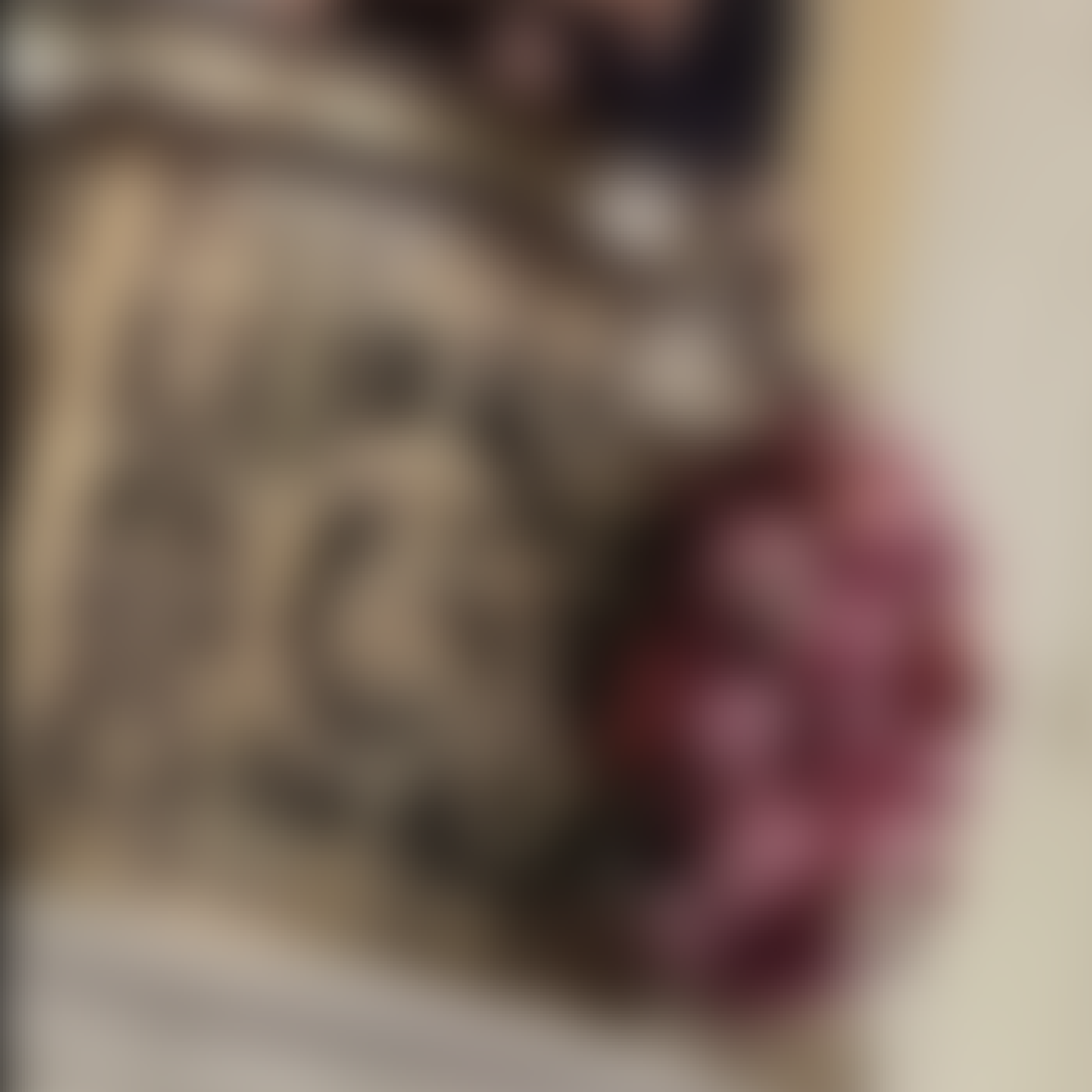}\\[1pt]
\mbox{\fontsize{9pt}{11pt}\selectfont Ground truth}\end{minipage} & \begin{minipage}[t]{2.500in}\centering \includegraphics[interpolate=false,height=2.500in]{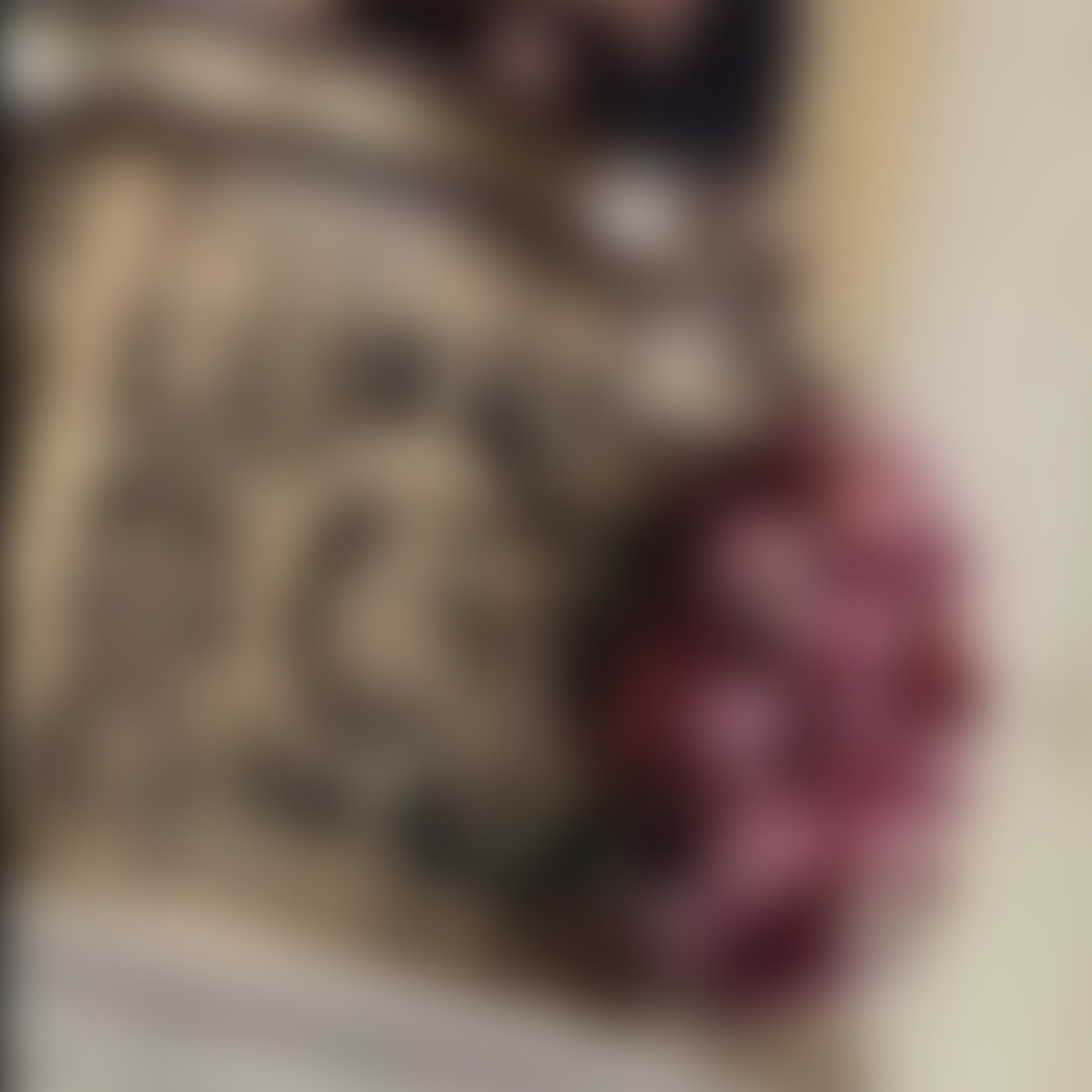}\\[1pt]
\mbox{\fontsize{9pt}{11pt}\selectfont Ours\,/\,67.9\,dB $\cdot$ 55,762 Mpix/s $\cdot$ fastest}\end{minipage} \\[12pt]
\begin{minipage}[t]{2.500in}\centering \includegraphics[interpolate=false,height=2.500in]{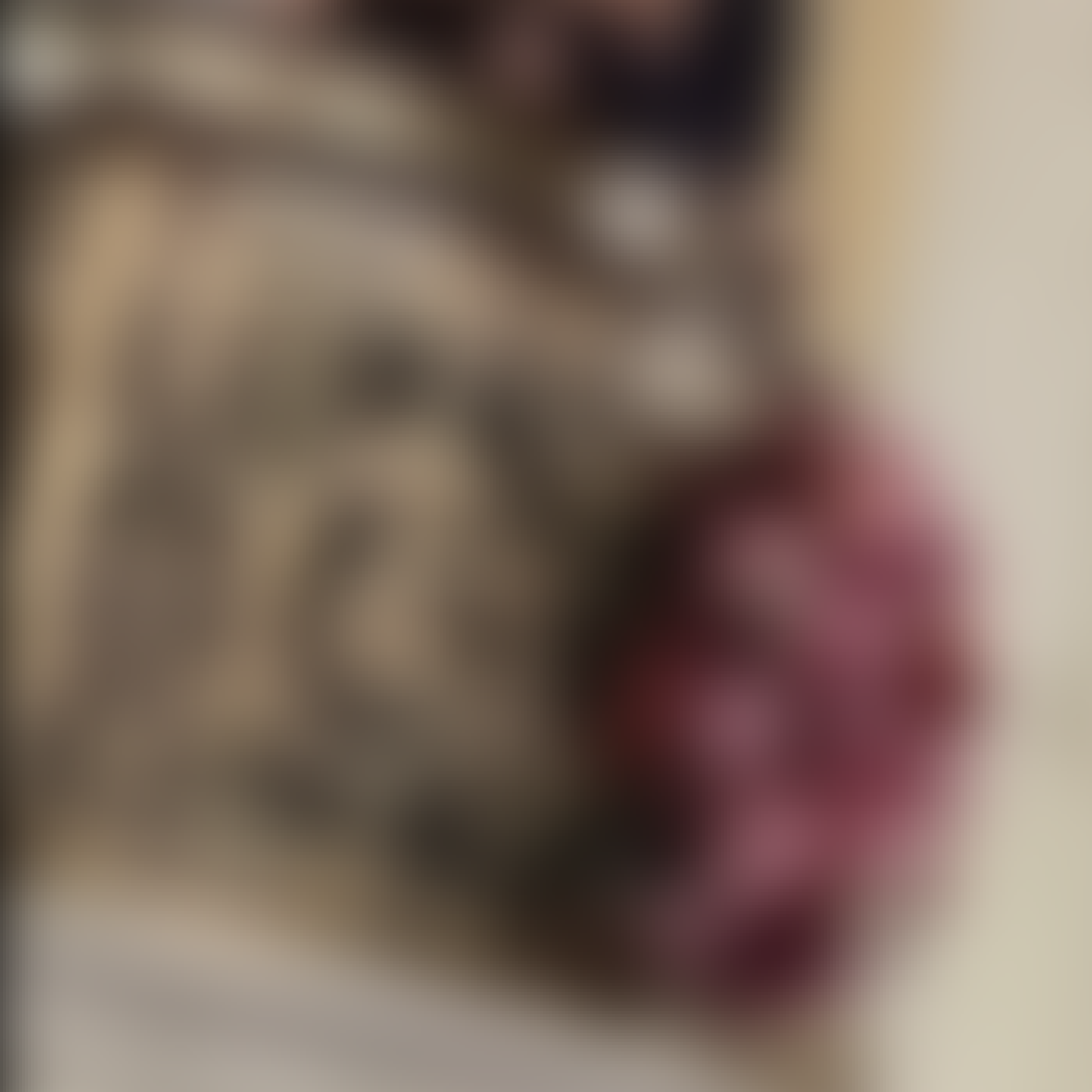}\\[1pt]
\mbox{\fontsize{9pt}{11pt}\selectfont Box3\,/\,42.7\,dB $\cdot$ 14,651 Mpix/s $\cdot$ 3.8$\times$ slower}\end{minipage} & \begin{minipage}[t]{2.500in}\centering \includegraphics[interpolate=false,height=2.500in]{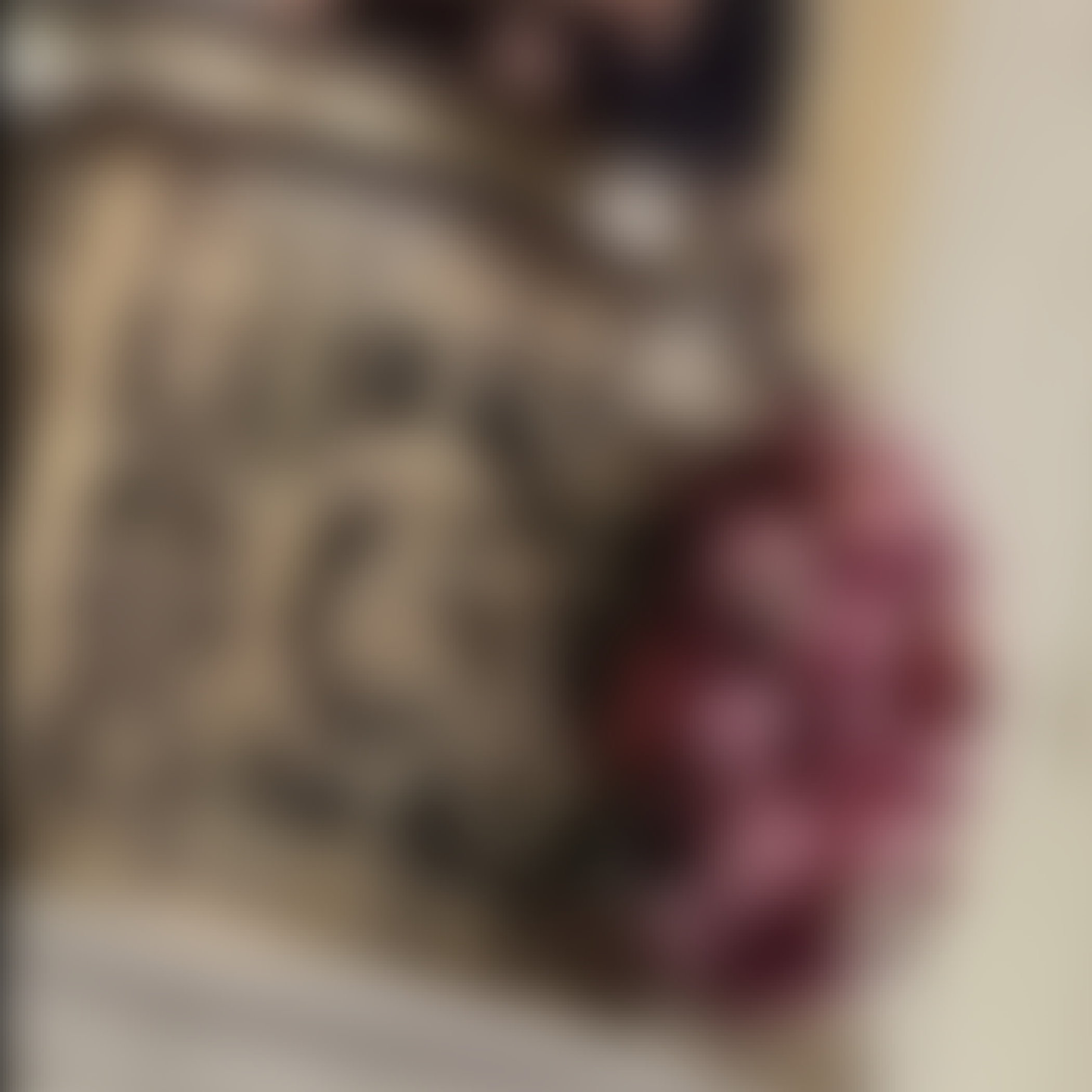}\\[1pt]
\mbox{\fontsize{9pt}{11pt}\selectfont CP\,/\,74.8\,dB $\cdot$ 13,566 Mpix/s $\cdot$ 4.1$\times$ slower}\end{minipage} \\[12pt]
\begin{minipage}[t]{2.500in}\centering \includegraphics[interpolate=false,height=2.500in]{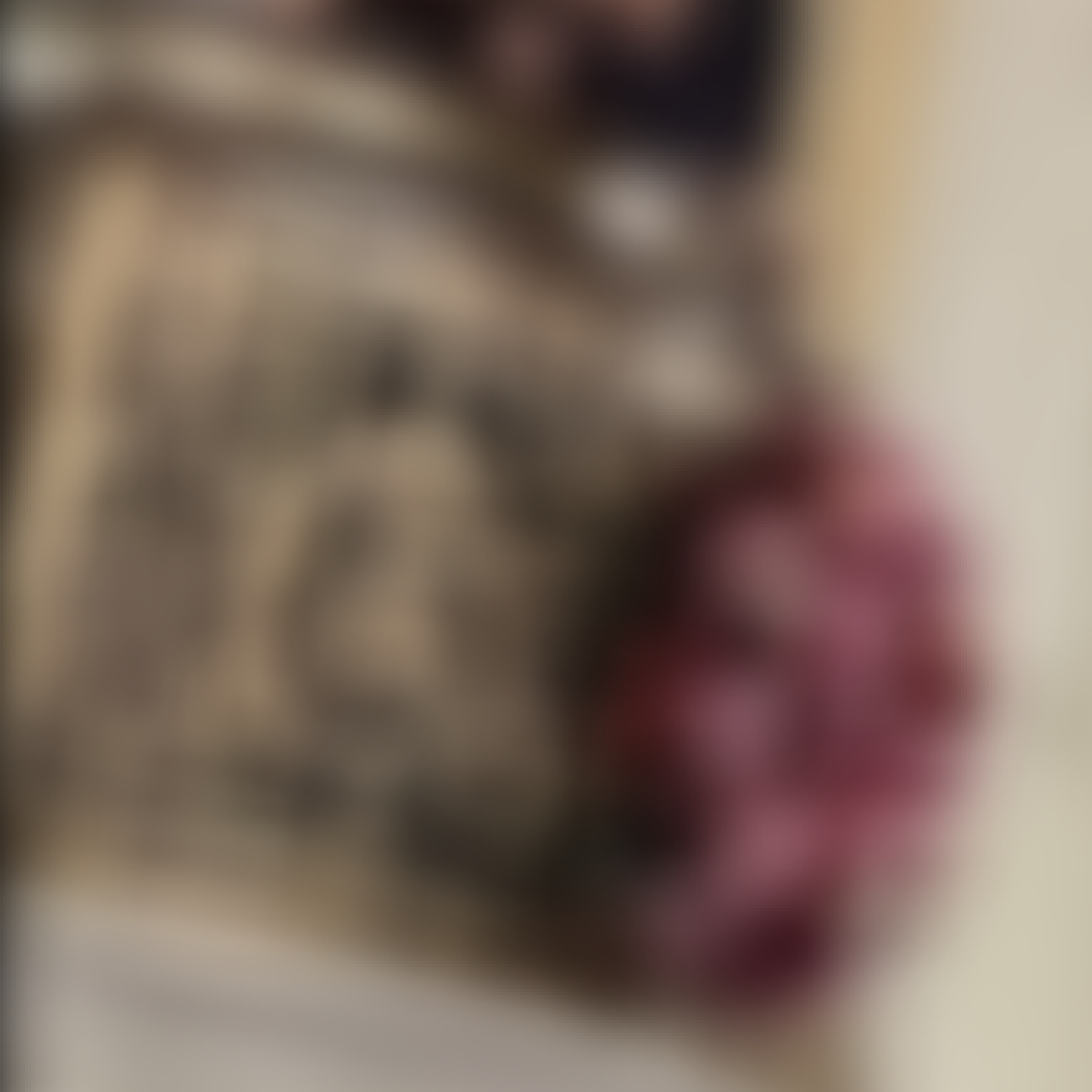}\\[1pt]
\mbox{\fontsize{9pt}{11pt}\selectfont FRM\,/\,43.4\,dB $\cdot$ 12,652 Mpix/s $\cdot$ 4.4$\times$ slower}\end{minipage} & \begin{minipage}[t]{2.500in}\centering \includegraphics[interpolate=false,height=2.500in]{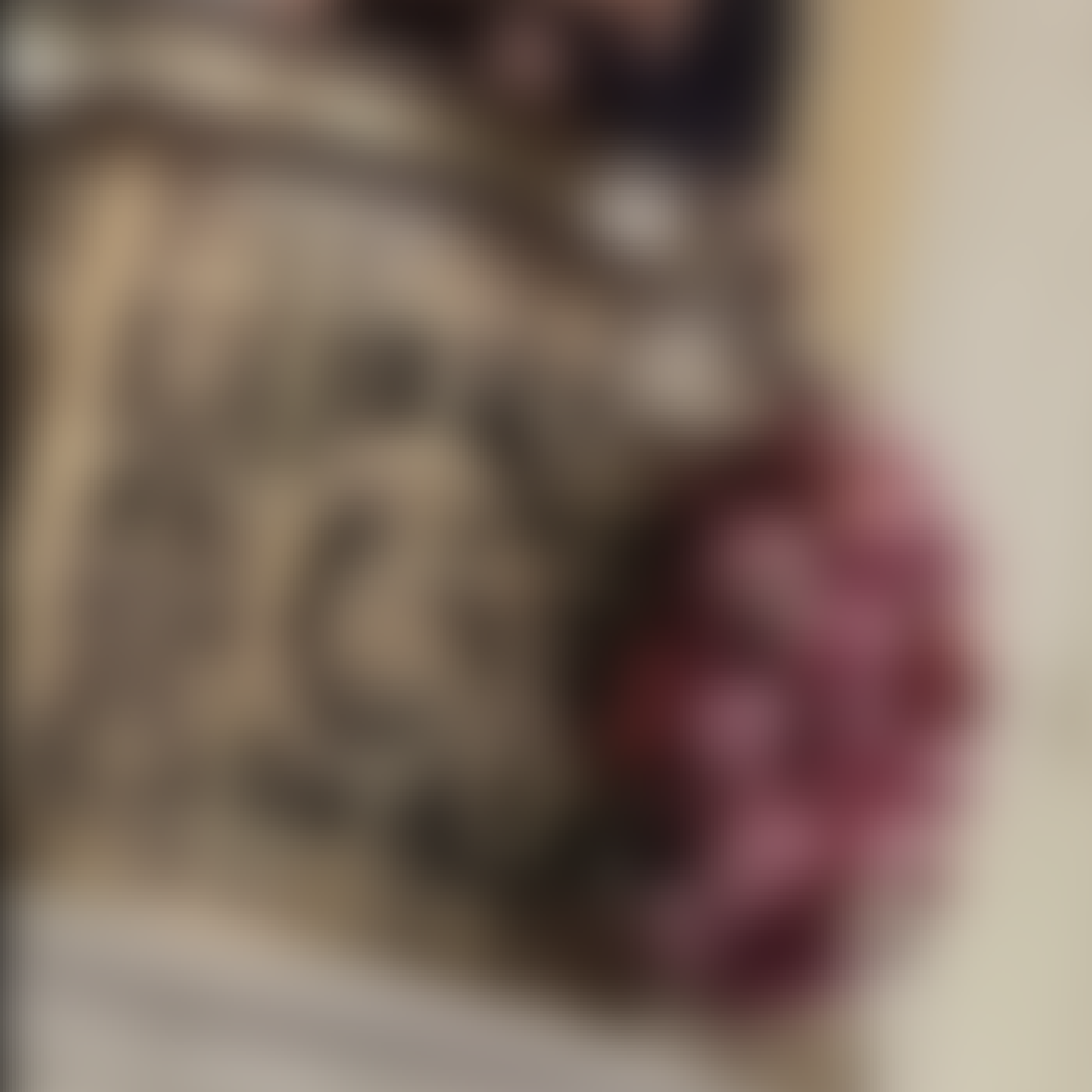}\\[1pt]
\mbox{\fontsize{9pt}{11pt}\selectfont YVV\,/\,46.9\,dB $\cdot$ 8,793 Mpix/s $\cdot$ 6.3$\times$ slower}\end{minipage} \\[12pt]
\end{tabular}
\caption{2D Gaussian filter, size 101 approximation outputs.}
\label{fig:patch_gaussian2D_101_IMG_5046_y2850_x50}
\end{figure*}

\begin{figure*}[p]\centering
\begin{tabular}{cc}
\begin{minipage}[t]{2.500in}\centering \includegraphics[interpolate=false,height=2.500in]{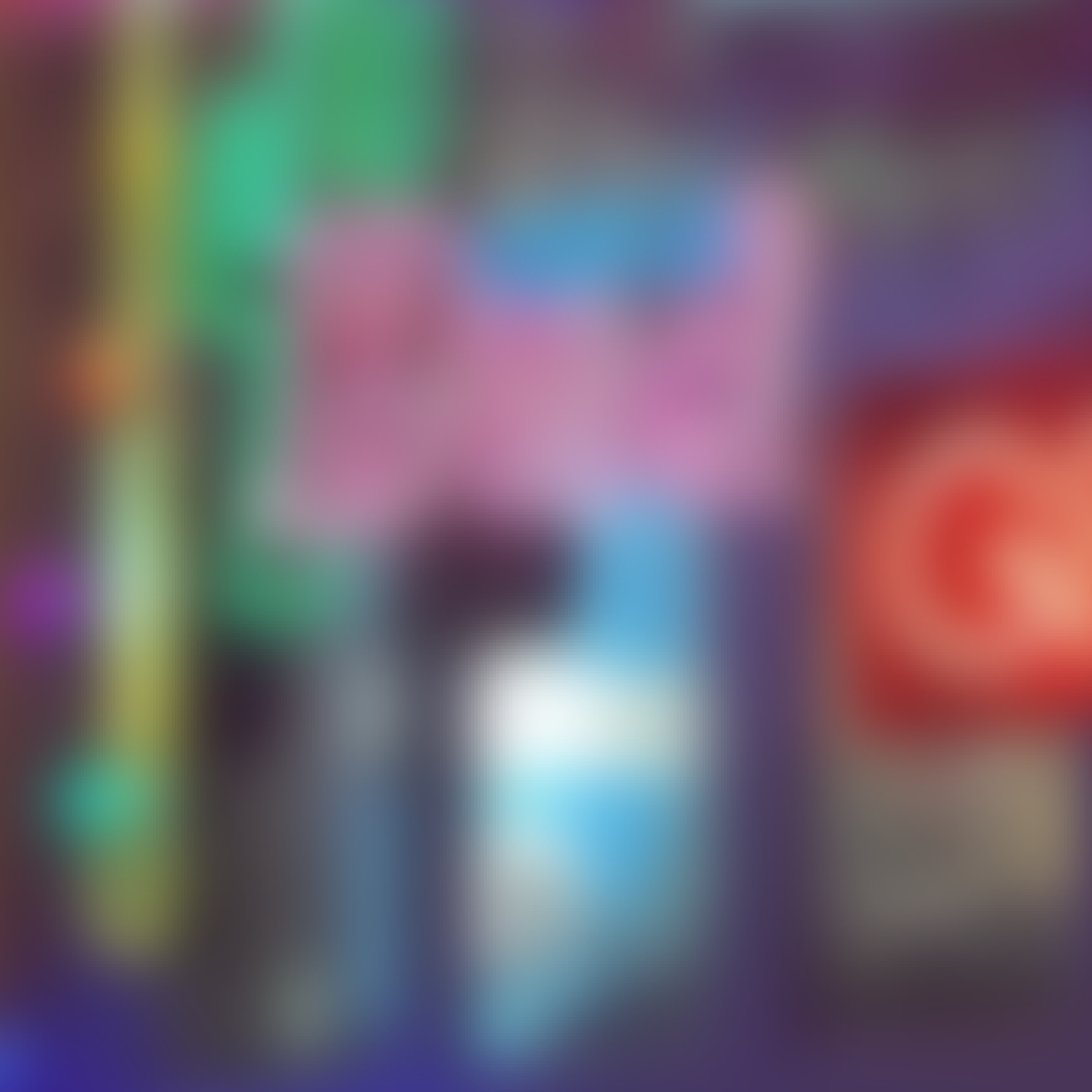}\\[1pt]
\mbox{\fontsize{9pt}{11pt}\selectfont Ground truth}\end{minipage} & \begin{minipage}[t]{2.500in}\centering \includegraphics[interpolate=false,height=2.500in]{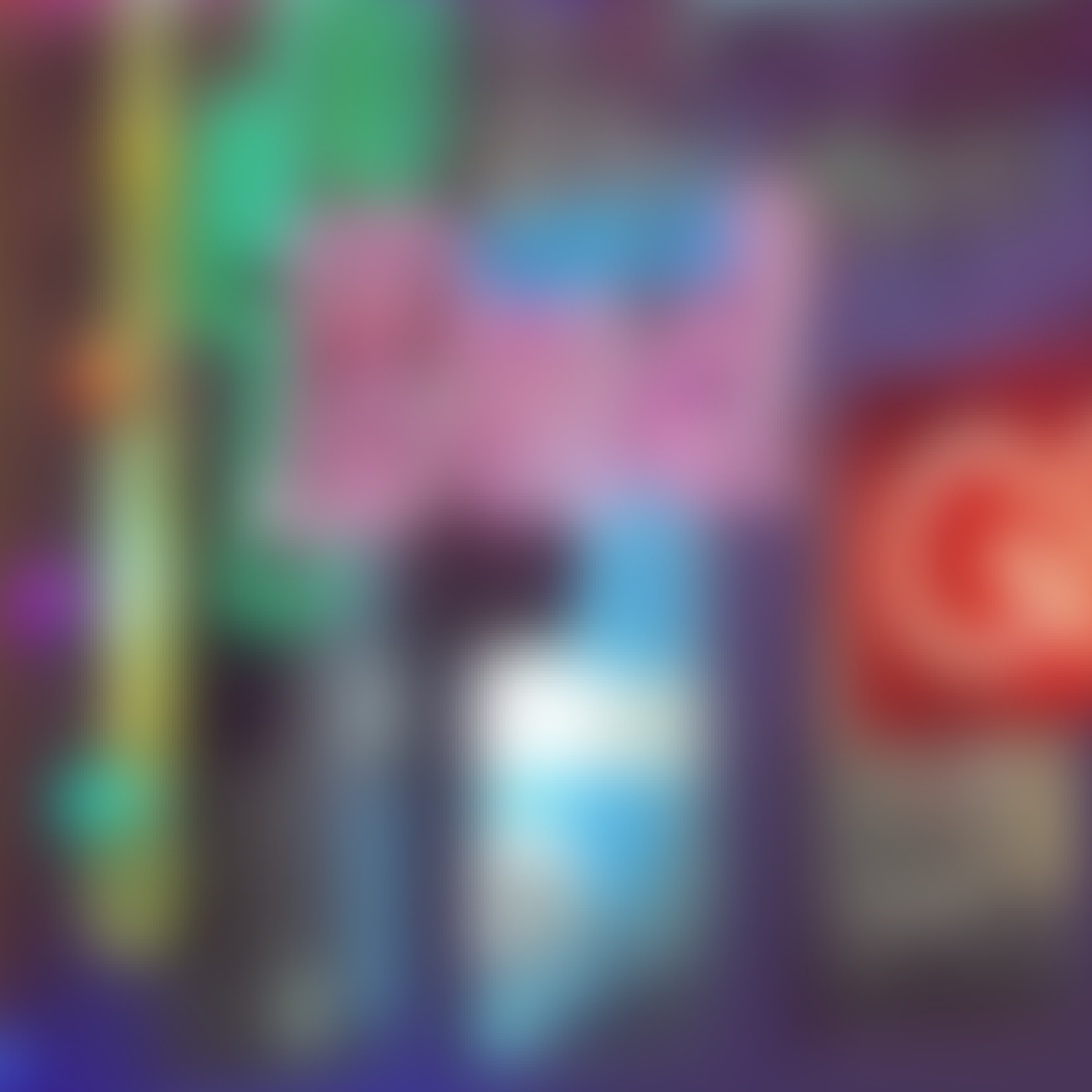}\\[1pt]
\mbox{\fontsize{9pt}{11pt}\selectfont Ours\,/\,66.8\,dB $\cdot$ 55,762 Mpix/s $\cdot$ fastest}\end{minipage} \\[12pt]
\begin{minipage}[t]{2.500in}\centering \includegraphics[interpolate=false,height=2.500in]{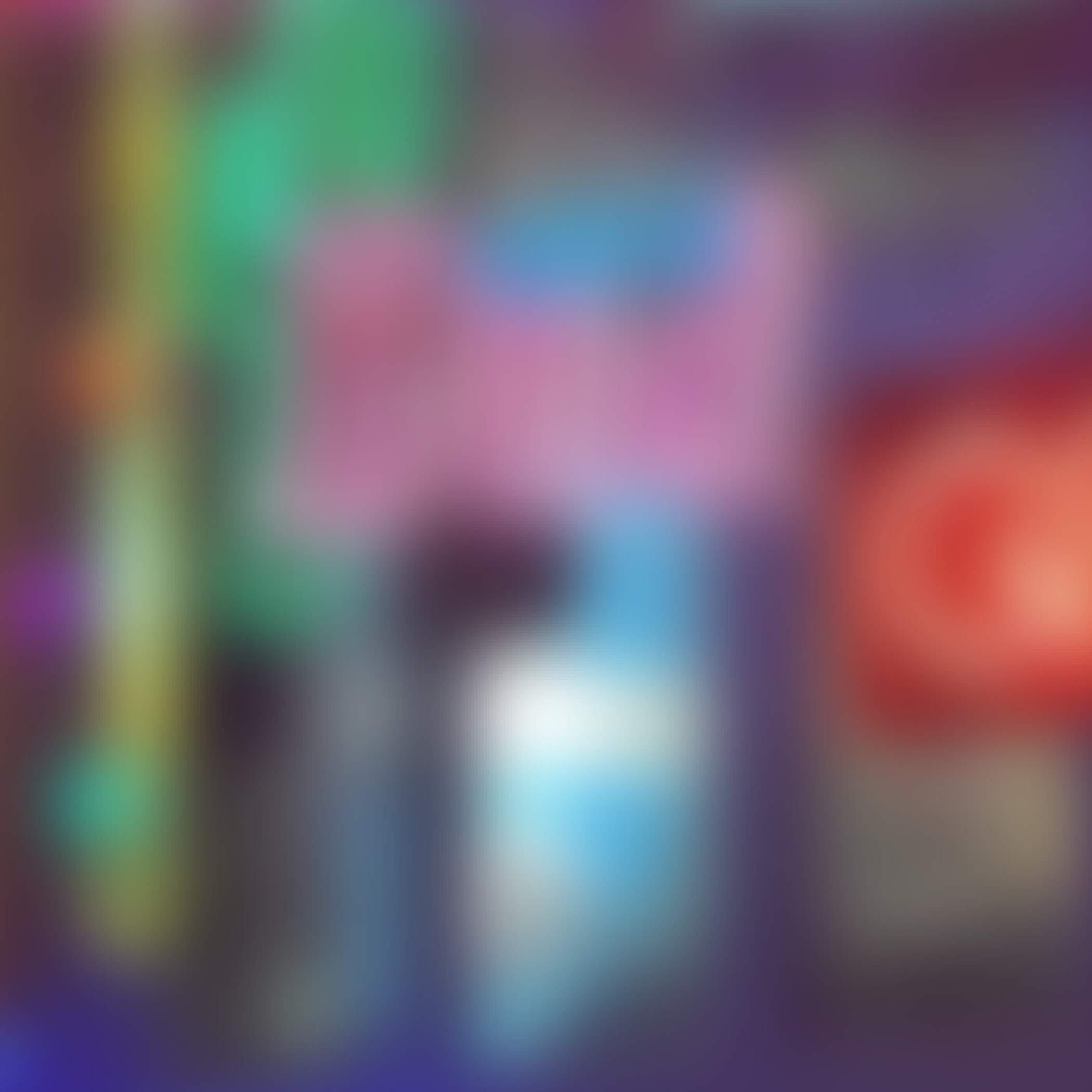}\\[1pt]
\mbox{\fontsize{9pt}{11pt}\selectfont Box3\,/\,43.8\,dB $\cdot$ 14,651 Mpix/s $\cdot$ 3.8$\times$ slower}\end{minipage} & \begin{minipage}[t]{2.500in}\centering \includegraphics[interpolate=false,height=2.500in]{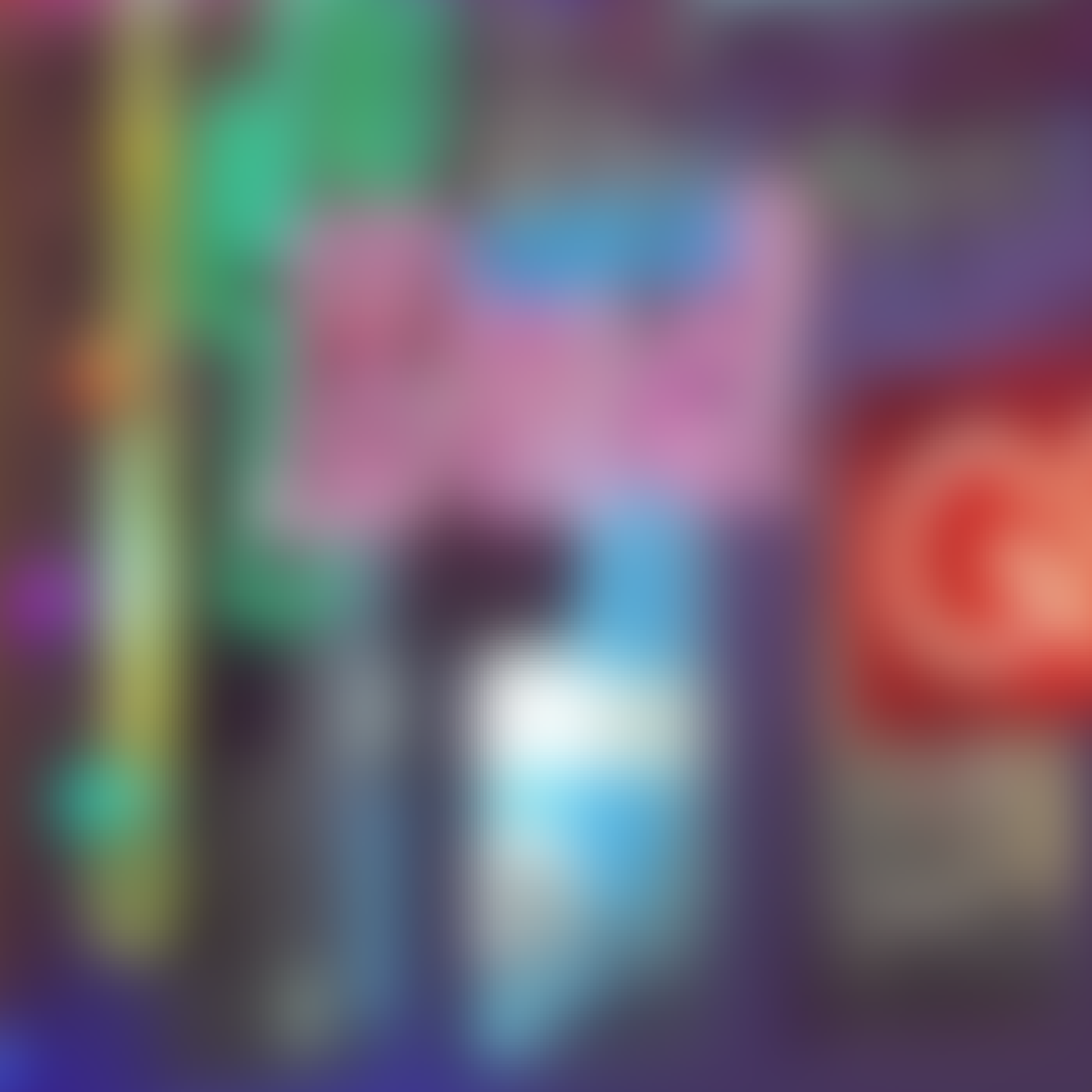}\\[1pt]
\mbox{\fontsize{9pt}{11pt}\selectfont CP\,/\,75.0\,dB $\cdot$ 13,566 Mpix/s $\cdot$ 4.1$\times$ slower}\end{minipage} \\[12pt]
\begin{minipage}[t]{2.500in}\centering \includegraphics[interpolate=false,height=2.500in]{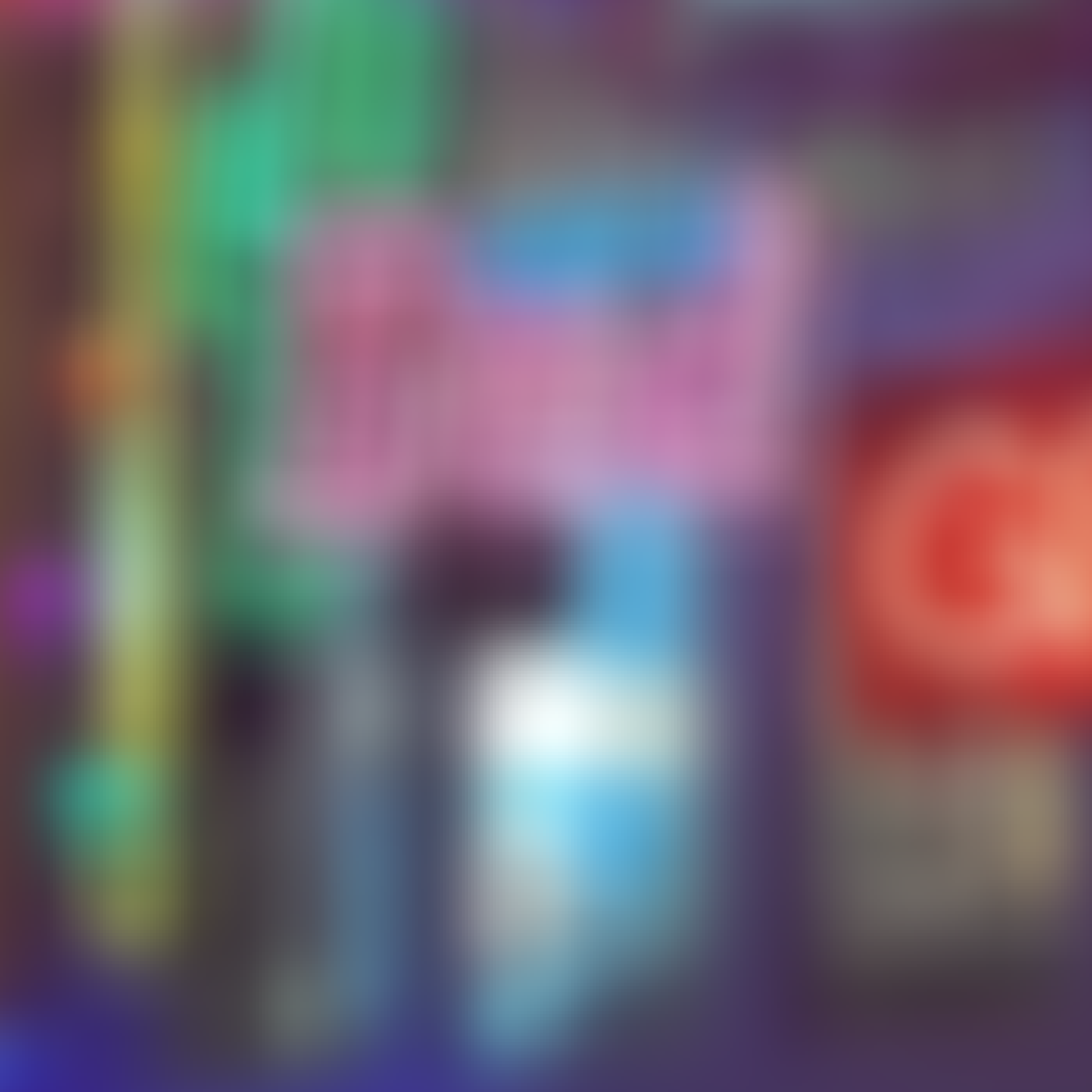}\\[1pt]
\mbox{\fontsize{9pt}{11pt}\selectfont FRM\,/\,44.3\,dB $\cdot$ 12,652 Mpix/s $\cdot$ 4.4$\times$ slower}\end{minipage} & \begin{minipage}[t]{2.500in}\centering \includegraphics[interpolate=false,height=2.500in]{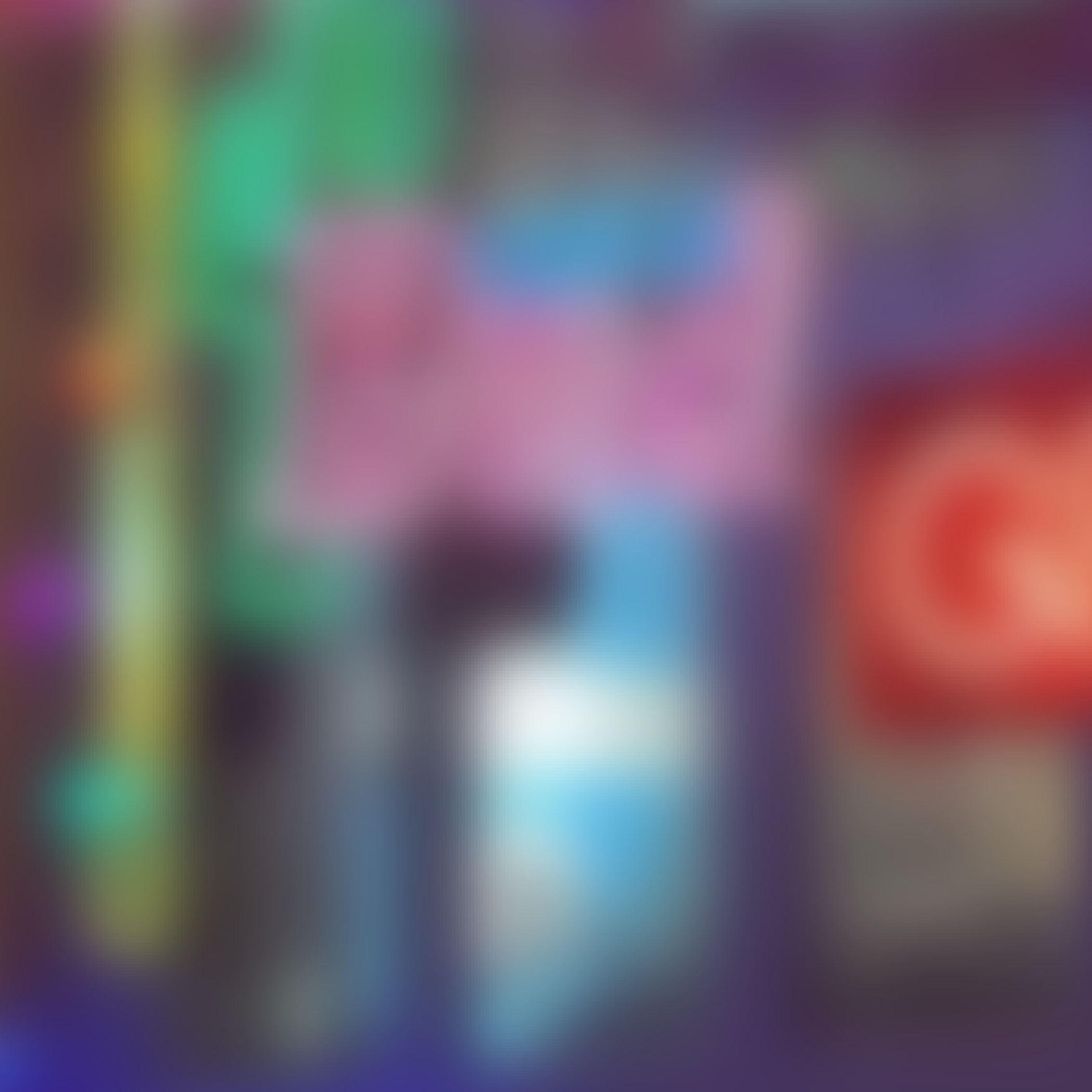}\\[1pt]
\mbox{\fontsize{9pt}{11pt}\selectfont YVV\,/\,46.9\,dB $\cdot$ 8,793 Mpix/s $\cdot$ 6.3$\times$ slower}\end{minipage} \\[12pt]
\end{tabular}
\caption{2D Gaussian filter, size 101 approximation outputs.}
\label{fig:patch_gaussian2D_101_IMG_5579_y2150_x1100}
\end{figure*}

\begin{figure*}[p]\centering
\begin{tabular}{cc}
\begin{minipage}[t]{2.500in}\centering \includegraphics[interpolate=false,height=2.500in]{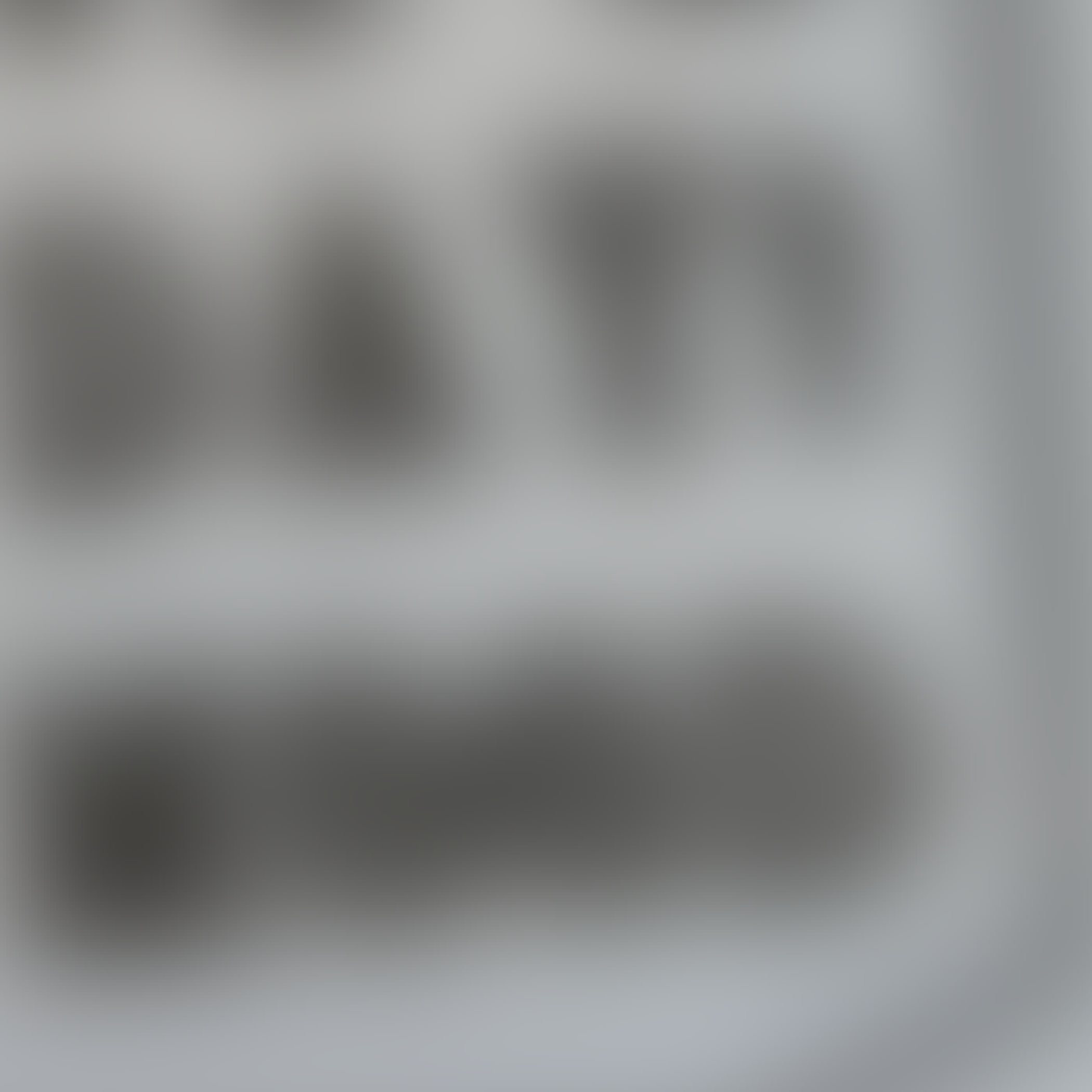}\\[1pt]
\mbox{\fontsize{9pt}{11pt}\selectfont Ground truth}\end{minipage} & \begin{minipage}[t]{2.500in}\centering \includegraphics[interpolate=false,height=2.500in]{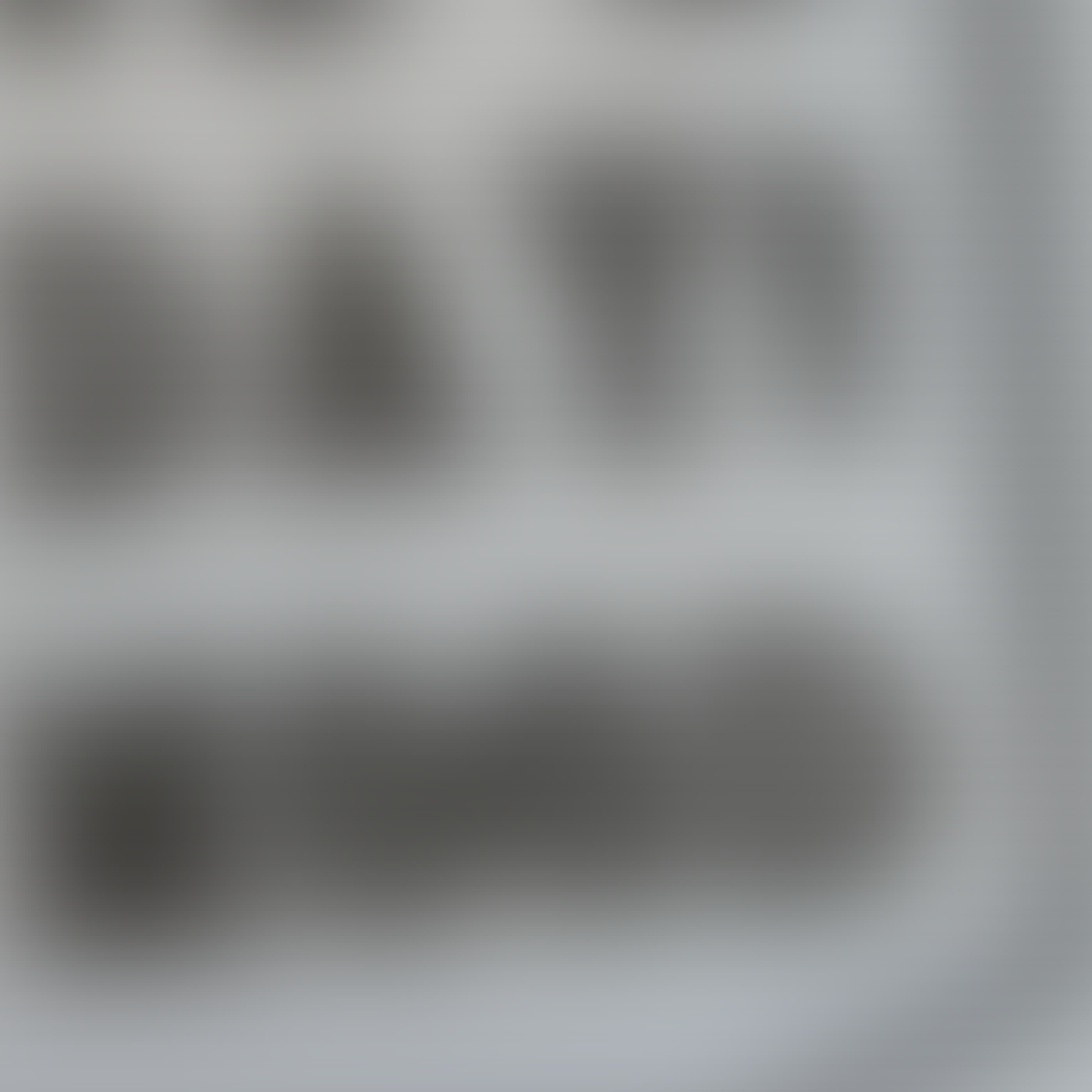}\\[1pt]
\mbox{\fontsize{9pt}{11pt}\selectfont Ours\,-\,58.2\,dB $\cdot$ 53,374 Mpix/s $\cdot$ fastest}\end{minipage} \\[12pt]
\begin{minipage}[t]{2.500in}\centering \includegraphics[interpolate=false,height=2.500in]{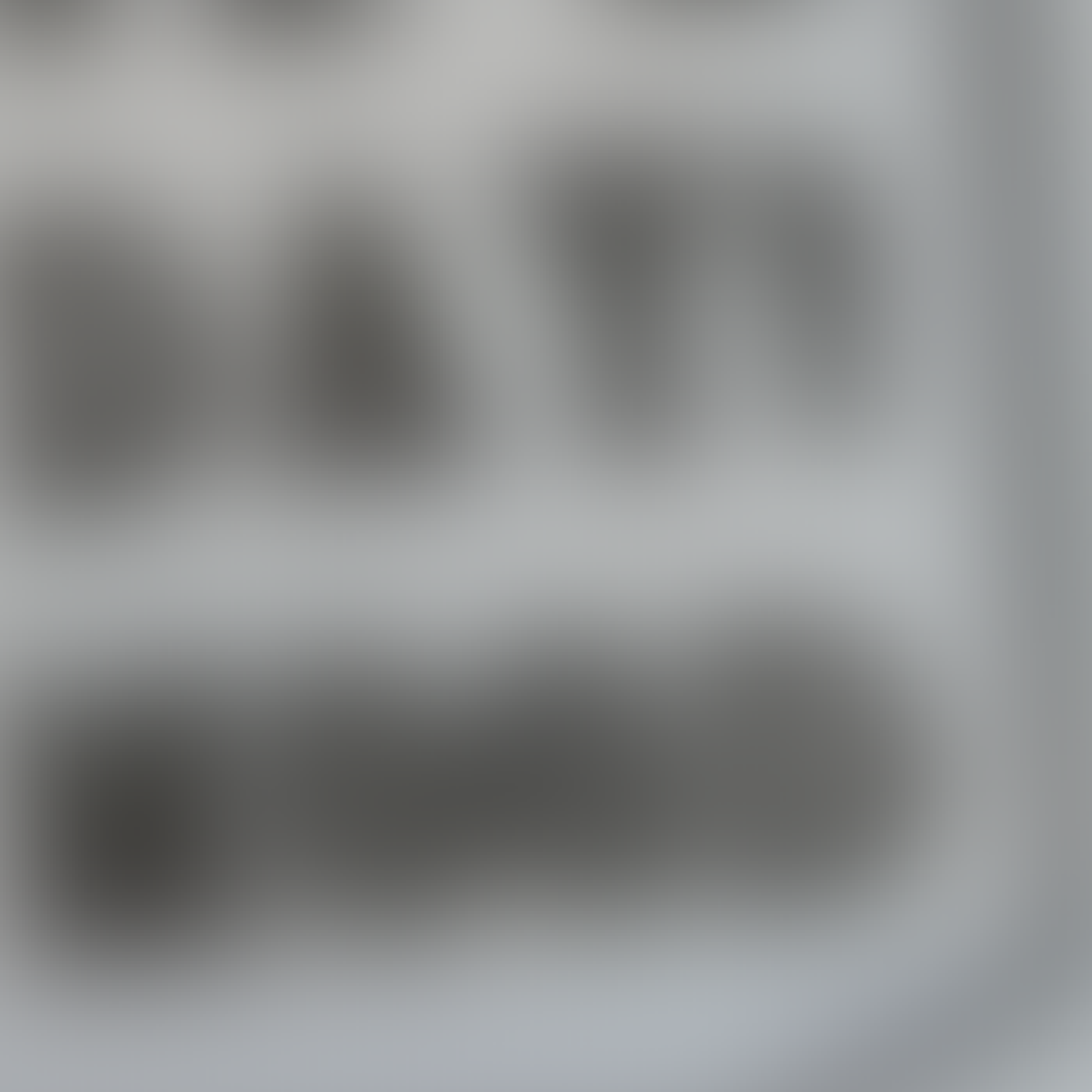}\\[1pt]
\mbox{\fontsize{9pt}{11pt}\selectfont CP\,-\,68.1\,dB $\cdot$ 12,251 Mpix/s $\cdot$ 4.4$\times$ slower}\end{minipage} & \begin{minipage}[t]{2.500in}\centering \includegraphics[interpolate=false,height=2.500in]{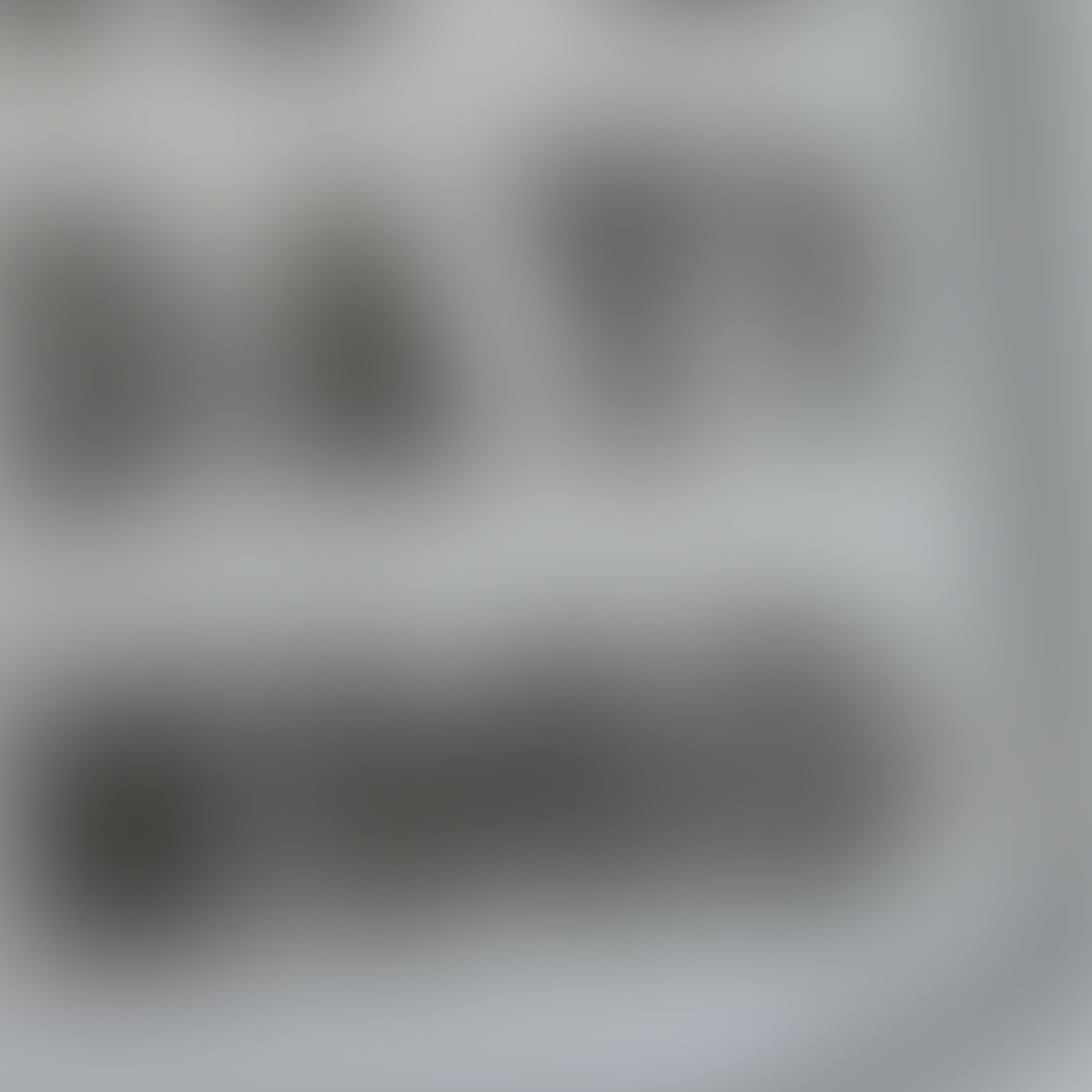}\\[1pt]
\mbox{\fontsize{9pt}{11pt}\selectfont Box3\,-\,43.3\,dB $\cdot$ 12,196 Mpix/s $\cdot$ 4.4$\times$ slower}\end{minipage} \\[12pt]
\begin{minipage}[t]{2.500in}\centering \includegraphics[interpolate=false,height=2.500in]{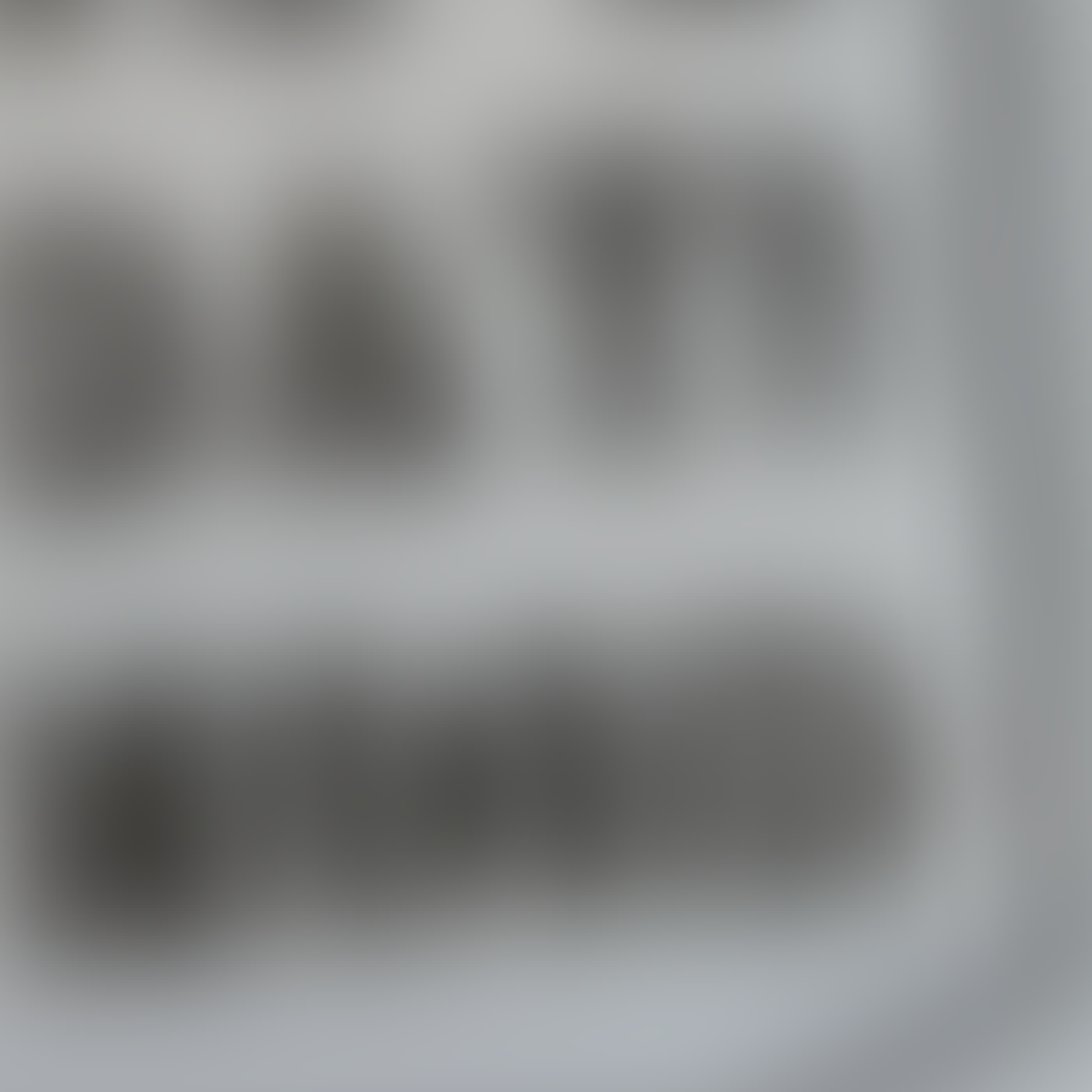}\\[1pt]
\mbox{\fontsize{9pt}{11pt}\selectfont FRM\,-\,44.4\,dB $\cdot$ 8,922 Mpix/s $\cdot$ 6.0$\times$ slower}\end{minipage} & \begin{minipage}[t]{2.500in}\centering \includegraphics[interpolate=false,height=2.500in]{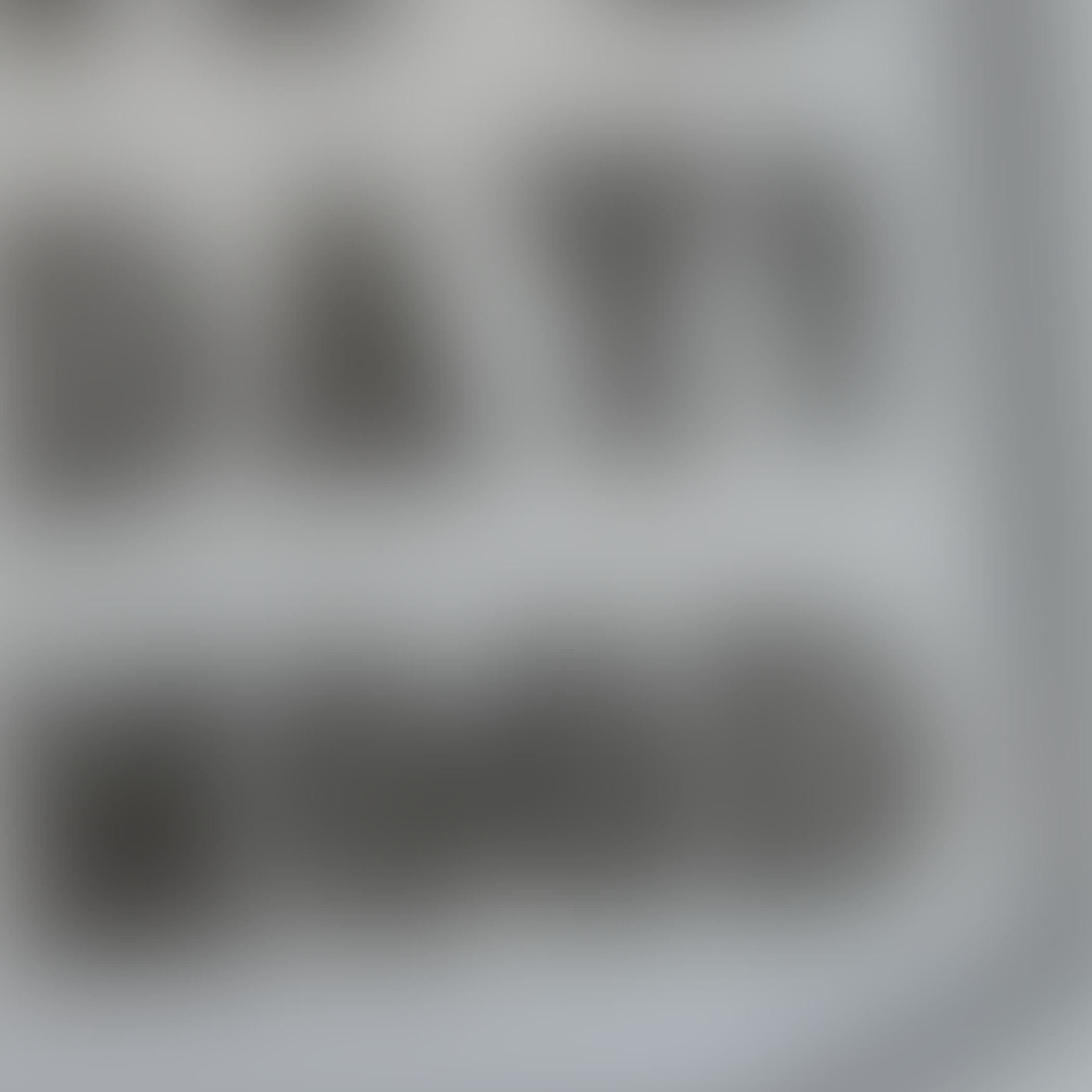}\\[1pt]
\mbox{\fontsize{9pt}{11pt}\selectfont YVV\,-\,44.2\,dB $\cdot$ 7,026 Mpix/s $\cdot$ 7.6$\times$ slower}\end{minipage} \\[12pt]
\end{tabular}
\vspace{-0.7em}
\caption{Figures 48 and 49 show outputs for 2D size 201 Gaussian filter approximations on patches from different images taken with the authors' iPhone 13 Pro. Our approximation and CP produce visually indistinguishable outputs from the ground truth, but our model is the fastest approximation and is 4.4x faster than CP. FRM has ringing artifacts, triple box over-blurs, and YVV under-blurs. The artifacts are less visually noticeable compared to the smaller 101 2D Gaussian.}
\label{fig:patch_gaussian2D_201_IMG_4390_y2900_x1500}
\end{figure*}

\begin{figure*}[p]\centering
\begin{tabular}{cc}
\begin{minipage}[t]{2.500in}\centering \includegraphics[interpolate=false,height=2.500in]{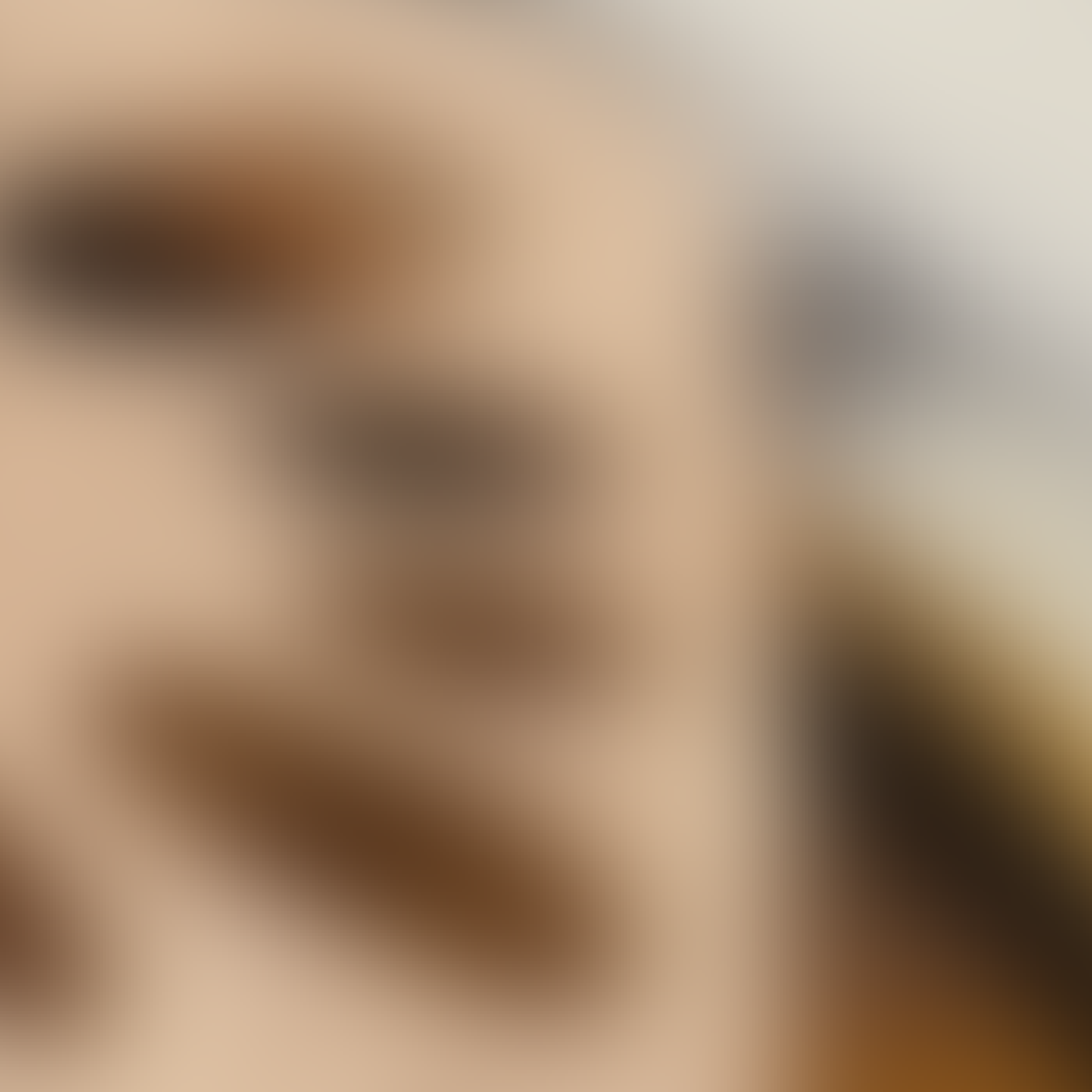}\\[1pt]
\mbox{\fontsize{9pt}{11pt}\selectfont Ground truth}\end{minipage} & \begin{minipage}[t]{2.500in}\centering \includegraphics[interpolate=false,height=2.500in]{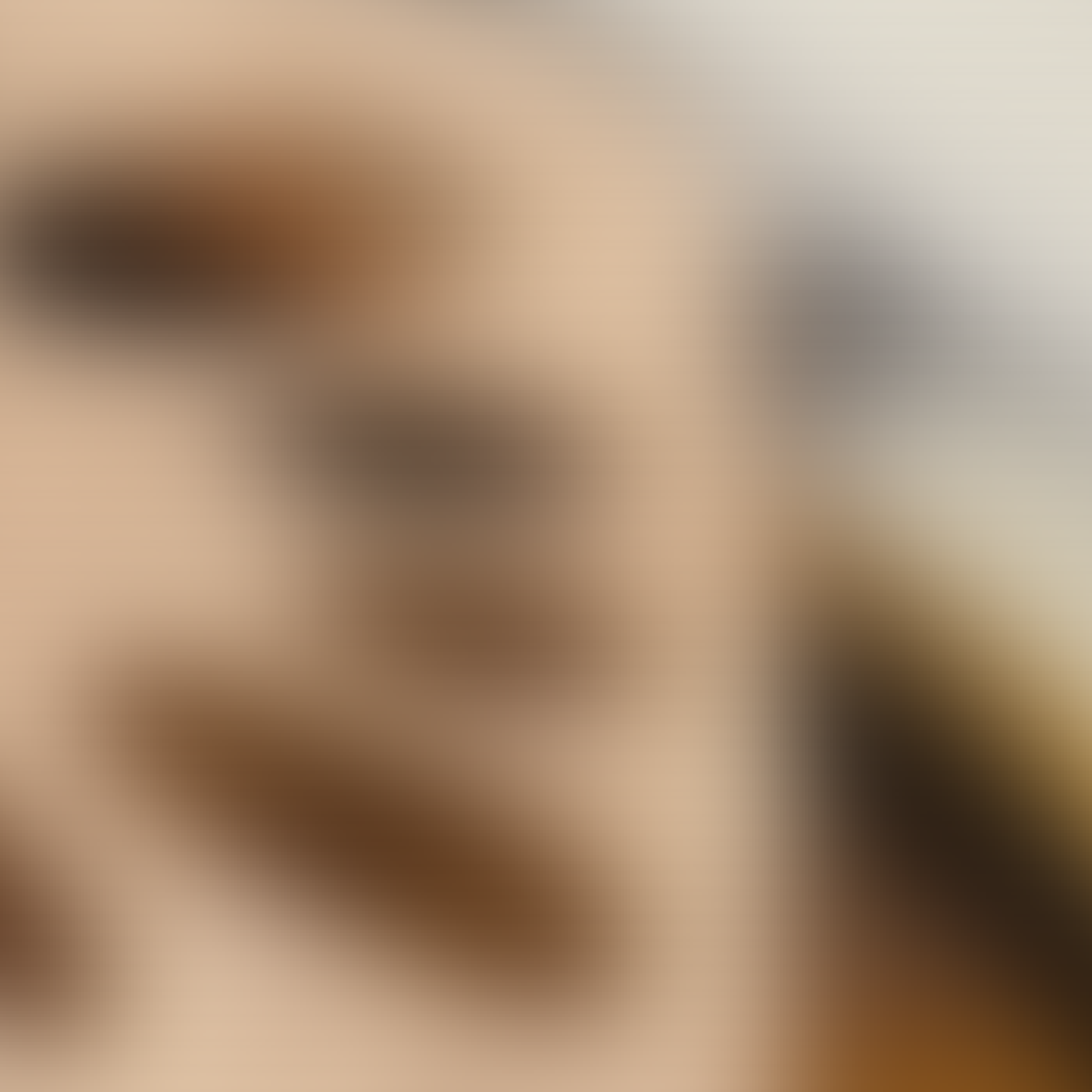}\\[1pt]
\mbox{\fontsize{9pt}{11pt}\selectfont Ours\,-\,58.3\,dB $\cdot$ 53,374 Mpix/s $\cdot$ fastest}\end{minipage} \\[12pt]
\begin{minipage}[t]{2.500in}\centering \includegraphics[interpolate=false,height=2.500in]{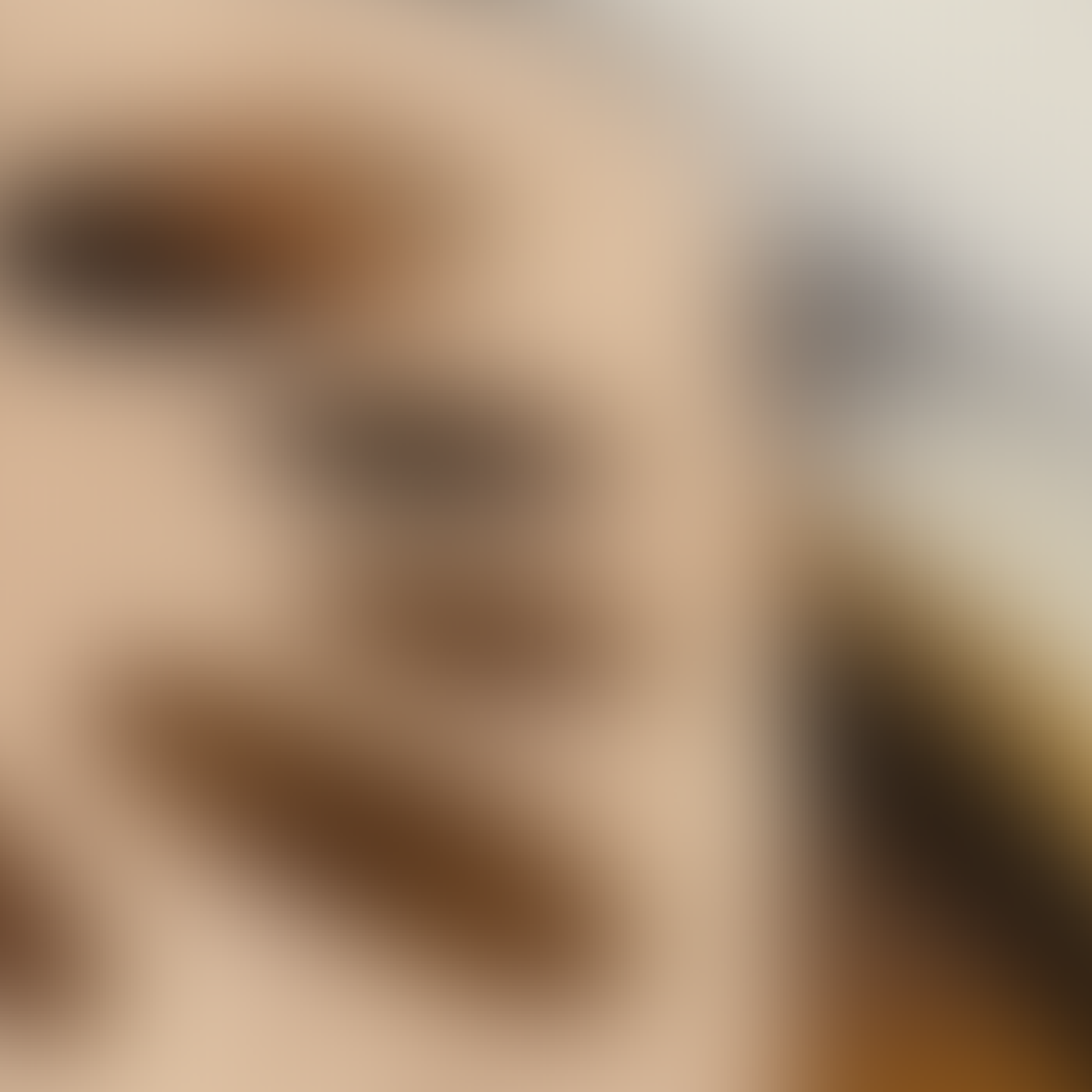}\\[1pt]
\mbox{\fontsize{9pt}{11pt}\selectfont CP\,-\,68.1\,dB $\cdot$ 12,251 Mpix/s $\cdot$ 4.4$\times$ slower}\end{minipage} & \begin{minipage}[t]{2.500in}\centering \includegraphics[interpolate=false,height=2.500in]{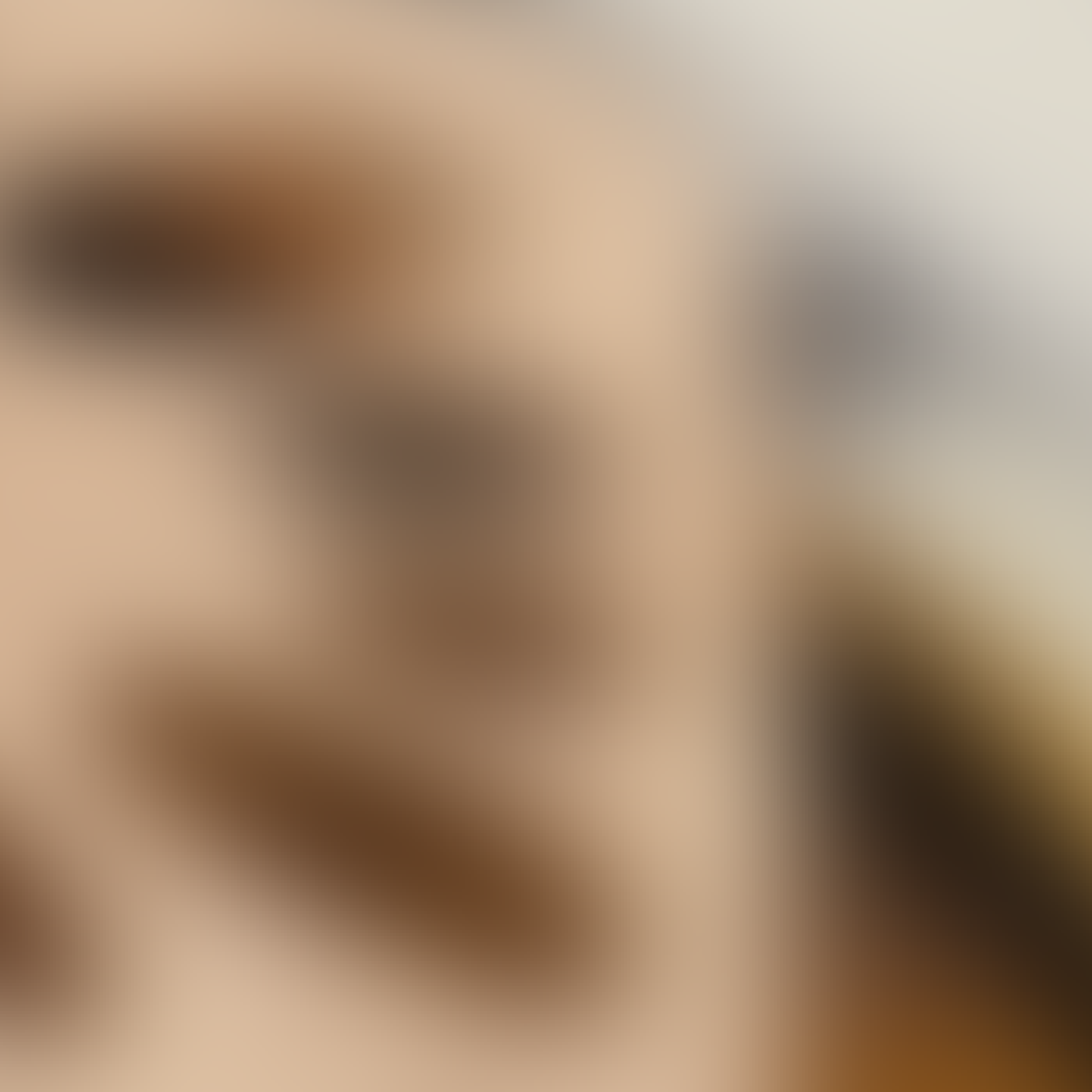}\\[1pt]
\mbox{\fontsize{9pt}{11pt}\selectfont Box3\,-\,43.5\,dB $\cdot$ 12,196 Mpix/s $\cdot$ 4.4$\times$ slower}\end{minipage} \\[12pt]
\begin{minipage}[t]{2.500in}\centering \includegraphics[interpolate=false,height=2.500in]{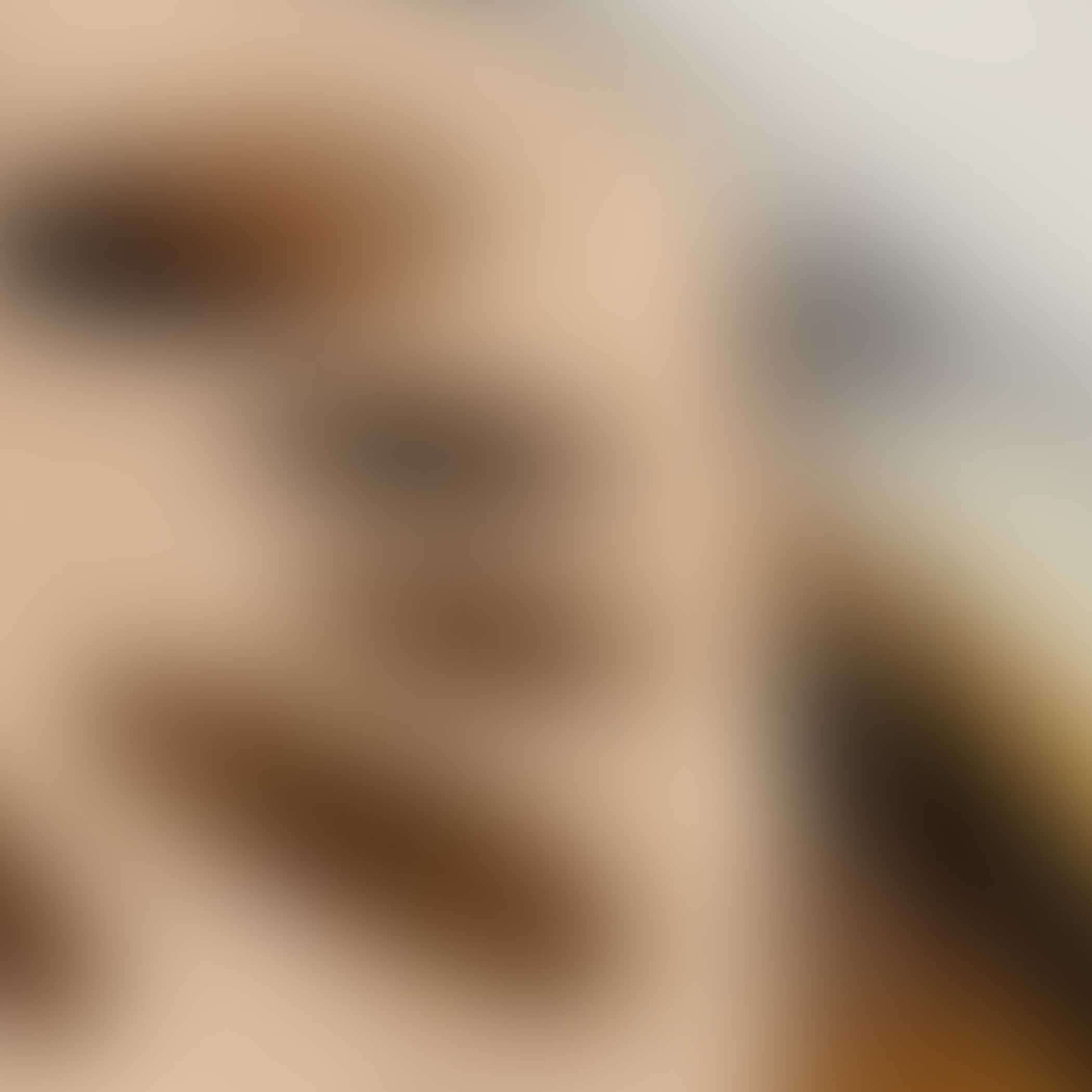}\\[1pt]
\mbox{\fontsize{9pt}{11pt}\selectfont FRM\,-\,49.6\,dB $\cdot$ 8,922 Mpix/s $\cdot$ 6.0$\times$ slower}\end{minipage} & \begin{minipage}[t]{2.500in}\centering \includegraphics[interpolate=false,height=2.500in]{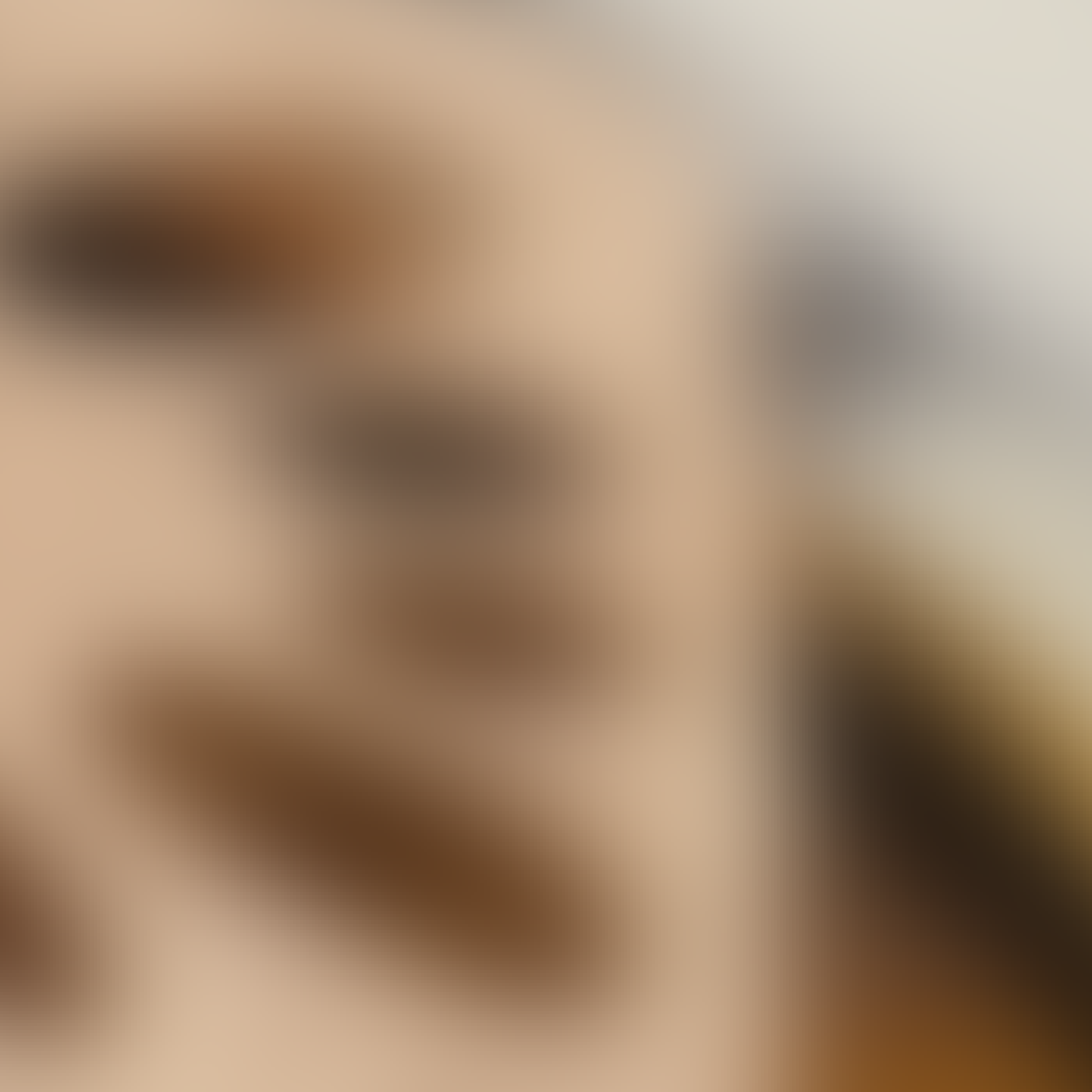}\\[1pt]
\mbox{\fontsize{9pt}{11pt}\selectfont YVV\,-\,43.8\,dB $\cdot$ 7,026 Mpix/s $\cdot$ 7.6$\times$ slower}\end{minipage} \\[12pt]
\end{tabular}
\caption{2D size 201 Gaussian filter approximation outputs.}
\label{fig:patch_gaussian2D_201_IMG_4303_y1850_x2200}
\end{figure*}

\begin{figure*}[p]\centering
\begin{tabular}{cc}
\begin{minipage}[t]{2.500in}\centering \includegraphics[interpolate=false,height=2.500in]{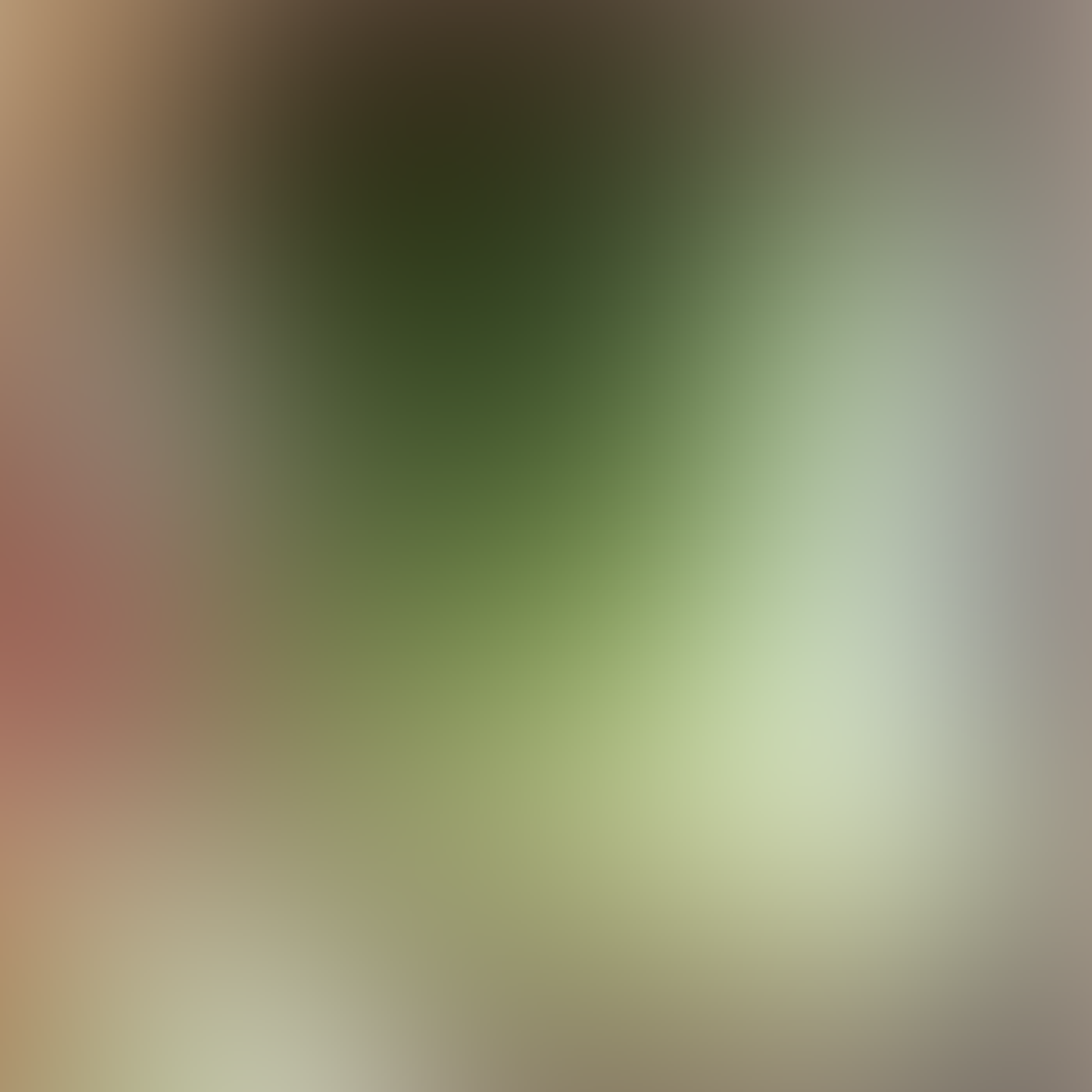}\\[1pt]
\mbox{\fontsize{9pt}{11pt}\selectfont Ground truth}\end{minipage} & \begin{minipage}[t]{2.500in}\centering \includegraphics[interpolate=false,height=2.500in]{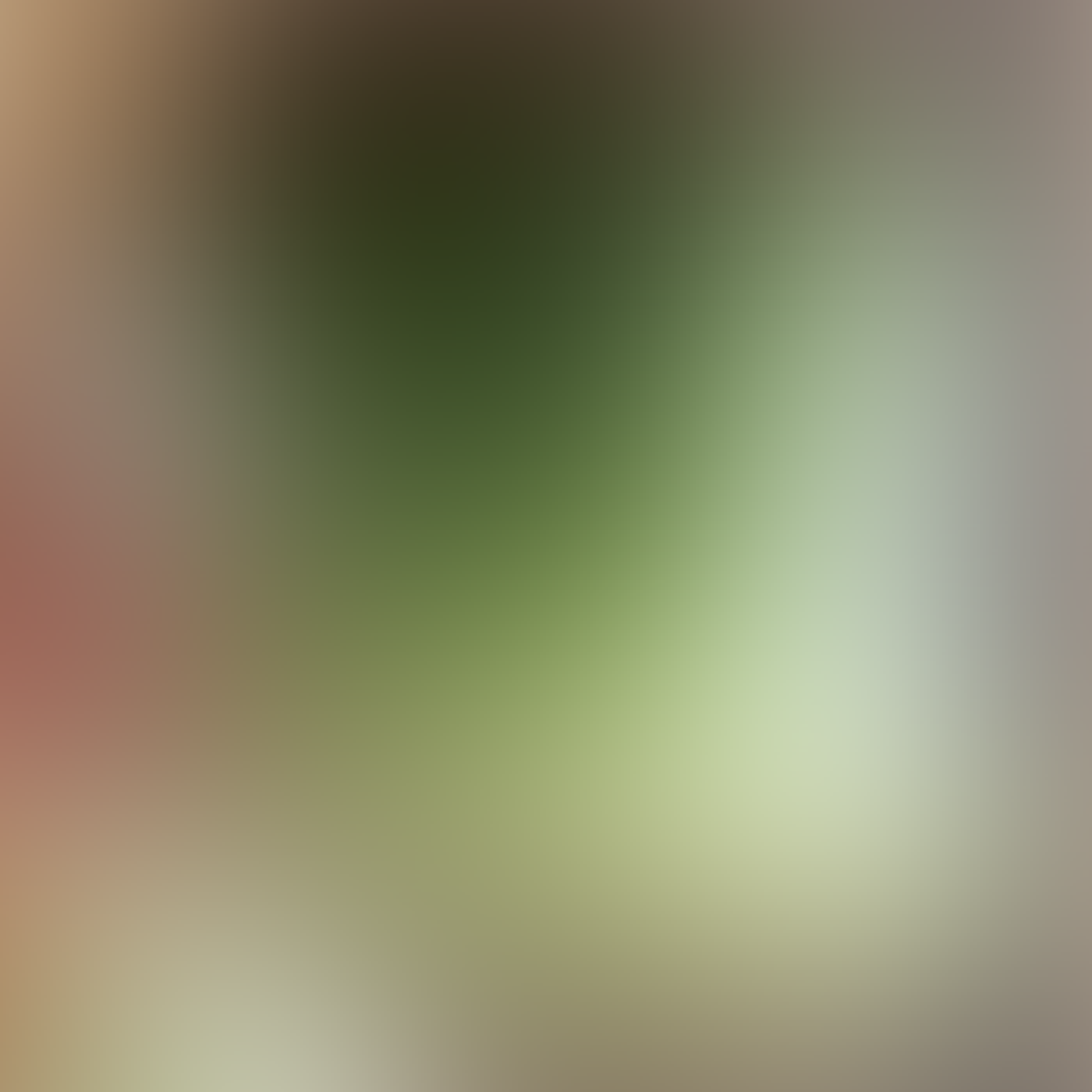}\\[1pt]
\mbox{\fontsize{9pt}{11pt}\selectfont Ours\,/\,71.5\,dB $\cdot$ 45,846 Mpix/s $\cdot$ fastest}\end{minipage} \\[12pt]
\begin{minipage}[t]{2.500in}\centering \includegraphics[interpolate=false,height=2.500in]{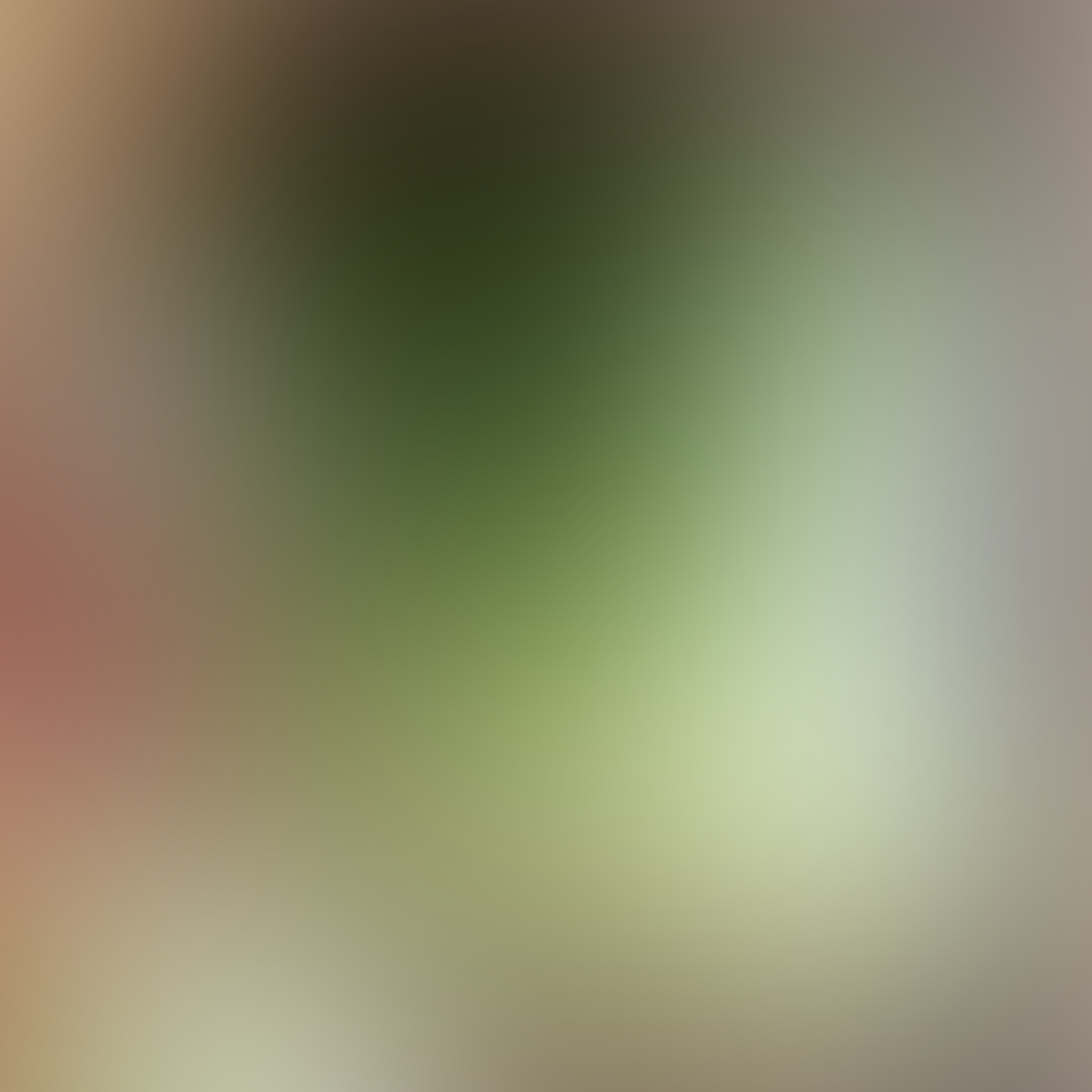}\\[1pt]
\mbox{\fontsize{9pt}{11pt}\selectfont Box3\,/\,44.3\,dB $\cdot$ 10,135 Mpix/s $\cdot$ 4.5$\times$ slower}\end{minipage} & \begin{minipage}[t]{2.500in}\centering \includegraphics[interpolate=false,height=2.500in]{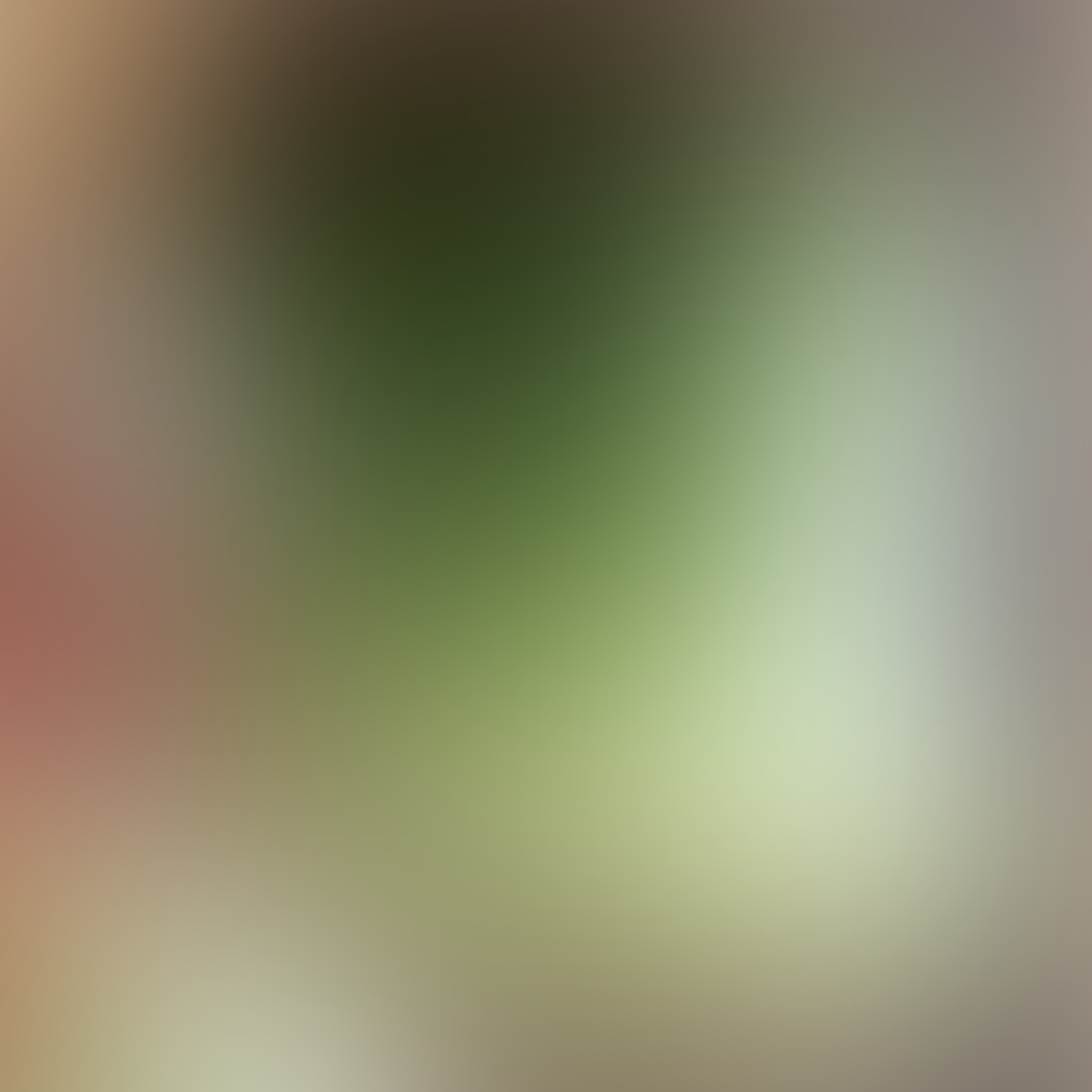}\\[1pt]
\mbox{\fontsize{9pt}{11pt}\selectfont CP\,/\,72.5\,dB $\cdot$ 9,353 Mpix/s $\cdot$ 4.9$\times$ slower}\end{minipage} \\[12pt]
\begin{minipage}[t]{2.500in}\centering \includegraphics[interpolate=false,height=2.500in]{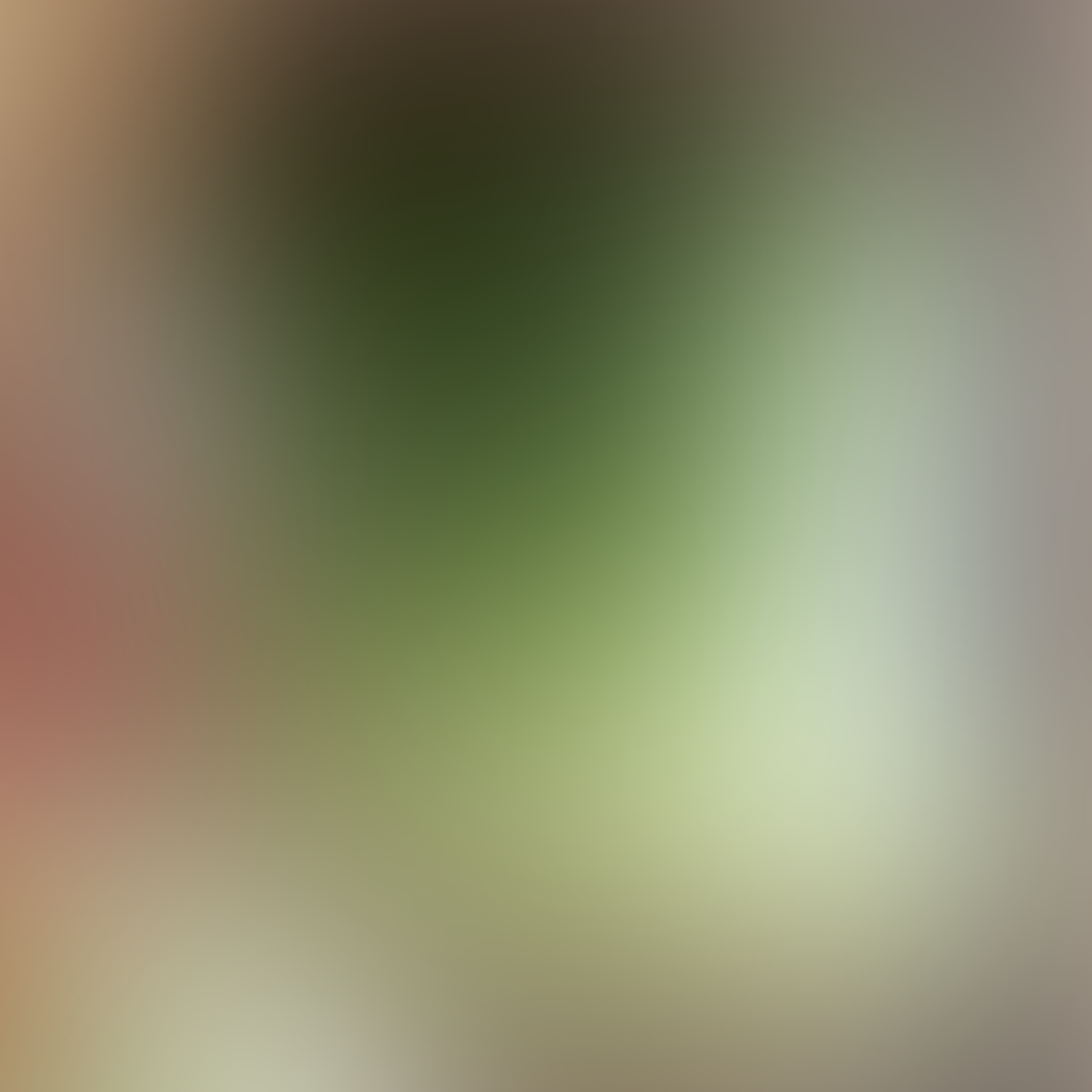}\\[1pt]
\mbox{\fontsize{9pt}{11pt}\selectfont FRM\,/\,70.0\,dB $\cdot$ 5,879 Mpix/s $\cdot$ 7.8$\times$ slower}\end{minipage} & \begin{minipage}[t]{2.500in}\centering \includegraphics[interpolate=false,height=2.500in]{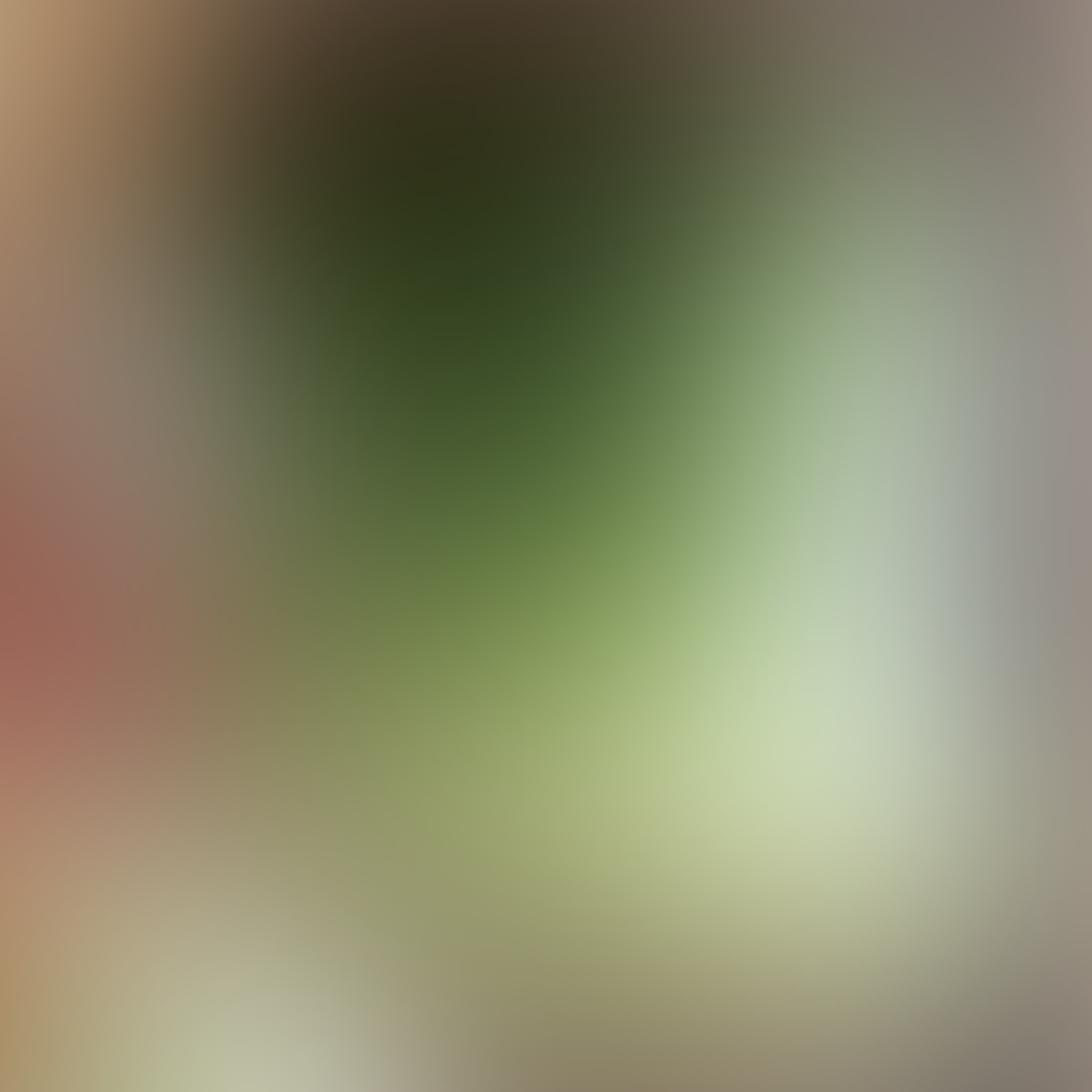}\\[1pt]
\mbox{\fontsize{9pt}{11pt}\selectfont YVV\,/\,44.3\,dB $\cdot$ 4,303 Mpix/s $\cdot$ 10.7$\times$ slower}\end{minipage} \\[12pt]
\end{tabular}
\vspace{-0.5em}
\caption{Outputs for 2D size 401 Gaussian filter approximations on a patch from an image taken with the authors' iPhone 13 Pro. For such large blur sizes the artifacts from triple box, FRM, and YVV are not noticeable. Our model is however still the fastest approximation. It is 4.5x faster than the fastest baseline, triple-box blur, and 10.7x faster than the slowest baseline, YVV.}
\label{fig:patch_gaussian2D_401_IMG_5781_y1600_x2300}
\end{figure*}

\urldef{\imglink}\url{https://raw.pixls.us/getfile.php/7738/nice/Nikon%20-%20Z5_2%20-%208bit%20compressed%20%283%3A2%29.NEF}
\begin{figure*}[p]\centering
\begin{tabular}{cc}
\begin{minipage}[t]{2.500in}\centering \includegraphics[interpolate=false,height=2.500in]{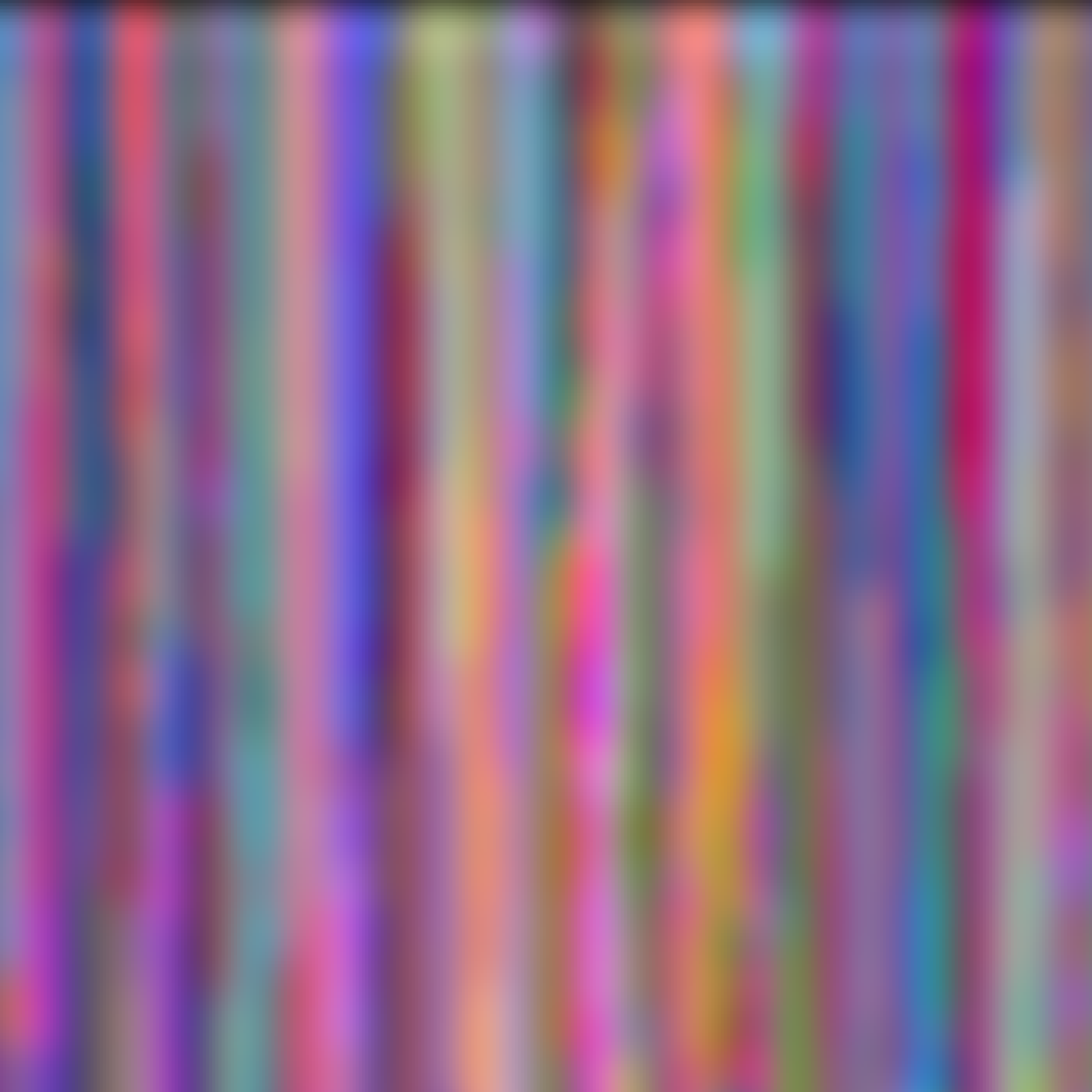}\\[1pt]
\mbox{\fontsize{9pt}{11pt}\selectfont Ground truth}\end{minipage} & \begin{minipage}[t]{2.500in}\centering \includegraphics[interpolate=false,height=2.500in]{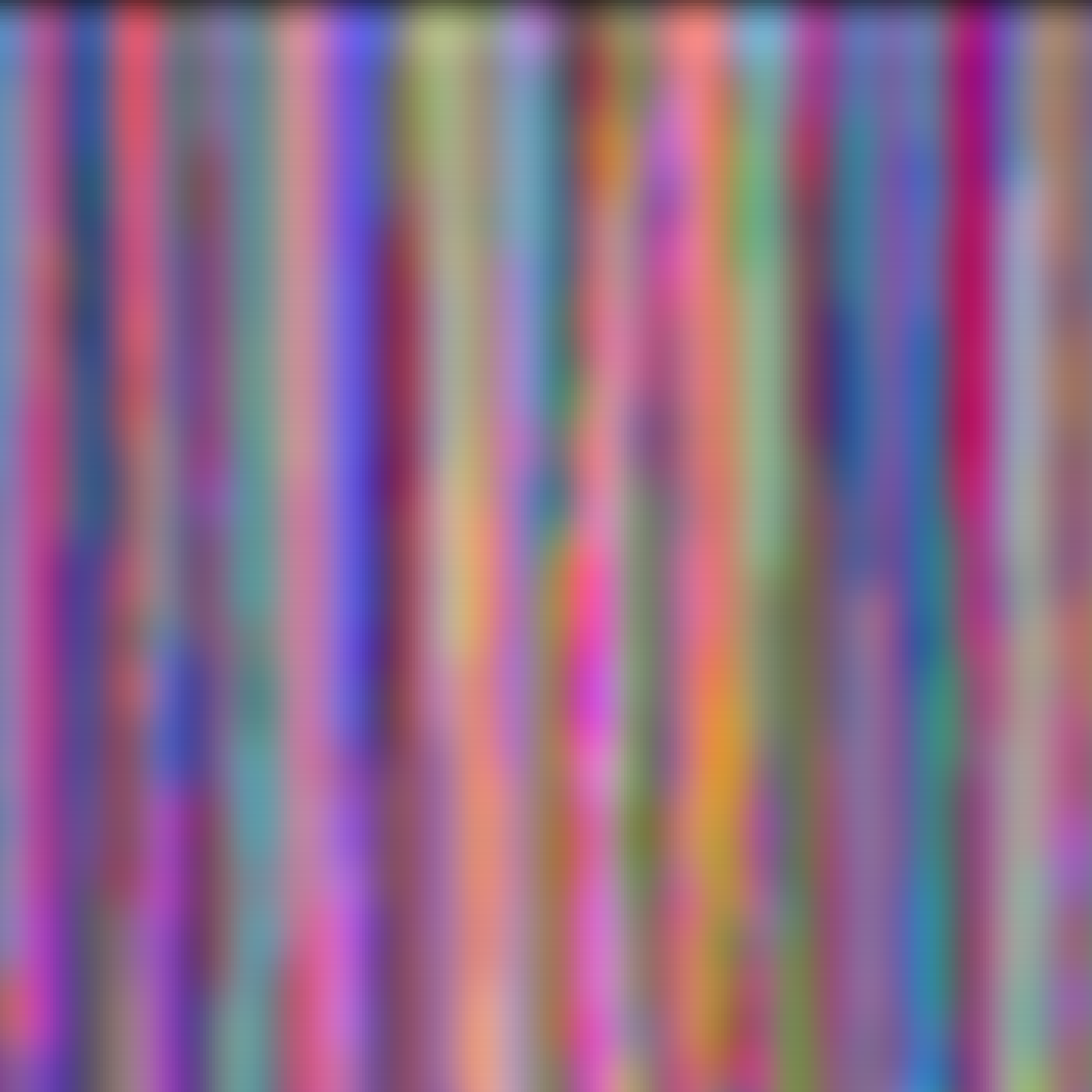}\\[1pt]
\mbox{\fontsize{9pt}{11pt}\selectfont Ours\,/\,55.2\,dB $\cdot$ 44,182 Mpix/s $\cdot$ fastest}\end{minipage} \\[12pt]
\begin{minipage}[t]{2.500in}\centering \includegraphics[interpolate=false,height=2.500in]{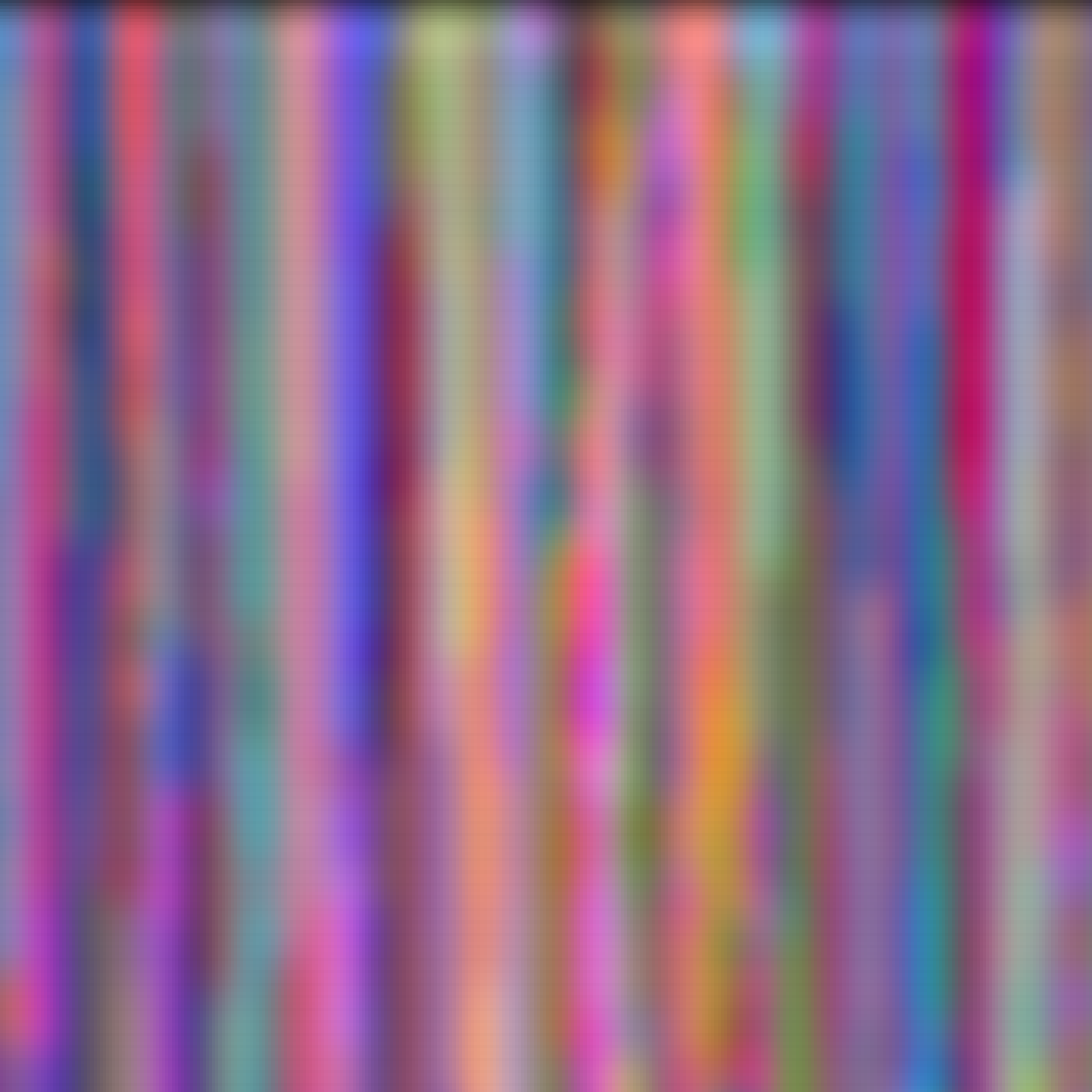}\\[1pt]
\mbox{\fontsize{9pt}{11pt}\selectfont CP\,/\,48.8\,dB $\cdot$ 13,634 Mpix/s $\cdot$ 3.2$\times$ slower}\end{minipage} & \begin{minipage}[t]{2.500in}\centering \includegraphics[interpolate=false,height=2.500in]{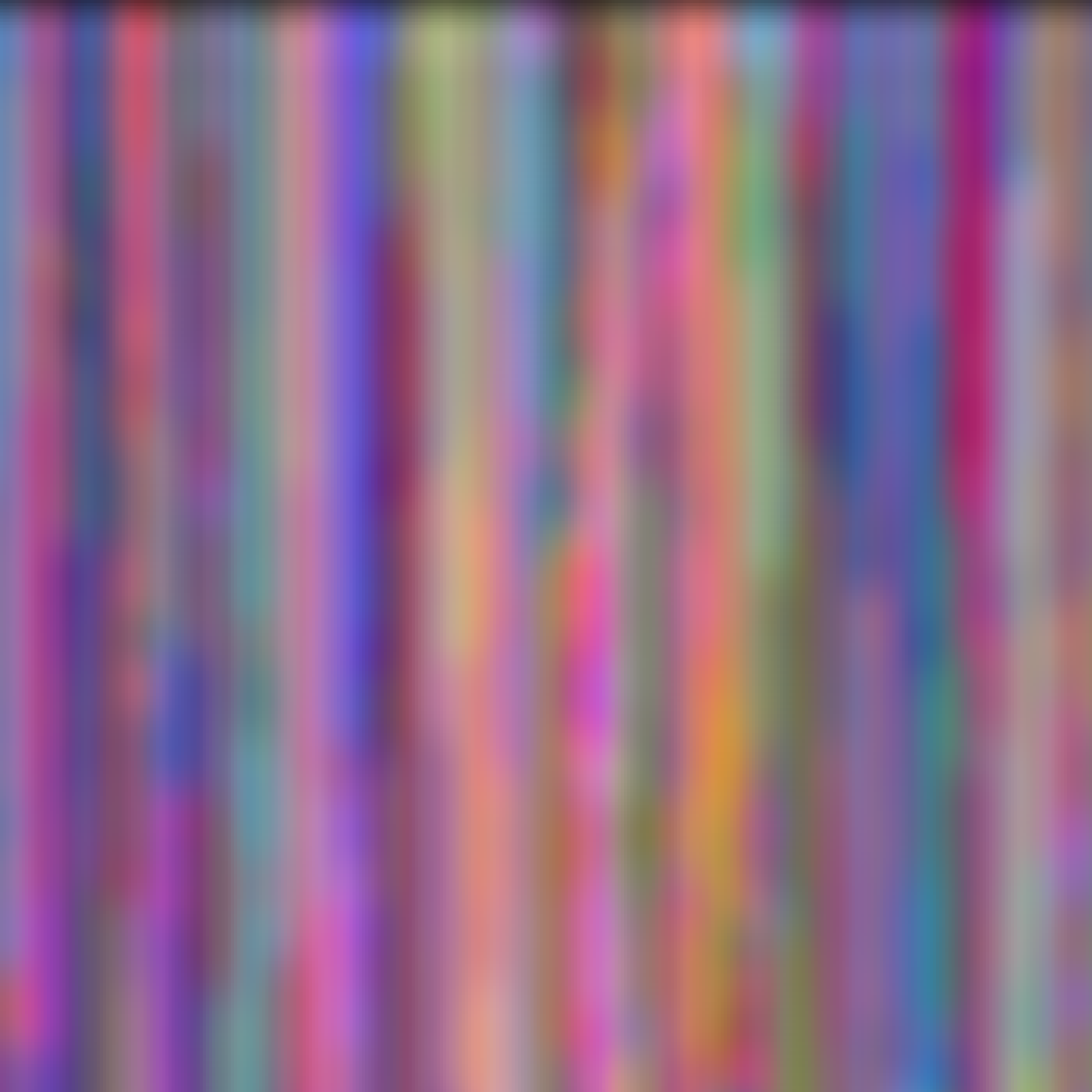}\\[1pt]
\mbox{\fontsize{9pt}{11pt}\selectfont FRM\,/\,38.9\,dB $\cdot$ 10,880 Mpix/s $\cdot$ 4.1$\times$ slower}\end{minipage} \\[12pt]
\end{tabular}
\caption{Figures 51 - 53 show outputs for 2D size 101 Lowpass filter approximations, on patches from image: \imglink. Both FRM's and CP's outputs have ringing artifacts, but FRM's outputs have much thicker and more noticeable bands. Our outputs are visually indistinguishable from the ground truth and are 3.2x faster than CP and 4.1x faster than FRM.}
\label{fig:patch_lowpass2D_101_Nikon___Z5_2___8bit_compressed_32_y50_x750}
\end{figure*}

\begin{figure*}[p]\centering
\begin{tabular}{cc}
\begin{minipage}[t]{2.500in}\centering \includegraphics[interpolate=false,height=2.500in]{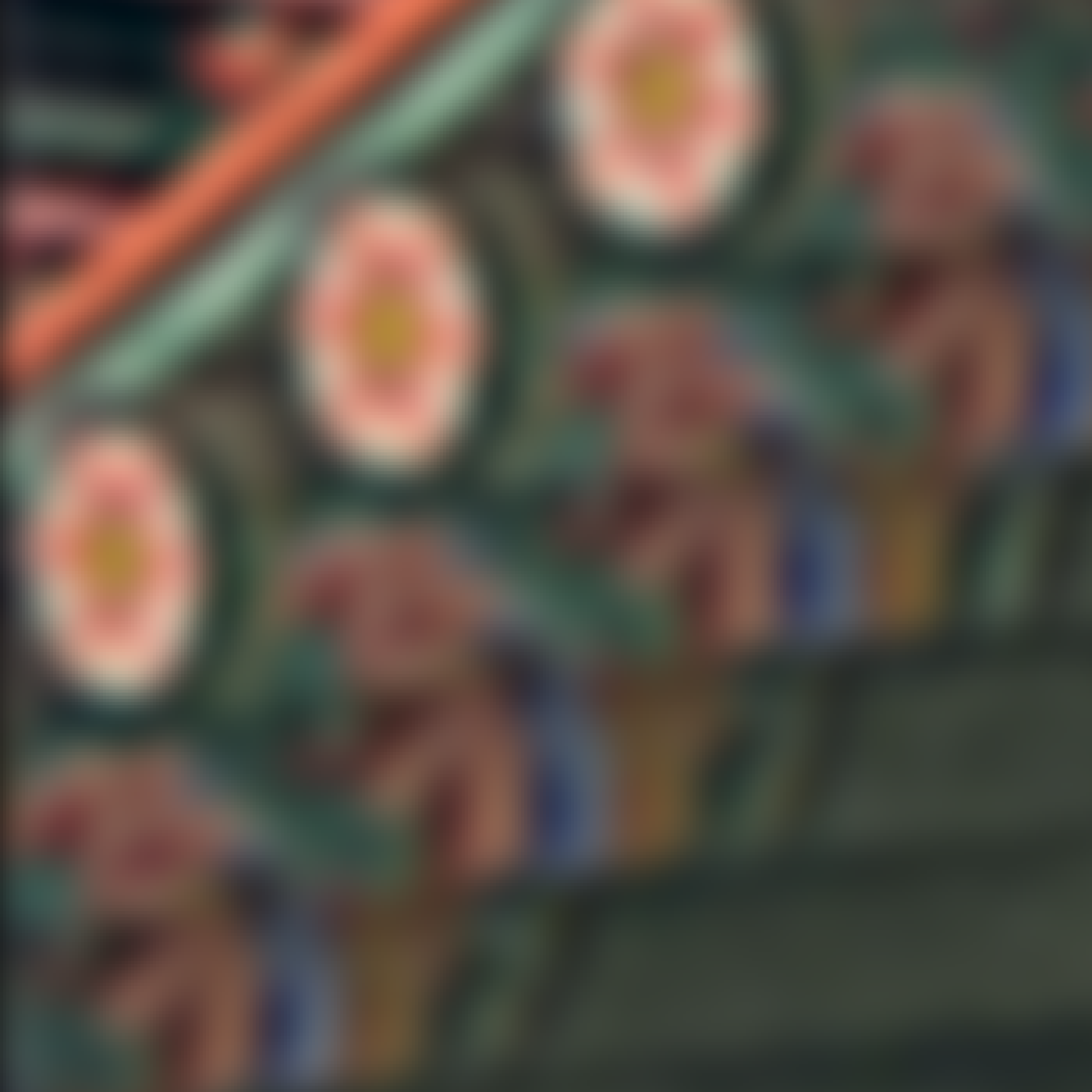}\\[1pt]
\mbox{\fontsize{9pt}{11pt}\selectfont Ground truth}\end{minipage} & \begin{minipage}[t]{2.500in}\centering \includegraphics[interpolate=false,height=2.500in]{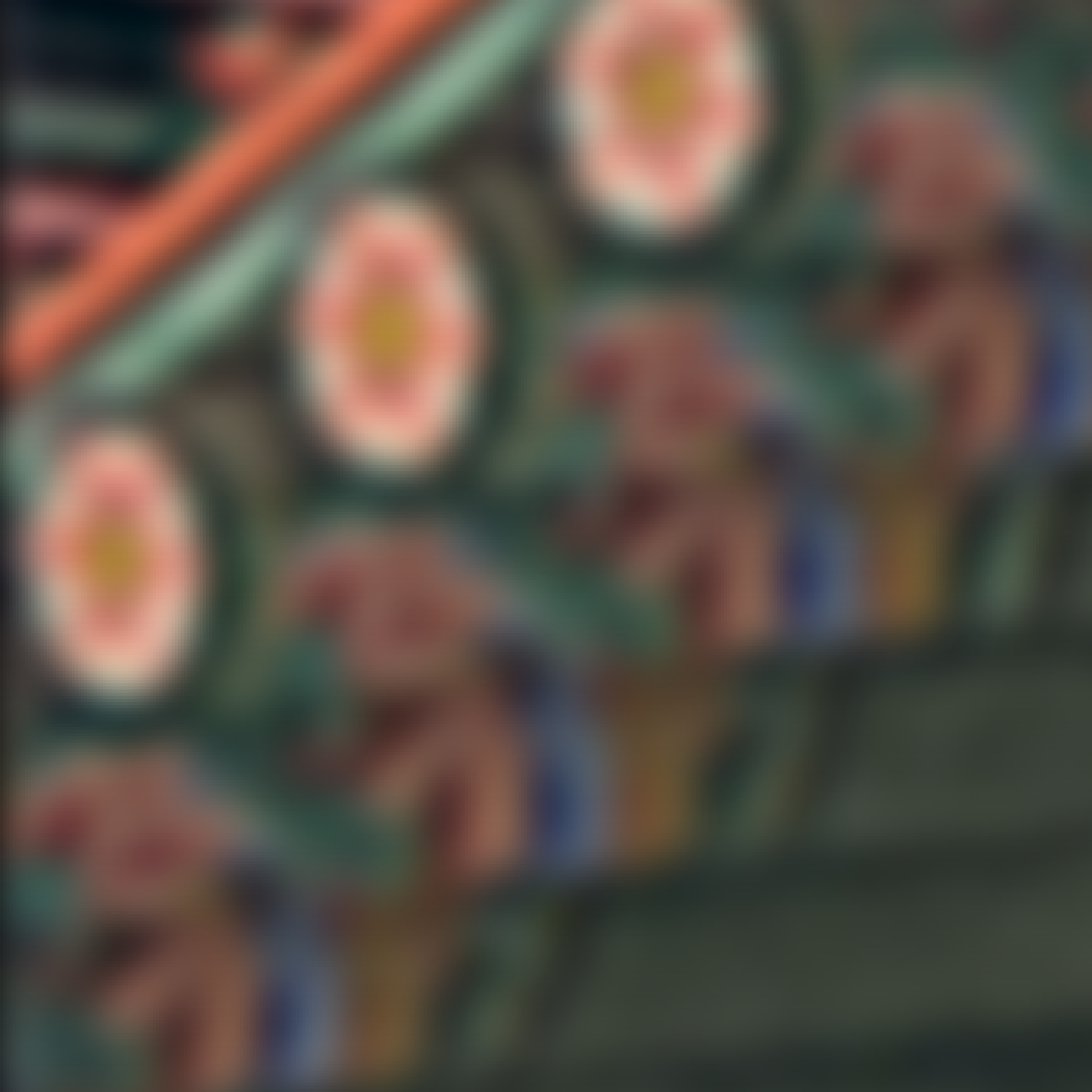}\\[1pt]
\mbox{\fontsize{9pt}{11pt}\selectfont Ours\,/\,59.5\,dB $\cdot$ 44,182 Mpix/s $\cdot$ fastest}\end{minipage} \\[12pt]
\begin{minipage}[t]{2.500in}\centering \includegraphics[interpolate=false,height=2.500in]{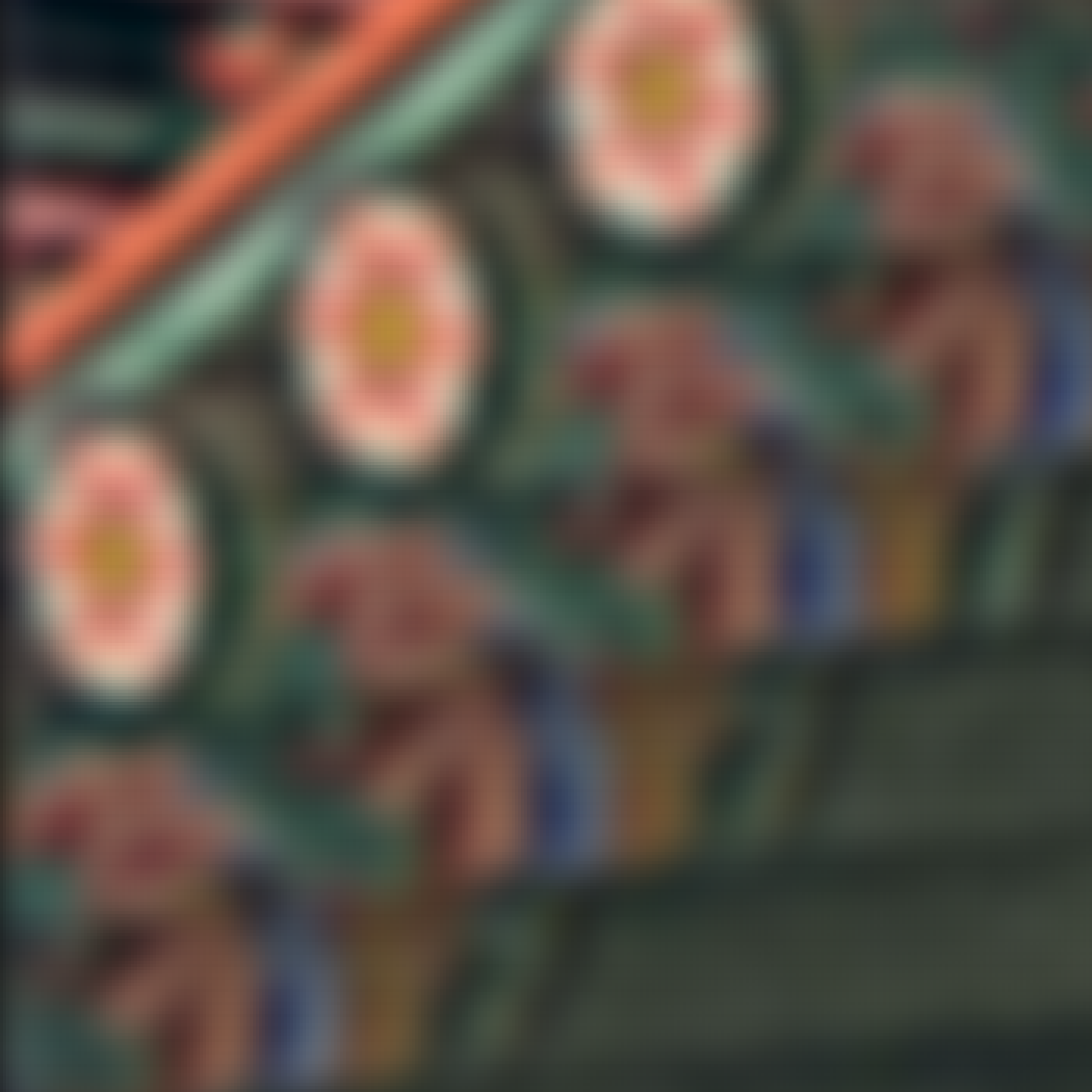}\\[1pt]
\mbox{\fontsize{9pt}{11pt}\selectfont CP\,/\,51.5\,dB $\cdot$ 13,634 Mpix/s $\cdot$ 3.2$\times$ slower}\end{minipage} & \begin{minipage}[t]{2.500in}\centering \includegraphics[interpolate=false,height=2.500in]{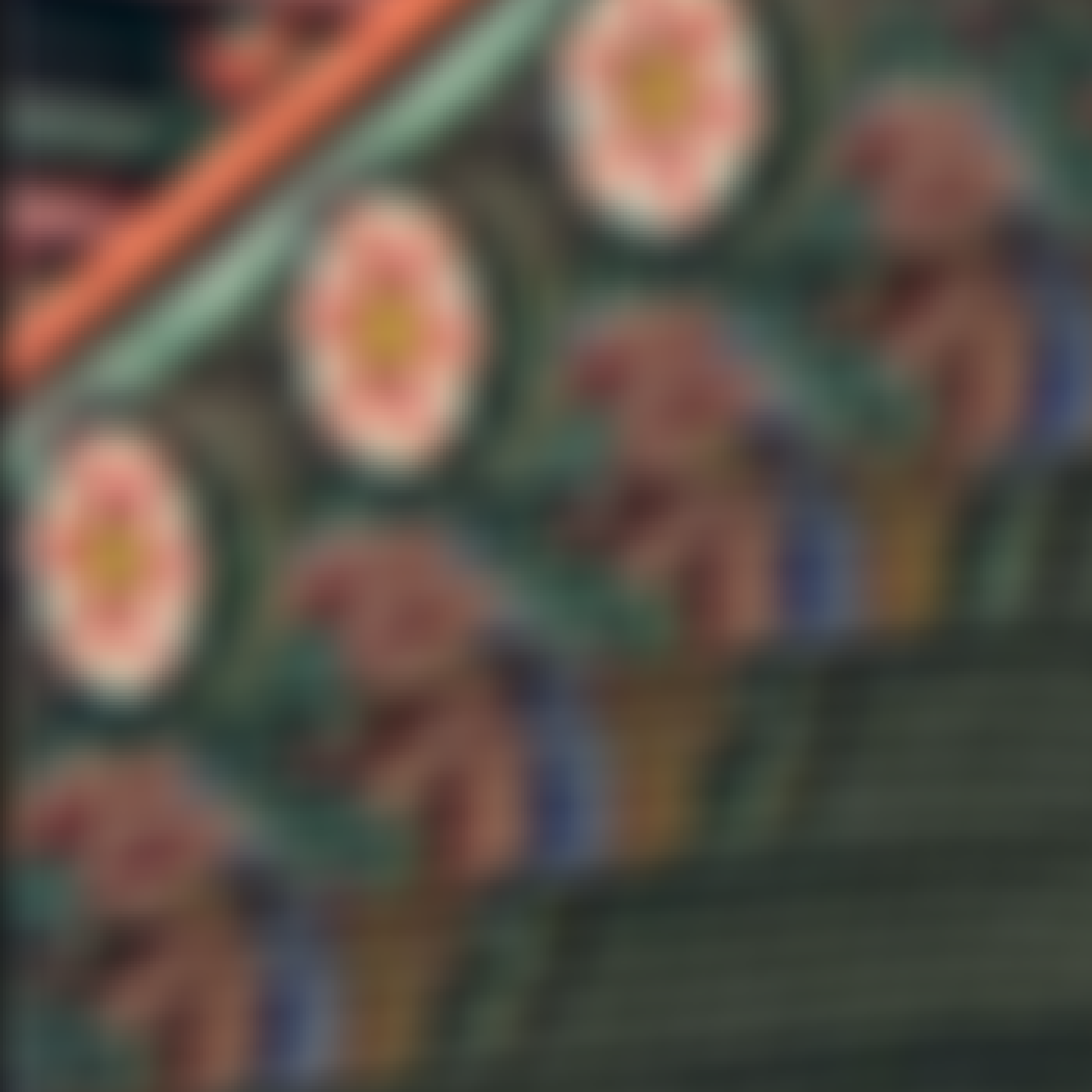}\\[1pt]
\mbox{\fontsize{9pt}{11pt}\selectfont FRM\,/\,45.4\,dB $\cdot$ 10,880 Mpix/s $\cdot$ 4.1$\times$ slower}\end{minipage} \\[12pt]
\end{tabular}
\caption{2D size 101 Lowpass filter approximation outputs.}
\label{fig:patch_lowpass2D_101_IMG_4455_y3200_x50}
\end{figure*}

\begin{figure*}[p]\centering
\begin{tabular}{cc}
\begin{minipage}[t]{2.500in}\centering \includegraphics[interpolate=false,height=2.500in]{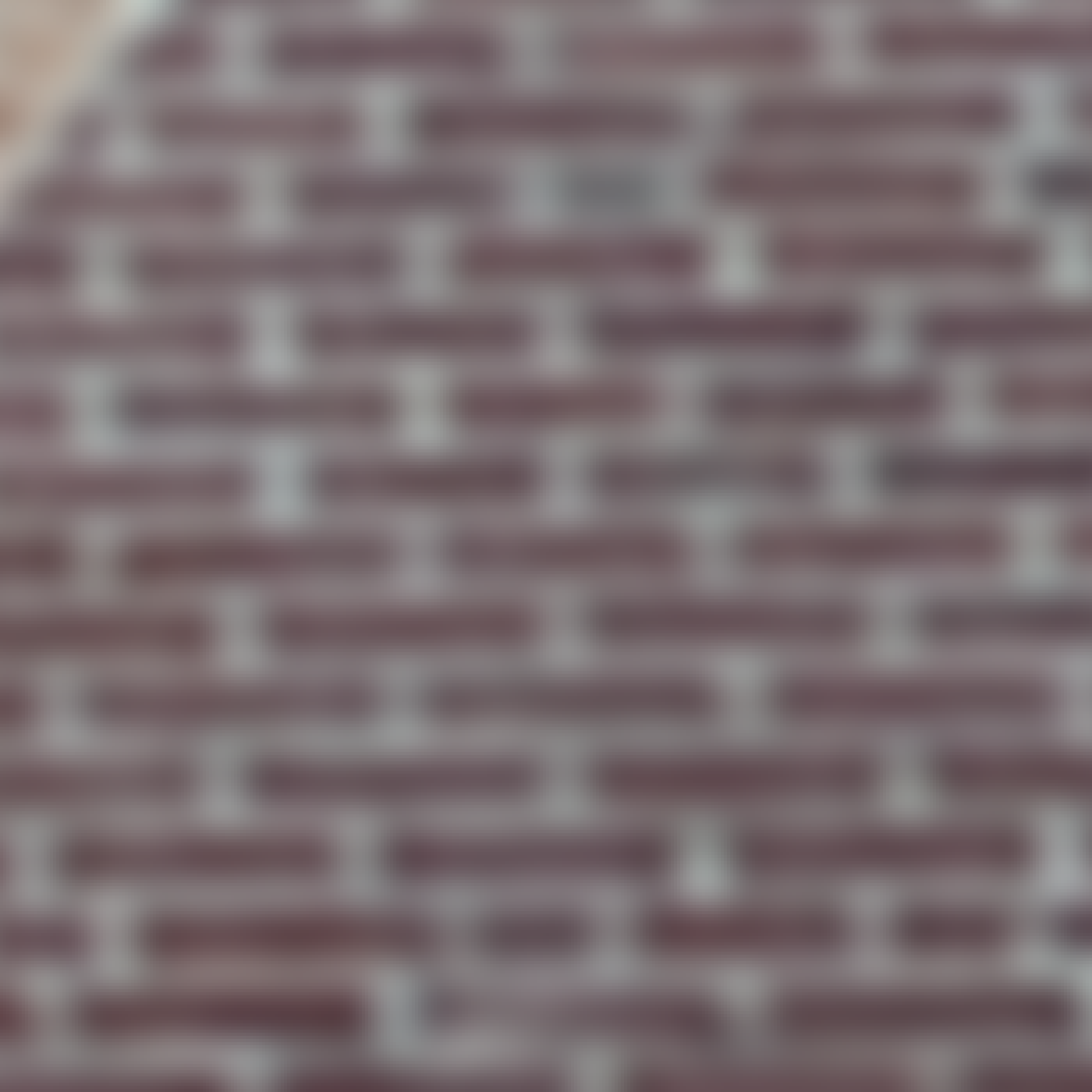}\\[1pt]
\mbox{\fontsize{9pt}{11pt}\selectfont Ground truth}\end{minipage} & \begin{minipage}[t]{2.500in}\centering \includegraphics[interpolate=false,height=2.500in]{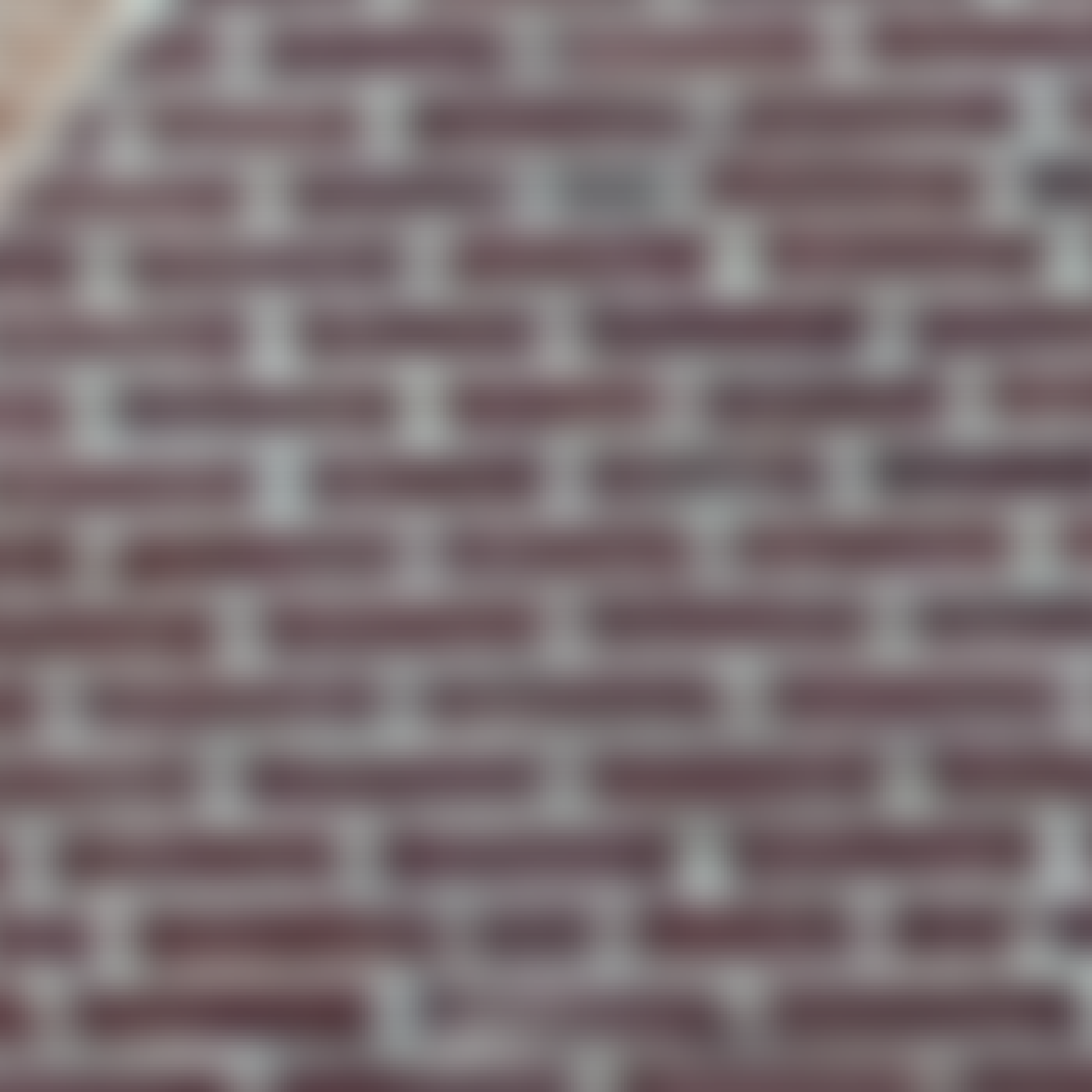}\\[1pt]
\mbox{\fontsize{9pt}{11pt}\selectfont Ours\,/\,57.9\,dB $\cdot$ 44,182 Mpix/s $\cdot$ fastest}\end{minipage} \\[12pt]
\begin{minipage}[t]{2.500in}\centering \includegraphics[interpolate=false,height=2.500in]{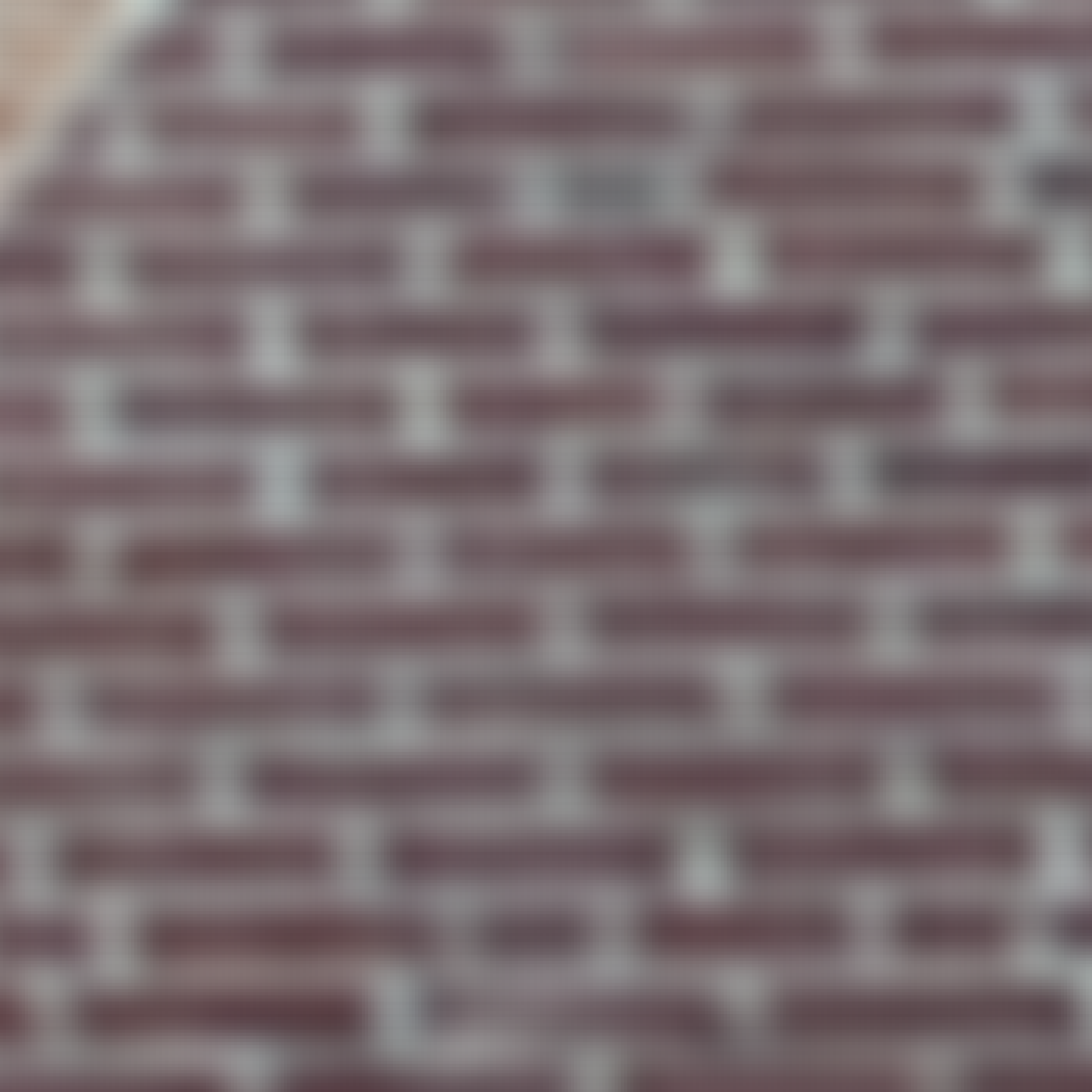}\\[1pt]
\mbox{\fontsize{9pt}{11pt}\selectfont CP\,/\,52.8\,dB $\cdot$ 13,634 Mpix/s $\cdot$ 3.2$\times$ slower}\end{minipage} & \begin{minipage}[t]{2.500in}\centering \includegraphics[interpolate=false,height=2.500in]{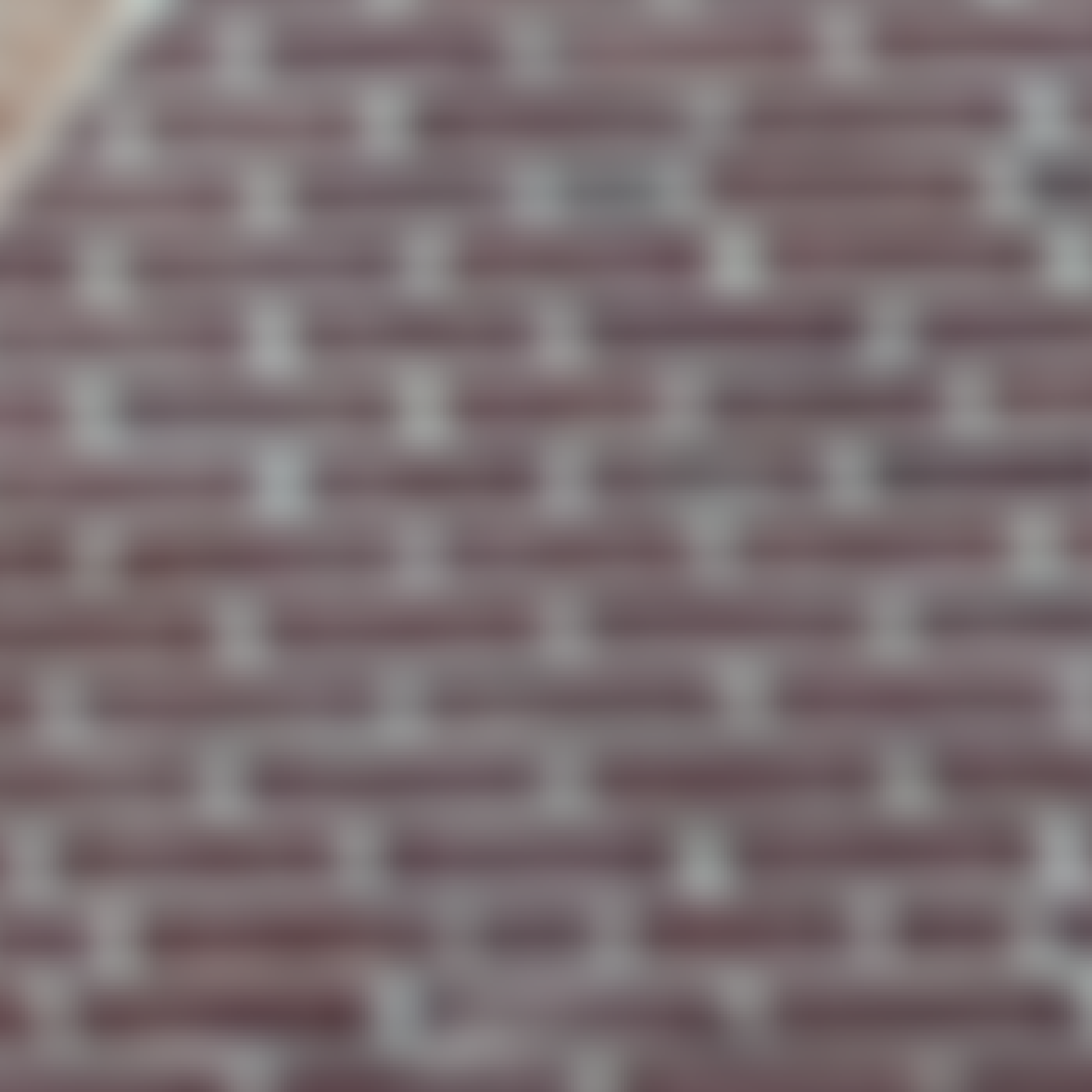}\\[1pt]
\mbox{\fontsize{9pt}{11pt}\selectfont FRM\,/\,41.2\,dB $\cdot$ 10,880 Mpix/s $\cdot$ 4.1$\times$ slower}\end{minipage} \\[12pt]
\end{tabular}
\caption{2D size 101 Lowpass approximation outputs.}
\label{fig:patch_lowpass2D_101_IMG_5700_y2500_x2500}
\end{figure*}

\urldef{\imglink}\url{https://raw.pixls.us/getfile.php/7738/nice/Nikon%20-%20Z5_2%20-%208bit%20compressed%20%283%3A2%29.NEF}

\begin{figure*}[p]\centering
\begin{tabular}{cc}
\begin{minipage}[t]{2.500in}\centering \includegraphics[interpolate=false,height=2.500in]{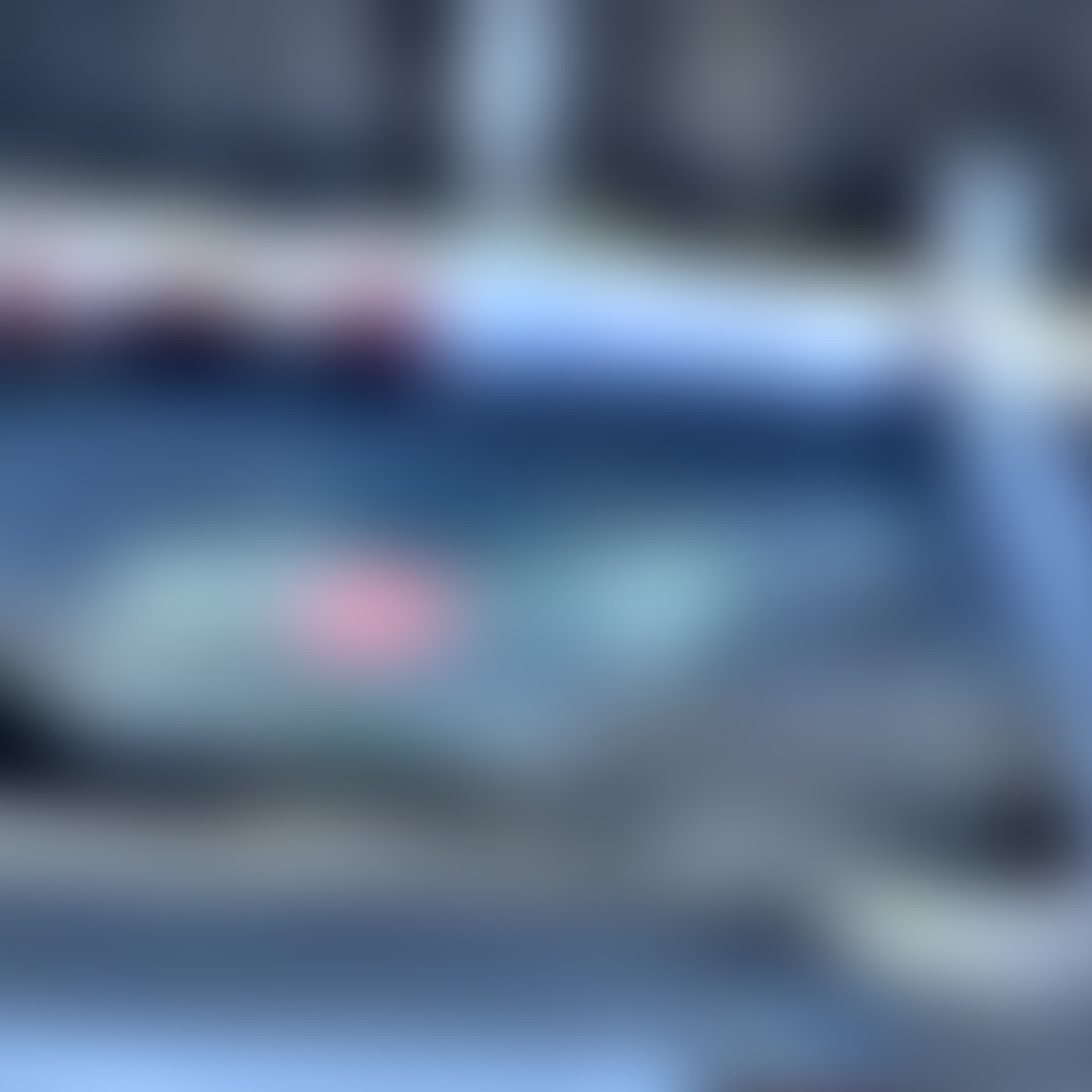}\\[1pt]
\mbox{\fontsize{9pt}{11pt}\selectfont Ground truth}\end{minipage} & \begin{minipage}[t]{2.500in}\centering \includegraphics[interpolate=false,height=2.500in]{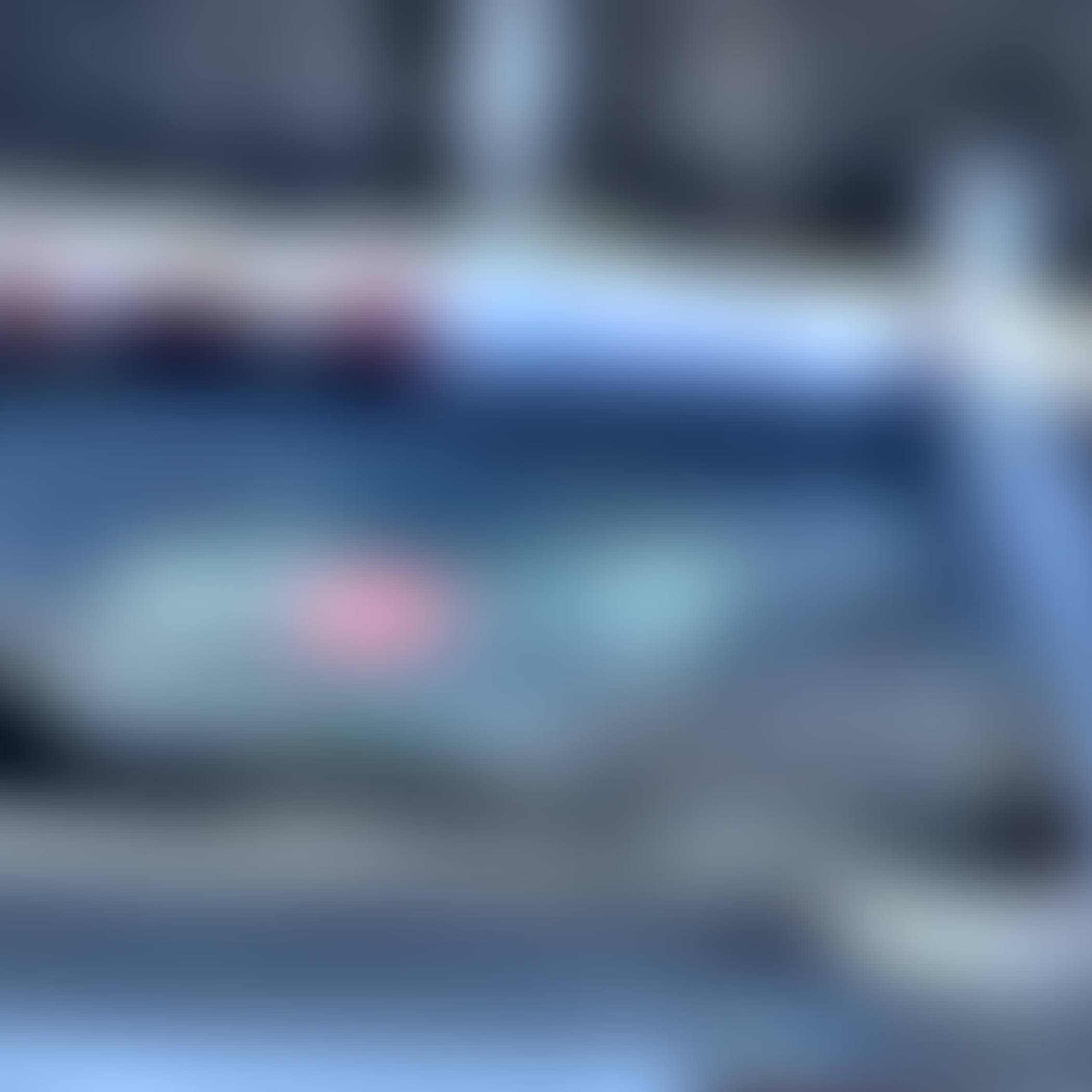}\\[1pt]
\mbox{\fontsize{9pt}{11pt}\selectfont Ours\,/\,70.8\,dB $\cdot$ 35,032 Mpix/s $\cdot$ fastest}\end{minipage} \\[12pt]
\begin{minipage}[t]{2.500in}\centering \includegraphics[interpolate=false,height=2.500in]{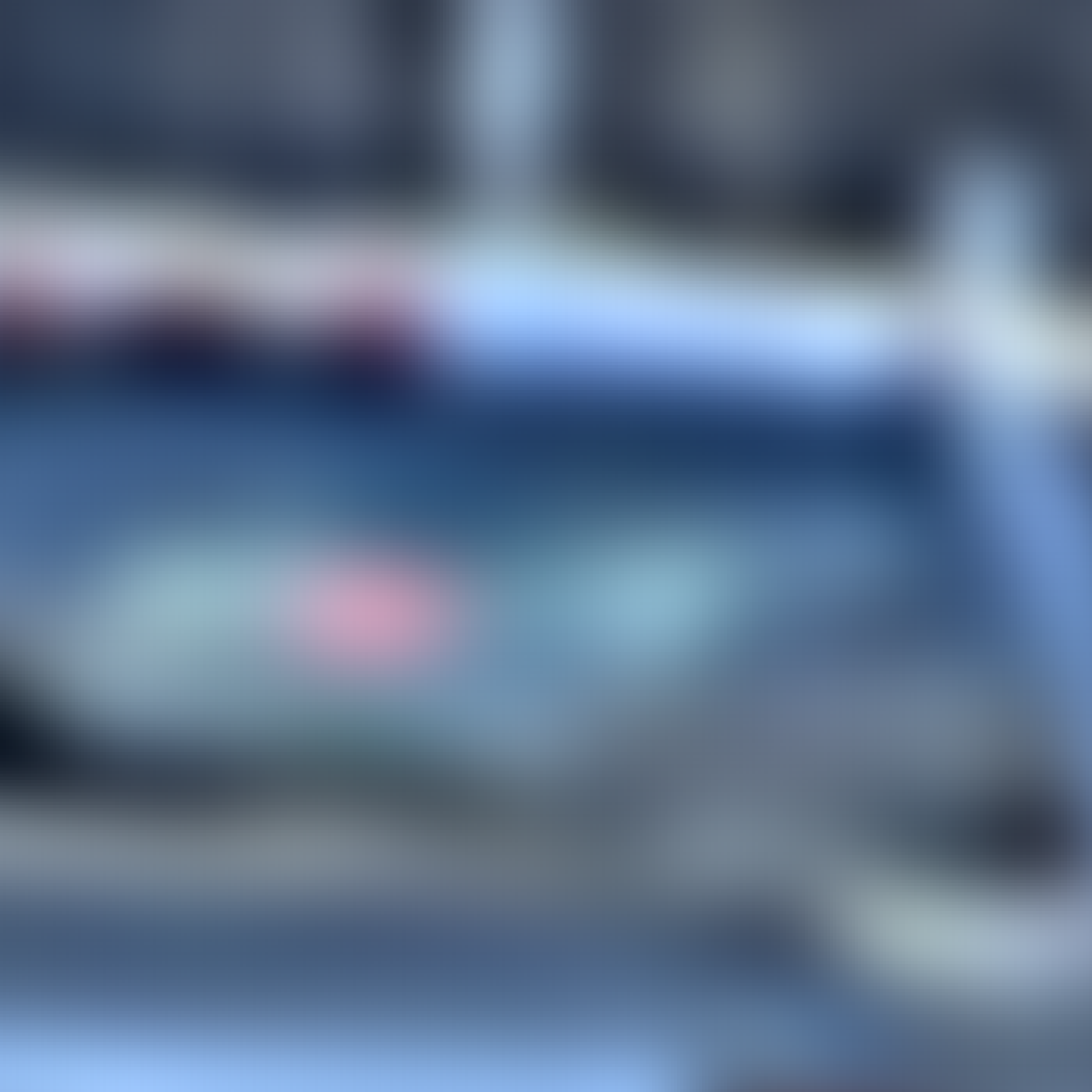}\\[1pt]
\mbox{\fontsize{9pt}{11pt}\selectfont CP\,/\,59.5\,dB $\cdot$ 11,899 Mpix/s $\cdot$ 2.9$\times$ slower}\end{minipage} & \begin{minipage}[t]{2.500in}\centering \includegraphics[interpolate=false,height=2.500in]{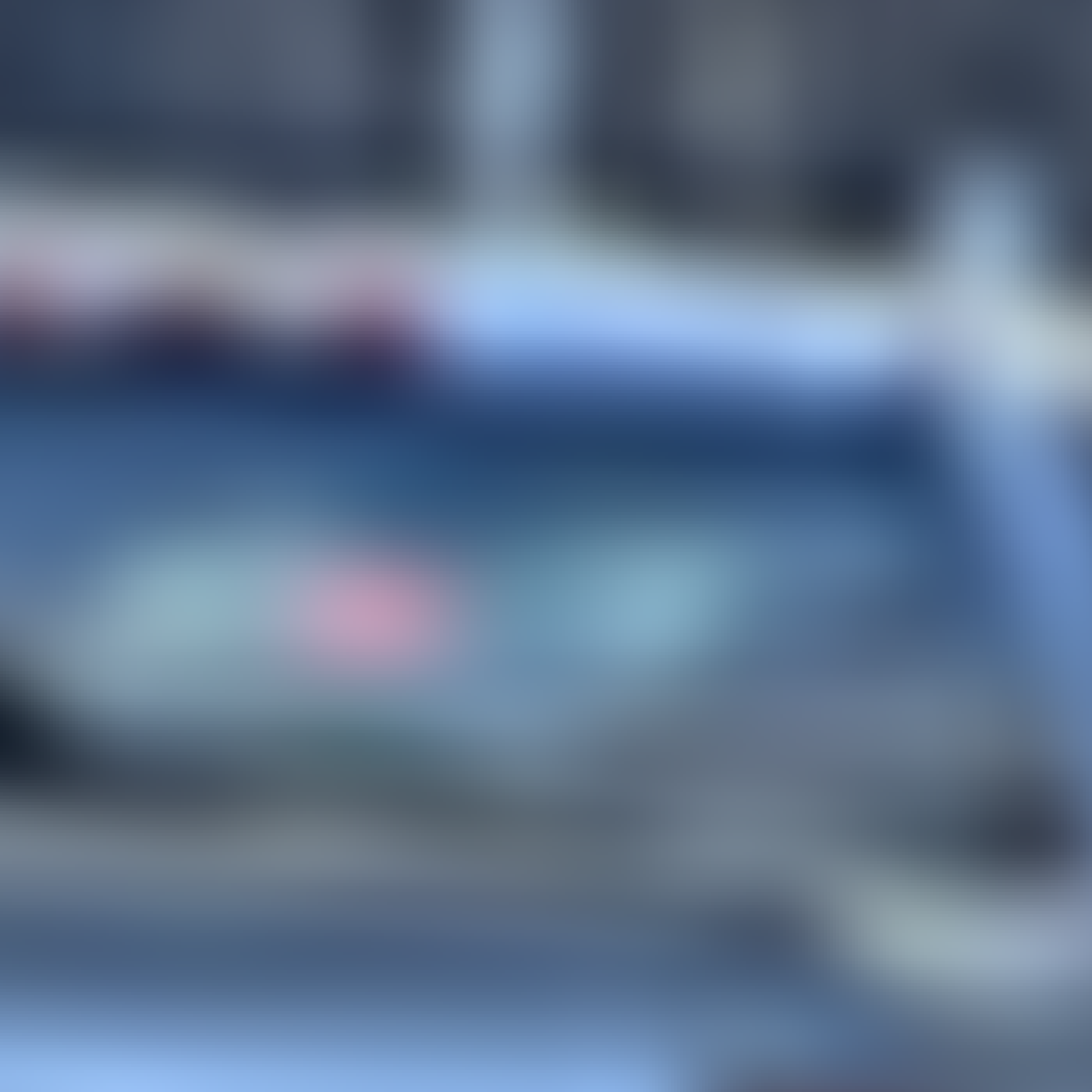}\\[1pt]
\mbox{\fontsize{9pt}{11pt}\selectfont FRM\,/\,39.9\,dB $\cdot$ 7,836 Mpix/s $\cdot$ 4.5$\times$ slower}\end{minipage} \\[12pt]
\end{tabular}
\caption{Figures 55 and 56 show outputs for 2D size 201 Lowpass filter approximations, on patches from different images taken with the authors’ iPhone 13 Pro. Compared to size 101, CP produces barely noticeable banding. FRM's outputs, however, still have visible ringing artifacts. Our approximation produces outputs that are visually indistinguishable from ground truth and 2.9x faster than CP and 4.5x faster than FRM.}
\label{fig:patch_lowpass2D_201_IMG_5736_y2550_x450}
\end{figure*}

\urldef{\imglink}\url{https://raw.pixls.us/getfile.php/7738/nice/Nikon%20-%20Z5_2%20-%208bit%20compressed%20%283%3A2%29.NEF}
\begin{figure*}[p]\centering
\begin{tabular}{cc}
\begin{minipage}[t]{2.500in}\centering \includegraphics[interpolate=false,height=2.500in]{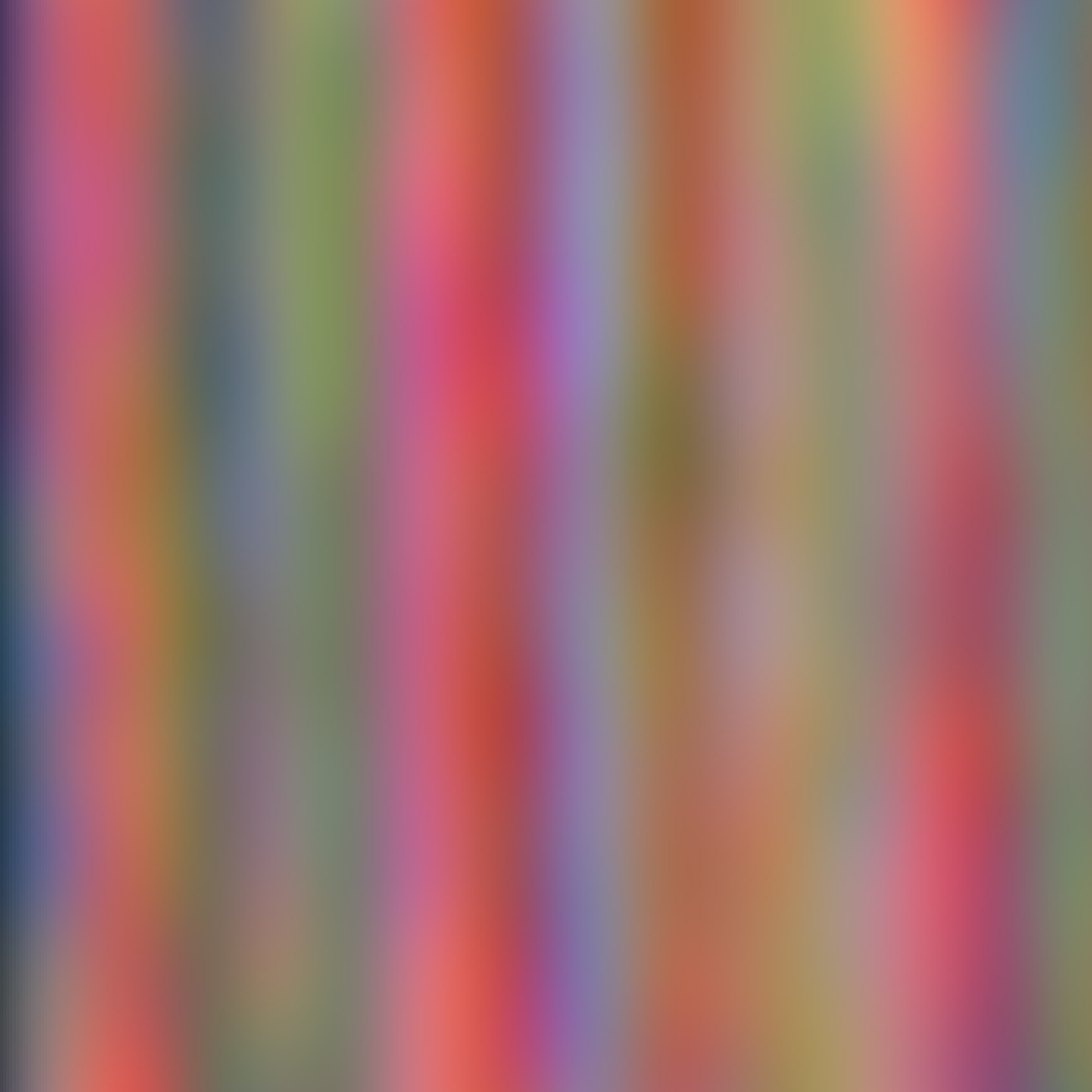}\\[1pt]
\mbox{\fontsize{9pt}{11pt}\selectfont Ground truth}\end{minipage} & \begin{minipage}[t]{2.500in}\centering \includegraphics[interpolate=false,height=2.500in]{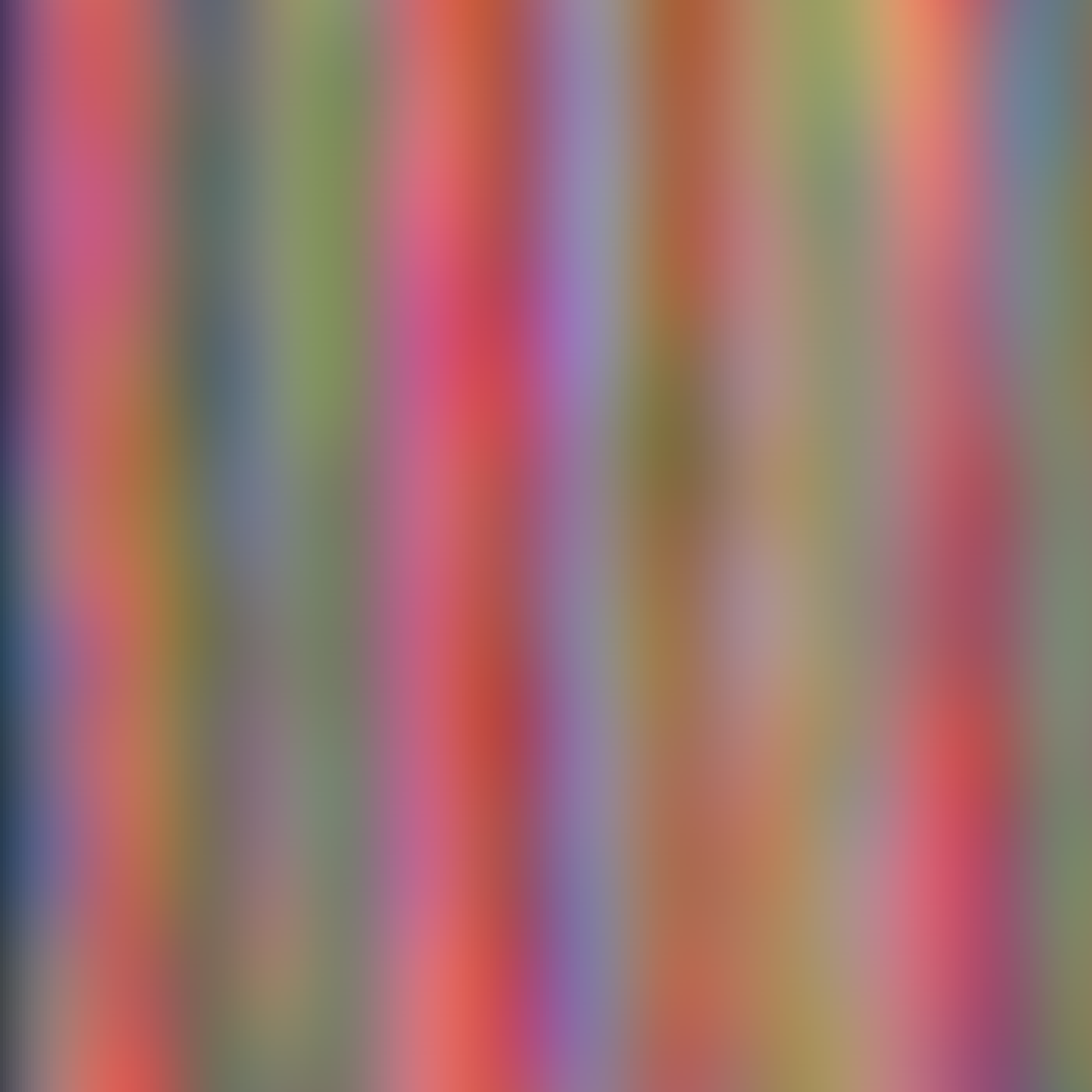}\\[1pt]
\mbox{\fontsize{9pt}{11pt}\selectfont Ours\,/\,71.1\,dB $\cdot$ 35,032 Mpix/s $\cdot$ fastest}\end{minipage} \\[12pt]
\begin{minipage}[t]{2.500in}\centering \includegraphics[interpolate=false,height=2.500in]{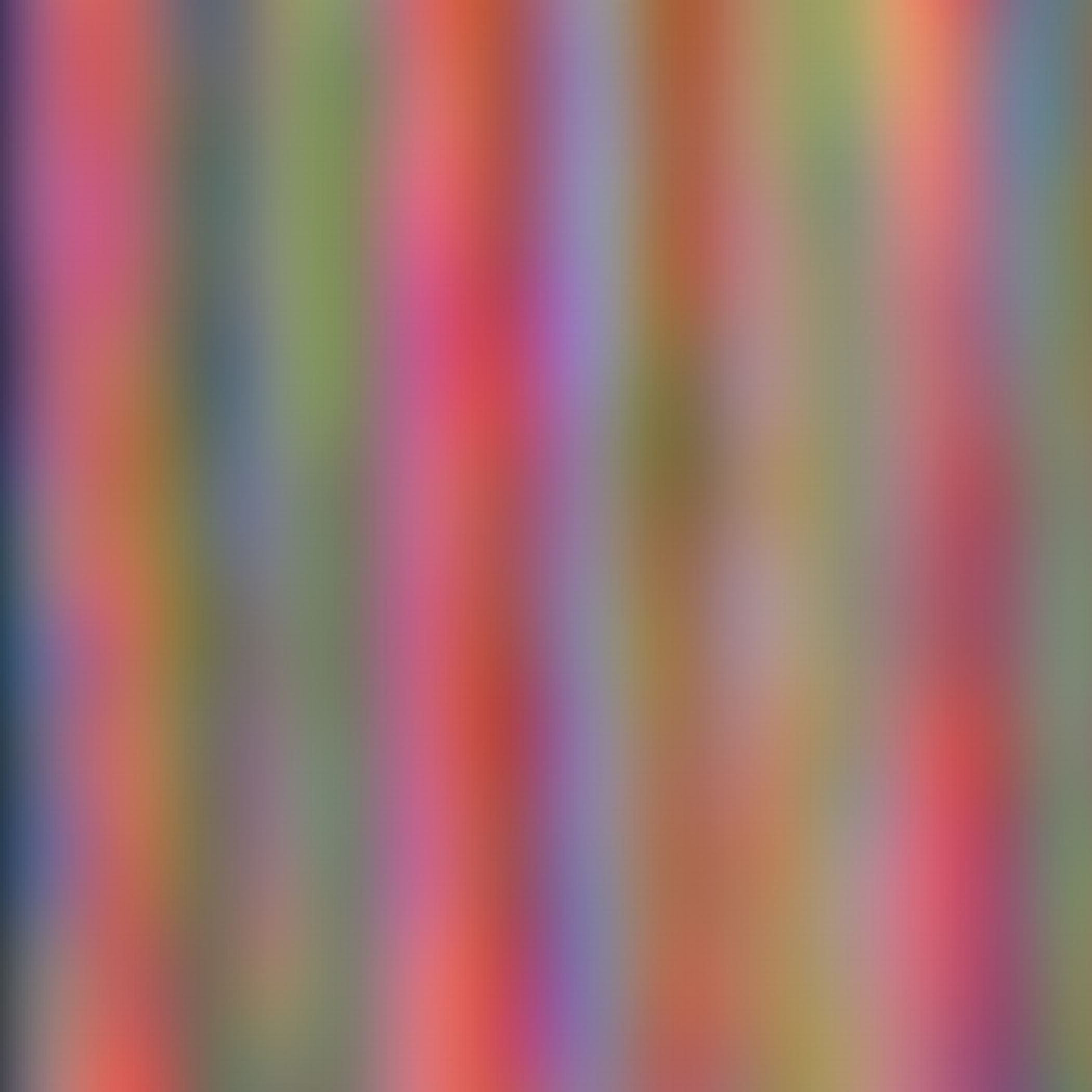}\\[1pt]
\mbox{\fontsize{9pt}{11pt}\selectfont CP\,/\,58.9\,dB $\cdot$ 11,899 Mpix/s $\cdot$ 2.9$\times$ slower}\end{minipage} & \begin{minipage}[t]{2.500in}\centering \includegraphics[interpolate=false,height=2.500in]{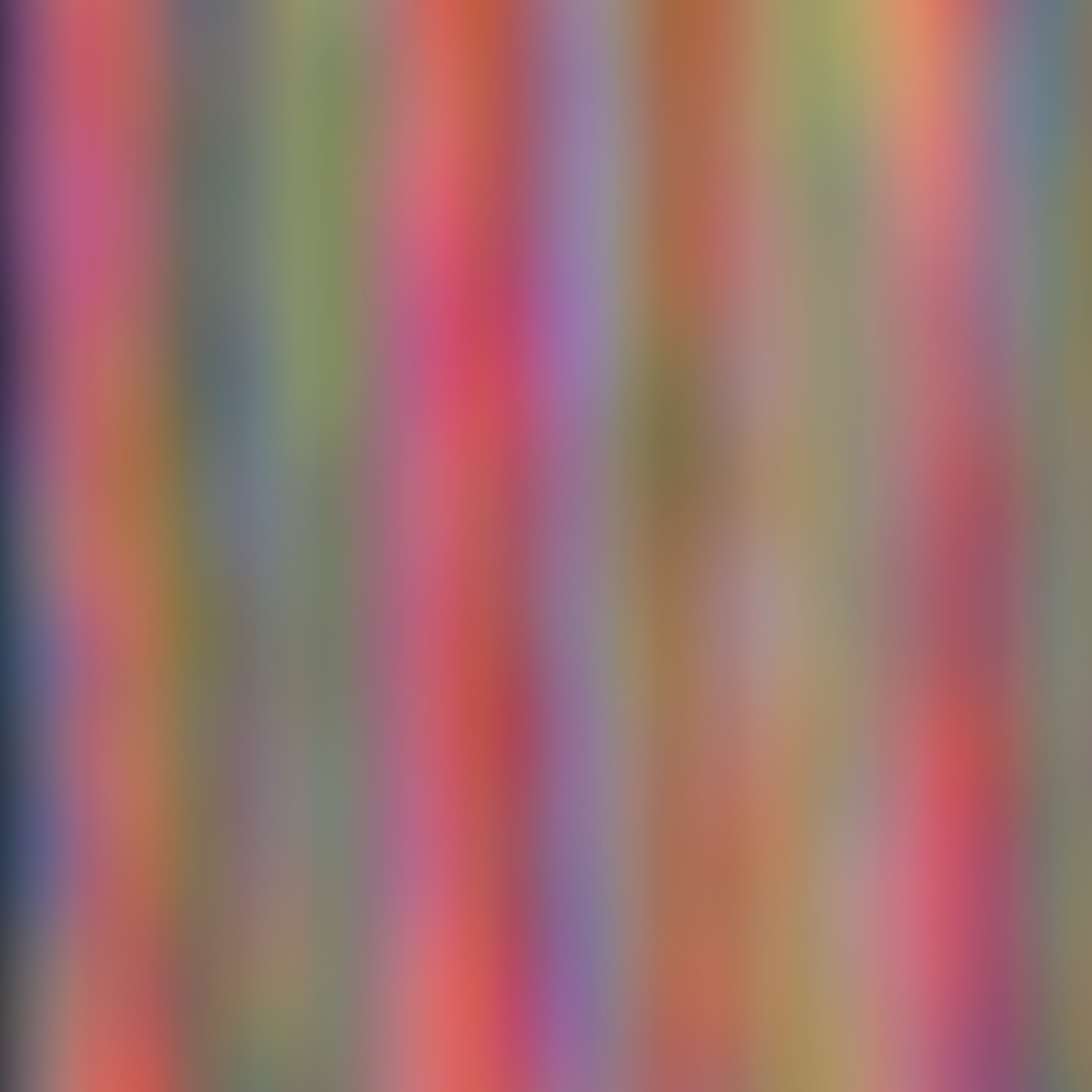}\\[1pt]
\mbox{\fontsize{9pt}{11pt}\selectfont FRM\,/\,40.5\,dB $\cdot$ 7,836 Mpix/s $\cdot$ 4.5$\times$ slower}\end{minipage} \\[12pt]
\end{tabular}
\caption{2D size 201 Lowpass filter approximation outputs.}
\label{fig:patch_lowpass2D_201_Nikon___Z5_2___8bit_compressed_32_y2200_x100}
\end{figure*}

\begin{figure*}[p]\centering
\begin{tabular}{cc}
\begin{minipage}[t]{2.500in}\centering \includegraphics[interpolate=false,height=2.500in]{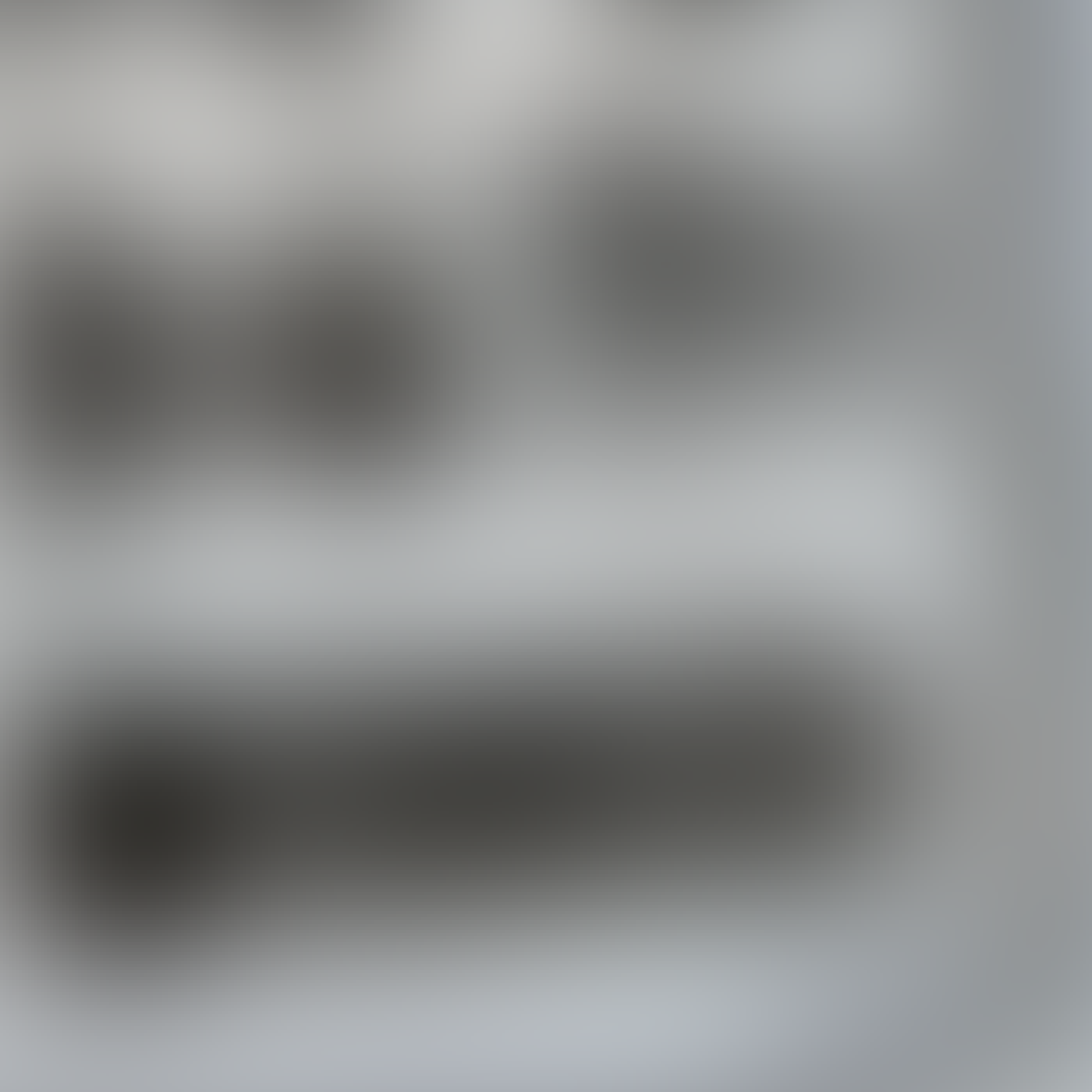}\\[1pt]
\mbox{\fontsize{9pt}{11pt}\selectfont Ground truth}\end{minipage} & \begin{minipage}[t]{2.500in}\centering \includegraphics[interpolate=false,height=2.500in]{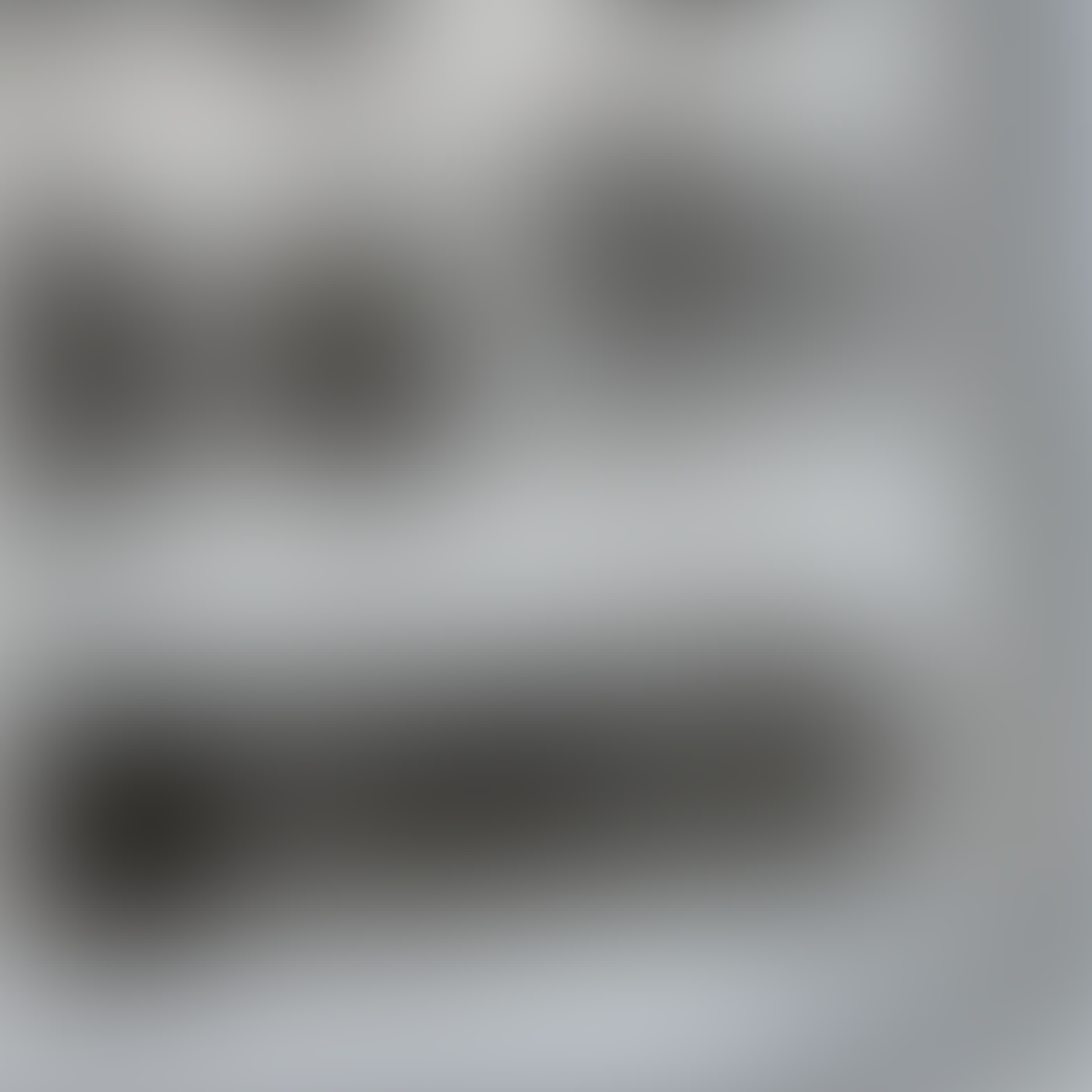}\\[1pt]
\mbox{\fontsize{9pt}{11pt}\selectfont Ours\,/\,71.4\,dB $\cdot$ 36,289 Mpix/s $\cdot$ fastest}\end{minipage} \\[12pt]
\begin{minipage}[t]{2.500in}\centering \includegraphics[interpolate=false,height=2.500in]{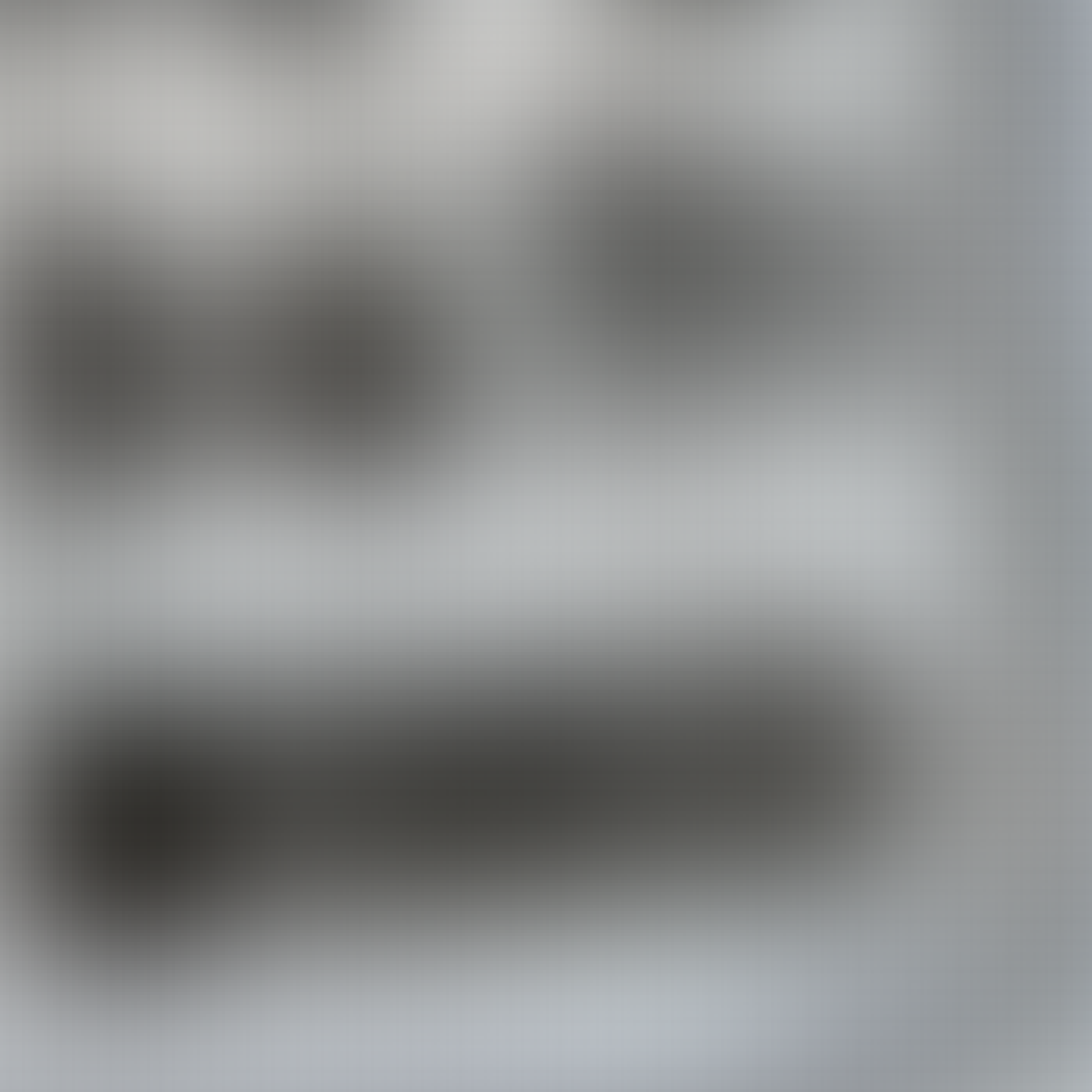}\\[1pt]
\mbox{\fontsize{9pt}{11pt}\selectfont CP\,/\,56.0\,dB $\cdot$ 9,141 Mpix/s $\cdot$ 4.0$\times$ slower}\end{minipage} & \begin{minipage}[t]{2.500in}\centering \includegraphics[interpolate=false,height=2.500in]{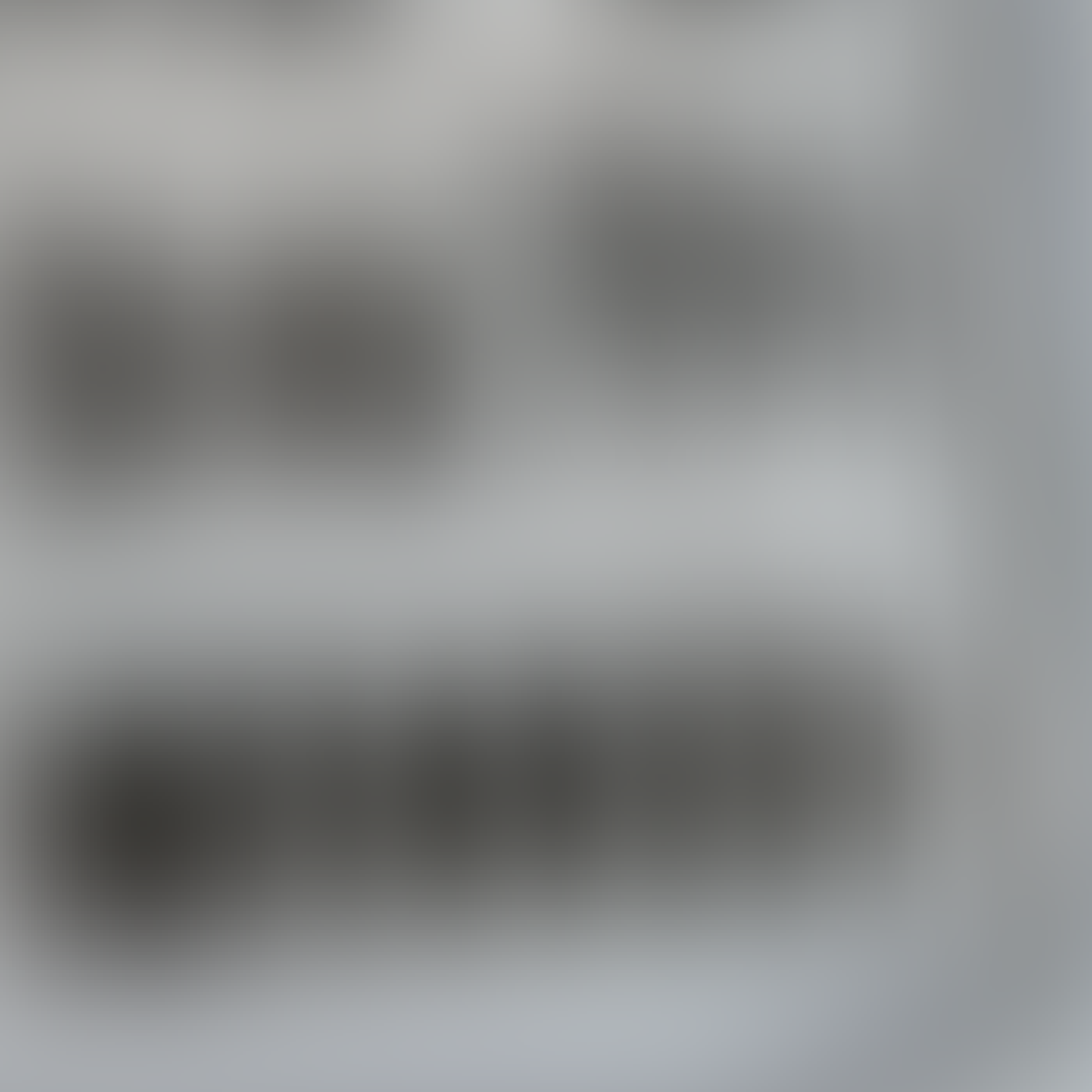}\\[1pt]
\mbox{\fontsize{9pt}{11pt}\selectfont FRM\,/\,39.2\,dB $\cdot$ 5,469 Mpix/s $\cdot$ 6.6$\times$ slower}\end{minipage} \\[12pt]
\end{tabular}
\caption{Outputs for 2D size 401 Lowpass filter approximations on patches from an image taken with the authors' iPhone 13 Pro. Our model produces outputs that are visually indistinguishable from ground truth. CP produces subtle thin ringing artifacts while FRM still produces noticeable thick ringing artifacts. Our model is 4x faster than CP and 6.6x faster than FRM.}
\label{fig:patch_lowpass2D_401_IMG_4390_y3000_x1600}
\end{figure*}
\urldef{\imglink}\url{https://raw.pixls.us/getfile.php/2980/nice/Fujifilm%20-%20GFX%2050R%20-%2014bit%2014bit%20uncompressed%20%283%3A2%29.raf}
\begin{figure*}[p]\centering
\begin{tabular}{cc}
\begin{minipage}[t]{2.500in}\centering \includegraphics[interpolate=false,height=2.500in]{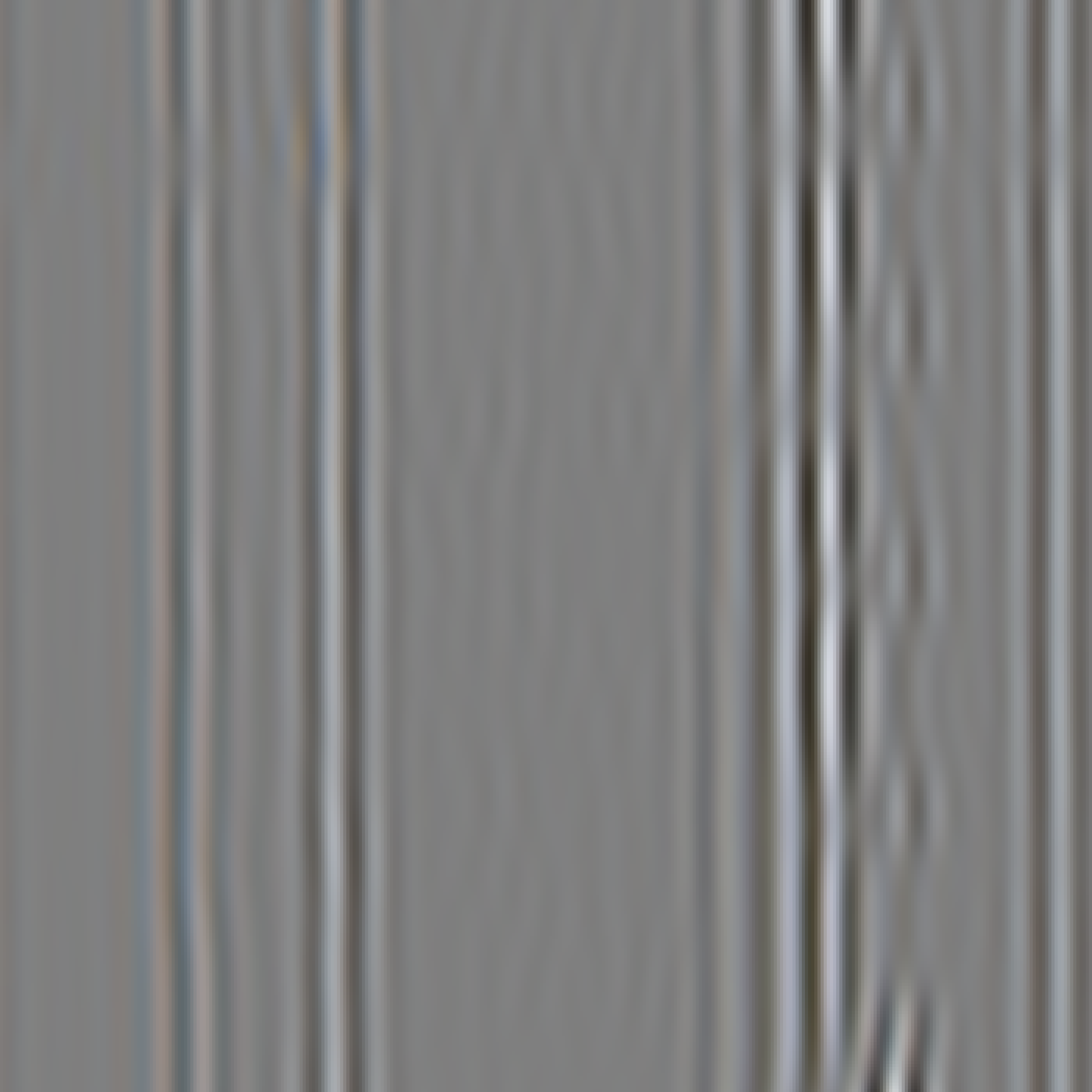}\\[1pt]
\mbox{\fontsize{9pt}{11pt}\selectfont Ground truth}\end{minipage} & \begin{minipage}[t]{2.500in}\centering \includegraphics[interpolate=false,height=2.500in]{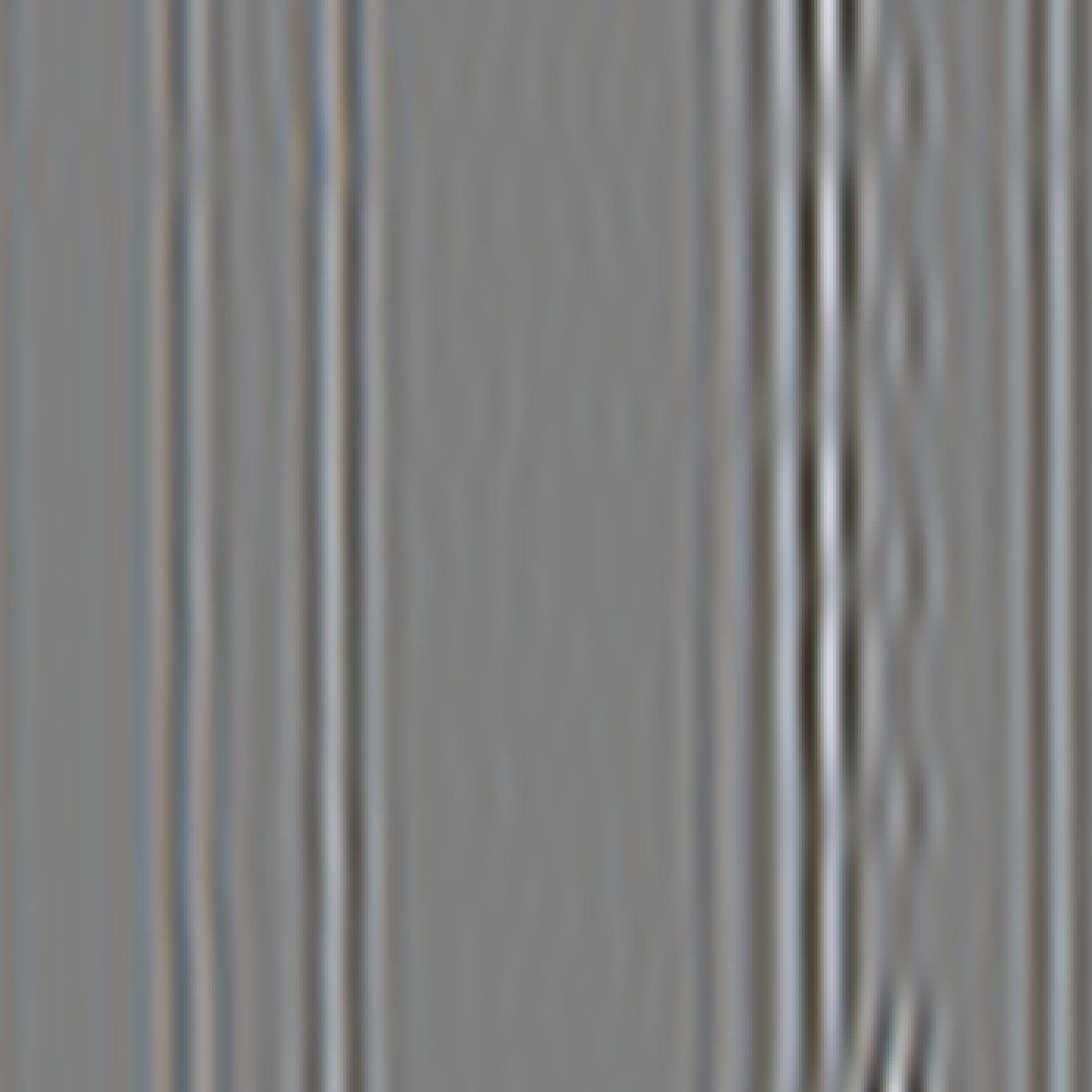}\\[1pt]
\mbox{\fontsize{9pt}{11pt}\selectfont Ours\,/\,64.0\,dB $\cdot$ 51,824 Mpix/s $\cdot$ fastest}\end{minipage} \\[12pt]
\begin{minipage}[t]{2.500in}\centering \includegraphics[interpolate=false,height=2.500in]{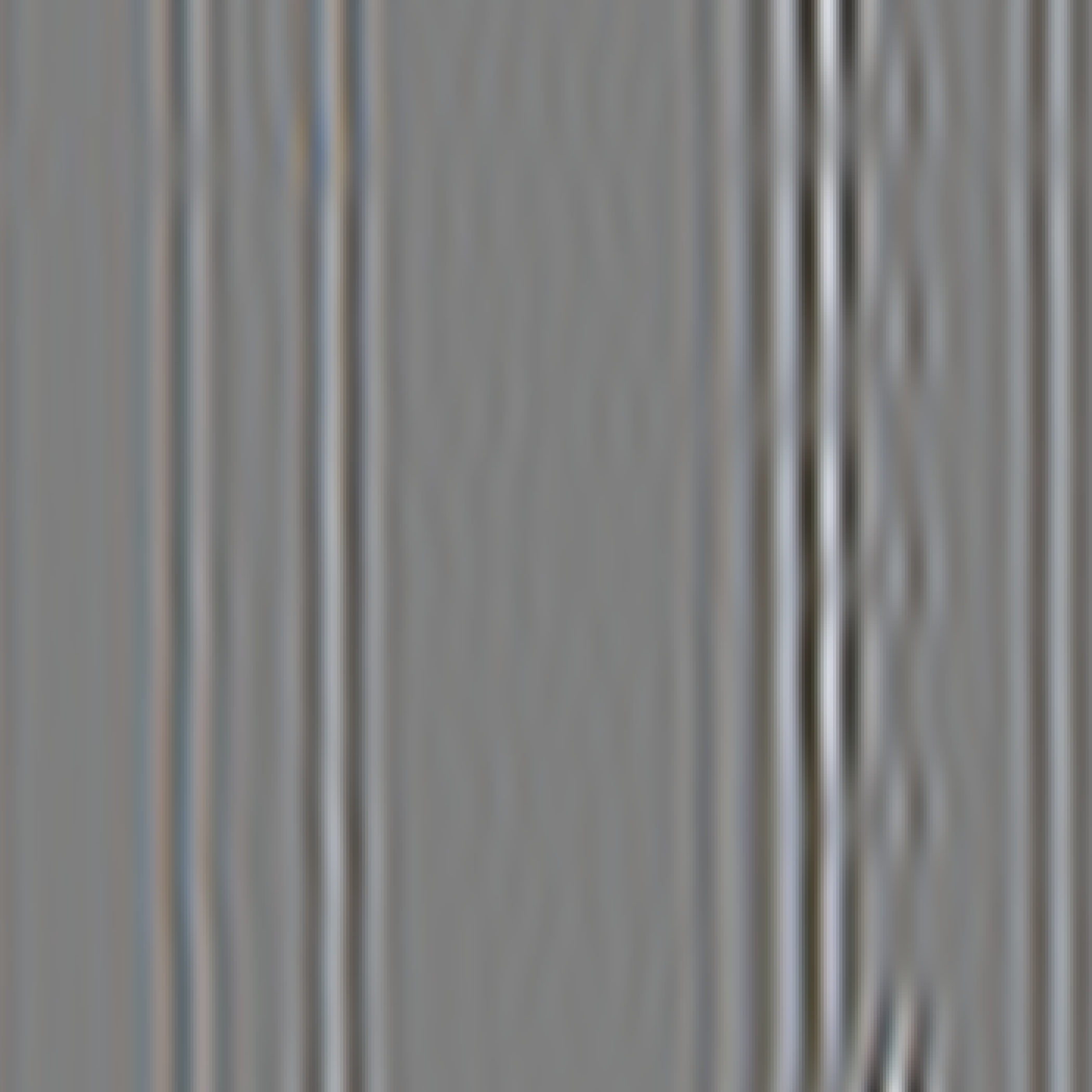}\\[1pt]
\mbox{\fontsize{9pt}{11pt}\selectfont CP\,/\,76.6\,dB $\cdot$ 12,986 Mpix/s $\cdot$ 4.0$\times$ slower}\end{minipage} & \begin{minipage}[t]{2.500in}\centering \includegraphics[interpolate=false,height=2.500in]{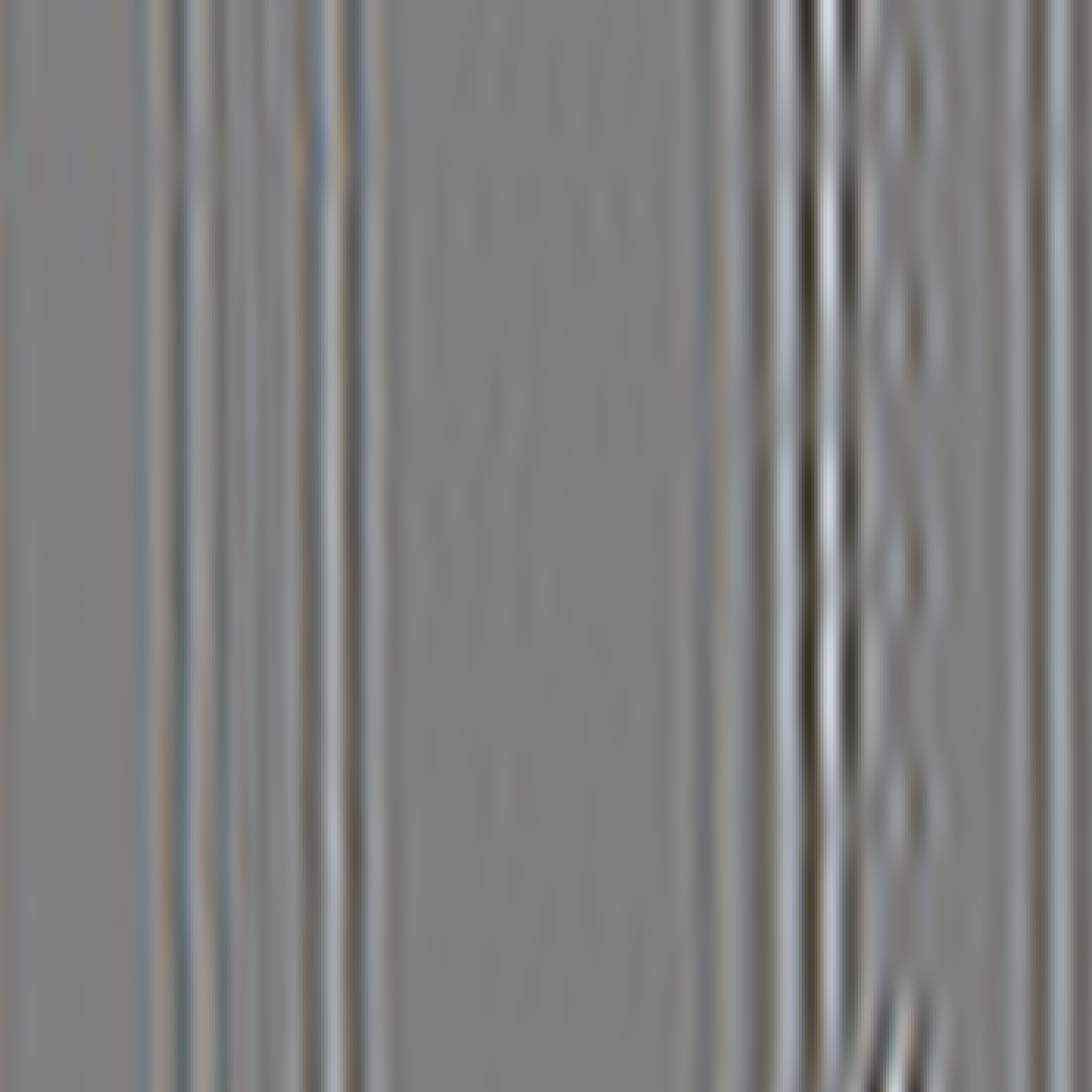}\\[1pt]
\mbox{\fontsize{9pt}{11pt}\selectfont FRM\,/\,52.4\,dB $\cdot$ 9,403 Mpix/s $\cdot$ 5.5$\times$ slower}\end{minipage} \\[12pt]
\begin{minipage}[t]{2.500in}\centering \includegraphics[interpolate=false,height=2.500in]{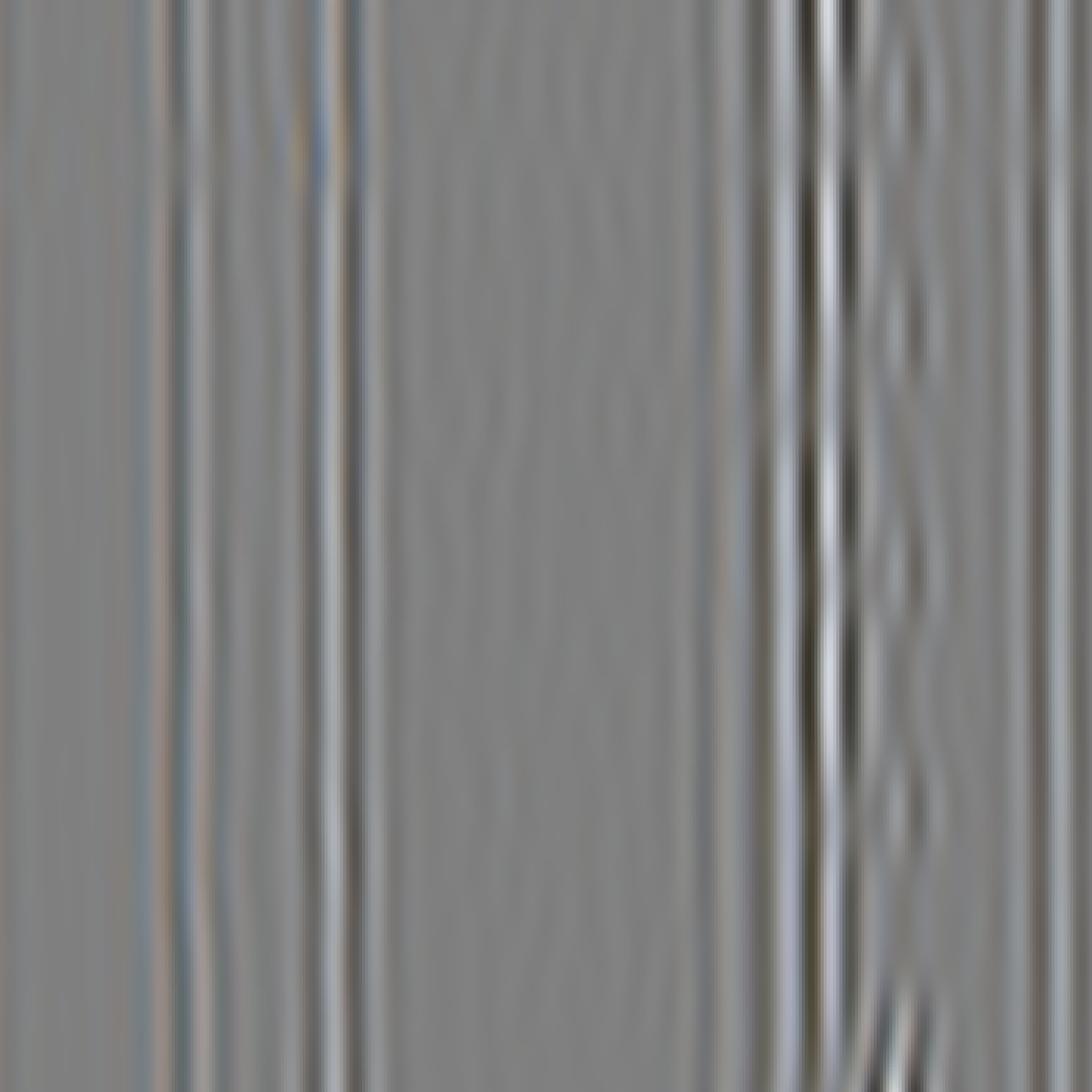}\\[1pt]
\mbox{\fontsize{9pt}{11pt}\selectfont YVV\,/\,54.0\,dB $\cdot$ 8,793 Mpix/s $\cdot$ 5.9$\times$ slower}\end{minipage} &  \\[12pt]
\end{tabular}
\vspace{-0.5em}
\caption{2D size 101 Gabor filter approximation outputs on an image patch taken from \imglink. Outputs look visually indistinguishable from the ground truth except for FRM which has vertical ringing artifacts. Our approximation is significantly faster than all the baselines, at 4.0x faster than CP, 5.5x faster than FRM and 5.9x faster than YVV.}
\label{fig:patch_gabor2D_101_Fujifilm___GFX_50R___14bit_14bit_uncompressed_32_y2850_x400}
\end{figure*}

\begin{figure*}[p]\centering
\begin{tabular}{cc}
\begin{minipage}[t]{2.500in}\centering \includegraphics[interpolate=false,height=2.500in]{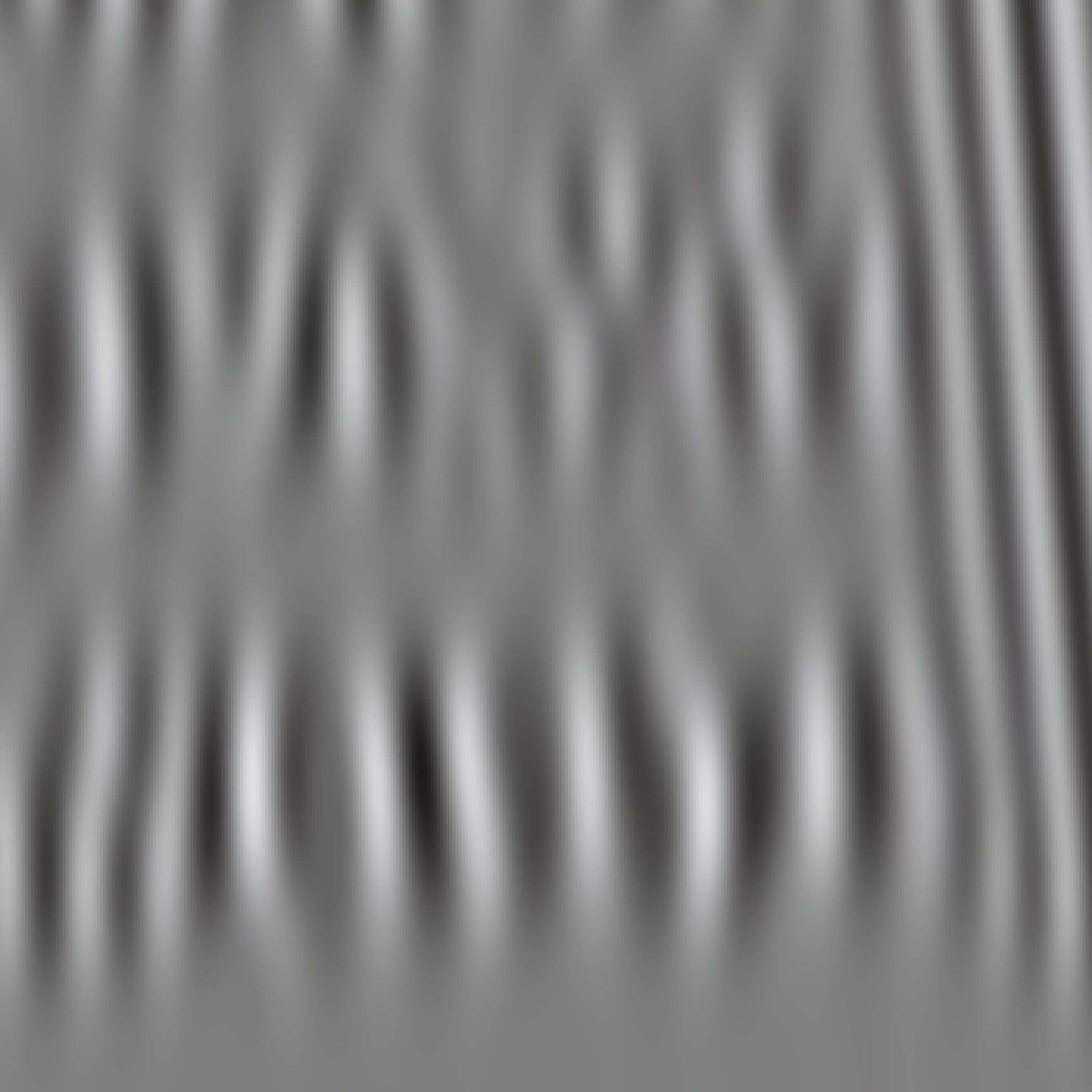}\\[1pt]
\mbox{\fontsize{9pt}{11pt}\selectfont Ground truth}\end{minipage} & \begin{minipage}[t]{2.500in}\centering \includegraphics[interpolate=false,height=2.500in]{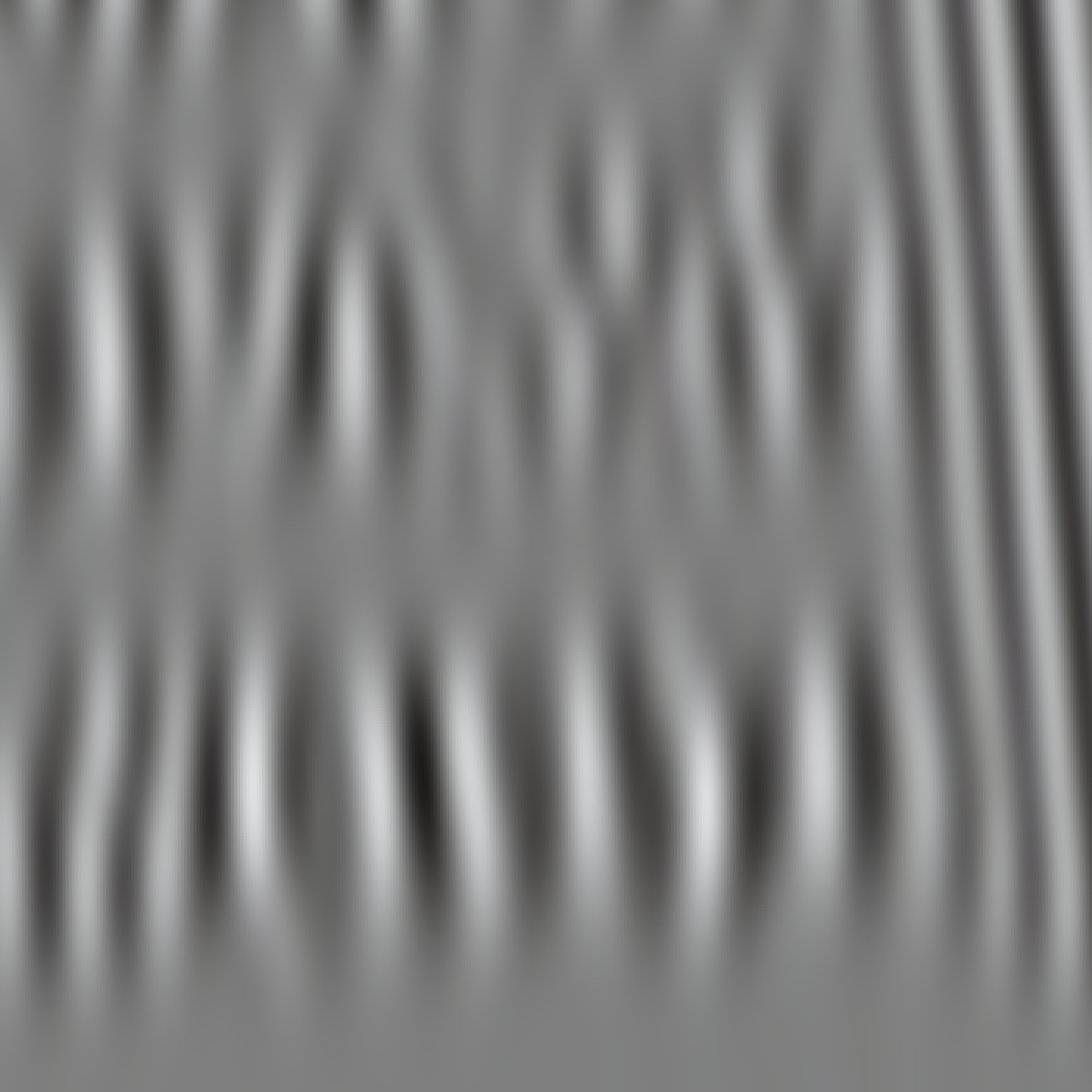}\\[1pt]
\mbox{\fontsize{9pt}{11pt}\selectfont Ours\,/\,56.6\,dB $\cdot$ 51,594 Mpix/s $\cdot$ fastest}\end{minipage} \\[12pt]
\begin{minipage}[t]{2.500in}\centering \includegraphics[interpolate=false,height=2.500in]{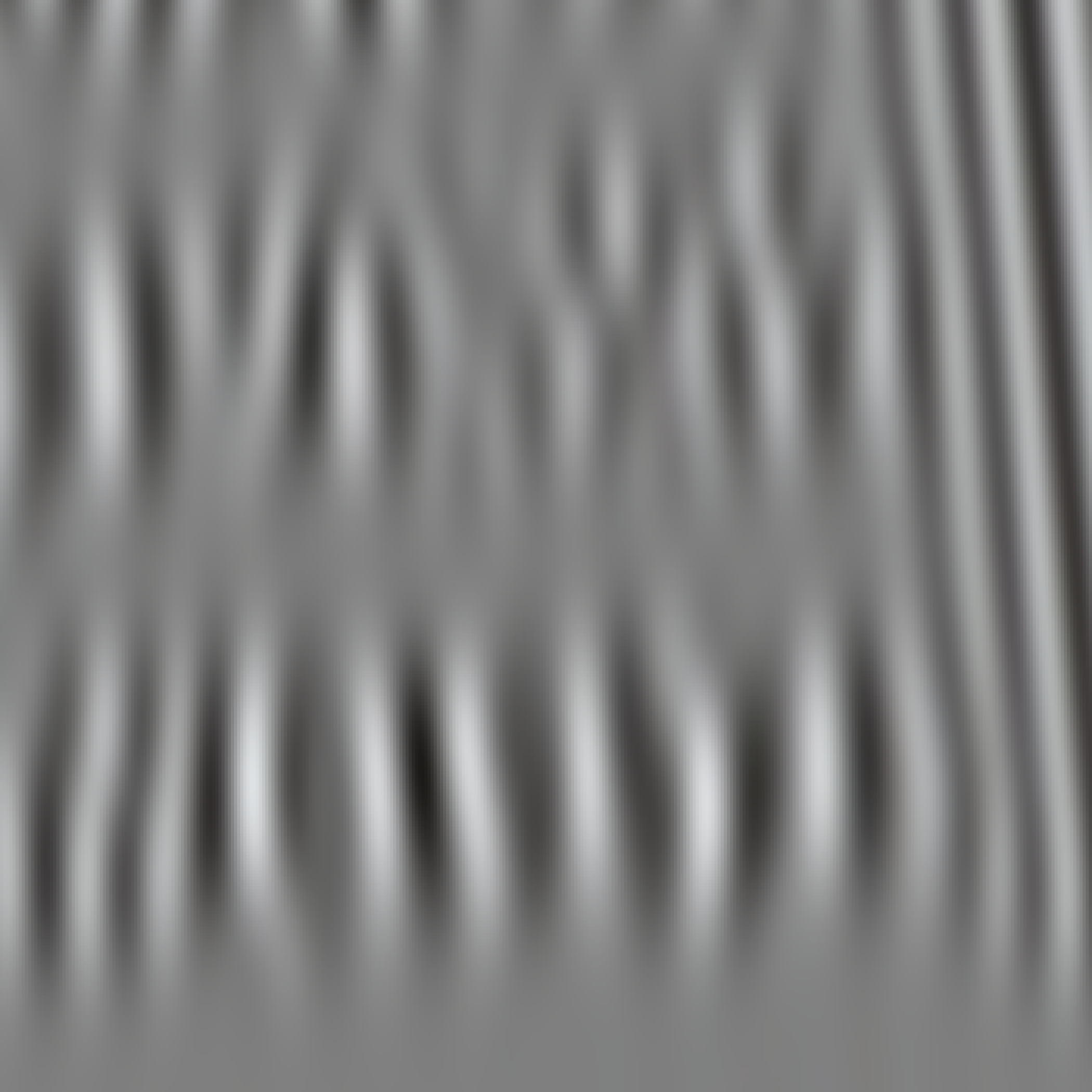}\\[1pt]
\mbox{\fontsize{9pt}{11pt}\selectfont CP\,/\,62.1\,dB $\cdot$ 11,857 Mpix/s $\cdot$ 4.4$\times$ slower}\end{minipage} & \begin{minipage}[t]{2.500in}\centering \includegraphics[interpolate=false,height=2.500in]{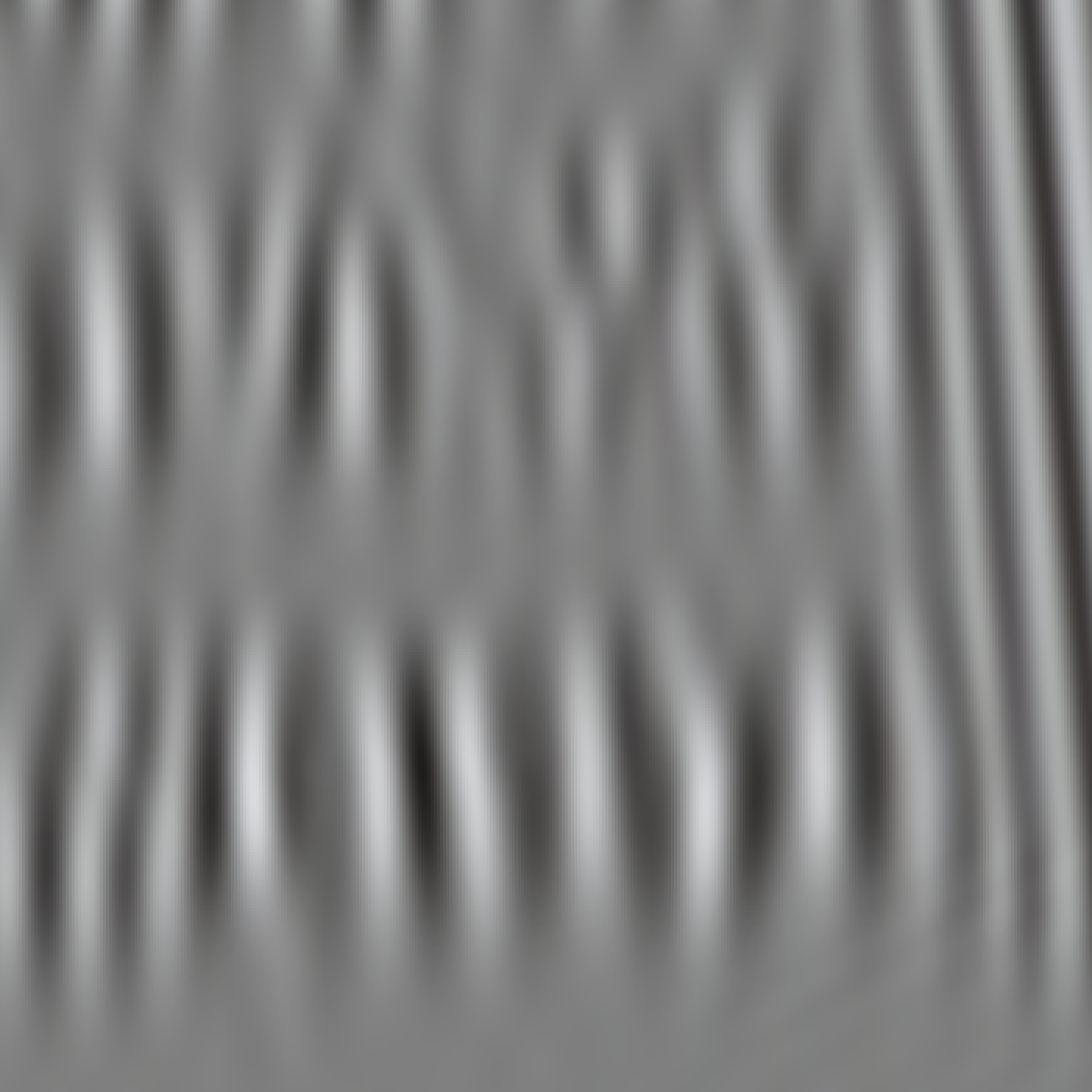}\\[1pt]
\mbox{\fontsize{9pt}{11pt}\selectfont FRM\,/\,48.0\,dB $\cdot$ 7,510 Mpix/s $\cdot$ 6.9$\times$ slower}\end{minipage} \\[12pt]
\begin{minipage}[t]{2.500in}\centering \includegraphics[interpolate=false,height=2.500in]{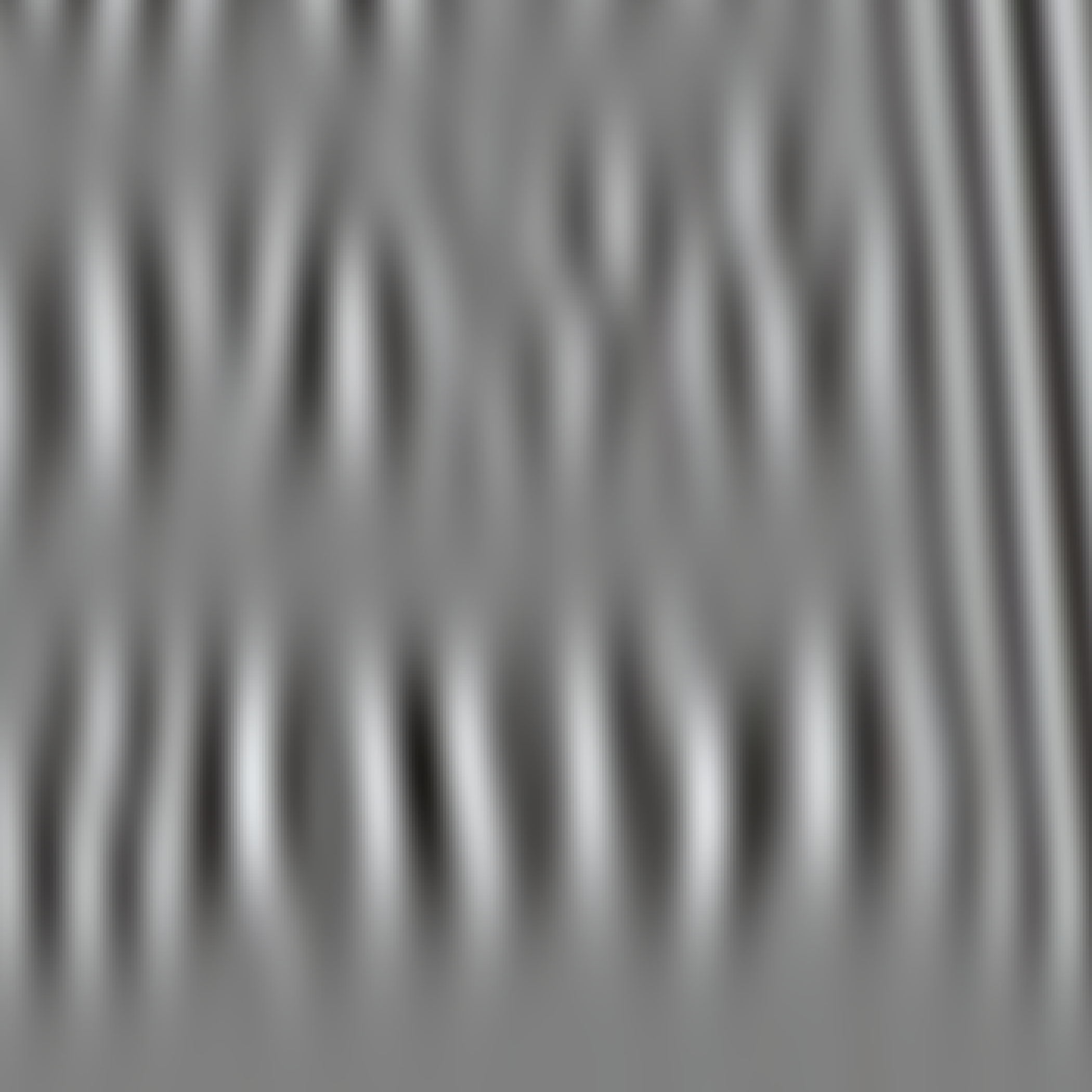}\\[1pt]
\mbox{\fontsize{9pt}{11pt}\selectfont YVV\,/\,51.2\,dB $\cdot$ 7,026 Mpix/s $\cdot$ 7.3$\times$ slower}\end{minipage} &  \\[12pt]
\end{tabular}
\caption{2D size 201 Gabor filter approximation outputs on image patches from a photo taken with the authors' iPhone 13 Pro. Outputs are visually indistinguishable from ground truth except for FRM, which again has ringing artifacts (diagonal banding in the bottom right corner). Our approximation is significantly faster than all the baselines, at 4.4x faster than CP, 6.9x faster than FRM, and 7.3x faster than YVV.}
\label{fig:patch_gabor2D_201_IMG_4390_y2900_x1500}
\end{figure*}

\begin{figure*}[p]\centering
\begin{tabular}{cc}
\begin{minipage}[t]{2.500in}\centering \includegraphics[interpolate=false,height=2.500in]{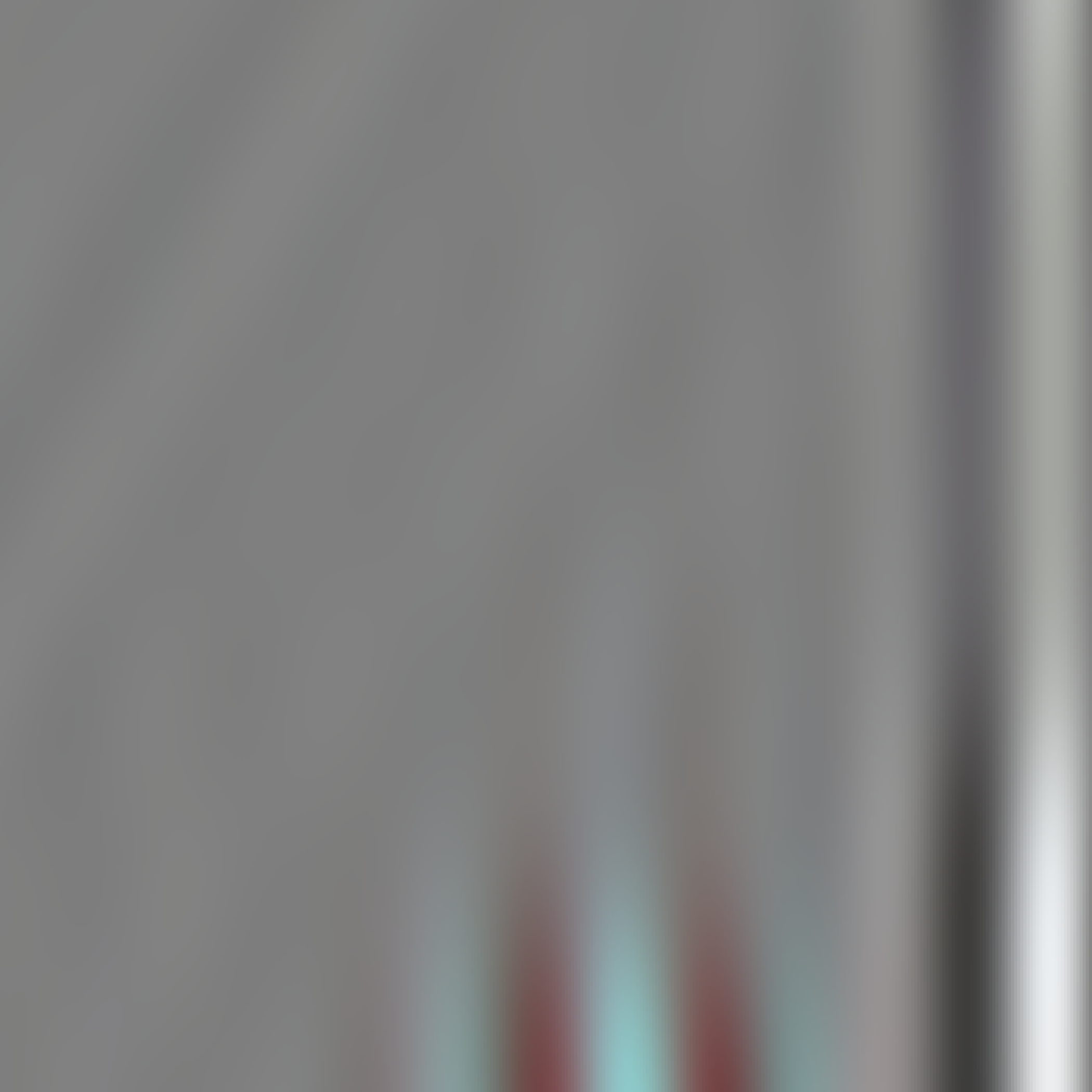}\\[1pt]
\mbox{\fontsize{9pt}{11pt}\selectfont Ground truth}\end{minipage} & \begin{minipage}[t]{2.500in}\centering \includegraphics[interpolate=false,height=2.500in]{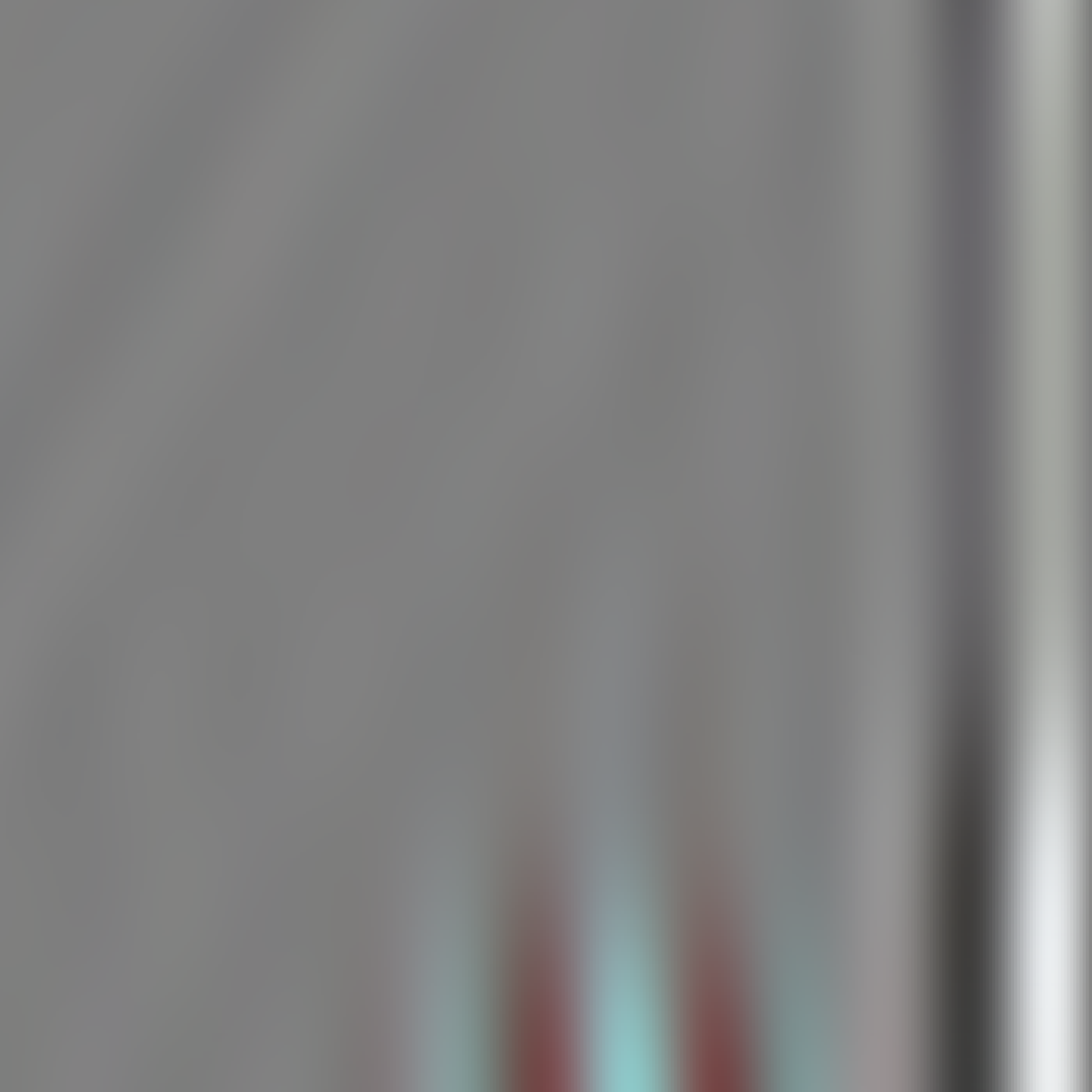}\\[1pt]
\mbox{\fontsize{9pt}{11pt}\selectfont Ours\,/\,81.8\,dB $\cdot$ 46,363 Mpix/s $\cdot$ fastest}\end{minipage} \\[12pt]
\begin{minipage}[t]{2.500in}\centering \includegraphics[interpolate=false,height=2.500in]{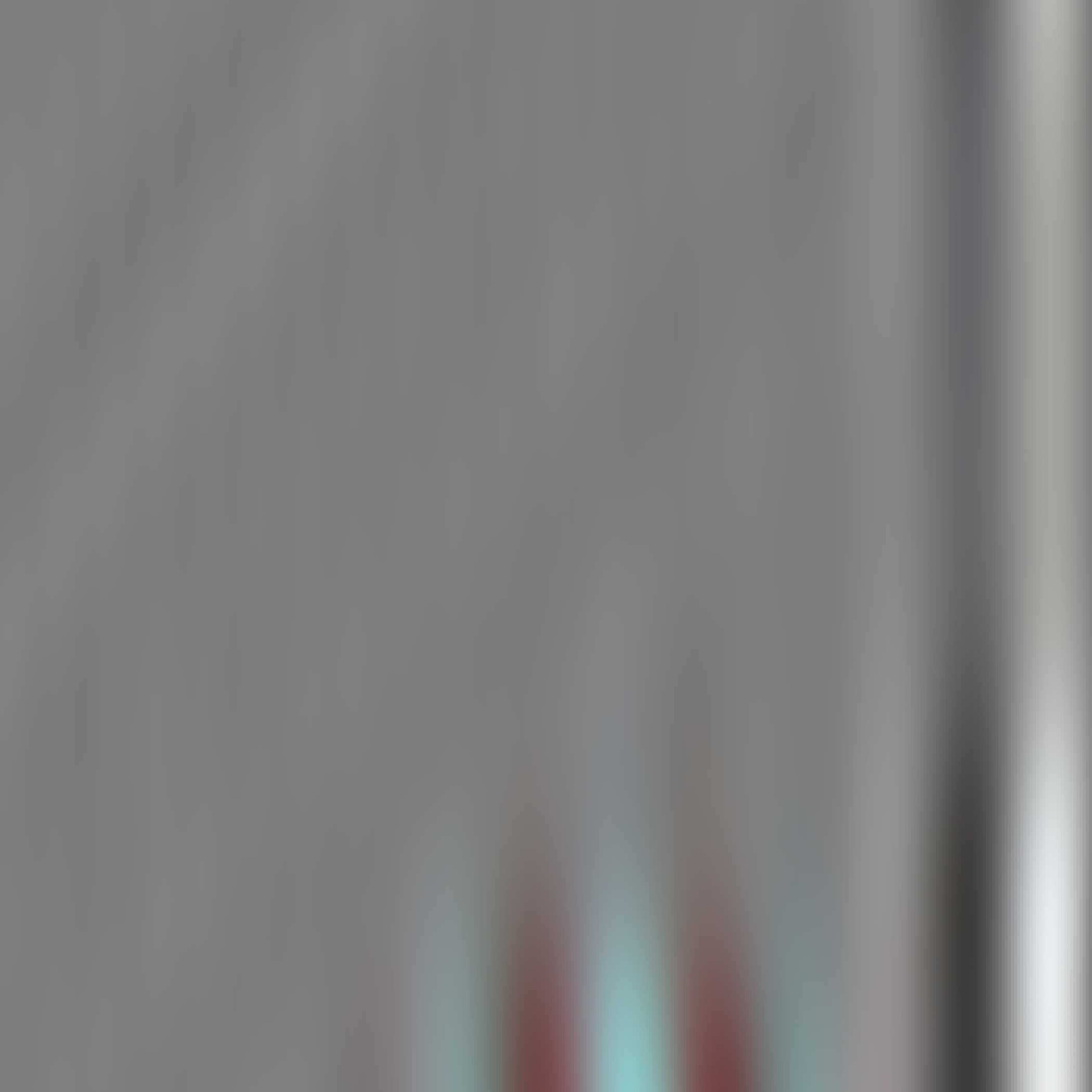}\\[1pt]
\mbox{\fontsize{9pt}{11pt}\selectfont CP\,/\,52.5\,dB $\cdot$ 9,421 Mpix/s $\cdot$ 4.9$\times$ slower}\end{minipage} & \begin{minipage}[t]{2.500in}\centering \includegraphics[interpolate=false,height=2.500in]{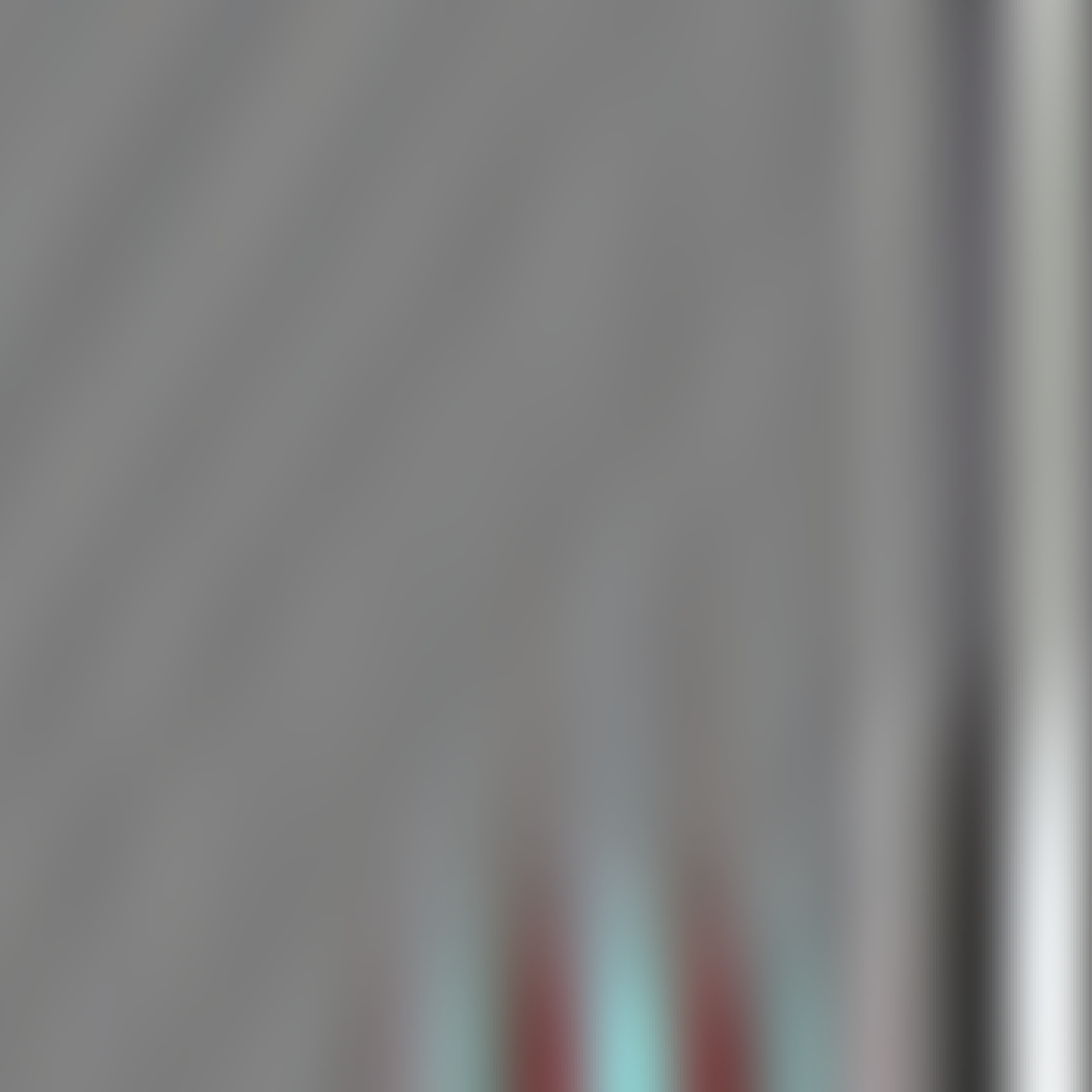}\\[1pt]
\mbox{\fontsize{9pt}{11pt}\selectfont FRM\,/\,52.7\,dB $\cdot$ 4,945 Mpix/s $\cdot$ 9.4$\times$ slower}\end{minipage} \\[12pt]
\begin{minipage}[t]{2.500in}\centering \includegraphics[interpolate=false,height=2.500in]{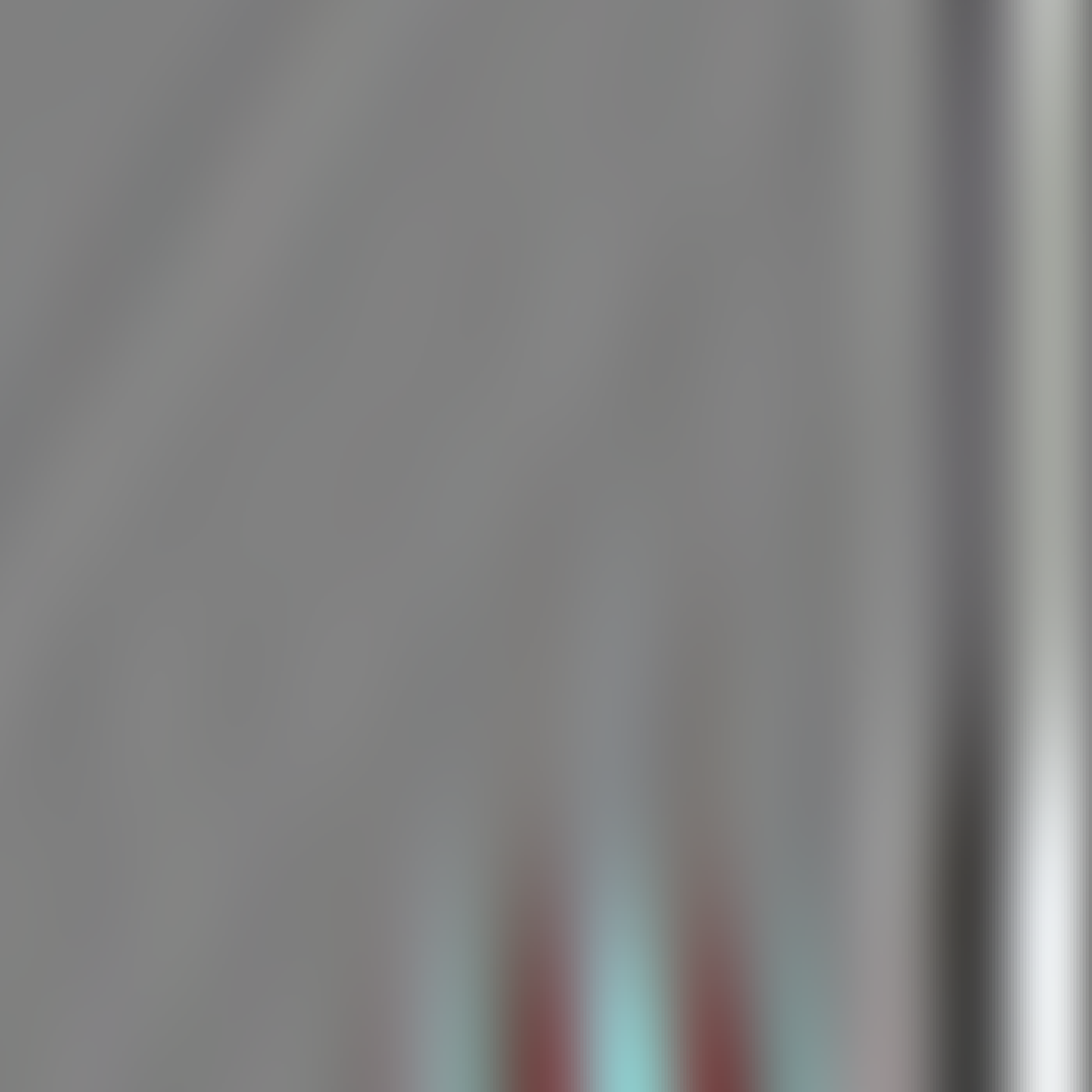}\\[1pt]
\mbox{\fontsize{9pt}{11pt}\selectfont YVV\,/\,53.1\,dB $\cdot$ 4,303 Mpix/s $\cdot$ 10.8$\times$ slower}\end{minipage} &  \\[12pt]
\end{tabular}
\vspace{-0.5em}
\caption{Figures 59 - 61 show 2D size 401 Gabor filter approximation outputs on patches from photos taken by the authors' iPhone 13 Pro. CP outputs have thin vertical banding artifacts. This figure's patch shows that the FRM approximation has too large of a response to diagonal frequencies, because it produces stronger diagonal bands than the ground truth image. Figure \ref{fig:patch_gabor2D_401_IMG_5740_y3000_x900} shows that FRM also produces vertical banding artifacts.}
\label{fig:patch_gabor2D_401_IMG_4460_y1950_x2874}
\end{figure*}

\begin{figure*}[p]\centering
\begin{tabular}{cc}
\begin{minipage}[t]{2.500in}\centering \includegraphics[interpolate=false,height=2.500in]{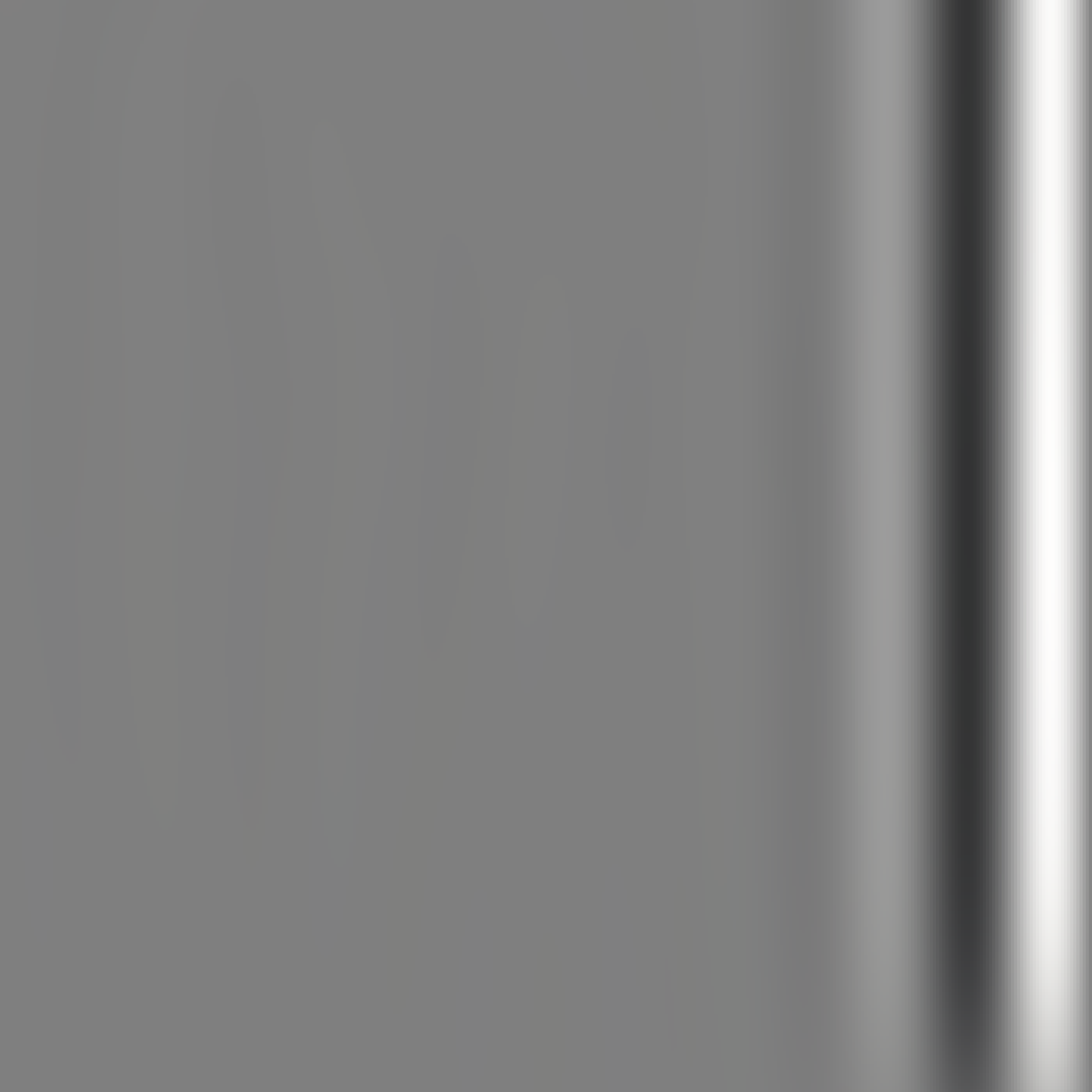}\\[1pt]
\mbox{\fontsize{9pt}{11pt}\selectfont Ground truth}\end{minipage} & \begin{minipage}[t]{2.500in}\centering \includegraphics[interpolate=false,height=2.500in]{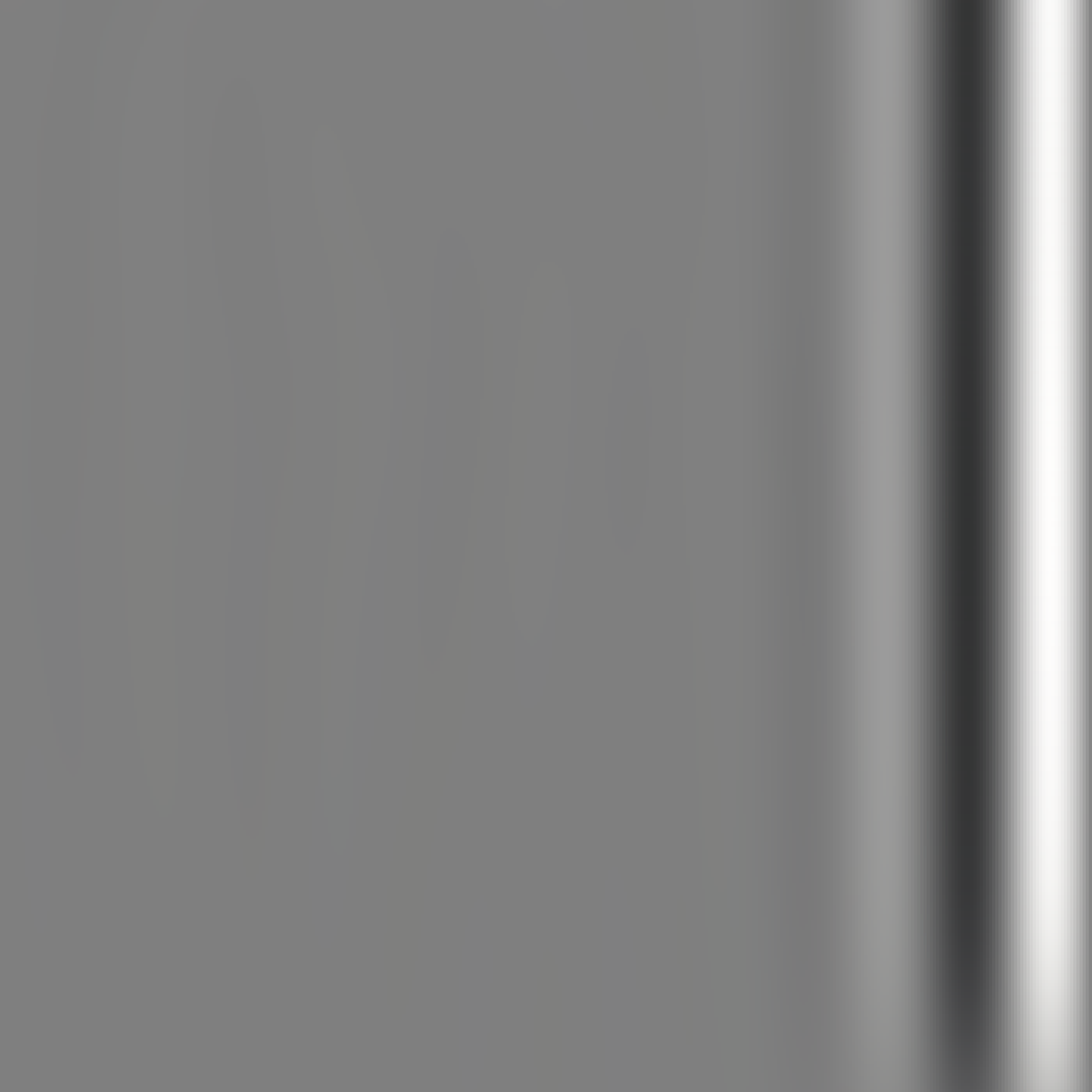}\\[1pt]
\mbox{\fontsize{9pt}{11pt}\selectfont Ours\,/\,81.0\,dB $\cdot$ 46,363 Mpix/s $\cdot$ fastest}\end{minipage} \\[12pt]
\begin{minipage}[t]{2.500in}\centering \includegraphics[interpolate=false,height=2.500in]{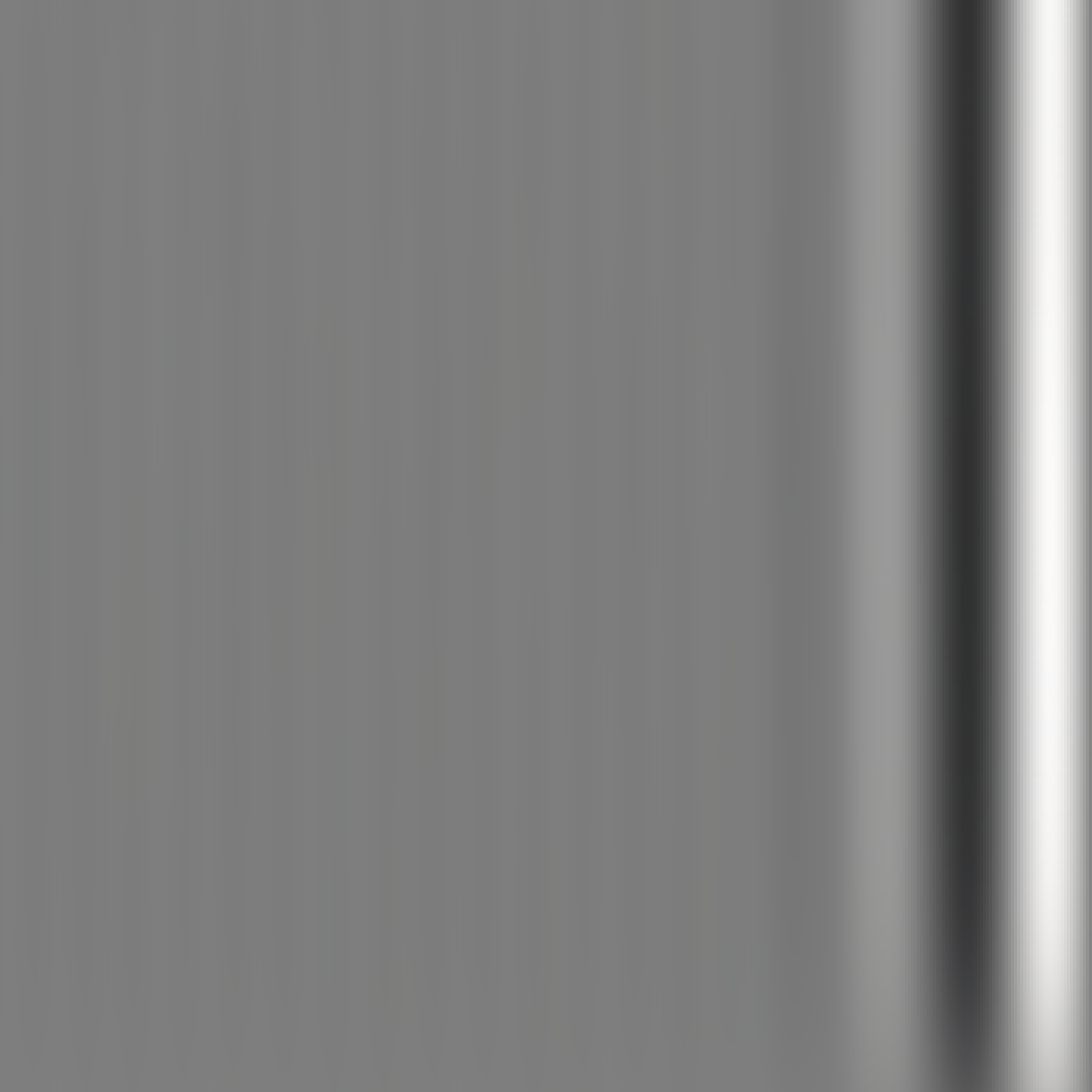}\\[1pt]
\mbox{\fontsize{9pt}{11pt}\selectfont CP\,/\,48.7\,dB $\cdot$ 9,421 Mpix/s $\cdot$ 4.9$\times$ slower}\end{minipage} & \begin{minipage}[t]{2.500in}\centering \includegraphics[interpolate=false,height=2.500in]{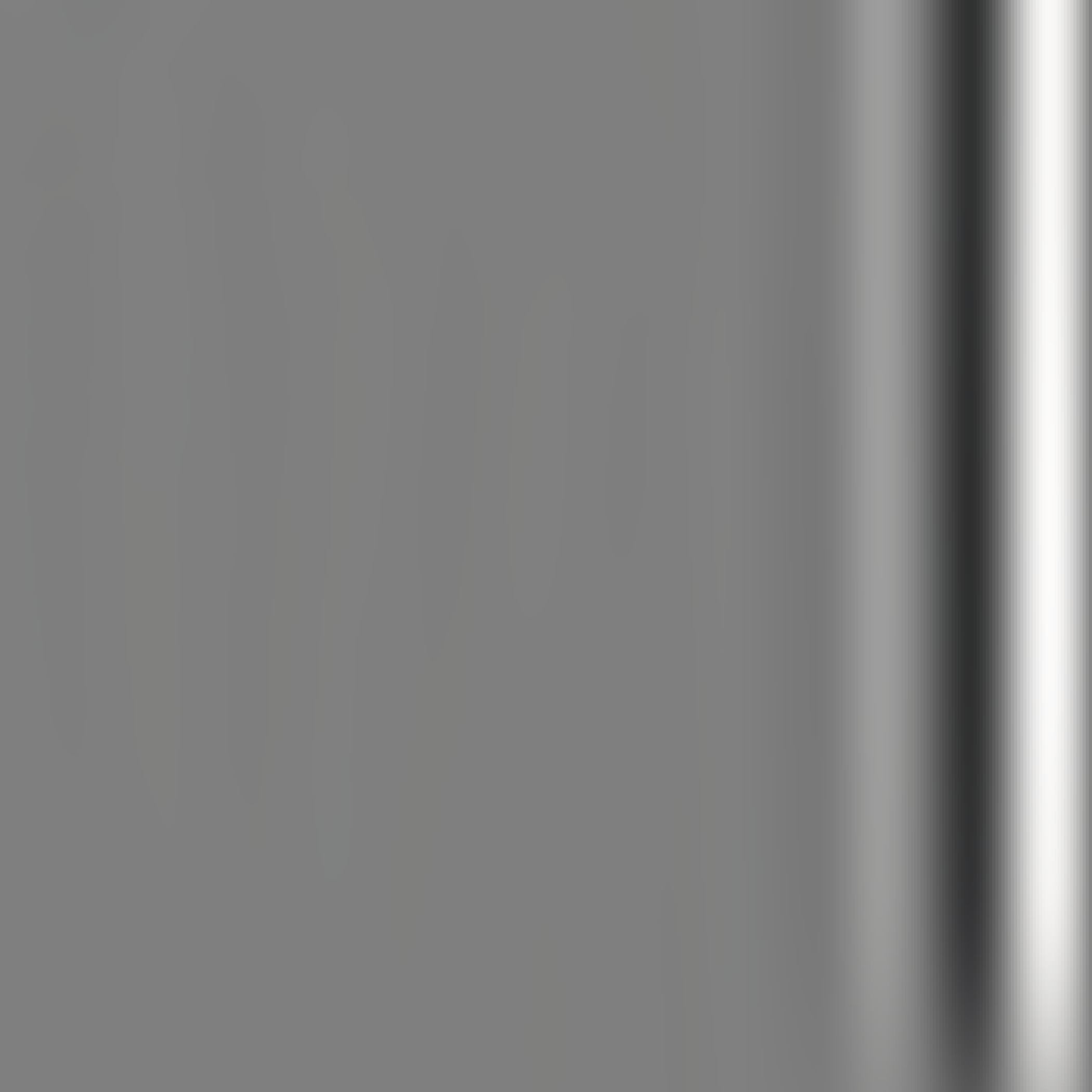}\\[1pt]
\mbox{\fontsize{9pt}{11pt}\selectfont FRM\,/\,54.8\,dB $\cdot$ 4,945 Mpix/s $\cdot$ 9.4$\times$ slower}\end{minipage} \\[12pt]
\begin{minipage}[t]{2.500in}\centering \includegraphics[interpolate=false,height=2.500in]{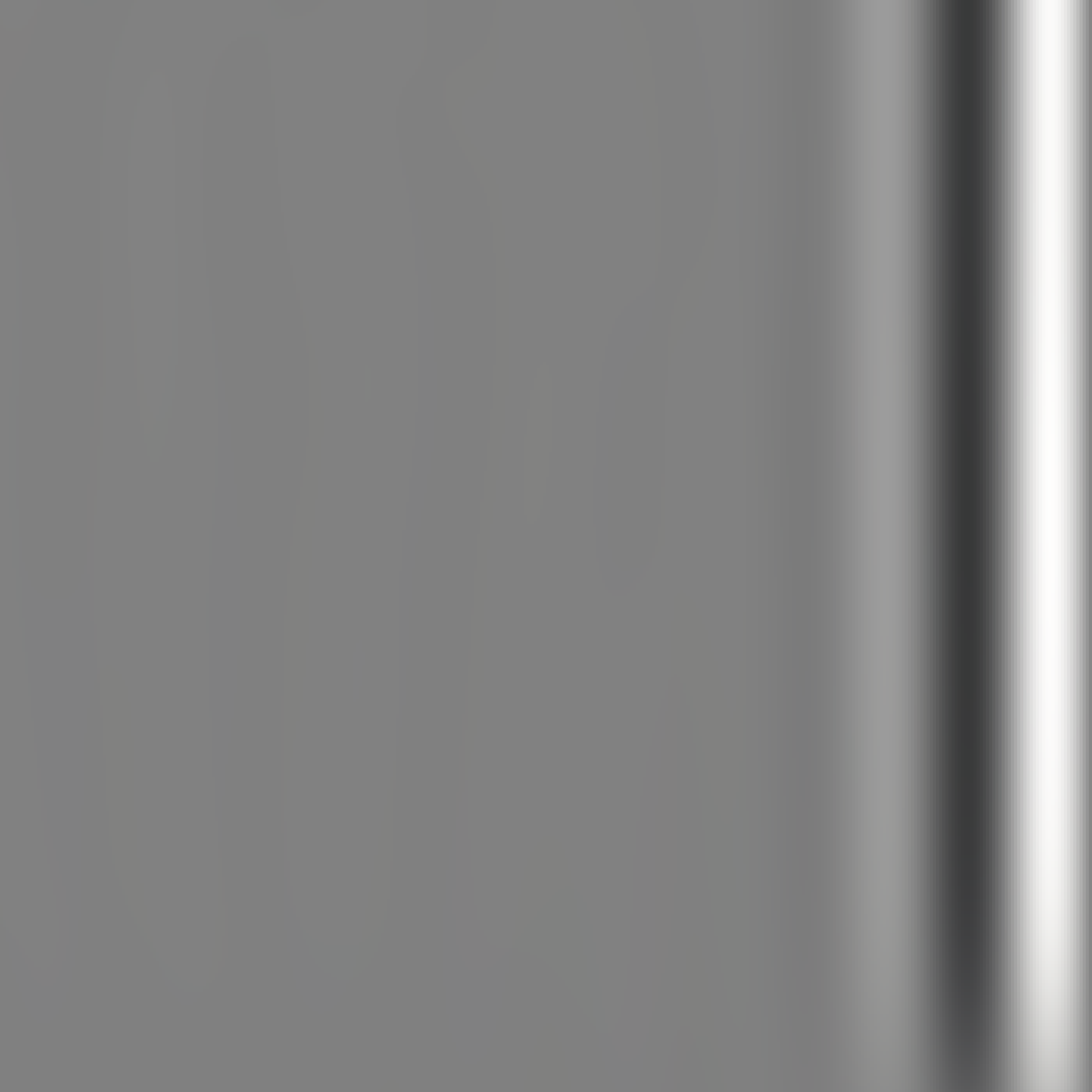}\\[1pt]
\mbox{\fontsize{9pt}{11pt}\selectfont YVV\,/\,50.3\,dB $\cdot$ 4,303 Mpix/s $\cdot$ 10.8$\times$ slower}\end{minipage} &  \\[12pt]
\end{tabular}
\caption{2D size 401 Gabor filter approximation outputs.}
\label{fig:patch_gabor2D_401_IMG_4460_y3882_x2874}
\end{figure*}

\begin{figure*}[p]\centering
\begin{tabular}{cc}
\begin{minipage}[t]{2.500in}\centering \includegraphics[interpolate=false,height=2.500in]{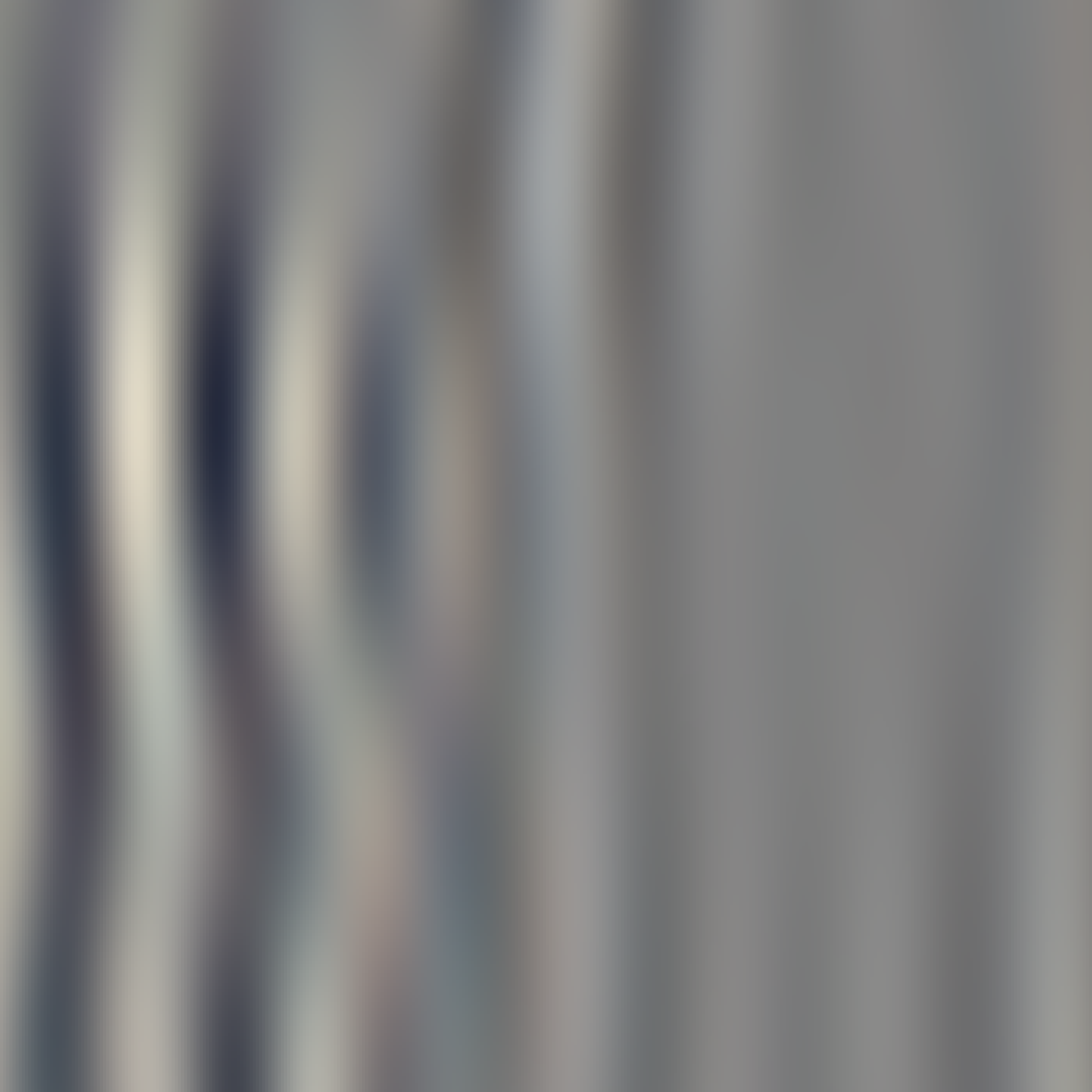}\\[1pt]
\mbox{\fontsize{9pt}{11pt}\selectfont Ground truth}\end{minipage} & \begin{minipage}[t]{2.500in}\centering \includegraphics[interpolate=false,height=2.500in]{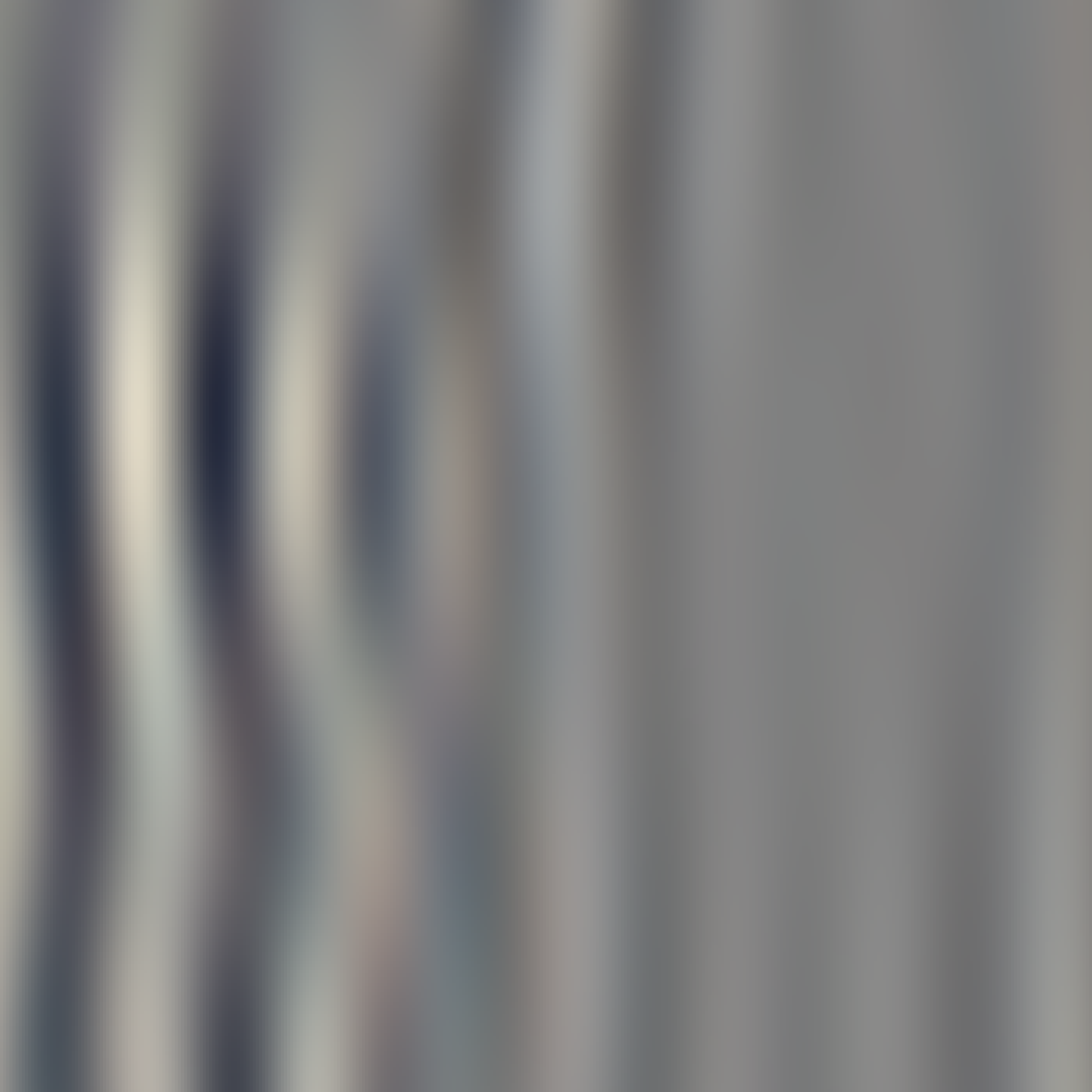}\\[1pt]
\mbox{\fontsize{9pt}{11pt}\selectfont Ours\,/\,75.6\,dB $\cdot$ 46,363 Mpix/s $\cdot$ fastest}\end{minipage} \\[12pt]
\begin{minipage}[t]{2.500in}\centering \includegraphics[interpolate=false,height=2.500in]{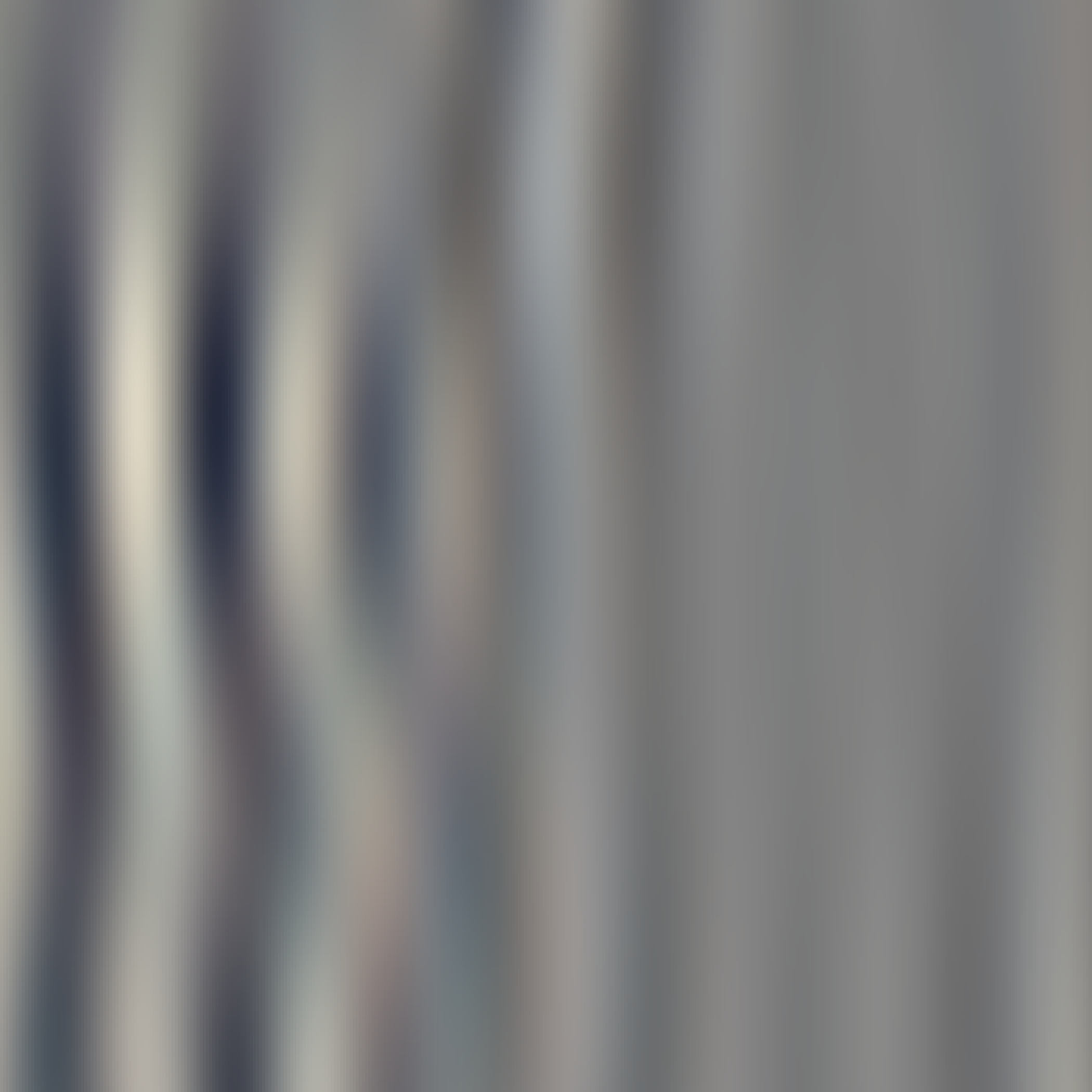}\\[1pt]
\mbox{\fontsize{9pt}{11pt}\selectfont CP\,/\,53.0\,dB $\cdot$ 9,421 Mpix/s $\cdot$ 4.9$\times$ slower}\end{minipage} & \begin{minipage}[t]{2.500in}\centering \includegraphics[interpolate=false,height=2.500in]{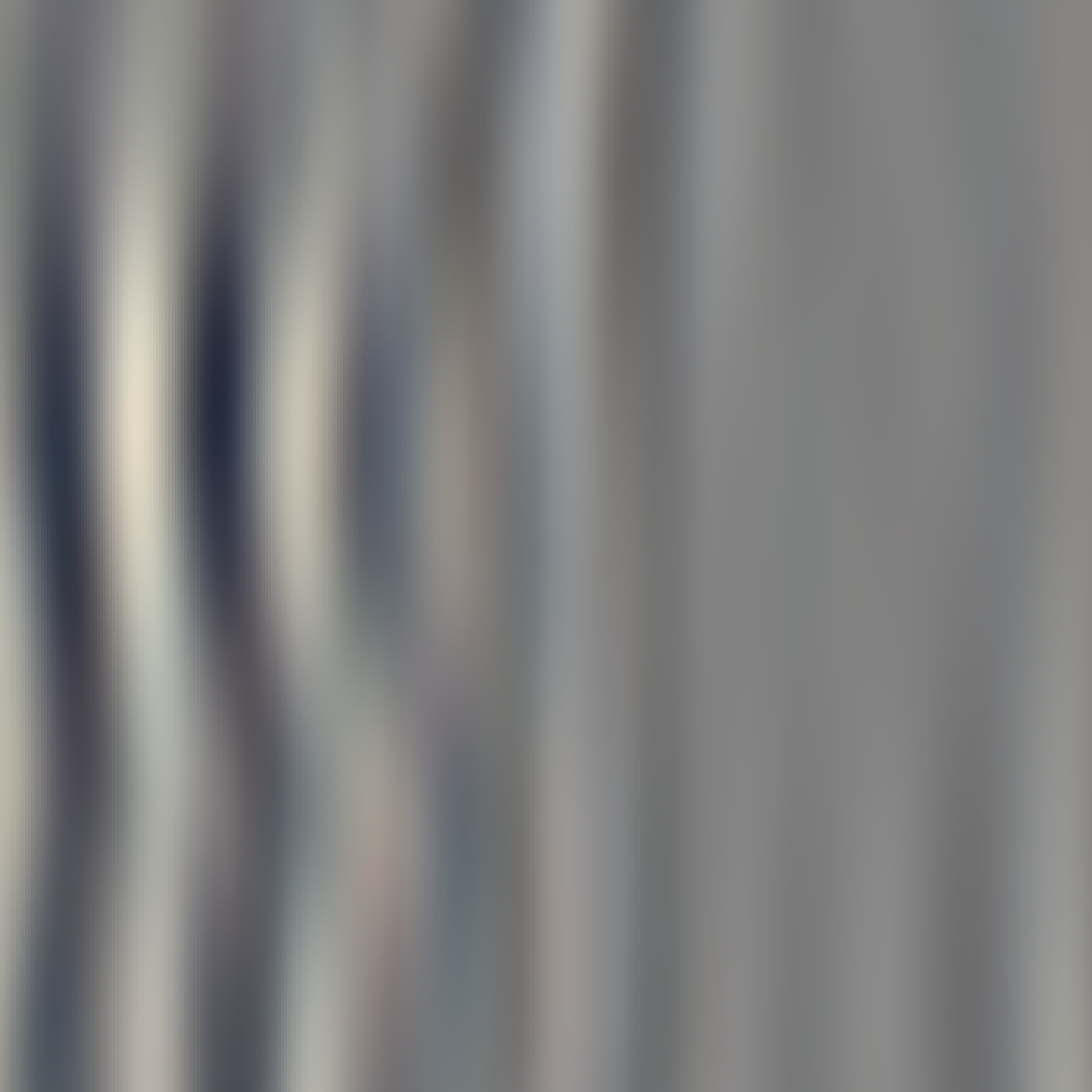}\\[1pt]
\mbox{\fontsize{9pt}{11pt}\selectfont FRM\,/\,49.2\,dB $\cdot$ 4,945 Mpix/s $\cdot$ 9.4$\times$ slower}\end{minipage} \\[12pt]
\begin{minipage}[t]{2.500in}\centering \includegraphics[interpolate=false,height=2.500in]{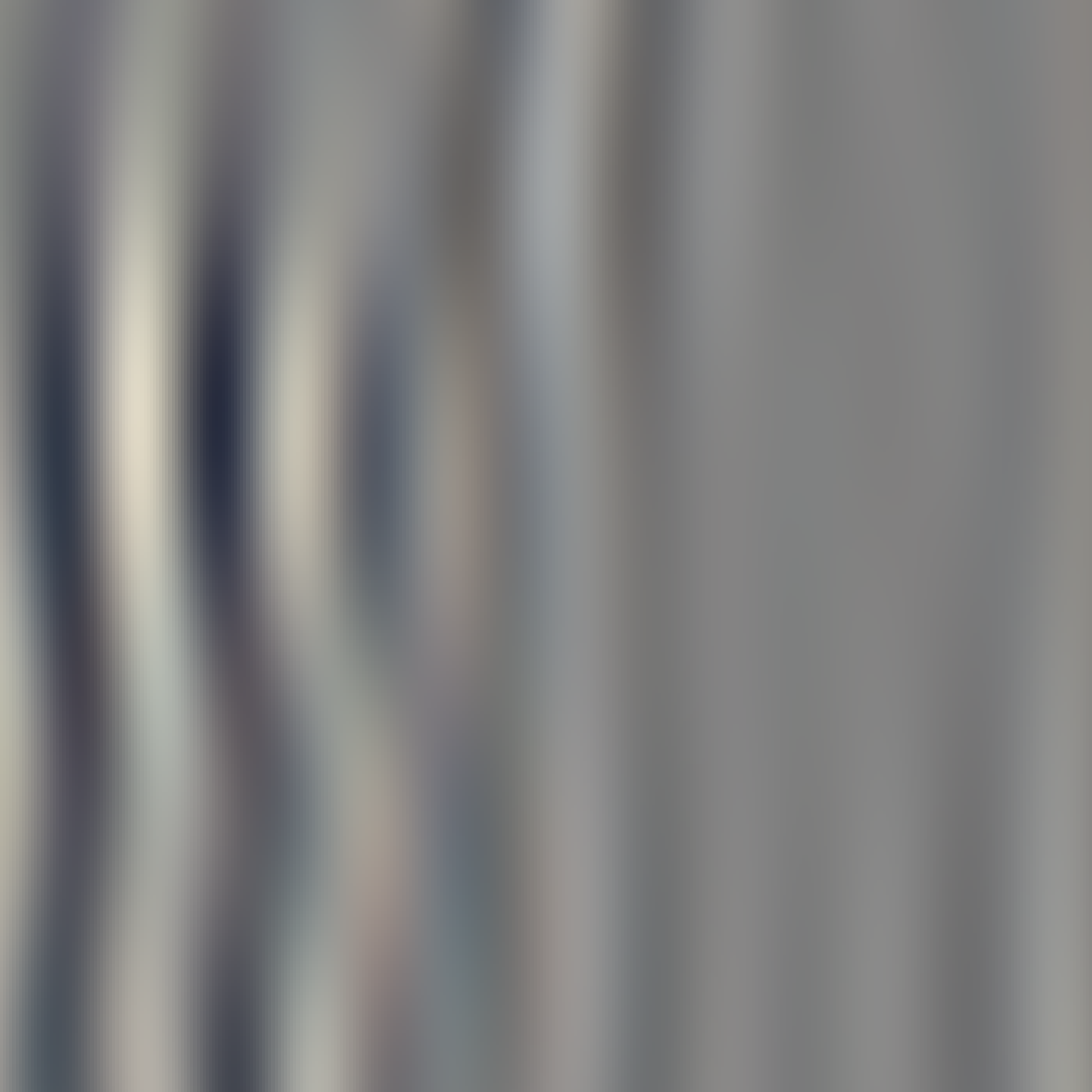}\\[1pt]
\mbox{\fontsize{9pt}{11pt}\selectfont YVV\,/\,54.5\,dB $\cdot$ 4,303 Mpix/s $\cdot$ 10.8$\times$ slower}\end{minipage} &  \\[12pt]
\end{tabular}
\caption{2D size 401 Gabor filter approximation outputs.}
\label{fig:patch_gabor2D_401_IMG_5740_y3000_x900}
\end{figure*}

\clearpage
\bibliographystyle{ACM-Reference-Format}
\bibliography{references}
\end{document}